\documentclass[pre,aps,final,twocolumn,showpacs,eqsecnum]{revtex4-1}
\newlength\figwidth\figwidth=0.5\textwidth

\usepackage{color,graphicx}
\usepackage{amsmath,amssymb,bm}
\usepackage{colordvi}
\usepackage{epsfig}

\begin{document}

\newcommand{\bel}[1]{\begin{equation}\label{#1}}

\def\ramaSM{\vadjust{\vbox to 0pt{\vss \hbox to \hsize
{\hskip\hsize \quad $\Leftarrow$\quad {\it SM}\hss}\vskip3.5pt}}}
\def\ramaJB{\vadjust{\vbox to 0pt{\vss \hbox to \hsize
{\hskip\hsize \quad $\Leftarrow$\quad {\it JB}\hss}\vskip3.5pt}}}
\def\ramaDG{\vadjust{\vbox to 0pt{\vss \hbox to \hsize
{\hskip\hsize \quad $\Leftarrow$\quad {\it PR}\hss}\vskip3.5pt}}}
\def\rama{\vadjust{\vbox to 0pt{\vss \hbox to \hsize
{\hskip\hsize \quad $\Leftarrow$\quad
{$\Longleftarrow$}\hss}\vskip3.5pt}}}

\def\eps{\varepsilon}
\def\press{{\cal P}}
\def\om{\omega }
\def\taubar{\tau_q}
\def\r{{\bf r}}
\def\p{{\bf p}}
\def\q{{\bf q}}
\def\u{{\bf u}}
\def\v{{\bf v}}
\def\chixz{\chi _{xz}}
\def\chizz{\chi _{zz}}
\def\chizero{\chi _0}
\def\chione{\chi _1}
\def\chitwo{\chi _2}
\def\chin{\chi _n}
\def\chicoll{\chi _{coll}}
\def\chibar{{\bar \chi }}
\def\chicollb{\chibar _{coll}}
\def\nbar{{\bar{\cal N}(T)}}
\def\cvbar{{\bar C}_V}
\def\gbarhp{{\Gamma}^{(0)}}
\def\gbarone{{\Gamma}^{(1)}}
\def\tbar{{\bar T}}
\def\tp{\pi^2 \tbar^2}
\def\taup{\pi^2 \taubar^3}
\def\azero{a^{(0)}}
\def\aone{a^{(1)}}
\def\omzero{\om^{(0)}}
\def\ups{\Upsilon}
\def\siml{\hbox{\kern.1em \lower.6ex \hbox{$\sim$} \kern-1.12em
          \raise.6ex \hbox{$<$} \kern.1em }}
\def\simg{\hbox{\kern.1em \lower.6ex \hbox{$\sim$} \kern-1.12em
          \raise.6ex \hbox{$>$} \kern.1em }}
\def\d{\hbox{d}}
\def\be{\begin{equation}}
\def\ee{\end{equation}}
\def\bea{\begin{eqnarray}}
\def\eea{\end{eqnarray}}
\def\l{\label}
\def\s{{\bf s}}
\def\Om{\Omega}
\def\chic{\chi_{coll}}
\def\chiin{\chi_{in}}
\def\dchic{\delta \chi_{coll}}
\def\dchiin{\delta \chi_{in}}
\def\dchi{\delta \chi}
\def\fhat{\hat{F}}
\def\f{{\cal F}}
\def\de{\delta E}
\def\hahat{\hat{H}}
\def\rhat{\hat{{\bf r}}}
\def\tn{\tilde n}
\def\dn{\delta n}
\def\k{\kappa}
\def\hahat0{\hat{H}_0}
\def\dk{\delta \kappa}
\def\dc{\delta C}
\def\db{\delta B}
\def\dg{\delta \gamma}
\def\df{\delta {\cal F}}
\def\tc{\tilde C}
\def\tb{\tilde B}
\def\tg{\tilde \gamma}
\def\tf{\tilde{{\cal F}}}
\def\tk{\tilde \kappa}
\def\tchic{{\tilde \chic}}
\def\tchi{{\tilde \chi}}
\def\a{{\cal A}}
\def\b{{\cal B}}
\def\bardf{{\delta \bar{{\cal F}}}}
\def\bs{\bigskip}
\def\ms{\medskip}
\def\bb{|}
\def\cos{\hbox{cos}}
\def\sin{\hbox{sin}}
\def\cosh{\hbox{cosh}}
\def\sinh{\hbox{sinh}}
\def\arccosh{\hbox{arccosh}}
\def\exp{\hbox{exp}}
\def\arcsinh{\hbox{arcsinh}}
\def\sinh{\hbox{sinh}}
\def\Im{{\mbox {\rm Im}}}
\def\Re{{\mbox {\rm Re}}}
\def\varx{x}
\newcommand{\lapp}{\mbox{\raisebox{0.7ex}{$<$} \hspace{- 1.5em}
\raisebox{- 0.7ex}{$\approx$}}}
\def\e{e}
\def\epsi{{\cal E}}
\def\siml{\hbox{\kern.1em \lower.6ex \hbox{$\sim$} \kern-1.12em
 \raise.6ex \hbox{$<$} \kern.1em}}
\def\simg{\hbox{\kern.1em \lower.6ex \hbox{$\sim$} \kern-1.12em
 \raise.6ex \hbox{$>$} \kern.1em}}
\newcommand{\Figurebb}[9]{
  \begin{figure}[H]\begin{center}
  \leavevmode
  \epsfysize=#7cm
  \epsfbox[#2 #3 #4 #5]{#6}
  \par
  \parbox{#8cm}{
  \caption[figure]{\renewcommand{\baselinestretch}{0.8} \small
                   \hspace{-0.3truecm}#9}\label{#1}}
  \end{center}
  \end{figure}
}
\newcommand{\beqar}{\begin{eqnarray}}
\newcommand{\eeqar}[1]{\label{#1} \end{eqnarray}}
\def\degdeg{$^\circ$}
\def\subsubsubsection#1{\smallskip\noindent{\it#1}\hfill\break}
\def\lt{\raisebox{0.2ex}{$<$}}
\def\gt{\raisebox{0.2ex}{$>$}}
\def\lambar{{\mathchar'26\mskip-9mu\lambda}}
\def\ja{{\cal J}}
\def\isospin{\rm q}
%

\title{A SEMICLASSICAL COLLECTIVE RESPONSE 
  OF HEATED, ASYMMETRIC\\ AND ROTATING NUCLEI} 
\author{A.\ G.\ Magner\thanks{magner@kinr.kiev.ua}}
\affiliation{
 Institute for Nuclear Research, 03680 Kiev, Ukraine}
\author{D.\ V.\ Gorpinchenko}
\affiliation{ Institute for Nuclear Research, 03680 Kiev, Ukraine, and 
National Technical University of Ukraine, ``KPI", Kyiv
03056, Ukraine}
\author{J.\ Bartel}
\affiliation{Institut Pluridisciplinaire Hubert Curien, CNRS/IN2P3,
Universit\'e de Strasbourg, F-67000 Strasbourg, France}

\begin{abstract}
The quasiparticle Landau Fermi-liquid and periodic orbit
theories are presented for the semiclassical 
description of 
collective excitations 
in nuclei, which are close to one of the main topics of the fruitful activity 
of S.\ T.\ Belyaev. 
Density-density response functions are studied at low temperatures
within the temperature-dependent collisional Fermi-liquid
theory in the relaxation time approximation. The
isothermal, isolated (static) and adiabatic susceptibilities for
nuclear matter show 
the ergodicity property. 
Temperature corrections to the response function, 
viscosity and thermal conductivity
coefficients have been derived, also in the long wave-length 
(hydrodynamic) limit. 
The relaxation and correlation functions are obtained through the
fluctuation-dissipation theorem and their properties are discussed in
connection to the static susceptibilities and ergodicity 
of the Fermi systems. Transport coefficients, 
such as nuclear friction and inertia as
   functions of the temperature for the hydrodynamic 
(heat-pole and
first sound) and Fermi-surface-distortion zero sound modes are 
derived within the Fermi-liquid droplet model. 
They are shown to be
in agreement with the semi-microscopical calculations based on the
nuclear shell model (SM) for large temperatures. 
This kinetic approach is extended to 
the study of the neutron-proton correlations in 
asymmetric neutron-rich 
nuclei. The surface symmetry binding-energy constants are presented 
as functions of the Skyrme force parameters
in the approximation of a sharp edged proton-neutron asymmetric
nucleus 
and applied to
calculations of the isovector giant dipole
resonance.  
The energies, sum rules and transition densities 
of these resonances  obtained by using 
analytical expression for these 
surface constants in terms of the Skyrme force parameters 
are in fairly good agreement with the experimental data.
An analysis of the experimental data, in  particular the specific 
structure of these resonances in terms of a main, and some satellite peaks, 
in comparison with our analytical approach and other theoretical 
semi-microscopical models, might turn out to be of capital importance for a 
better understanding of the values of the fundamental surface symmetry-energy 
constant.
The semiclassical collective moment of inertia is derived 
analytically beyond the quantum perturbation approximation of the
cranking model for any potential well as a mean field. 
It is shown that
this moment of inertia can be approximated by 
its rigid-body value for
the rotation with a given frequency within the ETF and more
general periodic orbit theories 
in the nearly local long-length approximation.  
Its semiclassical shell-structure components are derived 
in terms of the periodic-orbit free-energy shell corrections. 
An enhancement of the moment of inertia near the 
symmetry-breaking bifurcation deformations was found. 
We obtained good agreement between the semiclassical and quantum shell-structure
components of the moment of inertia for several critical
bifurcation deformations for the completely analytically
solved example of the harmonic oscillator mean field.

\end{abstract}

\pacs{21.60.Ev,21.60.Sc,24.30.Cz,21.10.Dr,03.65.Sq}

\date{\today}

\maketitle


\section{INTRODUCTION}
\label{introd}

Nuclear collective motion, such as fission, or multipole
vibration and rotation excitation modes,
was successfully studied by using several
microscopic-macroscopic approximations to
the description of the finite Fermi systems of the strongly interacting
nucleons \cite{migdal,myswann69,fuhi,bohrmot,mix,belyaevzel}.
Many significant phenomena deduced from experimental data on nuclear
fission, vibrations and rotations were explained within the
theoretical approaches based
 mainly on the cranking model \cite{inglis,bohrmotpr,valat,bohrmot}
and its extensions to the pairing correlations
\cite{belyaevfirst,belyaev61,belzel,belsmitolfay87},
including shell and temperature effects \cite{brackquenNPA1981},
and to non-adiabatic effects
\cite{fraupash,zelev,mix,zshimansky,marshalek,belyaevhighspin,afanas,belyaevbif},
which were originally applied for the
rotational modes (see also \cite{sfraurev} for the review paper
and references therein).

For the nuclear collective excitations
within the general response-function theory
\cite{bohrmot,siemjen,hofbook}, the basic idea is to parametrize
the complex dynamical problem of the collective motion of many
strongly interacting particles in terms of a few collective
variables found from the physical
meaning of the considered dynamical problem, for example
the nuclear surface itself \cite{strtyap,strutmagbr,strutmagden}
or its multipole deformations \cite{bohrmot}.
We can then study the response to an external field of the dynamical
quantities describing the nuclear collective motion in terms of
these variables. Thus, we get important information on the transport
properties of nuclei. For such a theoretical description of the
collective motion it is very important to take into account the
temperature dependence of the
dissipative nuclear characteristics as the friction coefficient, as
shown in \cite{hofivyam,ivhofpasyam,hofmann,hofbook}.
The friction depends strongly on the temperature and its
temperature dependence can therefore not be ignored
in the description of the collective excitations in nuclei.
Concerning the temperature dependence of
the nuclear friction, one of the most important problems is
related to the properties of the static susceptibilities and
ergodicity of the Fermi systems like nuclei.

However, the quantum description of dissipative phenomena in
nuclei is rather complicated because we have to take into account
the residual interactions beyond the mean-field approximation.
Therefore, more simple models
\cite{strutmagbr,kolmagpl,magkohofsh,kolmagsh}
accounting for some
macroscopic properties of the many-body Fermi-system are helpful
to understand the global average properties of the collective
motion.
Such a model is based on the Landau Fermi-liquid theory
\cite{landau,abrikha,pinenoz},  applied for the nuclear interior
and simple
macroscopic boundary conditions on the nuclear surface
\cite{strutmagbr,strutmagden,magstrut,magboundcond,kolmagsh,magsangzh,BMRV}
(see
also macroscopic approaches with different boundary conditions
\cite{bekhal,ivanov,abrditstrut,komagstrvv,abrdavkolsh}).
In \cite{magkohofsh}, the response-function theory can be applied to
describe collective nuclear excitations as the isoscalar quadrupole
mode. The
transport coefficients, such as friction and inertia, are simply
calculated within
the macroscopic Fermi-liquid droplet model (FLDM)
\cite{kolmagpl,magkohofsh,kolmagsh} 
and
their temperature dependence  can be clearly discussed
(see also earlier works
\cite{strutmagden0,galiqmodgen,galiqmod,strutmagden,magstrut,denisov}). 
The asymmetry of
heavy nuclei near their stability line
and the structure of the isovector dipole resonances
are studied in \cite{kolmag,kolmagsh,BMV,BMR}
(see also \cite{abrIVGDR,abrdavpl}). In this way,
the giant multipole resonances were described, and, with
increasing temperature \cite{kolmagpl,magkohofsh}, a transition
from zero sound modes to the hydrodynamic first sound.
The friction in \cite{kolmagpl,magkohofsh}
is due to the collisions of particles, which were taken into
account in the relaxation-time approximation
\cite{abrikha,pinenoz,sykbrook,brooksyk,heipethrev,baympeth} with a
temperature and
frequency dependence (retardation effects)
\cite{landau,kolmagpl}.

The most important
results obtained in \cite{magkohofsh,hofivmag} are related to the
overdamped surface excitation mode for the low energy region and
its dissipative characteristics as friction.
For the low excitation energy region these investigations can be
completed by the additional sources of the friction related to a
more precise description of the heated Fermi liquids presented in
\cite{heipethrev,baympeth} for the infinite matter. Following
\cite{heipethrev}, we should take into account the thermodynamic
relations along with the dynamical Landau--Vlasov equation and
introduce the local equilibrium distribution instead of the one of global
statics, used earlier in \cite{magkohofsh,hofivmag} for the
linearization procedure of this equation. These new developments
of the Landau theory are especially important for the further
investigations of the temperature dependence of the friction.
For the first step we have to work out in more details the theory
\cite{heipethrev} of the heated Fermi liquids for nuclear matter
to apply then it for the dynamical description of the collective
motion in the interior of nuclei in the macroscopic FLDM
 \cite{kolmagpl,magkohofsh}. Our purpose is also to find the
relations to some general points of the response function theory
and clarify them taking the example of the analytically solved
model based on the non-trivial temperature-dependent Fermi-liquid
theory. One of the most important questions which would be better
to clarify is the above mentioned ergodicity property, temperature
dependence of the friction and coupling constant.

Another important
extension of this macroscopic theory is to study the structure
of the isovector giant
dipole resonance (IVGDR) as a splitting phenomenon due to the nuclear
symmetry interaction between neutrons and protons
 near the stability line \cite{kolmag,abrIVGDR,abrdavpl,kolmagsh,BMV,BMRV,BMR}.
The neutron skin of exotic nuclei with a large excess of neutrons
is also still one of the exciting subjects of
    nuclear physics and nuclear astrophysics
\cite{myswann69,myswnp80pr96,myswprc77,myswiat85,danielewicz1,pearson,danielewicz2,vinas1,vinas2,vinas3,vinas4}.
Simple and accurate solutions for the isovector  particle density
distributions were obtained within the nuclear effective surface (ES)
approximation
\cite{strtyap,strutmagbr,strutmagden,magsangzh,BMRV}.
It exploits the saturation of nuclear matter and a
narrow diffuse-edge region in finite heavy nuclei. The ES is defined as
the location of points of the maximum density gradient. The coordinate
system, connected locally with
the ES, is specified by the distance  from the
given point to the surface and by tangent coordinates  at the ES.
The variational condition for the nuclear energy with some
additional fixed integrals of motion in the local energy-density theory
\cite{rungegrossPRL1984,marquesgrossARPC2004} is
significantly
simplified in these coordinates. In particular,  in the extended
Thomas--Fermi (ETF) approach \cite{brguehak,sclbook}
(with Skyrme forces
\cite{chaban,reinhard,bender,revstonerein,ehnazarrrein,pastore})
 this can be done for any deformations by using an expansion  in a small
leptodermic parameter. The latter
is of the order of the diffuse edge
thickness of heavy enough nucleus over its mean curvature radius, or
the number of nucleons in power one third
under the distortion constraint in the case of
deformed nuclei. The accuracy of the ES  approximation in the ETF
approach without spin-orbit (SO) and asymmetry terms was checked
\cite{strutmagden} by comparing results of  Hartree--Fock (HF)
\cite{brink,ringschuck} and ETF calculations  \cite{brguehak,sclbook}  for some
Skyrme forces. The ES approach (ESA) \cite{strtyap,strutmagbr,strutmagden}
was then extended by taking SO and asymmetry effects into account
\cite{magsangzh,BMRV}.
Solutions for the
isoscalar and isovector particle densities and
energies at the quasi-equilibrium
in the ESA of the ETF approach were applied to
analytical calculations of the neutron skin and isovector stiffness
coefficients in the leading order of the leptodermic parameter
and to the derivations of the macroscopic boundary conditions \cite{BMRV}.
Our results are compared with the fundamental researches
\cite{myswann69,myswnp80pr96,myswprc77,myswiat85} in the
liquid droplet model (LDM). These analytical expressions for the
energy surface constants can be used
for IVGDR calculations within the FLDM
\cite{denisov,kolmag,kolmagsh,BMV,BMR}.

A further interesting application
of the semiclassical response theory would consist in the study of the
properties of collective rotation bands in heavy deformed nuclei.
One may consider nuclear collective rotations
within the cranking model
as a response to the Coriolis external-field perturbation.
The moment of inertia (MI)
can be calculated as a susceptibility with respect to this external field.
The rotation frequency of the rotating Fermi system in the cranking model
is determined for a given nuclear
angular momentum through a constraint, as
for any other integral of motion, as in particular the particle number
conservation \cite{ringschuck}. In order to simplify
such a rather complicated problem, the Strutinsky shell correction method (SCM)
\cite{strut,fuhi} was adjusted to the collective nuclear rotations in
\cite{fraupash,mix}. The collective MI
is expressed as function of the particle number and temperature
in terms of a smooth part
and an oscillating shell correction. The smooth component can be described
by a suitable macroscopic model, like the dynamical ETF approach
\cite{bloch,amadobruekner,rockmore,jenningsbhbr,sclbook,brguehak,bartelnp,bartelpl}
 similar to the FLDM,
which has proven to be both simple and precise.
For the definition of the MI shell correction, one can apply the Strutinsky
averaging procedure to the single-particle (s.p.) MI, in the same way as for
the well-known free-energy shell correction.

For a deeper understanding of the quantum results and the correspondence
between classical and quantum physics of the MI shell components, it
is worth to analyze these shell components in terms of periodic orbits
(POs), what is now well established as the semiclassical periodic-orbit
theory (POT) \cite{gutz,bablo,strutmag,bt,creglitl,sclbook,migdalrev}
(see also its
extension to a given angular momentum projection along with the
energy of the particle \cite{magkolstr} and to the particle densities
\cite{strutmagvvizv1986,brackrocciaIJMPE2010} and
pairing correlations
\cite{brackrocciaIJMPE2010}).
Gutzwiller was the first who developed the POT for
completely chaotic Hamiltonians with only one integral of motion
 (the particle energy) \cite{gutz}.
The Gutzwiller approach of the POT extended
to potentials with continuous symmetries for the description of the
nuclear shell structure can be found in \cite{strutmag,smod,creglitl,sclbook}.
The semiclassical shell-structure corrections to the level density
and energy have been tested for a large number of s.p.\ Hamiltonians
in two and three dimensions (see, for instance,
\cite{sclbook,magosc,ellipseptp,spheroidpre,spheroidptp,maf,magNPAE2010,magvlasar}).
For the Fermi gas the
 entropy shell corrections of the POT as a sum of periodic orbits were
derived in \cite{strutmag}, and with its help,
simple analytical expressions for the
shell-structure energies in cold nuclei were obtained there following
a general semiclassical theory \cite{sclbook}.
These energy shell corrections are in good agreement with the
quantum SCM results, for
instance for elliptic and spheroidal cavities, including the
 superdeformed bifurcation region
\cite{ellipseptp,spheroidptp}.
In particular in three dimensions, the superdeformed bifurcation
nanostructure leads as function of deformation to the
double-humped shell-structure energy with the first and second
potential wells in heavy enough nuclei
\cite{smod,migdalrev,spheroidptp,sclbook,magNPAE2010},
which is well known as the double-humped fission barriers in the region
of actinide nuclei.
At large deformations the second well can be understood semiclassically,
for spheroidal type shapes, through the bifurcation of equatorial orbits
into equatorial and the shortest 3-dimensional periodic orbits, because of
the enhancement of the POT amplitudes of the shell correction to the level
density near the Fermi surface at these bifurcation deformations.

For finite heated fermionic systems, it was also shown
\cite{strutmag,kolmagstr,magkolstrutizv1979,richter,sclbook,brackrocciaIJMPE2010}
 within the POT that the
shell-structure of the entropy, the thermodynamical (grand-canonical) potential
 and the free-energy shell
corrections can be obtained by multiplying the terms of the POT expansion
by a temperature-dependent factor, which is exponentially decreasing
with temperature. For the case of the so called
``{\it classical rotations}$\,$'' around the symmetry $z$ axis of the nucleus,
the MI shell correction is obtained, for any rotational frequency and at finite
temperature, within the extended Gutzwiller POT through the averaging of the
individual angular momenta aligned along this symmetry axis
\cite{magkolstr,kolmagstr,magkolstrutizv1979}.
A similar POT problem, dealing with the magnetic susceptibility of fermionic
systems like metallic clusters and quantum dots, was worked out in
\cite{richter,fraukolmagsan}.

It was suggested in \cite{dfcpuprc2004} to use the spheroidal cavity and
the classical perturbation approach to the POT by Creagh \cite{creagh,sclbook}
to describe the collective rotation of deformed nuclei around an axis ($x$
axis) perpendicular to the symmetry $z$ axis.
The small parameter of the POT perturbation approximation turns out to be
proportional to the rotational frequency, but also to the classical action
(in units of $\hbar$), which causes an additional restriction to Fermi
systems (or particle numbers) of small enough size, in contrast to the usual
semiclassical POT approach.

In \cite{mskbg,mskbPRC2010}, the nonperturbative extended Gutzwiller POT
was used for the calculation of the MI shell corrections within the
mean-field cranking model for both the collective and the alignment
rotations.
In these works, for the statistical equilibrium nuclear rotations,
the semiclassical MI shell corrections were obtained in good agreement with the
quantum results
in the case of the harmonic-oscillator potential.
We extend this approach for collective rotations
perpendicular to symmetry axis to the analytical calculations of
the MI shell corrections for the  case of different mean fields,
in particular with spheroidal
shapes and sharp edges. The main purpose is to study semiclassically
the enhancement effects in the MI within the improved stationary
phase method (improved SPM or shortly ISPM)
\cite{ellipseptp,spheroidptp,maf,migdalrev,magvlasar},
due to the bifurcations of the
periodic orbits in the superdeformed region.

In the present review in Section \ref{eqmotion} we present some
basic formulas of the temperature-dependent Fermi-liquid theory
\cite{heipethrev}. We consider in Sec.\ \ref{conserveqs} the
particle number and momentum conservation equations and derive
from them the energy conservation and general transport equations,
in particular,
the expressions for the viscosity, shear modulus and thermal conductivity
coefficients. In Sec.\ \ref{respfunsec} we determine the
density-density and density-temperature response functions with
the low temperature corrections. Section
\ref{longwavlim} shows the long wave-length (LWL, or hydrodynamic) limit
for the response functions, and the specific expressions for the
transport coefficients.
In Sec.\ \ref{suscept}, one obtains the static
isolated, isothermal, and adiabatic susceptibilities to clarify
some important points of the general response function theory,
mainly, the ergodicity property of the Fermi systems
\cite{hofmann,kubo}. We study the relaxation and correlation
functions on the basis of the fluctuation-dissipation theorem and
establish their relations to the ergodicity of the Fermi-liquid
system in section \ref{relaxcorr}.
General aspects of the response function theory for the collective
motion in nuclei are presented in
Sec.\ \ref{basdef}
in line with \cite{hofmann,hofbook}. Section
\ref{fldm} shows the basic ingredients and
the collective response function of the nuclear FLDM.
Section \ref{transprop} is devoted to the
derivation of the temperature dependence of the
transport coefficients, such as friction, inertia,
and stiffness for the density modes
for slow collective motion. The numerical illustrations are given in Sec.\
\ref{discuss}.  In Sec.\ \ref{npcorivgdr}, the semiclassical theory is extended
to neutron-proton asymmetric nuclei and applied for the
calculations of IVGDRs.
In Sec.\ \ref{semshellmi}, the smooth
ETF and fluctuating shell-structure components of the moments of inertia
are derived for collective rotations of heavy nuclei.
The MI shell component is analytically presented in terms of the
periodic orbits and their bifurcations within the POT. This component
is compared with the quantum results for the simplest case of the deformed
harmonic oscillator Hamiltonian.
Comments and conclusions are finally given in Sec.\ \ref{concl}.
Some details of the thermodynamical, FLDM
(in the LWL limit) and POT calculations, such as
the analytical derivations
of the in-compressibility, viscosity, thermoconductivity,
coupling, and surface symmetry-energy constants, as well as the
semiclassical MI
are presented in Appendices A-E.


\section{The quasiparticle kinetic theory}
\label{kinapp}

\subsection{Equations of motion for the heated Fermi liquid}

\label{eqmotion}

In the semiclassical approximation the dynamics of a Fermi liquid
may be described by the distribution function $f(\r,\p,t)$ in the
one body phase-space. Restricting to small deviations of particle
density $\rho(\r,t)$ and temperature $T$, from their values in a
thermodynamic  equilibrium one may apply the linearized
Landau--Vlasov equation \cite{abrikha,heipethrev}:
\bea\l{landvlas}
\frac{\partial}{\partial t}
    \delta f(\r,\p,t) &+&
    \frac{\partial \eps_{\p}^{\rm g.e.}}{\partial \p}
    {\bf \nabla}_r \delta f(\r,\p,t) - 
    - {\bf \nabla}_r \left[\delta \eps(\r,\p,t) \right.\nonumber\\ 
    &+&\left.V_{\rm ext}\right]
    {\bf \nabla}_p f_{\rm g.e.}(\eps_\p^{\rm g.e.}) =
    \delta St. 
\eea
The right hand side (r.h.s.) represents the dynamic component
of the integral collision term $\delta St$, and
$V_{\rm ext}$ stands for an external field. We introduce here
the Fermi distribution
\bel{fgeq}
f_{\rm g.e.}(\eps_\p^{\rm g.e.})  =
    \left[1 + \exp \left(\frac{\eps_\p^{\rm g.e.}-\mu}{T}\right)\right]^{-1}
\ee
of the {\it global equilibrium} (g.e.), with  $\mu$ being the chemical
potential,  the temperature $T$ is given, as usually
in nuclear physics, in the energy
(MeV) units (without Boltzmann's constant), and
$\delta f(\r,\p,t)$ measures the deviation
\bel{dfgeqrpt} \delta f(\r,\p,t) =  f(\r,\p,t) -
f_{\rm g.e.}(\eps_\p^{\rm g.e.}).
\ee
For the sake of simplicity, the s.p.\ energy $\eps_\p^{\rm g.e.}$
will be assumed to be of the form
$~\eps_\p^{\rm g.e.} =
p^2/2m^*~$ with $m^*$ being the effective nucleonic
mass. In (\ref{landvlas}),
$\delta \eps(\r,\p,t)$ stands for the variation of the
quasiparticle energy $\eps(\r,\p,t)$,
\bea\l{deleps}
&&\delta
\eps(\r,\p,t) = \eps(\r,\p,t)- \eps_\p^{\rm g.e.}  \nonumber\\
&=&    \frac{1}{\mathcal{N}(T)} \int
\frac{2 d\p^\prime}{(2 \pi {\hbar})^3}\;
    \mathcal{F}(\p, \p^\prime)\;
    \delta f(\r, \p^\prime, t).
\eea
The quasiparticles' density of
states $\mathcal{N}(T)$ at the chemical potential $\mu$ is given
by
 \bel{enerdensnt}
 \mathcal{N}(T)=\int
  \frac{2 d\p^\prime}{(2 \pi {\hbar})^3}
    \left(
-\frac{\partial f_{\p^{\prime}}} {\partial \eps_{\p^{\prime}}} \right)_{g.e}.
\ee
Evidently, because of our linearization the density
$\mathcal{N}(T)$ here is the one of equilibrium. In the sequel such a
convention will inherently be applied to any coefficient of
quantities of order $\delta f$.
The factor 2 accounts for the spin degeneracy.
The amplitude of the
quasiparticle interaction, $\mathcal{F}(\p,\p^\prime)$,
commonly is written in terms of
the Landau parameters $\mathcal{F}_0$ and $\mathcal{F}_1$, according to
\bel{intampfpp}
\mathcal{F}(\p, \p^\prime) =
    \mathcal{F}_0 + \mathcal{F}_1 {\hat p}
    \cdot {\hat p}^\prime\;, \qquad   {\hat p} =
    \p/p.
\ee
These two constants may be related to the two properties of
nuclear matter, namely the isothermal in-compressibility $K^{T}$
(see Appendix A.1),
\bel{isotherk}
K^{T} =
9 \rho\mathcal{G}_0/\mathcal{N}(T),
\ee
and the effective mass $m^*$,
\bel{effmass}
m^*=  \mathcal{G}_1 m, \quad
\mathcal{G}_n=\left(1+\frac{\mathcal{F}_n}{2n+1}\right)
\ee
($n=0, 1$). The equation for the effective
mass $m^*$ is known \cite{abrikha,heipethrev} to be valid
for systems obeying Galileo invariance, which shall be assumed
here.

In principle, the Landau parameters $\mathcal{F}_0$ and $\mathcal{F}_1$
might vary with the momenta $p$ and $p'$. Such a dependence will
be neglected henceforth.  This approximation appears to be
reasonable as we are going to stick to small excitations near the
Fermi surface and to temperatures $T$, which are small as compared to
the chemical potential $\mu$. Likewise, we shall discard any
temperature dependence of the effective mass. Notice that
in addition to the ratio $({T /\mu})^2$, this dependence would be
governed by the additional factor $|{m^* / m}-1|$ which is small
for nuclear matter. These assumptions will allow us to simplify
further the theory \cite{heipethrev} and to get
more explicit results by making use of the temperature expansion
for the response functions in the small parameter $T/\mu$, as well
as of the standard perturbation approach to eigenvalue problems
needed later for the hydrodynamic (long-wave
length) limit. We will follow \cite{heipethrev} in neglecting
higher order terms in the expansion (\ref{intampfpp}) in Legendre
polynomials.

Later on we want to study motion of the system which can be
classified as an excitation on top of the {\it local equilibrium}.
Following  \cite{abrikha,heipethrev}, the
collision term $\delta St$ can  be considered in the relaxation
time approximation,
\bea\l{intcoll}
&&\delta St =
-\frac{\delta f_{\rm l.e.}(\r,\p,t)}{\tau}\;, \qquad
    f_{\rm l.e.}\left(\eps_\p^{l.e.}\right) =  \nonumber\\
    &=&\left[1 + \exp\left(\frac{\eps_\p^{\rm l.e.}-\mu(\r,t)
    -\p\u(\r,t)}{T(\r,t)}\right)\right]^{-1}.
\eea
Here, $f_{\rm l.e.}(\eps_\p^{\rm l.e.})$  is the distribution function
of a {\it local equilibrium} (l.e.), and $\eps_\p^{\rm l.e.}$ is the associated
quasiparticle energy. $\mu(\r,t)$ represents the chemical
potential, $\u(\r,t)$ the mean velocity field, and $T({\bf r},t)$
the temperature, all defined in the local sense.  Like in
\cite{magkohofsh}, the relaxation time $\tau$ is assumed to be
independent of the quasiparticle momentum $\p$. However, it  will be
allowed to depend $\tau$ on $T$  as well as on the frequency of the
motion (thus, accounting for retardation effects in collision
processes).   In
(\ref{intcoll}), $\delta f_{\rm l.e.}(\r,\p,t)$
is defined as
\bel{dfleqrpt} \delta
f_{\rm l.e.}(\r,\p,t) =  f(\r,\p,t) - f_{\rm l.e.}(\eps_\p^{\rm l.e.}).
\ee
It differs from $\delta f(\r,\p,t)$ of (\ref{dfgeqrpt}) by the
variations of local quantities.  For the latter, we may write
\bel{dfgeqdfleq} \delta f(\r,\p,t) =
    \delta f_{\rm l.e.}(\r,\p,t) +
    \delta f_{\rm l.e.}\left(\eps_\p^{l.e.}\right)
\ee
with
\bea\l{dfleq}
&&\delta f_{\rm l.e.} \left(\eps_\p^{\rm l.e.}\right)
    = f_{\rm l.e.}\left(\eps_\p^{\rm l.e.}\right) -
    f_{\rm g.e.}\left(\eps_\p^{\rm g.e.}\right)  \qquad \nonumber\\
&=&
   \left(\frac{\partial f_\p}{\partial \eps_\p}\right)_{\rm g.e.}
    \left(\delta \eps_\p^{\rm l.e.} - \delta \mu -\p \u -
    \frac{\eps_\p^{\rm g.e.}-\mu}{T}\delta T\right).\qquad
\eea
For the l.e. quasiparticle energy $\eps_\p^{\rm l.e.}$, one has
\bel{delelocglo} \eps_\p^{\rm l.e.}=\eps_\p^{\rm g.e.}+
    \delta \eps_\p^{\rm l.e.},
\ee
where $\delta \eps_\p^{\rm l.e.}$ is defined like in
(\ref{deleps}) with only $\delta f(\r,\p,t)$ replaced by
$\delta f_{\rm l.e.}(\r,\p,t)$. According to (\ref{dfgeqdfleq}) and
(\ref{dfleq}), for the simplified interaction
(\ref{intampfpp}), one gets
\bel{delepsleq} \delta \eps_\p^{\rm l.e.}=
    \delta \eps (\r,\p,t)=
    \frac{\mathcal{F}_0}{\mathcal{N}(T)} \delta \rho(\r,t)
    + \frac{\mathcal{F}_1 m \rho}{\mathcal{N}(T) p_{{}_{\! {\rm F}}}^2} \p\u,
\ee
 where $\delta \rho$ is the dynamical component of the particle
density
 \bel{densit} \rho(\r, t) =
    \int \frac{2 d\p}{(2 \pi \hbar)^3} \;
    f(\r, \p, t) = \rho_\infty + \delta \rho(\r,t)
\ee
 with $\rho_\infty$ being its g.e. value
associated to $f_{\rm g.e.}(\eps_\p^{\rm g.e.})$ for the infinite Fermi liquid.
The vector of the mean
velocity $\u$ can be expressed in terms of the first moment of the
distribution function (current density) and the particle density
(\ref{densit}),
\bel{veloc} \u(\r,t) =    \frac{1}{\rho}
    \int \frac{2  d \p}{(2 \pi \hbar)^3} \>
\frac{\p}{m} \delta f(\r, \p, t).
\ee

The definition of the collision term in the form (\ref{intcoll})
is incomplete without posing conditions for the conservation of
the particle number, momentum, and energy (for simplicity
of notations, we shall omit index $\infty$ in
the static nuclear-matter density component $\rho$ at
second order terms in the energy density variations).
Notice that to the order considered, in the equation
for energy conservation, $\eps$ may be replaced by
$\eps_\p^{\rm g.e.}$ (see also \cite{baympeth}).
Incidentally, for the
quasiparticle interaction (\ref{intampfpp}), this substitution
even becomes {\it exact}, as the dynamical part $\delta \eps$
would drop out of the last integral (as follows from
(\ref{delepsleq}), (\ref{delelocglo}), and two first equations
in the following set of conditions
\cite{heipethrev},
\begin{eqnarray}
    \int {\rm d}\p\; \delta f_{\rm l.e.}(\r, \p, t)\; = \;0\;, \qquad
    \int {\rm d}\p\;\p\;\delta f_{\rm l.e.}(\r,\p, t)
    \; = \;0\;, \nonumber\\
    \int {\rm d}\p\;
    \eps\;
    \delta f_{\rm l.e.}(\r, \p, t) \; = 0. \qquad\qquad\qquad\qquad
\label{consereq}
\end{eqnarray}
These equations
mimics conservation of the
corresponding quantities in each collision of quasiparticles and
ensures that of the same quantities calculated for the total
system (without external fields). Together with the basic equation
(\ref{landvlas}), one thus has 6 equations for the 6 unknown
quantities $\delta \rho(\r,t)$, $\delta \mu(\r,t)$, $\u(\r,t)$ and
$\delta T(\r,t)$. They allow one to find unique solutions as
functionals of the external field $V_{\rm ext}(t)$. Below we shall
solve these equations in terms of response functions.
It may be noted that, due
to the conditions (\ref{consereq}), the first variation of the
distribution function $\delta f(\r,\p,t)$
(\ref{dfgeqdfleq}) disappears from the dynamical component
$\delta \rho(\r,t)$ of the density $\rho(\r,t)$ and of the
velocity field $\u(\r,t)$. As one knows (see, e.g.,
\cite{heipethrev,abrikha,baympeth}), the equation for the velocity field
reduces to an identity if one takes into account the definition of
the effective mass $m^*$ given by (\ref{effmass}).

\subsection{The conserving equations}
\label{conserveqs}

In this section, we like to deduce conserving equations for the
particle number, momentum, and energy, which later on will turn out
helpful to find appropriate solutions of the Landau--Vlasov
equation (\ref{landvlas}). The procedure, which basis on a moment
expansion, is well known from textbooks
\cite{pinenoz,baympeth,forster}.
We will follow more
closely the version of \cite{kolmagpl,magkohofsh} (see also 
\cite{galiqmod}).

\subsubsection{THE MOMENT EXPANSION}
Whereas particle number conservation implies to have
 \bel{conteq}
\frac{\partial \rho}{\partial t} +
    {\bf \nabla} \left(\rho \u\right) =  0,
\ee
the  momentum conservation is reflected by the following set of
equations
 \bel{momenteq}
m \rho
\frac{\partial u_\alpha}{\partial t} +
    \sum_{\beta}
\frac{\partial \Pi_{\alpha \beta}}{\partial r_{\beta}}  =
    -\frac{\partial V_{\rm ext}}{\partial r_\alpha}.
\ee
 Besides quantities introduced before, they involve
\bel{momentflux}
\Pi_{\alpha\beta}=
    \int \frac{2  d \p}{(2 \pi \hbar)^3 } \>
    \frac{p_\alpha p_\beta}{m^*} \delta f(\r, \p, t)
    + \frac{\mathcal{F}_0}{\mathcal{N}(T)} \delta \rho\left(\r,t\right)
    \;\delta_{\alpha\beta}.
\ee
 Substituting for  $\delta f(\r, \p, t)$
(\ref{dfgeqdfleq})  into (\ref{momentflux}), one gets
\bel{momentflux1} \Pi_{\alpha\beta} = - \sigma_{\alpha\beta} +
\delta \press ~\delta_{\alpha\beta}.
\ee
The first component $\sigma_{\alpha\beta}$,
which results from the first term $\delta
f_{\rm l.e.}(\r, \p, t)$ on the right of (\ref{dfgeqdfleq}),
determines the dynamic shear stress tensor,
\bel{presstens}
\sigma_{\alpha\beta}(\r, t)=
   - \int \frac{2  d \p}{(2 \pi \hbar)^3}\;
    \frac{p_{\alpha} p_{\beta}}{m^*}
    \delta f_{\rm l.e.}(\r, \p, t)\;,
\ee
whose trace vanishes.
For a linearized dynamics, the non-diagonal components of the
momentum flux tensor $\Pi_{\alpha\beta}$ equal the corresponding
stress tensor (but with the opposite sign), with correction terms
being proportional to $u_\alpha u_\beta$ in $\delta f$, and thus,
of higher order, see (\ref{veloc}) for $u_{\alpha}$.

The second component of the momentum flux tensor of
(\ref{momentflux1}) can be derived from the variation  $\delta
f_{\rm l.e.}(\eps_\p^{\rm l.e.})$ as given by (\ref{dfleq}). It
represents the compressional part of the momentum flux tensor,
\bea\l{pressdef}
&&\int \frac{2  \d \p }{(2 \pi \hbar)^3}~
    \frac{p_\alpha p_\beta }{m^*}~
    \delta f_{\rm l.e.}\left(\eps_\p^{\rm l.e.}\right) +
\frac{\mathcal{F}_0}{\mathcal{N}(T)}~\delta \rho~
\delta_{\alpha\beta}= \delta \press \delta_{\alpha\beta}
\nonumber\\
&&{\rm with}\quad \delta \rho \equiv \delta \rho(\r,t)= \int
\frac{2\d \p}{(2 \pi \hbar)^3}~
    \delta f_{\rm l.e.}\left(\eps_\p^{\rm l.e.}\right)
\eea
[mind (\ref{dfgeqdfleq}) and (\ref{consereq})]. Notice, that here
only the diagonal parts survive. The only non-diagonal ones could come
from the terms in (\ref{dfleq}) involving  $\p\u$; but they vanish
when integrating over angles in momentum space. Traditionally,
$\delta \press$ in (\ref{pressdef}) is referred to as the
scalar pressure, see \cite{brenig}.
Using
(\ref{dfleq}) for the distribution  $\delta
f_{\rm l.e.}(\eps_\p^{\rm l.e.})$ and its properties mentioned above,
after some simple algebraic transformations, one
gets
 \bea\l{pressureq}
\delta \press &=&\frac{2}{3}~ \int \frac{2 \d \p}{(2 \pi \hbar)^3}
    \frac{p^2}{2 m^*}
    \delta f_{\rm l.e.}\left(\eps_\p^{\rm l.e.}\right) +
\frac{\mathcal{F}_0 }{\mathcal{N}(T)}~\delta \rho \nonumber\\
&=&{K^T \over 9}
\delta \rho + \rho\left(\varsigma-
\frac{\mathcal{M} }{\mathcal{N}}\right)~\delta T,
\eea
 with $K^T$ being the isothermal
in-compressibility (\ref{isotherk}). For the derivation of the
second equation in (\ref{pressureq}),  one can use (i)
the transformation of $\delta \mu$ to the
variations of $\delta \rho$ and $\delta T$
[see (\ref{dmu})], and (ii) the relations
 (\ref{entropydef}), (\ref{densstat}), and (\ref{mcapt}) for
the entropy per particle $\varsigma$, the particle density $\rho$
as well as for the quantity $\mathcal{M}$ (\ref{mcapt}),
respectively. Inspecting (\ref{dpdtr}) and
(\ref{incomprTdef}), it becomes apparent that the expression on
the very right of (\ref{pressureq}) may indeed be interpreted as
an  expansion of the static pressure to the first order in
$\delta \rho$ and $\delta T$. It is thus seen that the truly
non-equilibrium component $\delta f_{\rm l.e.}(\r,\p,t)$ only appears
in the shear stress tensor $ \sigma_{\alpha\beta} $ given in
(\ref{presstens}).

Note, here and below within
Sec.\ \ref{conserveqs}, we omit immaterial constants related to the
global equilibrium (static) components of the moments to simplify
the notations and adopt them to the ones of the standard textbooks
when it will not lead to misunderstanding. We should
emphasize that the Landau quasiparticle theory which is a basis of
our derivations is working in a self-consistence way with small
deviations from (small excitations near) the Fermi surface
which are denoted by symbol "$\delta$" and takes
above mentioned static components as those of the external phenomenological
(experimental) data. Therefore, all relations discussed below in
this section should be understood as the ones between such
close-to-Fermi-surface quantities within our linearized
Landau--Vlasov phase space dynamics after exclusion of all above
mentioned immaterial constants. Nevertheless, we keep the symbol
$\delta$ with the scalar pressure $\delta \mathcal{P}$ to avoid
possible misunderstanding related to the linearization procedure,
see more comments below after (\ref{deformtens}).

\subsubsection{THE STRESS TENSOR}

It may be worthwhile to relate the stress tensor  $
\sigma_{\alpha \beta}$ given in (\ref{presstens}) to the standard
form  in terms of the
coefficients of the shear modulus $\lambda$  and the viscosity
$\nu$,
 \bel{prestensone}
\sigma_{\alpha\beta}=
    \sigma_{\alpha\beta}^{(\lambda)}
    + \sigma_{\alpha\beta}^{(\nu)}.
\ee
 Here, the first term $ \sigma_{\alpha\beta}^{(\lambda)}$
is the conservative part of the stress tensor $
\sigma_{\alpha\beta}$,
\bel{presslamb}
\sigma_{\alpha\beta}^{(\lambda)}=
    \lambda\left(\frac{\partial w_{\alpha}}{\partial r_{\beta}} +
    \frac{\partial w_{\beta}}{\partial r_{\alpha}} -
    \frac{2}{3}{\bf \nabla}{\bf w}\;\delta_{\alpha \beta}\right)
\ee
with $~\u = \partial {\bf w}/\partial t$
and ${\bf w}$ being the displacement field.
The second term in (\ref{prestensone}) can be written as
\bel{pressnu}
\sigma_{\alpha\beta}^{(\nu)}=
    \nu\left(\frac{\partial u_{\alpha}}{\partial r_{\beta}} +
     \frac{\partial u_{\beta}}{\partial r_{\alpha}} -
    \frac{2}{3}{\bf \nabla}\u\;\delta_{\alpha \beta}\right),
\ee
 where $\nu$ is the coefficient of the shear viscosity (or the
first viscosity). For more details see Appendix A.2,
in particular for expressions of the coefficients $\lambda$
(\ref{shearmod}) and $\nu$ (\ref{viscos}) in terms of
Fermi liquid interaction parameters.

To obtain microscopic expressions for the shear modulus $\lambda$
and the viscosity $\nu$, one needs to exploit the solution $\delta
f_{\rm l.e.}(\r,\p,t)$ of the Landau--Vlasov equation (\ref{landvlas})
for the stress tensor $\sigma_{\alpha\beta}(\r,t)$ (\ref{presstens}),
reducing the latter to the form (\ref{prestensone}). Such a
calculation of $\lambda$ and $\mu$ in terms of the
Landau Fermi-liquid parameters is discussed in Appendix A.2,
in which Fourier transforms are exploited \cite{kolmagpl}.
Equivalently, one may
express functions of space and time by plane waves, which for the
distribution function reads \cite{strutmagden0,kolmagpl}
\bel{planewave}
\delta f(\r, \p, t)=
\delta {\tilde f}\left(\q,\p,\om\right)
  \exp\left[{i(\q\r-\om t)}\right]
\ee
 with $\q$ being the wave vector and $\om$ the frequency of the
vibrational modes of nuclear matter. Such a plane-wave
representation is to be applied to both sides of (\ref{presstens})
and (\ref{prestensone}). The amplitudes for the velocity
$\u$ and the displacement ${\bf w}$ field then satisfy ${\bf
{\tilde w}}={\tilde\u}/(-i\om)$.

Using (\ref{presstens}),
(\ref{prestensone}), (\ref{presslamb}) and (\ref{pressnu}) for
the stress tensor $\sigma_{\alpha\beta}$
and (\ref{pressureq}) for the scalar pressure $\delta \mathcal{P}$,
  \bel{incomprtotdef}
  \delta \press=\frac{K_{\rm tot}}{9}~\delta \rho ,
\ee
one
finally may write down a general expression for the momentum flux
tensor $\Pi_{\alpha\beta}(\r,t)$ (\ref{momentflux1}),
\bea\l{momentfluxtot}
&&\Pi_{\alpha\beta}=
    -\lambda\left(\frac{\partial w_{\alpha}}{\partial r_{\beta}} +
    \frac{\partial w_{\beta}}{\partial r_{\alpha}} -
    \frac{2}{3}{\bf \nabla}{\bf w}\;\delta_{\alpha \beta}\right) ~~\nonumber\\
    &-&\nu\left(\frac{\partial u_{\alpha}}{\partial r_{\beta}} +
    \frac{\partial u_{\beta}}{\partial r_{\alpha}} -
    \frac{2}{3}{\bf \nabla}\u\;\delta_{\alpha \beta}\right)
    + \frac{K_{\rm tot}}{9}~\delta \rho~\delta_{\alpha\beta}.~~~
\eea
A total effective in-compressibility $K_{\rm tot}$ includes the change of
the pressure due to
variations of the temperature with density.
With the help of (\ref{dpdtr}), the in-compressibility $K_{\rm tot}$
can be expressed through the specific
heat per particle  ${\tt C}_{\mathcal{V}}$
(\ref{specifheatpdef}),
 \bel{incomprtot}
K_{\rm tot}=K^{T}+ 6 {\tt C}_{\mathcal{V}}\; \rho
    ~\frac{\delta {\tilde T}}{\delta {\tilde \rho}}.
\ee
Again, $\delta {\tilde T}$ and $\delta {\tilde \rho}$ are the
Fourier components of $\delta T(\r,t)$ and $\delta \rho(\r,t)$.
Like all other kinetic coefficients, such as
$\lambda$ and $\nu$ given in (\ref{shearmod}) and (\ref{viscos}),
respectively, this effective, total in-compressibility modulus
$K_{\rm tot}$, too, depends on $\om$ and $q$.  Later on we shall
discuss in more detail these quantities in the LWL
limit. In this limit, the total in-compressibility $K_{\rm tot}$ will
be seen to become identical to the adiabatic one $K^{\varsigma}$
given in (\ref{incompradexp}).

\subsubsection{ENERGY CONSERVATION AND THE GENERAL TRANSPORT\\ EQUATION}

So far we have not looked at the energy conservation. For this purpose,
one needs to consider thermal aspects as they appear in equations
for the change of entropy and temperature. To do this we will
follow standard procedures. We first built the
scalar product of the mean velocity $\u$ with the vector equation,
whose component $\alpha$ is given by (\ref{momenteq}). Making
use of the continuity equation (\ref{conteq}), after some
manipulations, one gets
\bea\l{enerconstwo}
&&\frac{\partial}{\partial t}
    \left(\frac{1}{2} m\rho u^2+\rho \epsi\right) = \nonumber\\
    &=& -\sum_{\alpha\beta}\frac{
\partial}{\partial r_{\beta}} \left[u_{\alpha}
    \left(\frac{1}{2} m \rho u^2 \delta_{\alpha\beta}
    +\rho W_{\alpha\beta}-
    \sigma_{\alpha\beta}^{(\nu)}
    -\kappa \frac{\partial T}{\partial r_{\alpha}}
    \delta_{\alpha\beta}\right)\right] \nonumber\\
    &+& \rho T \left(\frac{\partial \varsigma}{\partial t}
    + \u{\bf \nabla} \varsigma\right)
   -{\bf \nabla}\left(\kappa {\bf \nabla}T\right) \nonumber\\
    &-&\frac{\nu}{2}\sum_{\alpha\beta}
    \left(\frac{\partial u_{\alpha}}{\partial r_{\beta} }+
    \frac{\partial u_{\beta}}{\partial r_{\alpha}} -
    \frac{2}{3}{\bf \nabla}\u\;\delta_{\alpha\beta}\right)^2
    -\rho \u{\bf \nabla} V_{\rm ext}.
\eea
On the left hand side, there appears the mean kinetic energy
density and the internal energy density $\rho \mathcal{E}$ per unit
volume (defined again up to an immaterial constant). The density
$\epsi$ itself may be split in three different components,
\bel{enerintr}
\epsi= \epsi^{(\lambda)}+\epsi_{\rm tot}^{(K)}
+ T \delta \varsigma\;.
\ee
  The first one,
\bel{shearen}
\epsi^{(\lambda)}
 =\frac{\lambda}{4 \rho}
    \sum_{\alpha \beta}\left(\frac{
\partial w_{\alpha}}{\partial r_{\beta}} +
    \frac{\partial w_{\beta}}{\partial r_{\alpha}} -
    \frac{2}{3}{\bf \nabla}{\bf w}\;\delta_{\alpha
    \beta}\right)^2\;,
\ee
 is related to shear deformations, which is known from the solid state
physics and for Fermi liquids as coming from
distortions of the Fermi surface \cite{landau}.
The second one may
be written as
\bel{enerintrk}
 \epsi_{\rm tot}^{(K)}= \frac{K_{\rm tot}}{18} \left(\delta \rho\right)^2 ;
\ee
 it represents the
compressional component, associated to the effective total
in-compressibility $K_{\rm tot}$,
which is
in line of the known thermodynamic relations.
Equation (\ref{enerintrk}) resembles the expression
found in \cite{strutmagbr,strutmagden}, except for a generalization
of the physical meaning of the in-compressibility modulus
$K_{\rm tot}$ as function of $\om$ and $q$
given in (\ref{incomprtot}), as compared to the quasistatic adiabatic
case. The third one in (\ref{enerintr}) represents the
change of heat part resulting from a change of entropy.
We keep here the dynamical variation symbol $\delta$ for the
entropy $\varsigma$ (also for the pressure $\mathcal{P}$ here and
below) to remember that all quantities of the Landau Fermi-liquid
theory are presented for small dynamical deviations near the
Fermi surface in the linear (or quadratic after multiplying
(\ref{momenteq}) by $\u$) form in $\delta f$. We avoid here a
misunderstanding with following transformations of the energy
$\epsi$ (\ref{enerintr}), say Legendre ones, to the
differential form  in line of a
general comment at the beginning of this section.
On the r.h.s. of (\ref{enerconstwo})
the enthalpy $W_{\alpha\beta}$ per particle has been introduced,
\bel{enthalp} W_{\alpha\beta}=
    \epsi \delta_{\alpha\beta}
    + \frac{1}{\rho}\left(\sigma_{\alpha\beta}^{(\lambda)} +
    \delta \press~ \delta_{\alpha\beta} \right)
\ee
 (see the comment above concerning $\delta \press$).
Furthermore, the thermodynamic relation for the dynamical
variations of the internal energy $\epsi$ in
terms of those $\varsigma$ for the entropy per particle $
\varsigma$, the density $\rho$ and the displacement tensor
$w_{\alpha\beta}$, is given by
\bel{thermener}
{\rm d} \epsi =T {\rm d}
\varsigma+ \frac{\delta \press}{\rho^2} {\rm d} \rho +
\frac{1}{\rho}
    \sum_{\alpha\beta}\sigma_{\alpha\beta}^{\lambda}
    {\rm d} w_{\alpha\beta}.
\ee
The displacement tensor $w_{\alpha\beta}$ is defined as
\bel{deformtens}
w_{\alpha\beta}=
    \frac{1}{2}\left(\frac{\partial w_{\alpha}}{\partial r_{\beta}}
    + \frac{\partial w_{\beta}}{\partial r_{\alpha}}\right)\;.
\ee
Note that equation (\ref{incomprtotdef}) for the pressure $\delta
\press$ is important to get
(\ref{enerintr}) by integration of (\ref{thermener}).
According to (\ref{enthalp}), we get the standard relation of
{\it linearized} thermodynamics  of
\cite{brenig}, for instance,
between the enthalpy (\ref{enthalp}) and entropy $\varsigma$ up to the
second order term in $\delta \press$.

In (\ref{enerconstwo}), we also added and subtracted the term
${\bf \nabla} {\bf j}_T$ containing the heat current,
\bel{currheat}
{\bf j}_{{}_{\! T}}= -\kappa {\bf \nabla} T\;,
\ee
with the coefficient $\kappa$ for the thermal conductivity.
We may now write the
equation for energy conservation as
\bea\l{enerconserv}
&&\frac{\partial}{\partial t}
    \left(\frac{1}{2} m\rho u^2+\rho \epsi \right)=
    -\sum_{\alpha\beta}\frac{
\partial}{\partial r_\beta} \left[u_\alpha
    \left(
   \frac{1}{2} m \rho u^2 \delta_{\alpha\beta} \right.\right.\nonumber\\
    &+&\left.\left.\rho W_{\alpha\beta}
       - \sigma_{\alpha\beta}^{(\nu)}
    -\kappa \frac{
\partial T}{\partial r_{\alpha}} \delta_{\alpha\beta}\right)\right]
    -\rho \u{\bf \nabla} V_{\rm ext}.
\eea
In this way, it is seen that from (\ref{enerconstwo}) and
(\ref{enerconserv}), together with the continuity equation
(\ref{conteq}) and the definition of the heat current ${\bf j}_T$
(\ref{currheat}), one
gets for the change of entropy:
\bea\l{entropyeq}
    \frac{\partial (\rho \delta \varsigma)}{\partial t}
    &=&-{\bf \nabla}\left(\rho \delta \varsigma\; \u
    + \frac{1}{T} {\bf j}_{{}_{\! T}}\right)
    + \frac{\kappa}{T^2}\left({\bf \nabla}T\right)^2 \nonumber\\
    &+&\frac{\nu}{2T} \sum_{\alpha\beta}
    \left(\frac{\partial u_{\alpha}}{\partial r_{\beta}} +
    \frac{\partial u_{\beta}}{\partial r_{\alpha}} -
    \frac{2}{3}{\bf \nabla}\u\;\delta_{\alpha \beta}\right)^2
\eea
[again the variation $\delta$ in $\delta \varsigma$ is not
omitted because of the following derivations of the Fourier
equation (\ref{fouriereq}) and thermal conductivity
(\ref{kappadef}) in Appendix A.2].
These two equations have a very clear physical meaning
for normal liquids and amorphous solids (a very
viscose liquids are associated to the amorphous
solids with some shear modulus $\lambda$ in our notations, i.e.,
solids without any crystal structure).
The first equation (\ref{enerconserv}) claims that the change of the
collective and internal energy, concentrated in unit volume per
unit of time and presented as the sum of the collective kinetic
and internal parts, equals the corresponding energy flux through
its surface and work of the external field. The second equation
(\ref{entropyeq}) is usually called as a general
heat transport equation.
This
equation states that the change of entropy in the unit volume
per unit of time equals the entropy flux through its surface
(heat energy flux). Two other terms
show the entropy increase related to the gradient of the
temperature and dissipation due to the shear ($\nu$) viscosity.
Note that there is no explicit dependence on the
external field in (\ref{entropyeq}). This dependence is
manifested only through the solutions of the dynamical equations
in terms of the moments. For zero external field (closed system)
the entropy is increasing  because of the basic thermodynamic law.
Therefore, according to (\ref{entropyeq}), the
shear viscosity $\nu$ (\ref{viscos}) and the thermal conductivity
$\kappa$ should be positive. The energy conservation equation for
the Fermi liquid (\ref{enerconserv}) differs from the one for
classical hydrodynamics  by the same
Fermi-surface distortions related to the shear modulus $\lambda$
(\ref{shearmod}) as discussed above. That is similar to the
amorphous solids  (in the above mentioned sense
of very viscose liquids). However, in
contrast to the latter, one obtains the energy conservation
condition (\ref{enerconserv}) for the dynamical variations of the
Fermi-liquid collective and internal energy with the specific
constants $\lambda$ (\ref{shearmod}), $\nu$ (\ref{viscos}) and
$K_{\rm tot}$ (\ref{incomprtot}) found from the relation to the
Landau--Vlasov equation (\ref{landvlas}). Our way of the
derivation of the energy conservation equation
(\ref{enerconserv}) for the Fermi liquids within the {\it
linearized} Landau--Vlasov dynamics (\ref{landvlas}),
as for normal liquids and
solids, leads to a more explicit form of the energy conservation
equation than that suggested in \cite{heipethrev,baympeth}. In this way,
we get rather simple expressions for the collective and internal
components of the energy out the hydrodynamical limit.

\subsubsection{POTENTIAL FLOW: FERMI LIQUID VERSUS HYDRODYNAMICS}
\label{potenflow}

Below, we shall be interested in the case of a viscous potential
flow, for which one has
\bel{velpoten}
\u={\bf
\nabla}\varphi,\quad {\bf w}={\bf \nabla}\varphi_w\quad
{\rm with} \quad \varphi=\dot{\varphi}_w
\ee
 [cf. the second
equation with (\ref{presslamb})]. With the help of these
definitions, the momentum equation (\ref{momenteq}) and the flux
tensor (\ref{momentfluxtot}) can be brought to the following
forms:
\bel{navstokeq0}
m \rho \frac{\partial \varphi}{\partial
t} - \frac{4}{3} \nu~ \Delta \varphi
 - \frac{4}{3} \lambda~ \Delta \varphi_w
+ \frac{K_{\rm tot}}{9} \delta \rho=-V_{\rm ext}
\ee
and
\bea\l{momentfluxpot}
\Pi_{\alpha\beta}&=& -2 \left(
\frac{\lambda}{-i\om} +\nu \right)
\left(\frac{
\partial^2\varphi}{\partial r_\alpha \partial r_\beta}- \triangle \varphi \;
\delta_{\alpha\beta} \right) \nonumber\\
&-&\left(m \rho \frac{
\partial \varphi}{\partial t}+ V_{\rm ext}\right) \delta_{\alpha\beta}.
\eea
The diagonal term given on the very right of
(\ref{momentfluxpot}) had been used to remove $\delta \rho$
which still appears in (\ref{momentfluxtot}). With the continuity
equation (\ref{conteq}) for the plane wave solutions
(\ref{planewave}), one has from (\ref{navstokeq0}) the equation
for the velocity potential $\varphi$ \cite{magkohofsh}:
\bel{navstokeq}
m \rho \frac{\partial^2 \varphi}{\partial t^2}
-\frac{\rho}{9}\left(K_{\rm tot} +12 \lambda/\rho \right) \Delta
\varphi
 - \frac{4}{3} \nu~ \Delta \frac{\partial \varphi}{\partial t}
 = - \frac{\partial V_{\rm ext}}{\partial t}.
\ee
The structure of
(\ref{navstokeq}) for the potential flow is similar to that of
the Navier-Stokes equation for the velocity potential $\varphi$.
The difference to the case of the common classical liquid, is seen in
the terms proportional to $\lambda$, viz in the presence of the
anisotropy term (\ref{presslamb}), which actually represents a
{\it reversible} motion. Such a term is known from the dynamics of
amorphous solids. We emphasize that for Fermi liquids, this term
arises only in the presence of the Fermi surface distortions, which
survive even in the non-viscous limit; they will turn out
important for our applications below. The shear modulus $\lambda$
may be interpreted as a measure of those distortions which are
related to a reversible anisotropy of the momentum flux tensor.
They disappear in the hydrodynamic limit, and so does
$\lambda$, in which case all formulas of this section turn into
those for normal liquids; for more details see section
\ref{longwavlim}.

At this place an important remark is in order. It should be noted
that in contrast to classical hydrodynamics
our system of equations for the moments is {\it not closed} to the
first few ones, namely particle density $\delta \rho$ and velocity
field $\u$. This is true in particular for (\ref{navstokeq})
for the  potential flow $\varphi$. Indeed, the coefficients
$\lambda$, $\nu$ and $K_{\rm tot}$ depend on the variable $\om/q$
which yet is unknown. The latter is determined from a
dispersion relation, which in turn has to be derived from the
Landau--Vlasov equation (\ref{landvlas}). Such a procedure goes
back to \cite{landau} where the dispersion relation was exploited
for the collisionless case at $T=0$. A collision term in the
relaxation approximation has been taken into account in
\cite{abrikha}. The extension to heated Fermi liquids and low
excitations, in the way, which we are going to use later on, has been
developed in \cite{heipethrev}. It may be noted that this version
of the dispersion relation, which we are aiming at, differs
essentially from the one obtained in the "truncated" (scaling model)
versions of
the Fermi-liquid theory of \cite{holzwarth,nixsierk}, where the
momentum flux tensor is not influenced by higher moments of the
distribution function.
We take into account all other  multipolarities (larger the quadrupole
one) of the Fermi-surface distortions when there is no
convergence in multipolarity expansion of the distribution
function for finite and large $\omega \tau$ or for finite
$K_{\rm tot}$, for instance for nuclear matter with small $\mathcal{F}_0$,
in contrast to the Fermi liquid $^3$He.

\subsection{Response functions}
\label{respfunsec}

\subsubsection{DYNAMIC RESPONSE}
\label{dynresp}

As mentioned earlier, we want to solve the linearized equations of
motion in terms of response functions. We concentrate on two
quantities, namely particle density $\rho(\r,t)$ and temperature
$T(\r,t)$ and examine how they react to the external field
$V_{\rm ext}(\r,t)$ introduced earlier. This may be quantified by the
following two response functions: The density-density response
$\chi_{DD}^{\rm coll}$ and the temperature-density response
$\chi_{TD}^{\rm coll}$ defined as
\bel{ddrespdef}
\chi_{DD}^{\rm coll}(q,\om)=
    - \frac{\delta \rho(q,\om)}{V_{\rm ext}(q,\om)}
\ee
  and
 \bel{dtrespdef} \chi_{TD}^{\rm coll}(q,\om)=
    - \frac{\delta T(q,\om)}{V_{\rm ext}(q,\om)},
\ee
respectively. To keep the notation simple, we will omit the
tilde characterizing the Fourier transform of the distribution
function (\ref{planewave}) (it should suffice to only show the
arguments $q,\om$). The definition of the response functions is
identical to the one of \cite{heipethrev}, except that we have
introduced the suffix "coll". This was done adopting a notation
used in the literature of nuclear physics when the dynamics of a
finite nucleus is expressed in terms of shape variables, to which
we will come below. Notice, however, that $V_{\rm ext}(q,\om)$ is only
proportional to the density,
$V_{\rm ext}(q,\om)=q_{\rm ext}(\om)\rho(q,\om)$, with $q_{\rm ext}(\om)$
being some externally determined function. Often, one therefore
defines response functions in a slightly modified way, in that the
functional derivatives are performed with respect to
$q_{\rm ext}(\om)$ instead of $V_{\rm ext}(q,\om)$ (see,
e.g., \cite{pinenoz}).

As will be seen below, these functions only depend on the wave
number $q$ but not on the angles of the wave vector $\q$. For this
reason, it is convenient to introduce the dimensionless quantities
$s$ and $\taubar$ (with $v_{{}_{\! {\rm F}}}=p_{{}_{\! {\rm F}}}/m^*$)
\bel{som}
s= \frac{\om}{v_{{}_{\! {\rm F}}} q},\quad
\taubar =\tau v_{{}_{\! {\rm F}}} q,
    \quad{\rm implying}\quad \om \tau = s  \taubar,
\ee
  instead of the frequency $\om$ and the wave number $q$.

To calculate the response functions (\ref{ddrespdef}) and
(\ref{dtrespdef}) we follow the procedure of \cite{heipethrev}. As
any further details may be found there, it may suffice to outline
briefly the main features. In short, the strategy is as follows.
Firstly, one rewrites the Landau--Vlasov equation (\ref{landvlas})
in terms of the Fourier coefficients introduced in
(\ref{planewave}). Evidently, in the spirit of the separation
specified in (\ref{dfgeqdfleq}), we need to evaluate explicitly
only the first component
$\delta f_{\rm l.e.}(\r,\p,t)$  which
enters the conditions (\ref{consereq}). By a
straightforward calculation, one may then express $\delta
f_{\rm l.e.}(\r,\p,\om)$ in terms of the unknown quantities $\delta
\rho$, $\u$, $\delta \mu$ and $\delta T$ for any given external
field $V_{\rm ext}$. The form is given in (\ref{basiceq}).
The continuity equation (\ref{conteq})
in the Fourier representation through
 (\ref{planewave}),
$\q\u=\om \delta
    \rho / \rho$,
may be used to eliminate the velocity field $\u$.
Furthermore, the thermodynamic relation [see (\ref{drhomt}),
(\ref{incomprTdef}) and (\ref{isotherk})]
\bel{dmu} \delta \mu =
    \left(\frac{\partial \mu }{\partial \rho}\right)_T \delta \rho +
    \left(\frac{\partial \mu }{\partial T}\right)_{\rho} \delta T =
    \frac{K^{T} }{9\rho}\;
    \delta \rho - \frac{\mathcal{M}(T)}{\mathcal{N}(T)}\;\delta T
\ee
allows one to express the chemical potential $\delta \mu$ in
terms of the two unknown variables $\delta \rho$ and $\delta T$.
Next, one may exploit the conditions (\ref{consereq}). As the
second (set of) equation(s) is just an identity, provided one
uses the appropriate definition of the effective mass
(\ref{effmass}), it is only
the first and the third equation which matter. They
may determine the remaining two
variables $\delta \rho$ and $\delta T$ in terms of the external
field,
\bel{drhodteqone}
    \left(\frac{i s \taubar }{i s \taubar -1}
    -\wp(s) \chizero \right) \delta \rho
    + \frac{1}{1-i s \taubar } \chione \delta T
    =-\chizero \delta V_{\rm eff}
\ee
 and
\bel{drhodteqtwo} -\wp(s) \chione \delta \rho
    + \frac{1}{1-i s \taubar } \left(\chitwo
    -i  s \taubar \rho \frac{{\tt C}_{\mathcal{V}}}{T} \right) \delta
T
    = -\chione \delta V_{\rm eff}.
\ee
 Here, the quantity
\bel{alphas}
   \wp(s) = \frac{1}{\mathcal{N}(T)}\frac{1}{i s \taubar -1}
    - \frac{3is}{\taubar \mathcal{N}(0)}
\ee
has been introduced with $\mathcal{N}(0)$ being the level density
(\ref{enerdensnt}) of the quasiparticles at $T=0$,
\bel{nzero} \mathcal{N}(0)= \frac{p_{{}_{\! {\rm F}}} m^*}{\pi^2 \hbar^3}=
    \frac{3}{2} \frac{\rho _0}{\eps_{{}_{\! {\rm F}}}},
\ee
and $\eps_{{}_{\! {\rm F}}}=p_{\rm F}^2/2m^*$.
The functions $\chin$ are given by
\bel{chinfun}
\chin = -\mathcal{N}(T) \left\langle
\frac{\q{\v_\p}}{\mathcal{D}_\p}
\left(\frac{\eps_{\p}-\mu}{T} -
\frac{\mathcal{M}(T)}{\mathcal{N}(T)}\right)^n\; \right\rangle
\ee
 with $n=0,1,2,...\;$,
\be\l{domindp}
\mathcal{D}_\p=
    \om-\q\v_\p+ i/\tau, \qquad\v_\p=\p/m^*.
\ee
 Furthermore, in (\ref{drhodteqone}) and
(\ref{drhodteqtwo}) a short hand notation $\delta V_{\rm eff}$ has
been used for the sum of two terms, namely
\bel{delueff} \delta
V_{\rm eff}=
    V_{\rm ext} + k(\om,T) \delta \rho
\ee
 with
\bel{couplconst} k(\om,T) =\frac{1}{\mathcal{N}(T)}
    \left[\mathcal{F}_0 + \frac{\mathcal{F}_1}{\mathcal{G}_1}
    \left(\frac{\om}{v_{{}_{\!{\rm F}}} q}\right)^2\right].
\ee
In (\ref{delueff}), $\delta V_{\rm eff}$ may be considered as an
{\it effective} field which includes the true external field
$V_{\rm ext}$ and the "screened" field $k\delta \rho$
\cite{heipethrev}. Our notation follows the one often used for
finite nuclei: The second term in (\ref{delueff}) plays the
role of the collective variable and $k$ of (\ref{couplconst})
represents the "coupling" constant (see, e.g.,
\cite{bohrmot,hofbook,hofmann}).

The response function $\chi_{DD}^{\rm coll}$ of (\ref{ddrespdef})
can be now obtained
from (\ref{drhodteqone}) and (\ref{drhodteqtwo}),
\bel{ddresp} \chi_{DD}^{\rm coll}( \taubar ,s)=
    \frac{\aleph( \taubar ,s)}{D( \taubar ,s)},
\ee
where
\bel{despfunc} D( \taubar ,s)=
    D_0( \taubar ,s)+ k( \taubar ,s) \aleph( \taubar ,s)
\ee
 with
\bel{chidomin} D_0( \taubar ,s)=
    \left(\frac{i s \taubar }{i s \taubar-1}
    -\wp(s)\chizero \right)
    \left(\chitwo - i s \taubar \rho
\frac{{\tt C}_{\mathcal{V}}}{T} \right)
    + \wp \chione ^2.
\ee
In (\ref{ddresp}), $\aleph(\taubar ,s)$ finally is given by
\bel{chinumer}
\aleph(\taubar ,s)=
    \chizero \left(\chitwo
    -i s \taubar \rho
\frac{{\tt C}_{\mathcal{V}}}{T} \right)- \chione ^2.
\ee

It is worth noticing that the collective response function for the
density-density mode, as given by (\ref{ddrespdef}) or
(\ref{ddresp}), can be expressed as
\bel{ddrespchiin}
\chi^{\rm coll}(q,\om)=
    \frac{\chi (q,\om)}{1+k(\om,T) \chi (q,\om)}.
\ee
This form is analogous to the form used to describe the
dynamics of shape variables \cite{bohrmot,hofbook,hofmann}. We omit
here the suffix $DD$ because the $TD$ response function takes on a
similar form
(with some modification of the numerator). It is here where the
"coupling constant" $k$ appears, as defined in
(\ref{couplconst}), together with the "intrinsic" (or
"un-screened" (see \cite{pinenoz,heipethrev}) response
function $\chi $,
\bel{chiindef} \chi ( \taubar ,s)=
    - \frac{\delta \rho}{\delta V_{\rm eff}}
    = \frac{\aleph( \taubar ,s)}{D_0( \taubar ,s)}.
\ee
 Both expressions can be found already in \cite{heipethrev}.
However, later we will find the form (\ref{ddresp}) more
convenient for our applications, in particular for the discussion
of the low frequency limit  $\om \tau \ll 1$.
When we shall expand first $\chi $ (\ref{chiindef}) in
(\ref{ddrespchiin}) in small $\om \tau$ near the poles of
$\chi ^{\rm coll}$ (\ref{ddresp}) (see next section), one should
assume that the singularities of $\chi$ related to zeros of $D_0$
in (\ref{chiindef}) are far away from zeros of $D$ in
(\ref{ddresp}), i.e., a smoothness of $\chi$ as function of
$\omega \tau$ near these poles. After the cancellation of a
possible singularity source $D_0$ in (\ref{ddresp}) we are free
from such an assumption.

Finally, let us turn to the temperature-density response function
$\chi_{TD}^{\rm coll}$ (\ref{dtrespdef}). It is determined by the same
system of equations (\ref{drhodteqone}) and (\ref{drhodteqtwo})
and can be written in the form (\ref{ddrespchiin}) but with
another ``intrinsic'' response function $\chi_{{}_{\! TD}}$ appearing in the
numerator,
\bel{chitindef} \chi_{{}_{\! TD}}( \taubar ,s)=
    - \frac{\delta T}{\delta V_{\rm eff}}.
\ee
 From (\ref{drhodteqone}), (\ref{drhodteqtwo}) and
(\ref{chitindef}), one obtains
\bel{chitbar}
\chi_{{}_{\! TD}}(\taubar,s)=
    -\frac{is\taubar~\chione}{D_0(\taubar ,s)},
\ee
where $D_0( \taubar ,s)$ is given by (\ref{chidomin}).
 As compared to the one printed in \cite{heipethrev},
this expression contains an additional factor $i s\taubar/\chi_1$,
which later on will turn out important, for instance, when
calculating susceptibilities and the in-compressibility $K_{\rm tot}$
(\ref{incomprtot}).  (We are grateful to H.
Heiselberg for confirming this misprint.)
Substituting (\ref{chitbar}) into the
numerator of (\ref{ddrespchiin})
instead of $\chi$, one gets the temperature-density response
function (\ref{dtrespdef}) in the form similar to (\ref{ddresp}),
\bel{dtresp} \chi_{TD}^{\rm coll}( \taubar,s)=
    -\frac{is\taubar~\chione}{D( \taubar ,s)}.
\ee
 Notice that according to (\ref{ddresp}) and
(\ref{dtresp}), both response functions (\ref{ddrespdef}) and
(\ref{dtrespdef}) have the same set of poles, which lie at the
roots of the equation
\bel{despeq} D( \taubar ,s)= 0.
\ee
This is identical to the condition of zero determinant
for the system of the linear equations (\ref{drhodteqone})
and (\ref{drhodteqtwo}).

\bigskip

\subsubsection{LOW TEMPERATURE LIMIT}
\label{lowtemlim}

The expressions for the collective response functions become much
simpler at low temperatures $T \ll \mu$. In
this case, one may calculate $\chin$ of (\ref{chinfun}) by
expanding in powers of ${T / \mu}$. For those applications to
nuclear physics we have in mind the temperature is sufficiently
small such that it suffices to mainly stick to order two. Fourth
order terms shall be shown only when necessary.

A basic element for the quantities which we need to evaluate is
the derivative $\partial f_{\p} / {\partial \eps_{\p}}$ taken at
global equilibrium:
\bel{dfpdep} \left(\frac{
\partial f_{\p}}{\partial \eps_{\p}}\right)_{\rm g.e.}=
    -\left[
4 T \cosh^2\left(\frac{\eps_\p-\mu}{2T}\right)\right]^{-1}_{\rm g.e.}.
\ee
 It appears in $\mathcal{N}(T)$ of (\ref{enerdensnt}) [see
also (\ref{averag})], which in turn is needed for $\chin$
of (\ref{chinfun}). For small $T$, this derivative is a sharp
bell-shaped function of $\eps_\p^{\rm g.e.}$, such that one may evaluate
the averaging integrals (\ref{chinfun}) and (\ref{averag}) by
expanding the smooth functions in terms of $\eps_\p^{\rm g.e.}$ near
$\eps_\p^{\rm g.e.}=\mu_{\rm g.e.}$. In this way, the Fourier-Bernoulli
integrals over the dimensionless variable
$\left[(\eps_\p-\mu)/T\right]_{\rm g.e.}$ appear (see, e.g.,
\cite{heipethrev})
which lead to
\bea\l{chitemzero}
\chi_0&=&
     \left[-Q_1(\zeta)+ \frac{\pi^2 {\bar T}^2}{12}
     \left(Q_1(\zeta)-\zeta Q_1^{\prime}(\zeta) \right.\right. \nonumber\\
   &-& \left.\left. \frac{1}{2} \zeta^2 Q_1^{\prime \prime}(\zeta)\right)
     + \mathcal{O}\left({\bar T}^4
\right)\right]
     \mathcal{N}(0),
\eea
 \bel{chitemone} \chi_1=
     \left[\frac{\pi^2 {\bar T}}{6} \zeta Q_1^{\prime}(\zeta)
     + \mathcal{O}\left({\bar T}^3
\right)\right]
     \mathcal{N}(0),
\ee
 and
\bea\l{chitemtwo}
\chi_2 &=&
    \left\{-\frac{\pi^2}{3} \left[Q_1(\zeta)+
\frac{\pi^2 {\bar T}^2}{120}
    \left(36 Q_1(\zeta)
 - 46 \zeta Q_1^{\prime}(\zeta) \right.\right.\right. \nonumber\\
 &-&\left.\left.\left.
    21 \zeta^2 Q_1^{\prime \prime}(\zeta)\right)\right]
    + \mathcal{O}\left({\bar T}^4
\right) \right\}
    \mathcal{N}(0).
\eea
 Here $Q_1(\zeta)$ is the Legendre function of second kind with
$\zeta= s+i/\taubar$,
 and
$\tbar=T/\eps_{{}_{\! {\rm F}}}$ is used also in Appendix A.3.
These quantities may now be used to calculate the response
functions (\ref{ddrespdef}) and (\ref{dtrespdef}), [or more
specifically (\ref{ddresp}) and (\ref{dtresp})].  For zero
temperature, one gets the standard solutions
\cite{abrikha,heipethrev}. So far no assumption has been made
concerning the parameter $\om \tau$ which specifies the importance
of  collision in various regimes of the collective motion
\cite{abrikha}. In particular, the formulas obtained in this
section are valid both for the regimes of zero sound ($\om \tau
\gg 1$) and  hydrodynamics ($\om \tau \ll 1$). For $\om \tau \gg
1$ our solutions agree with those of \cite{abrikha,heipethrev}.
However, below we shall be interested mainly in  collective
excitations of low frequencies. The notion "low frequencies" is
meant to indicate that the corresponding excitation energies are
smaller than those of the
giant resonances. Next we will turn to the hydrodynamic regime
where $\om \tau \to 0$. As we shall see, at low temperatures our
solutions approach the ones of normal classical liquids, in
agreement with \cite{forster,brenig}.

\bigskip

\subsection{Hydrodynamic regime}
\label{longwavlim}

\medskip

\subsubsection{DISPERSION RELATION}
\label{disrel}

The response functions can be
simplified significantly in the long-wave length
limit. Using
$\taubar$ introduced in (\ref{som}), this (LWL) limit may be defined as
$\taubar \ll 1$. It can be reached in two ways, namely for small
wave numbers $q$ and finite collision time $\tau$ or for small
$\tau$ but finite $q$. Both cases imply that the dimensionless
parameter $\om \tau=s \taubar$, which determines the collision
rate in comparison to the frequency of the modes, becomes small
for any finite value $s$ of (\ref{som}) ($|s| \siml 1$).
As will be shown below for nuclear matter at low temperatures, this
quantity $s$ is not enough large, in distinction to the
case of liquid $^3He$. Therefore, a small $\taubar$ implies
hydrodynamic behavior, in contrast to the zero sound regime; where
$\taubar \gg 1$, or $\om \tau \gg 1$.

The Landau--Vlasov equation (\ref{landvlas}) is an integral
equation. Its solution may be sought for in terms of an eigenvalue
problem with the distribution function $\delta f$ being the
eigenfunctions and the sound velocity $s$ (\ref{som}) being the
eigenvalues, see also \cite{abrikha,pinenoz}. This eigenvalue
problem may be solved perturbatively with $\taubar$ being the
smallness parameter
\cite{sykbrook,brooksyk}.
It may be noted in passing that this method may be applied to some extent
as well
to the eigenvalue problem of the Schr\"odinger equation.
We shall use it to get the hydrodynamic sound
velocities from the kinetic equation, see
\cite{sykbrook,brooksyk}. To this end, we expand the solutions for
$s$ and $\delta f$ into  power series with respect to $\taubar$,
but restricted to linear order. Thus, we may write
\bel{somexp} s =
s_0 + i s_1 \taubar,
\ee
 where $s_0$ and $s_1$ are independent of
the expansion parameter $\taubar$. In Appendix A.2,
it is
shown how the density-density response function may be calculated
in the LWL limit.
There two non-linear equations for the
coefficients $s_0$ and $s_1$ are obtained from the dispersion
relation (\ref{despeq}), namely (\ref{eqzero}) and (\ref{eqone})).
The first equation [see (\ref{eqzero})] has one obvious solution
$s_0=s_0^{(0)}=0$ and two others $s_0=\pm s_0^{(1)}$ with the
same modulus,
\bea\l{sfirst0}
    s_0^{(1)} &=& \sqrt{\frac{\mathcal{G}_0 \mathcal{G}_1
\mathcal{N}(0)}{3 \mathcal{N}(T)}
    \left(1 + \frac{\tp}{3 \mathcal{G}_0}\right)} \nonumber\\
&\approx&
    \sqrt{\frac{\mathcal{G}_0 \mathcal{G}_1}{3}
    \left[1 +
    \frac{\tp \left(4 + \mathcal{G}_0\mathcal{G}_1\right)}{12 \mathcal{G}_0}
\right]}.
\eea
Substituting $s_0^{(0)}$ and $s_0^{(1)}$ into the second equation (\ref{eqone}),
one finds the two solutions for $s_1$, $s_1^{(0)}$ and $s_1^{(1)}$,
respectively. These solutions for $s$ (\ref{somexp})
can be written in terms of the dimensional
frequency $\om$ by means of (\ref{som})
in the following form:
\bea\l{shp}
\om^{(0)} &=& -i
\frac{\gbarhp}{2}\;, \qquad \gbarhp = s_1^{(0)}v_{{}_{\! {\rm F}}} q \qquad\nonumber\\
&=&
\frac{2 \taubar v_{{}_{\! {\rm F}}} q}{3}
\left[1 - \frac{\tp \left(80-29 \mathcal{G}_0\right)}{120
\mathcal{G}_0}\right]\qquad
\eea
and
\bel{sfirst}
\om_{\pm}^{(1)} =
            \pm \om_0^{(1)} -i \frac{\gbarone}{2}
\quad {\rm with}\quad
            \om_0^{(1)}=s_0^{(1)} v_{{}_{\! {\rm F}}} q ,
\ee
where
\bea\l{gambarone}
\gbarone &=&
     s_1v_{{}_{\! {\rm F}}} q =
     \frac{4}{15} \taubar v_{{}_{\! {\rm F}}} q \mathcal{G}_1
    \left[1 \!+\! \frac{5 \tp}{6}
    \left(\frac{1}{\mathcal{G}_0 \mathcal{G}_1}\right)\right] \nonumber\\
   &\approx&  \frac{4}{15} \taubar v_{{}_{\! {\rm F}}} q \mathcal{G}_1
    \left[1 \! +\! \frac{5 \tp}{12}
    \left(1\!+\! \frac{1}{\mathcal{G}_0 \mathcal{G}_1}\right)\right].
\eea
The first root $\om^{(0)}$ given in (\ref{shp}) is purely
imaginary and corresponds to the overdamped excitations of the
hydrodynamic Raleigh mode \cite{kubo,forster}. The second
and third ones $\om_{\pm}^{(1)}$ correspond to the usual first
sound mode, expressed in terms of the (macroscopic) parameters of
viscosity and thermal conductivity of normal liquids
 \cite{forster,brenig}.
In (\ref{sfirst0}) and
(\ref{gambarone}), small corrections  of the order of the product
of the two small quantities $\tbar^2$ and $\mathcal{F}_0$ have been
neglected, along with $\tbar^2 |(m^*-m)/m|=\tbar^2|\mathcal{F}_1|/3$.
This procedure should be valid for nuclear matter; where
the relevant parameters are small, both $\mid\mathcal{F}_0\mid$ and
$|(m^*-m)/m|$ being of order $ \approx 0.2$. Discarding such
small corrections, our results for the sound frequencies
$\om_\pm^{(1)}$ (\ref{sfirst}) are in agreement with
\cite{heipethrev}. In particular, up to these small corrections,
the volume (or second) viscosity disappears, as it is the case in
\cite{heipethrev}. In the expressions (\ref{shp}) to
(\ref{gambarone}) more explicit temperature corrections are given
for $\om^{(0)}$ and $\om^{(1)}$ than those discussed in
\cite{heipethrev}. This will turn out important for the thermal
conductivity $\kappa$, which we shall address in
Sec.\ \ref{viscosthermcond} [see (\ref{kappaexp})]. The "widths"
$\gbarhp$ and $\gbarone$ are proportional to $\taubar$, and thus, to
the relaxation time $\tau$ which represents the effects of
two-body collisions. For nuclear matter, the Landau parameters
$\mathcal{F}_0$ and $\mathcal{F}_1$ are small [$\mathcal{G}_0$ and
$\mathcal{G}_1$ are close to unity, see (\ref{effmass})]. For this reason,
according to
the last equation in (\ref{som}), the sound velocities cannot be
large [see the approximation
(\ref{sfirst0})]. So, the LWL limit
($\taubar \ll 1$)
may be identified with the hydrodynamic collision regime
$\om \tau =s \taubar \siml \taubar \ll 1$.
Note that for the Fermi liquid $^3He$, for instance, the parameters
$\mathcal{F}_0$ and $\mathcal{F}_1$ are large and second order equation of
(\ref{sfirst0}) can not be applied. Moreover, according to the
first line in (\ref{sfirst0}), the sound velocity is large.
Therefore, in this case a smallness $\taubar$ does not mean yet that
$\om \tau$ is also small, i.e., the LWL
condition is not
enough for the hydrodynamical collision regime.

\bigskip

\subsubsection{RESPONSE FOR INDIVIDUAL MODES}
\label{respmodes}

In the following, we are going to examine the collective response
function $\chi_{DD}^{\rm coll}$ (\ref{ddresp}), in particular its
behavior in the neighborhood of the individual modes given by
(\ref{shp}) and (\ref{sfirst}). To simplify the notation, we shall
at times omit the lower index "DD" and move down the upper index
"coll". Near any of the sound poles $\om_{\pm}^{(1)}$ given in
(\ref{sfirst}), the collective response function $\chi_{\rm coll}$
(\ref{ddresp}) may be written as
\bea\l{chicollone}
\chi_{\rm coll}^{(1)}\left(q,\om\right) &=&
    \aone \left( \frac{1}{\om-\om_{-}^{(1)}} -
    \frac{1}{\om-\om_{+}^{(1)}} \right)\quad {\rm with} \nonumber\\
\aone &=&
    \frac{
\om_0^{(1)} \mathcal{N}(T)}{2 \mathcal{G}_0
\left[1 + \tp /(3 \mathcal{G}_0) \right]}.
\eea
 Here, we have made use of (\ref{ampsexp}), (\ref{dszexpzero}),
(\ref{couplconst}) as well as of (\ref{somexp}).
It will turn out convenient to present separately the dissipative
and reactive parts, $\chi_{\rm coll}^{(1)\;\prime\prime}$ and
$\chi_{\rm coll}^{(1)\;\prime}$, respectively,
\bea\l{chiconeqompp}
\chi_{\rm coll}^{(1)\;\prime\prime}(q,\om) &=&
    \frac{1}{2} \aone \left[
\frac{\Gamma^{(1)}}{\left(\om-\om_0^{(1)} \right)^2 +
    \left(\Gamma^{(1)}\right)^2 / 4 } \; \right.\nonumber\\
   &-&\left. \frac{\Gamma^{(1)}}{\left(\om + \om_0^{(1)} \right)^2 +
    \left(\Gamma^{(1)}\right)^2 / 4} \right]
\eea
 and
\bea\l{chiconeqomp}
\chi_{\rm coll}^{(1)\;\prime}(q,\om)&=&
    \aone \left[\frac{
\om + \om_0^{(1)}}{\left(\om + \om_0^{(1)}\right)^2 +
    \left(\Gamma^{(1)}\right)^2 / 4} \right.\nonumber\\
   &-&\left.\frac{\om-\om_0^{(1)}}{\left(\om-\om_0^{(1)}\right)^2 +
    \left(\Gamma^{(1)}\right)^2 / 4}\right].
\eea
 Notice that for $\taubar=+0$ the Lorentzians in (\ref{chiconeqompp}) turn into
$\delta$-functions.

The relaxation time $\tau$, which determines the dimensionless
quantity $\taubar =\tau \, v_{{}_{\! {\rm F}}} q$, might depend on temperature and
frequency. A useful form is found in
\bel{tautom} \tau = \frac{
\tau_o}{T^2 + c_o \left(\hbar \om \right)^2} \quad
    \approx \quad \frac{\tau_o}{T^2} \quad {\rm for} \quad
c_o(\hbar \om)^2 \ll T^2,
\ee
 with some parameters $\tau_o$ and
$c_o$ independent of $T$ and $\om$; see, e.g., \cite{magkohofsh}. As
indicated on the very right, for our present purpose we may
neglect the frequency dependence, simply because we are interested
in describing low frequency modes at larger temperatures (with
respect to $\hbar \om$). Indeed,
it is such a condition which
helps justifying the assumption of local equilibrium. We shall
return to this question later, when we are going to apply the Landau
theory to a finite Fermi-liquid drop. Substituting (\ref{tautom})
into the damping coefficient ${\Gamma}^{(1)}$ (\ref{gambarone}),
one has
 \bel{gambaronet} \gbarone =
     \frac{4 \tau_o v_{{}_{\! {\rm F}}}^2 q^2 \mathcal{G}_1}{15 T^2}
    \left[1 +
    \frac{5 \tp}{12}\left(1+ \frac{1}{\mathcal{G}_0
\mathcal{G}_1}\right)\right].
\ee
 To leading order, this gives the expected
dependence on temperature commonly associated to hydrodynamics,
namely $\gbarone \propto 1/T^2$.

Finally, we may note that in the long-wave limit the effective,
total in-compressibility $K_{\rm tot}$ (\ref{incomprtot}) becomes
identical to the adiabatic in-compressibility $K^{\varsigma}$
(\ref{incompraddef}) specified in Appendix A,
\bel{Kadiabat}
K^{\varsigma}=
K^{T}\left(1 + \frac{4T{\tt C}_{\mathcal{V}}}{K^{T}}\right),
\ee
see (\ref{isotherk}) for the isothermal in-compressibility $K^{T}$
and (\ref{incompradexp}) at small temperatures.
For the derivation
of this identity, it is more easy to
consider variations $\delta \tilde{T}$ and $\delta \tilde{\rho}$
as caused formally by some external field $V_{\rm ext}$. Then,
one can represent $\delta {\tilde T}/\delta {\tilde \rho}$ in
(\ref{incomprtot}) in terms of the ratio of the
temperature-density $\chi_{{}_{\! TD}}(\taubar,s)$ (\ref{chitbar}) to
density-density $\chi_{{}_{\! DD}}(\taubar,s)$ (\ref{chiindef}),
\bel{dtdrhoexp0}
\frac{\delta {\tilde T}}{\delta {\tilde \rho}}
\equiv \frac{\delta {\tilde T}/V_{\rm ext}}{\delta {\tilde
\rho}/V_{\rm ext}}= \frac{
\chi_{{}_{\! TD}}(\taubar,s)}{\chi_{{}_{\! DD}}(\taubar,s)}\;.
\ee
 Using then the LWL
expansions
(\ref{chiexpone}) and (\ref{ampsexp}) up to the third order
terms in $\taubar$, from (\ref{dtdrhoexp0}) we get
\bel{dtdrhoexp}
\frac{\delta {\tilde T}}{\delta {\tilde \rho}}
\approx \frac{\tbar}{\mathcal{N}(0)}\;\left(1-
\frac{i \taubar}{3
s_0^{(1)}}\right) \quad {\rm for}\quad \taubar \rightarrow 0,
\ee
 where $s_0^{(1)}$ was defined in (\ref{sfirst0}). As the
specific heat ${\tt C}_{\mathcal{V}}$ (\ref{specifheatv}) is
proportional to $\tbar$, we need in (\ref{dtdrhoexp}) only
linear terms to get the temperature correction of the second
order in $\tbar$ in the total in-compressibility $K_{\rm tot}$
(\ref{incomprtot}). Substituting (\ref{dtdrhoexp}),
(\ref{specifheatv}) and (\ref{isotherkexp}) into
(\ref{incomprtot}) for the total in-compressibility $K_{\rm tot}$,
one obtains identically the same as in (\ref{incompradexp}) for
the adiabatic in-compressibility $K^\varsigma$. The same result
(\ref{dtdrhoexp}) in the LWL limit can be obtained
also from (\ref{dtdrho}).

Let us address now the pole at $\omzero$ [see (\ref{shp})]. Near
the latter, the collective response function $\chi_{\rm coll}$
(\ref{ddresp}) becomes [as may be checked with the help of
(\ref{ampsexp}), (\ref{dszexpzero}), (\ref{couplconst}) and
(\ref{somexp})]
\bea\l{chicollhp} \chi_{\rm coll} ^{(0)}
&=&
    \frac{i \azero}{\om-\om^{(0)}}=
    \frac{i \azero}{\om + \frac{i \gbarhp}{2}}
\qquad {\rm with} \nonumber\\
 \azero &=&
    \frac{
\pi^4 \tbar^2 \taubar (8-3 \mathcal{G}_0)}{108
\mathcal{G}_0^2}~v_{{}_{\! {\rm F}}} q ~\mathcal{N}(0),
\eea
and  $\gbarhp$ being defined in (\ref{shp}). It may
be rewritten in a more traditional form, see \cite{kubo},
\cite{forster} and \cite{brenig}. Introducing the "diffusion
coefficient"
\bel{diffusd}
{\tt D}_T = \frac{\kappa}{{\tt C}_\press \rho}
    =\frac{\tau v_{{}_{\! {\rm F}}}^2}{3}
\left[1 - \frac{\tp \left(80-29 \mathcal{G}_0\right)}{120
\mathcal{G}_0}\right]
\ee
$[{\tt D}_T =
\Gamma^{(0)}/(2 q^2)]$,  one gets
\bel{chichpqom} \chi_{\rm coll}^{(0)} =
\frac{i {\tt D}_T q^2}{\om + i {\tt D}_T q^2} ~\chi_{\rm coll}^{(0)} (q, \om = 0).
\ee
Note that according to (\ref{shp}) and (\ref{tautom}), the
temperature dependence of $\Gamma^{(0)}$ becomes similar to the
one found in (\ref{gambaronet}),
\bel{gambarhpt} \gbarhp
= \frac{2 \tau_o v_{\rm F}^2 q^2}{3 T^2} \left[1 - \frac{\tp \left(80-29
\mathcal{G}_0\right)}{120 \mathcal{G}_0}\right].
\ee
 For the dissipative and reactive parts of the response function
$\chi_{\rm coll}^{(0)}$ (\ref{chichpqom}), from (\ref{chichpqom}) one gets
\bea\l{chichpqompp}
\chi_{\rm coll}^{(0)\;\prime\prime}(q,\om)&=&
     \azero \frac{\om}{\om^2 + \left(\Gamma^{(0)}\right)^2 / 4},\nonumber\\
\chi_{\rm coll}^{(0)\;\prime}(q,\om)&=&
     \azero \frac{
\Gamma^{(0)}/2}{\om^2 + \left(\Gamma^{(0)} \right)^2 / 4}.
\eea
 The strength distribution $\chi_{\rm coll}^{(0)\;\prime\prime}$
has a maximum at $\om=\Gamma^{(0)}/2$ and a width $\Gamma^{(0)}/2
\propto \taubar$. In the LWL
limit $\taubar \ll 1$
this distribution becomes quite sharp with the maximum lying close
to $\om=0$. As may be inferred with the help of (\ref{chicollhp})
and (\ref{shp}), the maximal value does not depend on $\taubar$
and is proportional to ${\bar T}^2$.  It will be demonstrated
shortly that the pole at $\omzero$ (\ref{shp}) is related to the heat
conduction, for which reason it sometimes is called "heat pole".
Notice that the reactive response function
$\chi_{\rm coll}^{(0) \; \prime}$ is finite at $\om=0$, with a value
independent of $\taubar$.

In the hydrodynamic regime with $\taubar \ll 1$, the response
function $\chi_{\rm coll}$ found for the {\it Fermi liquid} becomes
identical to the one for {\it normal liquids}
\cite{forster,brenig}. This can be made more apparent after
introducing the dimensional sound velocity $c$, a width parameter
$\Gamma$, determined as
\bel{liqparam} c=v_{{}_{\! {\rm F}}} s_0^{(1)}\;, \qquad\qquad
\Gamma=\Gamma^{(1)}/q^2 \;,
\ee
  as well as the diffusion
coefficient ${\tt D}$ (\ref{diffusd}) and the specific heats. The
sum of the two contributions discussed above may then be written
as
\bea\l{forstereq} \chi_{\rm coll}^{\prime\prime} &=&
    \rho \left(\frac{\partial \rho}{\partial \press}\right)_T
    \left[\frac{\left({\tt C}_{\mathcal{V}}/{\tt C}_\press\right)
    c^2 q^4 \Gamma\; \om}{\left(\om^2-c^2q^2\right)^2
+ \left(\om q^2
\Gamma\right)^2} \right.\nonumber\\
    &+& \left.\frac{\left(1-{\tt C}_{\mathcal{V}}/{\tt C}_\press\right)
q^2 {\tt D}_T\; \om}{\om^2 +\left(q^2 {\tt D}_T\right)^2}\right].
\eea
 Traditionally, the peaks related to the first and second terms
are called Brillouin and Rayleigh (or Landau--Placzek) peak,
respectively. The ratio of the specific heats ${\tt C}_\press$ and
${\tt C}_{\mathcal{V}}$ per particle is discussed in Appendix A.1,
see (\ref{cpcvkakt}) and (\ref{cpcvexp}). Note
that the sound speed $s_0^{(1)}$, see (\ref{sfirst0}), is identical
to the adiabatic sound velocity found in Appendix A.1,
see
(\ref{velocad}) ($c$ in dimensional units for normal liquids), as
it should be for normal liquids
\cite{brenig,forster}.
The structure of (\ref{forstereq}) is identical to that discussed
in the literature (see, e.g., (4.44a) of \cite{forster}), if one
only expresses the quantities appearing here in terms of viscosity
and thermal conductivity. As a matter of fact, the alert reader
might expect a third term (as in (4.44a) of \cite{forster}), but
this one is of the order of $\taubar^2$ and thus is neglected here.
The specific temperature dependence of these parameters (in the
LWL limit) will be discussed in the next subsection,
with respect to the specific heats, see also Appendix A.1.

Note that in the derivation of the both amplitudes $\azero$
(\ref{chicollhp}) and $\aone$ (\ref{chicollone}) we took
$D(s)$ (\ref{despfunc}) at low temperatures using
(\ref{chitemzero}) to (\ref{chitemtwo}); and then, expand first
it near the poles (\ref{shp}) and (\ref{sfirst0}), respectively;
and second, in small $\taubar$ of the LWL
limit. This way
of the calculation is much more simpler because the two last
operations can be exchanged only when we shall take into account
next order terms in $\taubar$, that takes much hard work. If we
exchange the last two operations, expanding first in $\tau_q$ in
the {\it linear} LWL
approximation (\ref{somexp}), and then,
doing expansion near the poles, some important terms will be lost.

\subsubsection{SHEAR MODULUS, VISCOSITY AND THERMAL CONDUCTIVITY}
\label{viscosthermcond}

As explained in Appendix A.2,
these coefficients may be
obtained by applying expansions to $\chin$ within the
perturbation theory mentioned above, for low temperatures (with
$\tbar \ll 1$); see  in particular (\ref{chiexpzero}), (\ref{chiexpone}) and
(\ref{chiexptwo}).
They specify the stress tensor $\sigma_{\alpha\beta}$
(\ref{prestensone})-(\ref{pressnu}) and the heat current ${\bf
j}_T$ (\ref{currheat}).

The shear
modulus $\lambda $ (\ref{shearmod}) in the time-reversible part $
\sigma_{\alpha\beta}^{(\lambda)}$ (\ref{presslamb}) of the
stress tensor $ \sigma_{\alpha\beta}$ (\ref{prestensone})
turns into zero in the long wave-length approximation linear in
$\taubar$ as in \cite{heipethrev}
up to immaterial corrections of the order of ${\bar T}^4$. By
another words, in this case, $\lambda $ is a small quantity of
the order of $\taubar^2$ because such corrections were neglected
everywhere. It means a disappearance of the Fermi-surface
distortions in our linear approach (\ref{somexp}) which are the
main peculiarity of Fermi liquids compared to the normal ones.

For the shear viscosity $\nu $
(\ref{viscos}) taken at the first sound frequency
$\om=\om_0^{(1)}$ (\ref{sfirst}), one obtains
\bel{shearvisonehp} \nu = \nu^{(1)}+\nu^{(2)}\;,
\ee
 where
\bel{shearvisone} \nu^{(1)} = \frac{2}{5} \rho \eps_{{}_{\! {\rm F}}} \tau
    \left(1 + \frac{5 \tp}{12} \right)~
\ee
 and
\bel{shearvishp} \nu^{(2)}=\frac{13 \pi^4}{720} \frac{\rho
\eps_{{}_{\! {\rm F}}} {\bar T}^4 }{v_{{}_{\! {\rm F}}}^2 q^2 \tau}.
\ee
 The first term
$\nu^{(1)}$ (\ref{shearvisone}) in (\ref{shearvisonehp}) is proportional to the
relaxation time $\tau$ and coincides mainly with that obtained
earlier for mono-atomic gases and
for a Fermi liquid by using another method
\cite{heipethrev},
except for the specific explicit dependence on temperature
presented here. The temperature dependence of the shear viscosity
$\nu^{(1)}$ (\ref{shearvisone}) is mainly the same as for the
rate of the sound damping $\Gamma ^{(1)}$ (\ref{gambaronet}),
$\nu^{(1)} \propto {1/T^2}$, with the temperature dependence of the
relaxation time $\tau$ (\ref{tautom}). Although the viscosity component
 $\nu^{(2)}$,
too, is related to the first sound solution $\om_0^{(1)}$, it is
proportional to $1/\tau$, similar to the viscosity of zero sound
but in contrast to the standard first sound viscosity
(\ref{shearvisone}). The $\nu^{(2)}$ component (\ref{shearvishp})
of the viscosity (\ref{shearvisonehp}) increases with temperature
as $T^6$, see also (\ref{tautom}) for the relaxation time $\tau$.
Although the second component $\nu^{(2)}$ of the shear viscosity
is proportional to ${\bar T}^4$, and thus may be considered small
under usual conditions, it may become important for small wave
numbers $q$ (or frequencies $\om$) [for more details, see the
discussion to come below in Sec.\ \ref{heatcorrfun}].
This component of the viscosity was not discussed
in \cite{heipethrev}.

Let us finally turn to the thermal conductivity $\kappa$ which
shows up in the equation for variations of temperature $T(\r,t)$
with $\r$ and $t$ (see Appendix A.2).
The form (\ref{kappadef}) [for
the heat mode $\om=\om^{(0)}$ of (\ref{shp})] may be rewritten as
\bel{kappaexp} \kappa =
    \rho \frac{{\tt C}_\press \Gamma^{(0)}}{2 q^2} \approx
   \frac{1}{3} \rho {\tt C}_\press v_{{}_{\! {\rm F}}}^2 \tau
   \left[1 - \frac{\tp \left(80-29 \mathcal{G}_0\right)}{120
\mathcal{G}_0}\right].
\ee
 We present here also explicitly the temperature
corrections up to the terms of the order of $\tbar^2$. Our
expression for the thermal conductivity $\kappa$ (\ref{kappaexp})
differs from the one found in \cite{forster} and
\cite{heipethrev} by small $\tbar^2$ corrections. However, they are not
important in the calculations of the damping coefficient
$\Gamma^{(1)}$ for the first sound mode defined in
\cite{forster}, and \cite{brenig},
see also the comment before (\ref{forstereq}),
\bel{gammaland}
\Gamma ^{(1)}=
    \frac{q^2}{m \rho }
    \left[
    \frac{4}{3} \nu^{(1)}
    + \frac{m \kappa}{{\tt C}_\press}
    \left(\frac{{\tt C}_\press}{{\tt C}_{\mathcal{V}}} -1\right)
     \right],
\ee
 Here, $\nu^{(1)}$ is the part of the shear viscosity
coefficient related to the first sound mode, see
(\ref{shearvisone}); ${\tt C}_\press/ {\tt C}_{\mathcal{V}}$ is the
adiabatic ratio of the specific heats, see
(\ref{specifheatpdef}) and (\ref{cvcpkSkT}). We omitted here
corrections related to the second viscosity in line of the second
approximation in (\ref{gambarone}).  In
(\ref{gammaland}),  $\kappa $ is multiplied by a small quantity of the
order of the $\tbar^2$ as follows from (\ref{cpcvexp}) and the
temperature corrections to $\kappa $ written explicitly in
(\ref{kappaexp}) can be neglected in (\ref{gammaland}) . The
expression for the damping coefficient $\Gamma^{(1)}$
(\ref{gammaland}) with the viscosity coefficient $\nu^{(1)}$
(\ref{shearvisone}), thermal conductivity $\kappa$
(\ref{kappaexp}) and specific heats from (\ref{cpcvexp}) and
(\ref{specifheatp}) for viscose normal liquids is in agreement
with our result for $\Gamma^{(1)}$ (\ref{gambarone}) including the
temperature corrections.

Thus, up to
the temperature corrections discussed above, we have agreement
with the results of
\cite{heipethrev} for the dispersion equation, viscosity and
thermal conductivity coefficients in the hydrodynamic limit. Our
derivations are more strict and direct within the perturbation
theory for the eigenvalue problem. We have the transition to the
hydrodynamics of normal liquids discussed in
\cite{forster,brenig,kubo}
in terms of
the  macroscopic parameters mentioned above.

\subsection{Susceptibilities}
\label{suscept}

In this section, we want to address the calculation of the static
susceptibilities, for which one distinguishes isolated, isothermal
and adiabatic ones \cite{kubo,forster,brenig}. Their comparison is
relevant for ergodicity properties, see \cite{kubo,brenig}. Here
we will concentrate on the density mode of nuclear matter
considered as an infinite Fermi-liquid system.

\subsubsection{ADIABATIC AND ISOTHERMAL SUSCEPTIBILITIES}
\label{susceptadt}

The isolated susceptibility $\chi_{{}_{\! DD}}(0)$ is defined as the
static limit of the response function $\chi_{{}_{\! DD}}(q,\om)$ (or
$\chi_{{}_{\! DD}}(\taubar,s)$ of (\ref{chiindef}) in dimensionless
variables), for which one first has to take the limit $q \to 0$
(or $\taubar \to 0$), and then, $\om \to 0$ (or $s \to 0$) (see, e.g.,
\cite{forster})
\bel{statresponse} \chi_{{}_{\! DD}}(0)=
 \lim_{\om \rightarrow 0}
    \left[\lim_{q \rightarrow 0}\;\chi_{{}_{\! DD}} \left(q,\om\right)\right]=
 \lim_{s \rightarrow 0}
    \left[\lim_{\taubar \rightarrow 0}\;\chi_{DD} \left(\taubar,s\right)\right].
\ee
 Apparently, $\chi_{{}_{\! DD}}(0)$ satisfies the relation
\bel{vardenschi0}
\delta \rho \equiv -\chi_{{}_{\! DD}}(0) \delta V_{\rm eff},
\ee
 where $\delta V_{\rm eff}$ and $\delta \rho$ are quasistatic
variations. They can be considered as independent of time, in
contrast to the ones discussed in Sec.\ \ref{dynresp}, see
(\ref{delueff}).

The isothermal susceptibility $\chi_{DD}^{T}$ is defined as the
density-density response at constant temperature $T$, and the
adiabatic one, $\chi_{DD}^{\varsigma}$, as that at constant
entropy (per particle $\varsigma$). Suitable variables for
studying the variations of the density $\rho$ are therefore
pressure $\press$ and temperature $T$ in the first case, and
pressure $\press$ and entropy per particle $\varsigma$ in the
second one. These two representations of $\delta \rho$ can be
written as
\bel{vardens} \delta \rho \equiv
    \left(\frac{\partial \rho }{\partial \press} \right)_T \delta \press
    + \left(\frac{\partial \rho }{\partial T} \right)_\press \delta T \equiv
    \left(\frac{\partial \rho }{\partial \press} \right)_\varsigma
    \delta \press
    + \left(\frac{\partial \rho }{\partial \varsigma} \right)_\press
    \delta \varsigma.
\ee
 For the isothermal and adiabatic susceptibilities
$\chi_{DD}^{T}$ and  $\chi_{DD}^{\varsigma}$, one thus gets the
following two relations:
\bel{vardenschiT} \delta \rho \equiv
    \left(\frac{\partial \rho }{\partial \press} \right)_T \delta \press
    =-\chi_{DD}^T \delta V_{\rm eff}
\ee
 and
\bel{vardenschiS} \delta \rho \equiv
    \left(\frac{\partial \rho }{\partial \press} \right)_\varsigma
    \delta \press
    =-\chi_{DD}^{\varsigma} \delta V_{\rm eff}.
\ee
 The variations of the density with pressure are related to the
(in-)compressibilities, see (\ref{incompraddef}) and
(\ref{incomprTdef}). As shown in Appendix A.1,
their ratio
can be expressed by that of the corresponding specific heats, see
(\ref{cvcpkSkT}). Building the ratio, one therefore gets
from (\ref{vardenschiT}) and (\ref{vardenschiS})
\bel{chitchiskSkT}
\frac{\chi_{DD}^{T}}{\chi_{DD}^{\varsigma}}=
\frac{K^{\varsigma}}{K^{T}}=
\frac{{\tt C}_\press }{{\tt C}_{\mathcal{V}}}.
\ee
 This is a
general relation from thermodynamics where we only have replaced
the system's total entropy \cite{brenig}
by the entropy per particle $\varsigma$ applied
for the intensive systems as normal and Fermi liquids.

We are interested more in the calculation of the differences
between the isothermal susceptibility $\chi_{DD}^{T}$ defined by
the relations in (\ref{vardenschiT}) (or adiabatic one
$\chi_{DD}^{\varsigma}$, see (\ref{vardenschiS})) and isolated
(static) susceptibility $\chi_{{}_{\! DD}}(0)$ presented by
(\ref{vardenschi0}) \cite{hofmann}. For this purpose, we
find first the ratio of the isothermal-to-isolated
susceptibilities $\chi_{DD}^{T}/\chi_{{}_{\! DD}}(0)$ in terms of the
ratio of the static "intrinsic" temperature-density response
function to the isolated one $\chi_{{}_{\! DD}}(0)$ (\ref{chiindef}). The
static temperature-density susceptibility $\chi_{{}_{\! TD}}(0)$ is
defined in the same way (\ref{statresponse}) as the static limit
of the "intrinsic" temperature-density response function
$\chi_{{}_{\! TD}}(\taubar,s)$ given by (\ref{chitindef}). Note that
the limits $\om \rightarrow 0$ (or $s \rightarrow 0$) and $q
\rightarrow 0$ (or $\taubar \rightarrow 0$)  which we consider to
get the static response functions are not commutative \cite{forster}.
Taking the second equations in (\ref{vardenschiT})  and
(\ref{vardenschi0}) for the intensive systems as liquids, one
gets
\bel{chitchi0}
\frac{\chi_{DD}^{T}}{\chi_{{}_{\! DD}}(0)}= 1 +
\left[\varsigma \left(
\frac{\partial \rho }{\partial \mu}\right)_T
- \left(\frac{\partial \rho }{\partial T}\right)_\mu\right]
\frac{\chi_{{}_{\! TD}}(0)}{\chi_{{}_{\! DD}}(0)}.
\ee
 We used here
the definitions (\ref{chiindef}) and (\ref{chitindef}) for the
density-density and temperature-density response functions and
(\ref{statresponse}) for their static limits $\chi_{{}_{\! DD}}(0)$
and $\chi_{{}_{\! TD}}(0)$. We then applied the thermodynamic relations of
Appendix A.1  for the transformations
of the derivative
$\left(\partial \rho / \partial \press\right)_T$. This derivative
 appears from the
definition of the isothermal susceptibility $\chi_{DD}^{T}$ in
(\ref{vardenschiT}) to another simpler thermodynamic
derivatives for the application to Fermi liquids, see below.
For this aim, we transform the variables $(T,\press)$ to the new
ones $(T,\mu)$. The derivatives of pressure $\press$ over
these two new variables can be then reduced to the ones of the
density $\rho$ shown in the r.h.s. of (\ref{chitchi0})
with the help of (\ref{gibbsduh}).

So, the calculations of the susceptibilities are resulted
in the
derivation of the static limits defined by
(\ref{statresponse}) for the temperature-density
$\chi_{{}_{\! TD}}(\taubar,s)$ and density-density $\chi_{{}_{\! DD}}(\taubar,s)$
response functions, see (\ref{chiindef}) and
(\ref{chitindef}), and their ratio $\chi_{{}_{\! TD}}(0)/\chi_{{}_{\! DD}}(0)$ for
the case of a heated Fermi liquid. We can then calculate the two
ratios of the susceptibilities (\ref{chitchi0}) and
(\ref{chitchiskSkT}) which both determine separately each
considered susceptibilities.

\subsubsection{FERMI-LIQUID SUSCEPTIBILITIES}
\label{flsuscept}

The expression for the ratio of the isothermal-to-static
susceptibilities (\ref{chitchi0}) can be simplified my making use
of the specific properties of Fermi liquids given by
(\ref{drhomt}) and second equation in (\ref{dpdtr}),
\bel{chitchi0fl}
\frac{\chi_{DD}^{T} }{\chi_{{}_{\! DD}}(0)} = 1
+\left(\frac{{\tt C}_\press }{{\tt C}_{\mathcal{V}}}-1\right)
\frac{\chi_{{}_{\! TD}}(0) }{\tbar \chi_{{}_{\! DD}}(0)}\;\mathcal{N}(0).
\ee
According to the definition (\ref{statresponse}) of the static
response functions applied to the ones in the ratio
$\chi_{{}_{\! TD}}(0)/\chi_{{}_{\! DD}}(0)$ of (\ref{chitchi0fl}), we shall
use (\ref{chitbar}) and (\ref{chiindef}) for the corresponding
intrinsic susceptibilities ($\mathcal{F}_{0}=\mathcal{F}_{1}=0$ there).
The static limit (\ref{statresponse}) of the
response functions $\chi_{{}_{\! DD}}(\taubar,s)$ (\ref{chiindef}) and
$\chi_{{}_{\! TD}}(\taubar,s)$ (\ref{chitbar}) in (\ref{chitchi0fl})
can be found by using the LWL
expansions over a
small parameter $\taubar \ll 1$ at low temperatures, see
Sec.\ \ref{lowtemlim} and Appendix A.2 for the
first limit ($\taubar \rightarrow 0$)
in (\ref{statresponse}). We substitute now the
perturbation theory expansions for small $\taubar$ for the
quantities $s$ (\ref{somexp}), $\chi_1$
(\ref{chiexpone}),
$\aleph$ (\ref{ampsexp}), and $D_0$ (\ref{dszexpzero}) into
(\ref{chiindef}) and (\ref{chitbar}). We get this limit as
functions of $s_0$ and $s_1$, and then, we shall take the second limit of
$s_{0} \to 0$ and $s_{1} \to 0$ [ $s \rightarrow 0$ in (\ref{statresponse})].
Finally, we
arrive then at the very simple result
 \bel{chitchid}
\frac{\chi_{{}_{\! TD}}(0) }{\chi_{{}_{\! DD}}(0)}=
\frac{\tbar }{\mathcal{N}(0)}
\ee
neglecting small cubic terms in $\tbar$, which correspond to $\tbar^4$
corrections in susceptibilities and do not matter in this section.
Note that the sequence of the limit transitions defined in
(\ref{statresponse}) and recommended in \cite{forster} is
important for the calculation of this ratio: We get zero for this
ratio if we take first $s \rightarrow 0$, and then, $\taubar
\rightarrow 0$.

Substituting now the ratio (\ref{chitchid}) of the
susceptibilities into (\ref{chitchi0fl}),
one obtains
\bel{chitchi0s}
\frac{\chi_{DD}^{T} }{\chi_{{}_{\! DD}}(0)} =
\frac{{\tt C}_\press }{{\tt C}_{\mathcal{V}}}
 = 1 + \frac{\tp }{3 \mathcal{G}_0},
\ee
see also (\ref{cpcvexp}) for the second equation. We compare
now this result with (\ref{chitchiskSkT}) and get that our
Fermi-liquid system satisfies the ergodicity property:
\bel{ergodicity}
\chi_{DD}^{(\varsigma)}=\chi_{{}_{\! DD}}(0) \;.
\ee
 This ergodicity property was proved at low temperatures, for which the
Landau Fermi-liquid theory  can be applied. It is related to
the adiabaticity of the velocity of the sound mode $s_0^{(1)}$,
see (\ref{sfirst0}) and discussion
after (\ref{forstereq}). Moreover, we got the normal liquid
(hydrodynamic) limit from the Fermi-liquid dynamics, and  therefore, the
ergodicity property is general for  heated Fermi liquids and
normal (classical) ones.

Another aspect of the discussed ergodicity property might be the
relation to the non-degeneracy of the excitation spectrum in the
infinite Fermi liquids beside of the spin degeneracy. We have only
the two-fold degenerate quasiparticle states, due to the spin
degeneracy. However, it does not influence on our results concerning the
ergodicity relations because we consider the density-density
excitations, which do not disturb  the spin degree of freedom. We
have only the multiplication factor two in all susceptibilities,
due to the spin degeneracy, that does not change the ratios of the
susceptibilities which are only important for the ergodicity
discussed here.

Our susceptibilities obtained above satisfy the Kubo relations,
see (4.2.32) of \cite{kubo}:
\bel{kuborel} \chi ^{T} \geq \chi
^{\varsigma} \geq \chi(0)
\ee
 with the equal sign for the second
relation because of the ergodicity property. To realize this, we
should take into account that
${\tt C}_\press > {\tt C}_{\mathcal{V}}$ (or $K^{\varsigma}>K^{T}$),
according to (\ref{cpcvkakt}),
because all quantities on the r.h.s. of this equation are positive
for the stable modes $\mathcal{G}_0=1+\mathcal{F}_0>0$. The equal sign
for the first relation in (\ref{kuborel}) becomes true in the
two limit cases: For the temperature $T$ going to 0 or for the
in-compressible matter when the interaction constant $\mathcal{F}_0$
tends to $\infty$. In both limit cases we made obvious equality
${\tt C}_\press={\tt C}_{\mathcal{V}}$ and all susceptibilities are
identical [equal signs in the both relations of
(\ref{kuborel})].

Note now that namely the specific Fermi-liquid expression of the
static susceptibility $\chi_{DD}(0)$, see (\ref{chiindef}) with
$\mathcal{F}_0=\mathcal{F}_1=0$ for the case of the intrinsic response
functions, depends on the sequence of the limit transitions
discussed near
(\ref{statresponse}), (\ref{chitchi0}), (\ref{chitchid}) above
and in \cite{forster}.
For the definition
(\ref{statresponse}) of \cite{forster}, one gets
\bel{statrespfl}
\chi_{{}_{\! DD}}(0)=\left(1 - \frac{5 \tp }{12}\right)
\mathcal{N}(0),\quad \chi_{DD}^{T}=\mathcal{N}(T).
\ee
 In the last
equation, we used also (\ref{chitchi0s}). Taking the opposite
sequence of the limit transitions, first $s \rightarrow 0$, one
has the result $\mathcal{N}(T)$ (\ref{enerdensnt})
for the isolated susceptibility
$\chi_{{}_{\! DD}}(0)$ like for the isothermal one $\chi_{DD}^{T}$. The
difference is in $\tbar^2$ corrections. Ignoring them, the both
versions of the limit transitions coincide, and we come to the
result independent on temperature discussed in \cite{pinenoz}~.
The ergodicity property (\ref{ergodicity}), Kubo's relations
(\ref{kuborel}) and relation of the isothermal susceptibility to
adiabatic one (\ref{chitchiskSkT}) do not depend on the specific
peculiarities of the static limit of the response function
discussed here in connection to Fermi liquids.

\subsection{Relaxation and correlation functions}
\label{relaxcorr}

\medskip

\subsubsection{RELAXATION FUNCTION}
\label{relaxfun}

Coming back to the dynamical problem, we note that
the dissipative part of the response function
$\chi^{\prime\prime}(\om)$ is related to the relaxation function
$\Phi^{\prime\prime} (\om)$ \cite{kubo} by
\bel{chiimpsi}
\chi
^{\prime\prime} =
    \om \Phi ^{\prime\prime} (\om )\;.
\ee
We follow the notations of \cite{hofmann,hofbook} and omit the
index "coll" in this section: For the comparison with the
microscopic results of \cite{hofmann} we need really the
relaxation and correlation functions related to the {\it
intrinsic} response functions. According to
(\ref{ddrespchiin}) and (\ref{couplconst}), all these intrinsic
functions can be formally obtained from the collective ones at the
zero Landau constants $\mathcal{F}_0$ and $\mathcal{F}_1$. Taking into
account also (\ref{chiconeqompp}) and (\ref{chichpqompp}), one has
\bea\l{relaxcom}
    \Phi^{\prime\prime}(\om) &=&
   \frac{\aone }{2 \om_0^{(1)}}
    \left[\frac{
\Gamma^{(1)} }{\left(\om-\om_0^{(1)} \right)^2
    + \left(\Gamma^{(1)}\right)^2/4} \right. \nonumber\\
    &+&\left.\frac{\Gamma^{(1)} }{\left(\om + \om_0^{(1)}\right)^2
    + \left(\Gamma^{(1)}\right)^2 /4}\right] \nonumber\\
    &+&
    \frac{\azero }{\gbarhp}\;
    \frac{\Gamma^{(0)} }{\om^2 + \left(\Gamma^{(0)} \right)^2 / 4}.
\eea
 This equation can be re-written in the same way like to
(\ref{forstereq}) in terms of the parameters $c$, $\Gamma$ and
${\tt D}_T$, see (\ref{liqparam}) and (\ref{diffusd}),
\bea\l{relaxbrenig}
\Phi^{\prime\prime}(\om) &=&
    \chi^T
    \left[\frac{
\left({\tt C}_{\mathcal{V}}/{\tt C}_\press\right) {\tt C}^2 q^4
\Gamma}{\left(\om^2-{\tt C}^2q^2\right)^2 +\left(\om q^2 \Gamma\right)^2}
\right.\nonumber\\
    &+&\left. \frac{
\left(1-{\tt C}_{\mathcal{V}}/{\tt C}_\press\right)
q^2 {\tt D}_T}{\om^2 +\left(q^2 {\tt D}_T\right)^2}\right].
\eea
 We used here the Jacobian relations
and (\ref{incomprTdef}), (\ref{isotherk})
for the transformation of the coefficient in front of the square
brackets in (\ref{forstereq}) to the one, the {\it intrinsic}
isothermal susceptibility $\chi^T$ (\ref{statrespfl})
($\mathcal{F}_0=0$). We also neglected
terms of the order of $\taubar^2$ as in the derivation of
(\ref{forstereq}). Equation (\ref{relaxbrenig}) for the relaxation
function $\Phi^{\prime\prime}(\om)$ is identical to the imaginary
part of the r.h.s. of (28.29) in \cite{brenig}
with the transparent physical meaning as (\ref{forstereq}). The
first term in the square brackets of (\ref{relaxcom}) and
(\ref{relaxbrenig}) is the
first sound Brillouin component with the poles (\ref{sfirst})
associated to the finite frequencies $\pm\om_0^{(1)}$  of the
time-dependent relaxation-function oscillations and their damping
rate $1/\Gamma^{(1)}$ ($\pm\om_s$ and $1/\gamma_s$ in the notation
of \cite{brenig}, respectively, see more complete discussion of
properties of the time-dependent relaxation function as a Fourier
transform of the relaxation function $\Phi(\om)$ in
\cite{brenig}). The second term in (\ref{relaxcom}) and
(\ref{relaxbrenig}) describes the pure
damped Raleigh mode corresponding to the overdamped pole
$\om^{(0)}$ (\ref{shp}) defined by the diffuseness coefficient
${\tt D}_T \propto \Gamma^{(0)}$  (or  $ \propto\gamma_{{}_{\! T}}$ in the
notation of \cite{brenig}). As noted in \cite{brenig}, the
strength of this peak is a factor $1-{\tt C}_{\mathcal{V}}/{\tt
C}_\press$ smaller than for the two first sound peaks. According
to (\ref{cpcvexp}), in the zero temperature limit $T
\rightarrow 0$, the Raleigh peak disappears but the
Brillouin ones become dominating because of $\Gamma \propto
\Gamma^{(1)} \propto 1/T^2$; see the second equation of
(\ref{liqparam}) for the relation of $\Gamma$ to
$\Gamma^{(1)}$ and (\ref{gambaronet}). Note also that the
coefficient in front of the square brackets in
(\ref{relaxcom}) is finite in the limit $T \rightarrow 0$.

\subsubsection{CORRELATION FUNCTION}
\label{corfun}

We like to present also the correlation function, partly for the
sake of completeness and partly to allow for comparisons with
calculations of the function in the nuclear SM approach of
\cite{hofivyam,ivhofpasyam}, see also
\cite{hofmann,hofbook}, to the collective motion of finite nuclei. Let us
use now the fluctuation-dissipation theorem \cite{kubo} to get the
correlation function $\psi^{\prime\prime}(\om)$,
\bel{fludiptheor}
\psi^{\prime\prime}(\om) \rightarrow
    \hbar \om \coth \left(\frac{\hbar \om }{2T} \right)
    \Phi ^{\prime\prime}(\om)=
     \hbar \coth \left(\frac{\hbar \om }{2T}\right)
     \chi ^{\prime\prime}(\om)\;.
\ee
 In the semiclassical limit $\hbar \rightarrow 0$ considered
here, one has
\bel{corrfun}
\psi^{\prime\prime}(\om) =
    ~\frac{2 T }{\om}~ \chi ^{\prime\prime}(\om)=~
    2T~ \Phi^{\prime\prime}\left(\om\right).
\ee
 According to (\ref{forstereq}),(\ref{relaxbrenig}), this
correlation function can be split into the two components as in
\cite{hofivyam,hofmann},
\bel{corrfunhof}
\psi^{\prime\prime}(\om) =
    \psi_0^{\prime\prime}(\om) + \psi_R^{\prime\prime}(\om)\;.
\ee
 Here, $\psi_0^{\prime\prime}$ is the heat pole part,
\bea\l{corrfunhp}
\psi_0^{\prime\prime}(\om) &=&
    \frac{2 T }{\om} \chi ^{(0)\;\prime\prime}(\om) \nonumber\\
  &=& 2T \chi^T~
\frac{\left(1-{\tt C}_{\mathcal{V}}/{\tt C}_\press\right)
q^2 {\tt D}_T}{\om^2 +\left(q^2 {\tt D}_T\right)^2},
\eea
 $\chi^{(0)\;\prime\prime}$ is given by the first equation in
(\ref{chichpqompp}) and is related to the second heat pole
terms in the square brackets of (\ref{forstereq}) and
(\ref{relaxbrenig}) [through (\ref{chiimpsi})]. This part is
singular at the zero frequency point $\om = 0$ for $\taubar
\rightarrow 0$, see (\ref{diffusd}) and (\ref{som}). The
other term $\psi_R^{\prime\prime}$ in (\ref{corrfunhof}) is
associated with the first sound component in the square brackets
of (\ref{forstereq}), (\ref{relaxbrenig}),
\bea\l{corrfun1}
\psi_R^{\prime\prime}(\om) &=&
    ~\frac{2 T }{\om}~ \chi ^{(1)\;\prime\prime}(\om) \nonumber\\
   &=& ~2T~ \chi^{T}~\frac{
\left({\tt C}_{\mathcal{V}}/{\tt C}_\press\right)
   c^2 q^4 \Gamma}{\left(\om^2-c^2q^2\right)^2 +\left(\om q^2 \Gamma\right)^2}.
\eea
 This component has no such singularity at $\om = 0$ for $\taubar \to 0$, as
seen from (\ref{liqparam}), (\ref{sfirst}) and
(\ref{gambaronet}) [see (\ref{chiconeqompp}) for
$\chi^{(1)\;\prime\prime}(\om)$ in the middle of
(\ref{corrfun1})]. According to the second equation in
(\ref{corrfunhp}), the heat pole part
$\psi_0^{\prime\prime}(\om)$ of (\ref{corrfunhof}) for the {\it
intrinsic} correlation function can be written as in
\cite{hofivyam,hofmann},
\bel{corrfunhphof}
\psi_0^{\prime\prime}(\om)= \psi^{(0)}
     \frac{\hbar {\it \Gamma}_T }{(\hbar \om)^2 + {\it \Gamma}_T^2 / 4},
\ee
 where
\bel{GammaT}
{\it \Gamma}_T=
2 \hbar  q^2 {\tt D}_T= \hbar
\Gamma^{(0)},
\ee
and
\bel{psi0}
(1 / T) \psi^{(0)}=\chi^T-\chi^\varsigma
 = \chi^T-\chi(0).
\ee
 We applied here (\ref{chitchiskSkT}) in the first equation
of (\ref{psi0}) and ergodicity condition (\ref{ergodicity})
for the second one. The specific expressions for the quantities
$\Gamma^{(0)}$, $\chi^T$ and $\chi(0)$ in the last two equations
(\ref{GammaT}) and (\ref{psi0}) can be found in (\ref{shp}),
(\ref{gambarhpt}) and (\ref{statrespfl}). Note that the
correlation function (\ref{corrfunhphof}), corresponding to the
heat pole, has the Lorentzian multiplier. This multiplier  approaches the
$\delta(\om)$ function in the hydrodynamic limit $\taubar
\rightarrow 0$ because of $\Gamma_T \rightarrow 0$, according to
(\ref{GammaT}) and (\ref{shp}) ($\Gamma^{(0)} \rightarrow
0$), i.e.,
 \bel{corrdeltafunlim}
    \psi^{(0)}(\om)
     \rightarrow 2 \pi \psi^{(0)} \delta (\om)\qquad\mbox{for}\qquad
     \taubar \rightarrow 0.
\ee
 The relations (\ref{corrfunhphof}), (\ref{psi0}) and
(\ref{corrdeltafunlim}) confirm the discussion in
\cite{hofmann} concerning the heat pole contribution to the
correlation function. The specific property of the Fermi liquid is
that this system is exactly ergodic, see (\ref{ergodicity}), as
used in the second equation of (\ref{psi0}).


\section{Nuclear response within the Fermi-liquid droplet model}
\label{respfuntheor}

\subsection{Basic definitions}
\label{basdef}

So far we considered the Fermi-liquid theory  for study of the
collective excitations at finite temperatures much smaller than
the Fermi energy $\eps_{{}_{\! {\rm F}}}$ in the {\it infinite} nuclear matter.
This theory can be also helpful for investigation of the
collective modes and transport properties of heavy heated nucleus
considered as a {\it finite} Fermi system within the macroscopic FLDM
\cite{strutmagbr,strutmagden0,galiqmodgen,galiqmod,magstrut,denisov,magboundcond,kolmagpl,magkohofsh,kolmagsh}.
Such a semiclassical nuclear model applied
earlier successfully to the giant multipole resonance description
\cite{strutmagden0,galiqmod,strutmagden,denisov,magpl,kolmagpl,kolmagsh,magkohofsh} is expected
to be also
incorporated in practice as an asymptotic high temperature limit
of the quantum transport theory \cite{hofmann} based on the shell model.
This theory takes into account the residue
interactions like particle collisions for study of the low energy
excitations in nuclei. The latter application  of the FLDM is very
important for understanding itself the dissipative processes
like nuclear fission at finite temperatures
(see, g.e., \cite{hofivyam,hofmann,hofbook,hofivmag,hofmag})

Following \cite{hofmann,hofbook}, let us describe the many-body
excitations of nuclei in terms of the response to an external
perturbation
\bel{extfield}
V_{\rm ext}=q_{\rm ext}(t)~{\hat F},
\ee
where
${\hat F}$ is some one-body operator,
\bel{qext}
q_{\rm ext}(t)=q_{\rm ext}^{\om}\exp[-i(\om + i\epsilon)],\qquad (\epsilon=+0)
\ee
 The linear
response function can be determined through the Fourier transform
$\langle {\hat F} \rangle_\om$ of the time-dependent quantum
average $\langle {\hat F} \rangle_t$ by
\bel{defresp}
\langle
{\hat F} \rangle_\om= -\chi_{FF}^{\rm coll}(\om)~q_{\rm ext}^\om.
\ee
 Here
and below we omit an unperturbed average value $\langle {\hat F}
\rangle_0$ and use the same notation as in \cite{hofmann}. In the
following, we shall consider the operators ${\hat F}$ neglecting the
momentum dependence in a phase space representation in the linear
approximation for an external field $V_{\rm ext}$ and writing
${\hat F}={\hat F}(\r)$. According to (\ref{defresp}), one can
then express explicitly $\chi_{FF}^{\rm coll}(\om)$ in terms of the
Fourier transform $\delta \rho_\om(\r)$ of the transition density
$\delta \rho(\r,t)$ \cite{magkohofsh} as
\bel{chicollrho}
\chi_{FF}^{\rm coll}(\om)= -\frac{1 }{q_{\rm ext}^\om} \int {\rm
d}\r~{\hat F}(\r)~\delta \rho_\om(\r).
\ee
 Note that in a
macroscopic picture the transition density is the dynamical part
$\delta \rho(\r,t)$ of the particle density,
\bel{partdens}
\rho(\r,t)=\rho_{\rm qs}+ \delta \rho(\r,t).
\ee
 Here, $\rho_{\rm qs}$ is
the quasistatic equilibrium particle density. We define now
${\hat F}$ as related to the variation of the self-consistent mean
field $V$ in the nuclear Hamiltonian:
\bel{hamil}
 {\hat H}={\hat H}_0 + V
= {\hat H}_0 + Q {\hat F} + \frac{1 }{2} Q^2 \left\langle \left(\frac{\partial^2
{\hat H}}{\partial Q^2}\right)_{Q=0}\right\rangle_0+...,
\ee
 where
$H_0$ is an unperturbed Hamiltonian. Introducing the
 collective variable $Q$  ($Q=0$ in equilibrium), one may write
 \bel{foper}
{\hat F}=\left(\frac{\partial {\hat H}}{\partial Q}\right)_{Q=0}=
\left(\frac{\partial V }{\partial Q}\right)_{Q=0}.
\ee
 The total Hamiltonian ${\hat H}_{\rm tot}$ is given by
\bel{hamiltot}
{\hat H}_{\rm tot}={\hat H}+q_{\rm ext}(t){\hat F}.
\ee

As shown in \cite{hofmann,hofbook}, a
conservation of the nuclear energy
$\langle {\hat H} \rangle$ for the Hamiltonian ${\hat H}$
(\ref{hamil}) leads to the equation of motion which is the
secular equation in the Fourier representation,
\bel{seculareq}
k^{-1} +\chi(\om)=0 .
\ee
 The coupling constant $k$ is given by
\bel{kstiffC0chi0}
 -k^{-1} = C(0) + \chi_{{}_{\! FF}}(0),
\ee
$C(0)=\left(\partial^2 E(Q,S)/ \partial Q^2\right)_{Q=0}$ is the stiffness
coefficient of the internal energy
$E(Q,S)$ for the constant nuclear entropy, $S_0$,
and $\chi_{FF}(0)$ is the static (isolated) susceptibility.
$\langle {\hat F} \rangle_\om$ and $Q_\om$ are related then each
other by the self-consistency condition
 \bel{selfconsist}
k \langle {\hat F} \rangle_\om = Q_\om
\ee
 with $Q_\om$ being the
Fourier component of the collective variable $Q(t)$.
The ergodicity condition,
\bel{ergodicity1}
\chi_{{}_{\! FF}}(0)=\chi_{FF}^{\rm ad},
\ee
 with $\chi_{FF}^{\rm ad}$ being the
adiabatic susceptibility was not used really in the derivation of
the self-consistency condition (\ref{selfconsist}) with the
coupling constant $k$ from (\ref{kstiffC0chi0}) in
\cite{hofmann} if the definition of slow variation of the
time-dependent $\langle {\hat F} \rangle$ is employed under the
certain physical conditions, see (3.7-14),(3.3-15) in
\cite{hofmann} and discussion there. The isolated susceptibility
$\chi_{{}_{\! FF}}(0)$ is the static limit $\om \rightarrow 0$ of the
intrinsic response function $\chi_{{}_{\! FF}}(\om)$ defined by
\bel{defrespintr}
\langle {\hat F}\rangle_\om = -\left(Q_\om +
q_{\rm ext}^\om\right) \chi_{{}_{\! FF}}(\om) .
\ee
 Thus, the intrinsic response
function $\chi_{{}_{\! FF}}(\om)$ is related to the collective response
function $\chi_{FF}^{\rm coll}(\om)$ through the relation
(\ref{ddrespchiin}) \cite{bohrmot,hofmann}.  Within the FLDM
formulated below, it is simpler to derive first the collective
response function $\chi_{FF}^{\rm coll}(\om)$ by making directly
use of the
definition (\ref{chicollrho}). For comparison with
the microscopic quantum theory \cite{hofmann} and for study of
the susceptibilities and of the ergodicity property, it is helpful
to present the intrinsic response function $\chi(\om)$ in terms
of the collective response function $\chi_{FF}^{\rm coll}(\om)$ found
from (\ref{ddrespchiin}) as
\bel{respintr}
\chi_{{}_{\! FF}}(\om)=
\frac{\chi_{FF}^{\rm coll}(\om) }{1 - k \chi_{FF}^{\rm coll}(\om)}.
\ee

\subsection{Fermi-Liquid Droplet Model}
\label{fldm}

In this section we follow \cite{magkohofsh} for the basic grounds
of the FLDM \cite{magstrut,kolmagpl} for heavy nuclei taking into account
the quasiparticle Landau--Vlasov theory for the collective
dynamics of the {\it heated} Fermi liquids described in
\cite{heipethrev} and developed in the previous sections in more
details for nuclear matter. The main idea is to apply this
semiclassical theory for the distribution function {\it inside the
nucleus} with the macroscopic {\it boundary conditions}
\cite{strutmagbr,magstrut} like for normal liquids {\it at its
moving surface}. These boundary conditions are used for the
solutions of the dynamical collisional Landau--Vlasov equation
(\ref{landvlas}) {\it coupled with the thermodynamic relations}
for motion in the Fermi-liquid-drop interior. Our derivations are
based on the conception of the linearized dynamics near the {\it
local} equilibrium instead of the global one considered earlier in
\cite{kolmagpl,magkohofsh}. This is important for a {\it low} frequency
region of the nuclear excitations which we are interested in this
review.

We shall consider below small isoscalar vibrations of the nuclear surface
near a spherical shape, which are induced by the external field
$V_{\rm ext}(t)$ (\ref{extfield}). To this end, we define
a collective variable $Q(t)$ in the usual way:
\bel{surface}
R = R_0
\left[1 + Q(t)Y_{L0}({\hat r})\right],
\ee
where $R_0$ is the
equilibrium radius of nucleus, and $Y_{L0}({\hat r})$ is the spherical
harmonics which represent the axially symmetric shapes as
functions of the radius vector angles ${\hat r}$. For $Q(t)$
we expect the form
\bel{collvarq}
Q(t) = Q_\om \exp\left(-i\om
t\right)
\ee
 with the same frequency $\om$ as for the external
field (\ref{extfield}).

\subsubsection{EQUATIONS OF MOTION INSIDE THE NUCLEUS}

 Quasiparticle conceptions
of the Landau Fermi-liquid theory  can be justified
in the nuclear volume, where variations of the density
$\rho(\r,t)$ (\ref{densit}) are small.  Therefore, in
the interior of sufficiently heavy nuclei, one may describe the
semiclassical phase-space dynamics
in terms of the distribution function $\delta f(\r,\p,t)$
(\ref{dfgeqdfleq}) which satisfies the
collisional Landau--Vlasov equation (\ref{landvlas}). We recall now
the equations of Sec.\ \ref{eqmotion} which present the
collective dynamics linearized with respect to the local
equilibrium (\ref{intcoll}). Our interior nuclear collective
dynamics is then described by 6 equations, see (\ref{landvlas}) and
(\ref{consereq}), for the 6 local quantities $\delta \rho(\r,t)$,
$\delta \mu(\r,t)$, $\u(\r,t)$ and $\delta T(\r,t)$ defined inside
of the nucleus as for the nuclear matter. The conserving
equations (\ref{conteq}), (\ref{momenteq}) [or (\ref{navstokeq})
for a potential flow], (\ref{enerconserv}) and (\ref{entropyeq})
are helpful to find them in the semiclassical approximation.

For the isoscalar multipole vibrations of the Fermi-liquid drop
surface (\ref{surface}), we  shall look for the solutions of
(\ref{landvlas}), (\ref{consereq})
in terms of  a superposition
of the plane sound waves (\ref{planewave}) over all  angles
${\hat q}$ of the unit wave vector ${\bf q}$ with the amplitude
$\mathcal{A}_L({\hat q})$,
\bea\l{superpos}
 \delta f(\r,\p,t) &=& \int {\rm d} \Omega_{\bf q}
\mathcal{A}_L({\hat q})
~\delta {\tilde f} (\q,\p,\om)\; \exp [i(\q\r-\om t)]
\nonumber\\
&&{\rm with} \qquad \mathcal{A}_L({\hat q})=
Y_{L0} ({\hat q}_z).
\eea
 Here $L$ is the multipolarity of collective vibrations,
${\hat q}_z$ is the projection of the unit
vector ${\hat q}=\q/q$ on the symmetry $z$-axis.
The Fourier amplitudes
$\delta f(\q,\p,\om)$ are presented as a spherical harmonic
expansion in momentum space,
\bel{tildef}
\delta {\tilde
f}(\q,\p,\om) = \left(\frac{\partial f_{\rm g.e.} (\eps_\p)}{\partial
\eps_\p} \right)_{\rm g.e.} \sum_{l^\prime} \mathcal{A}_{l^\prime}(\om,q)
Y_{l^\prime 0} ({\hat p} \cdot {\hat q}),
\ee
 where $\mathcal{A}_{l^\prime}$ are small vibration amplitudes.
For such
solutions, the velocity field $\u$ corresponds to the
potential flow (\ref{velpoten}).

The relaxation time $\tau$ in (\ref{intcoll}) is assumed to be
frequency and temperature dependent as in (\ref{tautom}).
Following \cite{hofmann,kolmagpl,magkohofsh}, we take the form:
\bel{relaxtime}
 \tau(\om,T)=\frac{\hbar}{{\it \Gamma}(\om,T)},
\ee
where
\bel{widthG}
 {\it \Gamma}(\om,T)=\frac{\pi^2 }{{\it
\Gamma}_0}~ \frac{T^2 + c_o (\hbar \om)^2 }{
1+\frac{\pi^2}{c^2}\left[T^2 + c_o (\hbar \om)^2\right]}.
\ee
 For $c_o$ one has several values. For instance, $c_o = 1 / 4\pi^2$,
according to \cite{landau,ayik}, $c_o = 1 /\pi^2$ follows
from \cite{pinenoz,hofmann}, $3 /
4\pi^2$ from \cite{kolmagpl} and several numbers near these
constants were suggested in \cite{sykbrook,brooksyk}. Formula
(\ref{widthG}) with the $c_o = 1/\pi^2$ and finite cut-off
constant $c$ which weakens the dependence on both
frequency $\om$ and
temperature $T$ at large values of these quantities may in some
sense be compared with the expressions suggested in \cite{hofmann} for
the imaginary part of the self-energy to be used in microscopic
computations \cite{hofmann,hofbook}
[$c$ in (\ref{widthG}) should not be
confused with the sound velocity $c$ used for the description of
normal liquids [see, g.e., (\ref{liqparam}) and (\ref{forstereq})].
In line
of these computations, we shall use ${\it \Gamma}_o=33.3$ MeV and
$c=20$ MeV in our FLDM calculations.
The value of the parameter $c_o=3/4 \pi^2$ is taken as in
\cite{kolmagpl,magkohofsh}. The specific value of this
parameter is not important for the following derivations and
results in this section because we shall apply the temperature-dependent
Fermi-liquid theory for low frequencies and large temperatures.
Note that for $c \to \infty$ the expression (\ref{widthG}) was derived in
\cite{landau,pinenoz,sykbrook,brooksyk,kolmagpl}.

\subsubsection{BOUNDARY CONDITIONS AND COUPLING CONSTANT}
\label{boundary}

The dynamics in the surface layer of nucleus can be described by
means of the macroscopic boundary conditions as in
\cite{magstrut} by using the effective surface approximation
\cite{strutmagbr,strutmagden,magboundcond}. For small vibration
amplitudes, they read:
\bel{bound1}
u_r \Big\vert_{r=R_0} = {\dot
R}(t) \equiv R_0 {\dot Q}(t) Y_{L0}({\hat r}),
\ee
 \bel{bound2}
\Pi_{rr} \Big\vert_{r=R_0} = P_{S} + P_{\rm ext},
\ee
 where $u_r$
and $\Pi_{rr}$ are the radial components of the velocity field
$\u$ (\ref{veloc}) and the momentum flux tensor
$\Pi_{\alpha\beta}$ (\ref{momentflux}) which are determined in
the nuclear volume, see
\cite{bekhal,ivanov,magboundcond,abrditstrut,komagstrvv,abrdavkolsh}
for other (mirror and diffused) boundary conditions used directly
for the distribution function as a solution of the Landau--Vlasov equation.
In the case of the potential flow (\ref{velpoten}),
 we shall use the specific expression for the
momentum flux tensor (\ref{momentfluxpot}) with the shear modulus
($\lambda$) and viscosity ($\nu$) coefficients given by
(\ref{shearmod}) and (\ref{viscos}), respectively. The
surface pressure $P_{S}$, which is due to the tension forces
for the isoscalar motion in symmetric nuclei, is
given by
\bel{surfpress}
P_{S} = \frac{\alpha }{R_0} (L-1)
(L+2)\; Q(t) Y_{L0} ({\hat r}),
\ee
 where $\alpha$ is the surface
tension coefficient, see Sec.\ \ref{npcorivgdr} and
Appendix D
for the isovector asymmetric modes. For the
tension coefficient $\alpha$, we used an expression found in
\cite{strutmagden} within  the  ESA.
This approximation is based on expansion of the nuclear
characteristics, such as the particle density and the total energy in
small parameter $a/R_{0} \sim A^{-1/3}$, where $a$ is the diffuseness
parameter and $R_{0}$ is the mean curvature radius of the nuclear surface
\cite{strutmagbr,strutmagden}, see also \cite{magsangzh,BMRV} and
Appendix D.
In this way, one derives the
nuclear energy expansion [Wiezs\"acker formula (\ref{EvEs}),
(\ref{Espm})], $E=E_{\mathcal{V}} +
E_{S} +...$,
with the volume part of the energy $E_{\mathcal{V}}$
proportional to the particle number $A$,
and the surface energy $E_{S}$, $E_{S}=b_{{}_{\! S}}A^{2/3}$
($b_{{}_{\! S}}=4\pi r_0^2 \alpha$ corresponds to the surface tension
constant $\alpha$, $b_{{}_{\! S}}\approx 20~\mbox{MeV}$,
$r_{{}_{\! 0}}=R_0/A^{1/3} \approx 1.1-1.2$ fm) and so on, see
\cite{strutmagbr,strutmagden,magsangzh,BMRV} and Appendices A.4
(symmetrical nuclei) and D (asymmetrical ones) for more details
(the suffix ``$+$'' is omitted here).
According to (\ref{sigma}) of \cite{strutmagden,BMRV},
\bel{tensionconst}
\alpha \approx 2
\mathcal{C}
\int\limits^\infty_0 {\rm d} r \left(\frac{
\partial \rho_{\rm qs}}{\partial r}\right)^2.
\ee
Here and below we neglect the relatively small
corrections of the order of $A^{-1/3}$ of the
ESA, which are in
particular related to the semiclassical $\hbar$ corrections and
external field. The coefficient $\mathcal C$
appears earlier in
front of the term which is proportional to $\left({{\bf \nabla}}
\rho_{\rm qs}(r)\right)^2$
in the nuclear energy-density formula [see (\ref{enerden})],
$\mathcal{C}=40-60~ {\rm MeV} \cdot {\rm fm}^5$ \cite{BMRV}.

An external pressure
$P_{\rm ext}$ appears in (\ref{bound2}),
where we make connection to the external
potential $V_{\rm ext}$ (\ref{extfield})
\cite{komagstrvv,magkohofsh},
\bel{extpress}
P_{\rm ext} =
-\int\limits_0^\infty dr \> \> \rho_{\rm qs} (r)
\frac{\partial V_{\rm ext}}{\partial r}.
\ee
 For the density in equilibrium,
one has
\bel{denseq}
\rho_{\rm qs} (r) =\rho_{{}_{\! 0}} w(\xi),\quad \xi=\frac{r-R}{a},\quad
a=\sqrt{\frac{\mathcal{C}_{+}\; \rho_\infty\; K}{30\;
b_{\mathcal{V}}^2}}.
\ee
This density is expressed in terms of the
profile function $w(\xi)$ with a sharp decrease from one to zero in
the narrow region of the order of the diffuseness parameter
$a$
near $\xi=0$ as in a step function ($w(\xi) \rightarrow \theta(R-r)$
for $a \rightarrow 0$), $b_{\mathcal{V}} \approx 16$ MeV
is the separation energy per
nucleon
\cite{strutmagbr,magstrut,strutmagden,BMRV}. The value
of equilibrium density inside the nucleus $\rho_{{}_{\! 0}}$
\cite{strutmagbr} is given by
\bel{rho0}
\rho_{{}_{\! 0}}=\rho_\infty\left(1+\frac{6
b_{{}_{\! S}} r_0}{K R_0}\right),
\ee
 where
$\rho_\infty$ is the particle density of the infinite nuclear
matter, $\rho_\infty=3/(4\pi r_0^3)$. The surface energy constant,
$b_{{}_{\! S}}$,  and in-compressibility modulus, $K$, in
(\ref{rho0}) depend on the condition of the constant temperature,
entropy and of the static limit, as shown in Appendix C.
In (\ref{rho0}) and below, we omit the index ${\tt X}$ of these
quantities which specifies one of these conditions, see Appendix C.
For instance, the in-compressibility in
(\ref{rho0})  is denoted simply as
$K=K_{\rm tot}(\om=0)=K^{\varsigma}$, as shown above through
(\ref{incomprtot}), (\ref{dtdrhoexp}) and (\ref{incompradexp}).
The surface energy constant $b_{{}_{\! S}}$ in (\ref{rho0}) is also
identical to the adiabatic one as the in-compressibility
(see Appendix C). The second term in the circle brackets of
(\ref{rho0}) is a small correction proportional to $A^{-1/3}$,
due to the surface tension.

Boundary conditions (\ref{bound1}) and (\ref{bound2}) were
re-derived here from (\ref{conteq}) and
(\ref{momenteq}) where all quantities are now extended to the
surface region with a sharp coordinate dependence of the particle
density as in the approach \cite{magstrut}.
However, we used the specific properties of the {\it
heated} Fermi-liquid drop following the same ESA
\cite{strutmagbr,magstrut,strutmagden}. For the derivation of (\ref{bound2}),
g.e.,
the key equation
(\ref{keybound}) for the Gibbs thermodynamic potential per
particle $g$, which satisfies the thermodynamic relations
(\ref{thermrelmu}), was applied instead of the energy per particle
$\eps$ of \cite{magstrut}. The result
(\ref{bound2}) has the same form as in
\cite{magstrut,magkohofsh} because in its derivation we have
simultaneously to use  (\ref{gradrelener}) of the
temperature-dependent Fermi-liquid theory (with the entropy term
$T d \varsigma$), in contrast to the adiabatic equation (17) of
\cite{magstrut}, see Appendix A.4 for details.

The external field $V_{\rm ext}$ (\ref{extfield}) in
(\ref{extpress}) is determined by the operator ${\hat F}(\r)$
(\ref{foper}), and hence, $V_{\rm ext}$ is concentrated in the surface
region of the nucleus. Indeed, for the operator ${\hat F}({\r})$
(\ref{foper}) in the FLDM, one gets the form
\bel{foperl}
{\hat F}({\r}) = \left(\frac{\delta V }{\delta \rho}~ \frac{\partial
\rho}{\partial Q}\right)_{Q=0}^{\rm qs}
=-R_0\left(\frac{\partial V }{\partial r}\right)_{R=R_0}^{\rm qs}
Y_{L0}({\hat r}),
\ee
 see (\ref{dhdq}). After substitution of
(\ref{foperl}) into (\ref{extpress}) we have
\bel{extpressl}
P_{\rm ext} = - \frac{1 }{kR_0^3}\; q_{\rm ext}(t) Y_{L0}
({\hat r}),
\ee
where
\bel{kfld}
k^{-1} = \frac{K \alpha R_0^4 }{18
\mathcal{C} \rho_\infty}
       \left[1 + \mathcal{O}\left(A^{-1/3}\right)\right]
       \approx \frac{K b_{{}_{\! S}} r_0^5 }{54 \mathcal{C}} A^{4/3}.
\ee
 The integration by parts in (\ref{extpress}) and the
equation (\ref{couplfld}) for the quasistatic coupling constant $k^{-1}$
were used in the derivation of (\ref{extpressl}), (\ref{kfld}), see
the second equation of (\ref{couplxchix}), and also applications to
calculations of the collective vibration modes in \cite{yaf,gzhmagfed}.

\subsubsection{COLLECTIVE RESPONSE FUNCTION}
\label{collresponse}

As shown in \cite{strutmagbr,strutmagden,galiqmod}, the 
linearized dynamic part of the
nucleonic density $\delta \rho (\r,t)$ for the isoscalar modes
can be represented as a sum
of the "volume" and the "surface" term,
\bel{densvolsurf}
\delta
\rho({\r},t) = \delta \rho^{\rm vol}({\r},t) w(\xi)
- \frac{\partial
w}{\partial r} \rho_{{}_{\! 0}} \delta R,
\ee
 where $\delta R$ is the
variation of nuclear radius (\ref{surface}), $\delta
R = R_0 Q(t) Y_{L0} ({\hat r})$, $w$ is defined around
(\ref{denseq})
and in Appendix D. For
isovector vibration modes of the odd multipolarity (dipole), one has to
account for the mass center conservation \cite{kolmagsh,BMR}
[see (\ref{trandenscl})].
The upper index "vol" in $\delta
\rho^{\rm vol}({\r},t)$ of (\ref{densvolsurf}) denotes that
the dynamical particle-density variation
is determined by the equations of motion in the nuclear
volume and is given in terms of the local part $\delta
f_{\rm l.e.}(\eps_{\rm l.e.})$ (\ref{dfleq}) of the distribution function
$\delta f({\r}, {\p},t)$ (\ref{dfgeqdfleq}) through
(\ref{densit}).

Solving (\ref{navstokeq}) with the first
boundary condition (\ref{bound1}), one gets the potential
$\varphi$ in the form
\bel{potensolut}
\varphi({\r},t) = \frac{1 }{q^2}
\frac{qR_0}{ j^\prime_L (qR_0)} {\dot Q}(t) j_{{}_{\! L}}(qr)  Y_{L0}
({\hat r}),
\ee
 where $j_{{}_{\! L}}(x)$ is the spherical Bessel function
and $j^\prime_L(x) = dj_{{}_{\! L}}(x)/dx$.  From the continuity equation
(\ref{conteq}) with (\ref{potensolut}), one has
\bel{densolut}
\delta \rho^{\rm vol}({\r},t) = \rho_{{}_{\! 0}} \frac{qR_0 }{ j^\prime_L(qR_0)}
Q(t) j_{{}_{\! L}}(qr) Y_{L0} ({\hat r}).
\ee
 Therefore, according to
(\ref{densvolsurf}) and (\ref{densolut}), one finds
\bel{trandensolut}
\delta \rho({\r},t) = \rho_{{}_{\! 0}} Q(t) Y_{L0}({\hat
r}) \left[\frac{qR_0 }{ j^\prime_L(qR_0)}  j_{{}_{\! L}}(qr) w(\xi)
- \frac{\partial w}{\partial r} R_0\right].
\ee

With this solution, we may now proceed to calculate the response
function $\chi_{FF}^{\rm coll}(\om)$ (\ref{chicollrho})
by expressing the integral over the coordinates $\r$ for the average
$\langle {\hat F}\rangle_\om$ (\ref{defresp}) in
the numerator of (\ref{chicollrho})
in terms of our collective variable $Q_\om$ given by (\ref{collvarq}).
Indeed, substituting the Fourier transform of
(\ref{trandensolut}) together with ${\hat F}$ from
(\ref{foperl}) into (\ref{chicollrho}), we obtain
\bel{chicollQq}
 \chi_{FF}^{\rm coll}(\om) = - \frac{Q_\om }{k q^\om_{\rm ext}}.
\ee

Using  (\ref{momentfluxpot}),
(\ref{surfpress}), (\ref{extpressl}) and (\ref{potensolut}),  one may
write the second boundary condition (\ref{bound2}) in terms of the
collective variable $Q(t)$ and periodic time dependence of the
external field $V_{\rm ext}$ in the form of the equation of motion
\bel{eqmotionQ}
\mathcal{B}_L(\varx) {\ddot Q} + \mathcal{C}_L(\varx)Q +
\mathcal{Z}_L(\varx) {\dot Q} = - q_{\rm ext}.
\ee
 We have introduced various new
quantities,
\bel{x}
\varx = \frac{\om}{\Om}=\frac{\om R_0 }{ v_{{}_{\! {\rm F}}} s}, \quad \mbox{with}
\quad
\Om=\frac{v_{{}_{\! {\rm F}}}}{R_0}\sim \frac{\eps_{{}_{\! {\rm F}}} }{A^{1/3} \hbar},
\ee
 which  is a complex function of $\om$ by
means of (\ref{despeq}) for the sound velocity $s$ with
(\ref{despfunc})-(\ref{chinumer}),
(\ref{chitemzero})-(\ref{chitemtwo}).
In (\ref{x}), $\Om$ is the characteristic frequency of the classical
particle rotation
in a mean potential well of the radius $R_0$ with the
energy near $\eps_{{}_{\! {\rm F}}}$, as a convenient frequency unit.
Other quantities are defined as
\bel{blx}
\mathcal{B}_L(\varx) = m \rho_{{}_{\! 0}} R_0^5 \frac{j_{{}_{\! L}}(\varx)}{\varx
j^\prime_L(\varx)},
\ee
 \bel{clx} \mathcal{C}_L(\varx) = C_L^{(S)} +
\mathcal{C}_L^{(\lambda)} (\varx),
\ee
 \bel{LDstiff} C_L^{(S)} =
{\alpha} R^2_0 (L-1) (L+2)= \frac{b_{{}_{\! S}} }{4\pi} A^{2/3} (L-1)
(L+2),
\ee
 \bel{clmu} \mathcal{C}_L^{(\lambda)}(\varx) = 2 \lambda R^3_0
\frac{\varx }{j^\prime_L(\varx)} (j^{\prime\prime}_L (\varx) + j_{{}_{\! L}}(\varx))\;,
 \ee
and
 \bel{zlx} \mathcal{Z}_L(\varx) = 2 \nu R^3_0 \frac{\varx }{ j^\prime_L(\varx)}
(j^{\prime \prime}_L (\varx)+ j_{{}_{\! L}}(\varx)).
\ee

From (\ref{eqmotionQ}), one has
\bel{Qqext}
\frac{Q_\om }{q^\om_{\rm ext}} = - \frac{1 }{k D_L(\om)}
\ee
 with
\bea\l{glx}
&&\mathcal{D}_L(\om) \equiv - \left[\mathcal{B}_L(\varx) \om^2 -
\mathcal{C}_L(\varx) + i \om
\mathcal{Z}_L(\varx)\right] ~~~\nonumber\\
&=& \frac{C_L^{(S)} }{j^\prime_{{}_{\! L}} (\varx)}
\left\{ j^\prime_L(\varx) - \frac{6 A^{1/3}\lambda\;
\varx }{b_{{}_{\! S}}(L-1) (L+2) \rho_{{}_{\! 0}} \eps_{{}_{\! {\rm F}}}} \left[
\left(i \nu \om -\lambda\right) \right.\right.~~~\nonumber\\
&\times& \left.\left.  j^{\prime \prime}_L (\varx) +
\left(\frac{s^2
\rho_{{}_{\! 0}} \eps_{{}_{\! {\rm F}}} }{\mathcal{G}_1} - \lambda + i \nu \om \right)
j_{{}_{\! L}}(\varx)\right] \right\}.~~~
\eea
In (\ref{Qqext}),
$k$ is the coupling constant (\ref{kfld}), see (\ref{blx})-(\ref{zlx}).
The kinetic
coefficients $\lambda$ and $\nu$ are the shear modulus $\lambda$
and viscosity $\nu$ given by (\ref{shearmod}) and
(\ref{viscos}), respectively. The two latter quantities enter
(\ref{glx}) in the following combination
\bel{lambvischixz}
\lambda -i \nu \om= s \chixz \rho_{{}_{\! 0}} \eps_{{}_{\! {\rm F}}}
\ee
through a function $\chixz$ defined by
(\ref{chixz}). Finally, the
response function $\chi_{FF}^{\rm coll}(\om)$
(\ref{chicollQq}) with
(\ref{Qqext}) writes
\bel{chicollfldm}
\chi_{FF}^{\rm coll}(\om) =  \frac{1 }{k^2 \mathcal{D}_L(\om)}.
\ee
The poles of this collective response function
are determined by the following equation,
see (\ref{glx}) for $\mathcal{D}_L(\om)$,
\bel{poleseq}
- \mathcal{B}_L(\varx) \om^2 + \mathcal{C}_L(\varx) -
i \om \mathcal{Z}_L(\varx) = 0,
\ee
 with $\,\varx\,$ defined by (\ref{x}).
The complex
solution of the dispersion equation (\ref{despeq}) for $s$
has two branches of the solutions. They are related
asymptotically to the Landau--Placzek heat $s^{(0)}$ and the sound
$s^{(1)}$ solutions considered all in the hydrodynamic limit for
the infinite nuclear matter in Sec.\ \ref{longwavlim}, see
(\ref{shp})-(\ref{gambarone}) for the corresponding
frequencies $\om^{(0)}$ and $\om^{(1)}$. For each branch denoted
below by the same upper index
$n=0,1$ as well in
Sec.\ \ref{longwavlim}, we have the roots of the secular equation
(\ref{poleseq}) written as
\bel{rootomega}
 \om^{(n)} =
\om_i^{(n)} - i \Gamma_i^{(n)}/2, \> \> \> i = 0,1,...,
\ee
 where
$i$ numbers $\om_i^{(n)}$ in order of their increasing magnitude.
We shall consider these roots with $\om_i^{(n)}$ in the frequency
region of about $\hbar \om \siml \hbar \Omega$ which overlaps the
low frequency energy region discussed below. We shall consider
enough large
temperatures $T \simg \sqrt{c_o} \hbar \Omega\; \simg \sqrt{c_o}
\hbar \om$ but smaller than the Fermi energy. This approximately
means $2 MeV \siml T \siml 10 MeV$ for $c_o=3/4\pi^2$ of
\cite{kolmagpl} ($A \sim 200$). (Low temperature limit is about
$1$ MeV for $c_o=1/4\pi^2$ of
\cite{landau,ayik}.)
For above mentioned frequencies $\om$ and temperatures $T$, for which
the quasiparticle and local-equilibrium conceptions of the theory
for the heated Fermi liquids can be applied, the only lowest
solutions have been found in the infinite sequence
(\ref{rootomega}). They are
associated with $i=0$, $1$ and $2$ for the "first sound"
branch $n=1$, and
that with $i=0$ for the "Landau--Placzek"
branch ($n=0$). (Quote marks show that the corresponding names
are realized in fact only asymptotically in the hydrodynamical
limit.) The total response function is the sum of the two
branches mentioned above. The response function
(\ref{chicollfldm}) contains all important information concerning
the excitation modes of the Fermi-liquid drop. One of the ways of
the receipt of this information is to analyze the response
function poles  (\ref{rootomega}) and their residua. However, this
way is often not so convenient and too complicate in the case when
a few poles are close to each other or they belong (or are close)
to the imaginary axis of the complex plane $\om$. More transparent
way which is free from such disadvantages is to describe the
response function in terms of the transport coefficients
\cite{hofmann}.

\subsection{Transport properties for a slow collective motion}
\label{transprop}

The macroscopic response of nucleus to an external field
is a good tool for calculations of
the transport coefficients. To achieve this goal we follow
the lines of \cite{hofmann,hofbook}. For instance, in cranking model type
approximations, one assumes the collective motion to be
sufficiently slow such that the transport coefficients can be
evaluated simply in the "zero-frequency" limit. For a such slow
collective motion we shall study here the transport coefficients
within the FLDM having a look at excitation energies smaller than the
distance between gross shells \cite{strutmag},
\be\l{hom}
\hbar \Omega = \frac{\hbar v_{{}_{\! F}}}{R} \approx
\frac{\eps_{{}_{\! F}}}{A^{1/3}} = 7-10\, {\rm MeV}
\ee
 in heavy nuclei [$\Omega$  is the particle rotation frequency
(\ref{x})] $\hbar\om
\siml \hbar\Omega$, i.e., less than or of the order of the giant
multipole resonance
energies. Within the low collective motion ($\om \siml \Omega$),
we shall deal with first more simple
case of the hydrodynamic approximation which can be applied for
frequencies much smaller than the characteristic "collisional
frequency" $1/\tau$ related to the relaxation time $\tau$
(\ref{relaxtime}), $\om \tau \ll 1$. Using this hydrodynamic
expansion of the macroscopic response function (\ref{chicollfldm})
in small parameter $\om \tau$, we shall look for in
Sec.\ \ref{zerofreqlim} the relation to the "zero frequency limit"
discussed in \cite{hofmann}. Another problem of our interest in
this section is related to the correlation functions, "heat pole
friction" and ergodicity property, see \cite{hofmann,hofbook}. We shall
consider in the next Sec.\ \ref{heatcorrfun} a smaller
frequency region where the nuclear heat pole like the Landau--Placzek mode
for the infinite matter appears within the hydrodynamic
approximation. This subsection will be ended by a more general
treatment of the transport coefficients in terms of the parameters
of the oscillator response function.
The method of \cite{hofmann,hofbook} can be applied for the low frequency
excitations, $\om \siml \Omega$, but also in the case when the
hydrodynamic approach fails, i.e. for $\om \tau \simg  1$.

Following \cite{hofmann,hofbook}, we shall study the "intrinsic"
response function $\chi_{{}_{\! FF}}(\om)$ related to the collective
one $\chi_{FF}^{\rm coll}(\om)$ (\ref{chicollrho}) by
the relation (\ref{respintr}). The collective response function
$\chi_{FF}^{\rm coll}(\om)$ (\ref{chicollfldm}) in the FLDM was
derived straightly from (\ref{chicollrho}) in terms of the
solution for the transition density $\delta\rho$
(\ref{densvolsurf}). The "intrinsic" response function can be
then got with help of (\ref{respintr}). This way is more
convenient in the FLDM with respect to the opposite one used
usually in the microscopic quantum calculations based on the shell
model \cite{hofmann}.

By making use of expansion of the denominator $\mathcal{D}_L(\om)$
(\ref{glx}) of the response function $\chi_{FF}^{\rm coll}(\om)$
(\ref{chicollfldm}) up to fourth order terms in small parameter
$\om \tau$ in the low frequency region $\om/\Omega \ll 1$, and then,
of (\ref{respintr}), one gets the response function in the $F$ mode
in the form:
\bea\l{respfunas}
\chi(\om)&=& k^{-2}\left(-M\om^2- i
\gamma \om + C_{\rm in} - \frac{i \ups C } {2 \om}\right)^{-1},\nonumber\\
C_{\rm in}&=&C - k^{-1},
\eea
 $M$, $C$ and $\gamma$ can be
defined as the $Q$-~mode mass, the stiffness and the friction coefficients
which are the values of $B_L(\varx)$ (\ref{blx}), $C_L(\varx)$
(\ref{clx}) and $\mathcal{Z}_L(\varx)$ (\ref{zlx}) for $\varx=0$ ($\om=0$).
Here and below we omit the low index "FF" in the
"FF"-response functions everywhere when it will not lead to
misunderstanding. Note that the formulas which we derive here and
below for the "intrinsic" response function $\chi(\om)$ can be
applied also to the collective response function
$\chi^{\rm coll}(\om)$ if we only omit the index "in" in $C_{\rm in}$
and in functions of $C_{\rm in}$ denoted by the same index (except
for some approximations based on the specific properties of
$C_{\rm in}$ compared to $C$ and noted below if necessary).
Another argument of the presentation of our results in terms of
the "intrinsic" response functions  is to compare them more
straightly with the discussed ones in \cite{hofmann} in connection
to correlation functions and ergodicity. For the
inertia  $M$ and stiffness $C$,  we obtain the parameters of the
classic hydrodynamic model, namely:
\bel{mass0}
M=\mathcal{B}_L(0) =
\frac{1 }{L} m \rho_{{}_{\! 0}} R^5_0 \equiv M_{\rm LD},
\ee
 the inertia of
irrotational flow, and
\bel{stiffness0}
 C=\mathcal{C}_L(0) =
C_L^{(S)} \equiv  C_{\rm LD}
\ee
 with $C_L^{(S)}$ being the
stiffness coefficient of the surface energy (\ref{LDstiff}). (We
introduced here more traditional notations labeled by index "LD"
which means the relation to the usual liquid-drop model of
irrotational flow). For friction $\gamma$, we arrive at
the temperature dependence typical for hydrodynamics,
\bea\l{friction0}
\gamma &=&\mathcal{Z}_L(0)=2 \nu_{{}_{\! {\rm LD}}} R_0^3 (L-1) \nonumber\\
&=& \frac{3 A (L-1)\eps_{{}_{\! {\rm F}}}
}{5 \pi}~ \frac{\nu_{{}_{\! {\rm LD}}}(T) }{\nu_{{}_{\! {\rm LD}}}(0)}~  \tau \equiv
\gamma_{{}_{\! {\rm LD}}}.
\eea
 Here, $\nu_{{}_{\! {\rm LD}}}$ is the classical hydrodynamic
limit $\nu^{(1)}$ (\ref{shearvisone}) for the viscosity
coefficient, $\tau$ the relaxation time
(\ref{relaxtime}), (\ref{widthG}) for $\om=0$,
\bel{wdthGT}
\tau\equiv\tau(0,T)=\frac{\hbar }{{\it \Gamma}(0,T)},\quad
{\it \Gamma}(0,T)= \frac{\pi^2 T^2 }{{\it \Gamma}_0 \left(1+
\pi^2 T^2 / c^2\right)}.
\ee
 However, our result
(\ref{friction0}) for the
classical liquid-drop model of irrotational flow, if only extended
to include the two-body viscosity, differs from the one found in
\cite{nixsierk,davinixsierk}
by an additional factor of
$(2L+1)/L$, see \cite{magkohofsh,ikmPRC}. We neglected the fourth order
terms in $\tbar$ (Sec.\ \ref{lowtemlim}) in
(\ref{mass0}), (\ref{stiffness0})
and (\ref{friction0}) because of the presence of more important
lower order terms there. For the coefficient $\ups$ in the term
proportional to $1/\om$ in (\ref{respfunas}), one obtains
\bel{gamma0hp}
\ups=\frac{13A^{1/3}\eps_{{}_{\! {\rm F}}} \pi^4 {\bar T}^4 s_0^2
}{60 b_{{}_{\! S}}(L-1)(L+2) \tau} =
\frac{24 \mathcal{G}_0 \eps_{{}_{\! {\rm F}}} A^{1/3}
}{b_{{}_{\! S}}(L-1)(L+2)}~ \nu^{(2)}.
\ee
 The expression in the middle
of these equations turns into zero for the Landau--Placzek kind
($n=0$) of the solutions (\ref{shp}) of dispersion equation
(\ref{despeq}) for the velocity $s$. It is, however, finite for the first
sound mode $n=1$ presented in (\ref{sfirst}) for
$s_0=s_0^{(1)}$. The second equation being true only for the first
sound mode was obtained by making use of (\ref{sfirst}) for
the first sound velocity $s_0^{(1)}$ and (\ref{shearvishp}) for
the viscosity component $\nu^{(2)}$  up to small temperature
corrections of the next order. The both equations (\ref{gamma0hp}) show
the main term in the temperature expansion of the coefficient
$\ups$ in front of $1/\om$ in (\ref{respfunas}). Note that it
appears in the order $\tbar^4$ and can not be neglected for enough
small frequencies $\om$. As seen from (\ref{respfunas})
considered for the case of the collective response, i.e., with
omitted index "in" in $C_{\rm in}$ of (\ref{respfunas}), for
enough small frequencies $\om$, there is the pole which equals
approximately $i\ups/2$. Therefore, the physical meaning of the parameter
$\ups$ (\ref{gamma0hp}) is a "width"  of the overdamped pole in
the asymptotic collective response function $\chi^{\rm coll}(\om)$ for
enough low frequencies. As shown below, this pole and
corresponding pole of the intrinsic response function
(\ref{respfunas}) is overdamped. It is similar to the Landau--Placzek
pole in the infinite nuclear matter and to the nuclear heat pole found in
\cite{hofmann}, see more detailed discussion in
Sec.\ \ref{heatcorrfun}. The "width" $\ups$ (\ref{gamma0hp}) of such
"heat pole" is inversely proportional to the relaxation time
$\tau$ and increases with temperature and particle number.
Note also that this "width" is proportional to the component
$\nu^{(2)}$ (\ref{shearvishp}) of the viscosity discussed in
Sec.\ \ref{viscosthermcond}. It is somewhat similar to the
viscose part of the standard expression for the first sound
"width" $\Gamma$ (\ref{gammaland}) in terms of the first component
$\nu^{(1)}$ of the viscosity coefficient (\ref{shearvisonehp}),
$\Gamma \propto \nu^{(1)}$. However, there is in
(\ref{gamma0hp}) the surface energy constant $b_{{}_{\! S}}$ and
particle number factor $A^{1/3}$ which are both the specific
parameters of a {\it finite} Fermi-liquid drop.

Thus, the denominator of the hydrodynamical response function
(\ref{respfunas}) contains the two friction terms. One of them is
proportional to the friction coefficient $\gamma$, $\gamma
\propto \nu^{(1)}$, and another one which is proportional to
$\Upsilon$ ($\Upsilon \propto \nu^{(2)}$). We shall consider in
the next two Secs. \ref{zerofreqlim} and \ref{heatcorrfun}
the two limit cases neglecting first the heat pole  $\Upsilon$
term for enough large frequencies $\om$ within the hydrodynamic
approximation $\om\tau \ll 1$, and then, the $\gamma$ friction one
for smaller frequencies with the dominating heat pole, respectively.

\subsubsection{HYDRODYNAMIC SOUND RESPONSE}
\label{zerofreqlim}

For enough large frequencies  $\om$ within the frequent
collisional (hydrodynamic) regime,
\bea\l{hydpolcond}
&&\om_{\rm crit} \tau \ll \om \tau \ll 1 \;, \qquad \om_{\rm crit}\tau
=\tau~\sqrt{C\ups /2\gamma}  \nonumber\\
&=&\frac{\pi^2\sqrt{13 \mathcal{G}_0 \mathcal{G}_1}
~s_0~\nu_{{}_{\! {\rm LD}}}(0) }{36(L-1)~\nu_{{}_{\! {\rm LD}}}(T)}~\tbar^2 ,
\eea
 one finds the
 first sound ($i=1$) solution $s$ (\ref{sfirst0}).
In this case, one can neglect
the last term proportional to $1/\om$ compared to the friction
term in the denominator of the asymptotic expression
(\ref{respfunas}). The critical frequency $\om_{\rm crit}$ is defined
in the second equation of (\ref{hydpolcond}) as a frequency
for which these two compared terms coincide,
$\om_{\rm crit}=\sqrt{C\ups/2\gamma}=\om_{{}_{\! {\rm LD}}}\sqrt{M\ups/2\gamma}~$
($\om_{{}_{\! {\rm LD}}}=\sqrt{C/M}$ is the frequency of the  surface
liquid-drop
vibrations). The critical value $\om_{\rm crit}\tau$ increases with
increasing temperature as $\tbar^2$ and does not depend on
particle number for the first sound mode $n=1$.  It equals zero
for the Landau--Placzek branch $n=0$, according to
(\ref{gamma0hp}) for $\ups$. For the $n=1$ mode,
$\om_{\rm crit}\tau$ is small for all temperatures $T ~\siml~
10~\mbox{MeV}$, $\om_{\rm crit}\tau \approx 0.6~ \tbar^2 ~\ll~ 1$ at
typical values of the parameters, $\eps_{{}_{\! {\rm F}}}=40~\mbox{MeV}$ and
$r_0=1.2~\mbox{fm}$, and
for a value $\mathcal{C}$ of the Skyrme forces considered in
\cite{strutmagden},
$\mathcal{C} =80~\mbox{MeV} \cdot \mbox{fm}^5$,
which is somewhat larger than those of \cite{BMRV}
(Sec.\ \ref{npcorivgdr} and Appendix D)
in the ESA, where $A^{-1/3}$ is assumed to be
small.  We took here and below $L=2$ for the quadrupole
vibrations, $\mathcal{F}_0=-0.2$, $\mathcal{F}_1=-0.6$ for the Landau
constants which are close to the values common used for the
calculations of the nuclear giant multipole resonances
\cite{spethwoud,hasse}, a little more "realistic" than in
\cite{kolmagpl,magkohofsh}. For frequencies $\om$ within the
condition (\ref{hydpolcond}), we arrive at the
oscillator-like response function,
\bel{oscresponse}
\chi(\om)\equiv k^{-2}\chi_{\rm osc}(\om) =k^{-2}\left(-M \om^2 -i
\gamma \om +C_{\rm in}\right)^{-1},
\ee
 with all hydrodynamic
transport coefficients presented in (\ref{mass0}),
(\ref{stiffness0}) and (\ref{friction0}). In the middle of (\ref{hydpolcond}),
$\chi_{\rm osc}(\om)$ is
the "intrinsic" oscillator response function which describes the
dynamics in terms of the $Q(t)$ variable for the collective
harmonic oscillator potential. As seen now, the constants $M$, $C$
and $\gamma$ were naturally called above as the transport
coefficients: The collective response function $\chi^{\rm coll}(\om)$
within the approximation (\ref{hydpolcond}) is the same
(\ref{oscresponse}) but with omitted index "in" in the
stiffness coefficient, as noted above. This remark is related also
to the oscillator $QQ$- response function $\chi_{\rm osc}^{\rm coll}(\om)$
useful for the following analysis of the response functions in
terms of the transport coefficients,
\bel{oscrespcoll}
\chi_{\rm osc}^{\rm coll}(\om) =\left(-M \om^2 -i \gamma \om
+C\right)^{-1}.
\ee
 We obtain the $QQ$- response functions from
the $FF$-ones, for instance, from $\chi(\om)$ (\ref{oscresponse}),
simply multiplying by the constant $k^2$ because of the
self-consistency condition (\ref{selfconsist}). Note also that
the condition (\ref{hydpolcond}) for the Landau--Placzek branch of
the solutions for the sound velocity $s$ [see (\ref{shp})], is
always fulfilled for $\om\tau \ll 1$.

In order to compare our results with those of previous
calculations \cite{hofmann},
we introduce the dimensionless quantity
\bea\l{eta}
\eta &=& \gamma / \left(2 \sqrt{M |C|}\right)  \nonumber\\
&=& \frac{2 \eps_{{}_{\! {\rm F}}}}{5 p_{{}_{\! {\rm F}}} r_0 A^{1/6}} ~
\sqrt{\frac{6 L (L-1)
\eps_{{}_{\! {\rm F}}} \mathcal{G}_1 }{(L+2) b_{{}_{\! S}}}}
~\frac{~\nu_{{}_{\! {\rm LD}}}(T)~\tau
}{~\nu_{{}_{\! {\rm LD}}}(0)},
\eea
 see
(\ref{friction0}), (\ref{mass0}) and (\ref{stiffness0}) for
$\gamma$, $M$ and $C$, respectively. The quantity $\eta$ in
(\ref{eta}) characterizes the effective damping rate of the
collective motion. Neglecting small temperature corrections of the
viscosity coefficient $\nu_{{}_{\! {\rm LD}}}=\nu^{(1)}$ in (\ref{eta}), see
(\ref{shearvisone}), and substituting
(\ref{relaxtime}), (\ref{widthG}) for the relaxation time
$\tau$ at $\om=0$, one writes
\bel{eta0}
\eta \approx \frac{2 \hbar
\eps_{{}_{\! {\rm F}}} {\it \Gamma}_0 }{5 \pi^2 p_{{}_{\! {\rm F}}} r_0 A^{1/6}} ~\sqrt{{6 L
(L-1) \eps_{{}_{\! {\rm F}}} \mathcal{G}_1 \over {(L+2)b_{{}_{\! S}}}}} ~
\frac{1 + \pi^2 T^2 /
c^2}{T^2}.
\ee
 This hydrodynamic effective friction $\eta$
mainly decreases with temperature as $1/T^2$. For large
temperatures and finite cut-off parameter $c$,
the dimensionless friction parameter
$~\eta$ (\ref{eta0}) approaches the
constant.

We have the two kind of poles of the response function
(\ref{oscresponse}) as roots of the quadratic polynomial in the
denominator, the overdamped poles, see
\cite{hofmann},
\bea\l{overdamp}
\om_{\pm}^{\rm over}&=&-i \Gamma_{\pm}^{\rm in}/2,\nonumber\\
\Gamma_{\pm}^{\rm in}&=&2\varpi_{\rm in}\left(\eta_{\rm in}\pm
\sqrt{\eta_{\rm in}^2-1}\right),\quad \eta_{\rm in}>1,
\eea
 and the
underdamped ones,
\bel{underdamp}
\om_{\pm}^{\rm under}=
\varpi_{\rm in}\left(\pm \sqrt{1-\eta_{\rm in}^2} - i
{\eta_{\rm in}}~\right), \quad \eta_{\rm in}<1\;.
\ee
 These solutions
depend on the two parameters,
\bel{varpietain}
\varpi_{\rm in}=\sqrt{\frac{|C_{\rm in}|}{M}}~ \qquad \mbox{and} \qquad
\eta_{\rm in}= \frac{\gamma }{2 \sqrt{M|C_{\rm in}|}} .
\ee
 Note also that
the two hydrodynamic poles in (\ref{oscresponse}) coincide
approximately for the both branches $n=0$ and $1$ of solutions to
the dispersion equation (\ref{despeq}) for the velocity $s$. The
difference between these two modes is related only to the last
term proportional to $\ups$ in the brackets of r.h.s. of
(\ref{respfunas}), and it was neglected under the condition
(\ref{hydpolcond}).

For the real and imaginary parts of the response function
$\chi(\om)$ (\ref{oscresponse}) with the help of
(\ref{varpietain}) for the overdamped case (\ref{overdamp}),
for instance, one gets for completeness
[see (\ref{overdamp})-(\ref{varpietain})]
\bea\l{oscrespre}
\chi^{\prime}(\om)&=& \frac{1 }{4 M k^2
\varpi_{\rm in}\sqrt{\eta_{\rm in}^2+1}}
\left(\frac{\Gamma_{-}^{\rm in}
}{\om^2+(\Gamma_{-}^{\rm in})^2/4} \right.\nonumber\\
&-& \left. {\Gamma_{+}^{\rm in} \over
{\om^2+(\Gamma_{+}^{\rm in})^2/4}}\right),
\eea
\bea\l{oscrespim}
\chi^{\prime\prime}(\om)&=& \frac{\om }{4 M
k^2\varpi_{\rm in}\sqrt{\eta_{\rm in}^2+1}} \left(\frac{1
}{\om^2+(\Gamma_{-}^{\rm in})^2/4} \right.\nonumber\\
&-& \left. \frac{1
}{\om^2+(\Gamma_{+}^{\rm in})^2/4}\right).
\eea

For a more simple case of the collective response in the FLDM, we
omit index "in" in formulas of this section [see the comment
after (\ref{respfunas})]. From (\ref{eta}) for $\eta$ with
the parameters used above for the estimate of $\om_{\rm crit} \tau$
of (\ref{hydpolcond}), and the "standard"
${\it \Gamma}_0=33.3$ MeV \cite{hofmann}; one has an overdamped motion, $\eta > 1$,
for all temperatures $T \siml 10~ {\rm MeV}$ and particle numbers
$A \siml 230$, as seen from Fig.\ \ref{fig1}. Moreover, for such
temperatures and particle numbers, one can expand the "widths"
$\Gamma_{\pm}$ in small parameter $MC/\gamma^2=(4\eta^2)^{-1}$,
see (\ref{overdamp}) omitting index "in". From
(\ref{overdamp}) (without index "in") one gets approximately
\bel{overdamp1}
\Gamma_{\pm}=4\varpi \eta \left\{{1-\left(4
\eta^{2}\right)^{-1} \atop {\left(4 \eta^2\right)^{-1}}}\right\}
= 2\left\{{\gamma/M \atop {C/\gamma}}\right\},\,\,
\eta^2~ \gg~ 1.
\ee
 Fig.\ \ref{fig1} shows that the above
mentioned parameter $1/(4 \eta^2)$ for the expansion in
(\ref{overdamp1}) is really small for all considered
temperatures. Using (\ref{mass0}),
(\ref{stiffness0}), (\ref{friction0}) for the transport
coefficients and  the definition of $\tau$ (\ref{wdthGT}) as in
the derivation of (\ref{eta}), (\ref{eta0}), one obtains from
(\ref{overdamp1})
\bea\l{Gammaplus}
\Gamma_{+}&=& \frac{16  \mathcal{G}_1
L(L-1)\eps_{\rm F}^2 }{5 \left(p_{{}_{\! {\rm F}}} r_0\right)^2~A^{2/3}}~
\frac{~\nu_{{}_{\! {\rm LD}}}(T) ~\tau }{ ~\nu_{{}_{\! {\rm LD}}}(0)} \nonumber\\
&\approx& \frac{16 \hbar^2
\mathcal{G}_1 L(L-1)\Gamma_0 \eps_{\rm F}^2 }{5
\pi^2\left(p_{{}_{\! {\rm F}}}r_0\right)^2~A^{2/3}}~ ~
\frac{1 + \pi^2 T^2 / c^2
}{T^2},
\eea
 \bea\l{Gammaminus}
\Gamma_{-}&=&\frac{5 b_{{}_{\! S}} (L+2)
}{6 \eps_{{}_{\! {\rm F}}}~A^{1/3}}~ \frac{~\nu_{{}_{\! {\rm LD}}}(0)
}{~\nu_{{}_{\! {\rm LD}}}(T) ~\tau} \nonumber\\
&\approx& \frac{5 \pi^2 b_{{}_{\! S}} (L+2) }{6 \hbar {\it \Gamma}_0
\eps_{{}_{\!{\rm F}}}~A^{1/3}}~ ~\frac{T^2 }{1 + \pi^2 T^2/c^2 }.
\eea
 One of
the "widths" specified by  $\Gamma_{+}$ (\ref{Gammaplus}) is
mainly the decreasing function of temperature, $\Gamma_{+}
\propto \tau\propto 1/T^2$ at low temperatures. It is typical for
the hydrodynamic modes as the first sound vibrations in normal
liquids; in contrast to another "width" $\Gamma_{-}$
(\ref{Gammaminus}), $\Gamma_{-} \propto 1/\tau \propto T^2$,
similar to the zero sound damping in relation to the $\tau $-
dependence. They both become about a constant for high
temperatures, due to the cut-off factor $c$.

Note, $\Gamma_{+}$ (\ref{Gammaplus}) decreases with particle
number as $A^{-2/3}$ while $\Gamma_{-} \propto A^{-1/3}$, see
(\ref{Gammaminus}). The different $A$-dependence of the "widths"
$\Gamma_{-}$ (\ref{Gammaminus}) and $\Gamma_{+}$
(\ref{Gammaplus}) can not be nevertheless
referred even formally to the
so-called  "one- and two-body
dissipation", respectively.  (Collisions with potential walls
without the integral collision term in the Landau--Vlasov equation
but with the mirror or diffused boundary conditions might lead to
the "widths" proportional to $\Omega$
in (\ref{x}), $\Omega
\propto A^{-1/3}$, as in equation (49) of \cite{komagstrvv} or
through the wall formula \cite{wall,blocki,BMY}.)
They both depend on the collisional
relaxation time $\tau$ and correspond to the "two-body"
dissipation. The latter means here the collisional damping of the
viscose Fermi liquid as in \cite{kolmagpl,magkohofsh}. The
physical source of the damping in the both cases is the same
collisions of particles in the nuclear volume, due to the integral
collision term (\ref{intcoll}) with the relaxation time $\tau$.
We would like to emphasize, however, that the collisional
$\Gamma_{-}$ (\ref{Gammaminus}) depends on the surface energy
constant $b_{{}_{\! S}}$ and disappears proportionally to $A^{-1/3}$
with increasing particle number $A$ like $\Omega$
of (\ref{x}) because we took into account a {\it finite} size of
the system through the boundary conditions
(\ref{bound1}), (\ref{bound2}). An additional overdamped pole with
the "width" $\Gamma_{-}$ (\ref{Gammaminus}) appears because of the
{\it finiteness of the system and collisions inside the nucleus.}
This looks rather in contrast to the wall friction \cite{blocki,BMY}
coming from
the collisions with the only walls of the potential well.

We shall come back now to the {\it intrinsic}
response function $\chi(\om)$ (\ref{oscresponse}).
For the "intrinsic stiffness" $C_{\rm in}$, one has
\bel{Cintr}
C_{\rm in}=-\left(1-kC\right)/k\approx -1/k.
\ee
 In the last equation,
we neglected a small parameter $kC$,
\bea\l{smallpar}
kC&=& \frac{54
(L-1)(L+2) \mathcal{C} }{4 \pi K r_0^5 A^{2/3}} \nonumber\\
&\approx&
\frac{9(L-1)(L+2)
\mathcal{C} }{4 \pi \mathcal{G}_0 \eps_{{}_{\! {\rm F}}} r_0^5 A^{2/3}}
 \approx \frac{3 }{A^{2/3}},
\eea
 for the typical values of the
parameters mentioned above before (\ref{oscresponse}). We
neglected also small temperature corrections of
(\ref{incompradexp}) for the in-compressibility modulus
$K$, $K=K^{\varsigma}$, in the second equation of (\ref{smallpar}).

Using a smallness of the parameter $kC$ (\ref{smallpar}), we shall
get now the relation of the coupling constant $k^{-1}$ with the
isolated susceptibility $\chi(0)$ and stiffness $C$ as in
equation (3.1.26) of \cite{hofmann}. For this purpose, we take the limit
$\om \rightarrow 0$ in (\ref{oscresponse}) for the "intrinsic"
response function $\chi(\om)$ and  expand then the obtained
expression for $\chi(0)$,
$\chi(0)=k^{-2}C_{\rm in}^{-1}=-k^{-2}(1-kC)^{-1}$, in powers of the
small parameter $kC$ (\ref{smallpar}) up to {\it second} order
terms. As result, we arrive at the relation
\bel{kstiffCchi0}
-k^{-1}=\chi(0) + C.
\ee

The liquid-drop transport coefficients $M$ (\ref{mass0}), $C$
(\ref{stiffness0}) and $\gamma$ (\ref{friction0}) can be now
compared with the ones in the "zero frequency limit" $M(0)$,
$C(0)$ and $\gamma(0)$, respectively, defined by
equations (3.1.84)-(3.1.86) in \cite{hofmann}:
\bel{C0}
C(0)=-\left(1/k
+ \chi(0)\right)=C,
\ee
\bel{gamma0} \gamma(0)=-i\left(\partial
\chi(\om)/\partial \om\right)_{\om=0}= \gamma,
\ee
 \bea\l{M0}
M(0)&=&\left(\frac{1}{2}\partial^2
\chi(\om)/\partial\om^2\right)_{\om=0}= M\left(1+\gamma^2
k/M\right)  \nonumber\\
&=&M\left(1+4\eta_{\rm in}^2\right).
\eea
 Expanding $\chi(\om)$
near the zero frequency $\om=0$ in the secular equation
(\ref{seculareq}), see \cite{hofmann}, we assumed here and will
show below that the "intrinsic" response function $\chi(\om)$ is
a smooth function of $\om$ for small frequencies $\om$ within the
hydrodynamic condition (\ref{hydpolcond}). The second and third
equations in (\ref{C0}), (\ref{gamma0}) and (\ref{M0}) were
got approximately in the ESA from (\ref{oscresponse}) up to
small corrections in the parameter $kC$ with help of
(\ref{kstiffCchi0}), second equation in
(\ref{varpietain}) and (\ref{Cintr}). As the liquid drop
stiffness $C$ equals approximately the stiffness in the "zero
frequency limit" $C(0)$, according to (\ref{C0}), the
equation (\ref{kstiffCchi0}) is identical to the relation
(\ref{kstiffC0chi0}) of the general response-function theory
\cite{hofmann} within the same ESA. As seen from
(\ref{C0})-(\ref{M0}), the stiffness $C(0)$ and friction
$\gamma(0)$  equal to the liquid drop parameters, but the inertia
$M(0)$ differs from the liquid-drop mass value $M$ by a positive
correction.

For the definition of transport coefficients in "the zero
frequency limit" (\ref{C0})-(\ref{M0}), we needed to know also
the properties of the "intrinsic" response function in the secular
equation (\ref{seculareq}),
concerning its pole
structure. For the "intrinsic" case the quantity
$\eta_{\rm in}$, see (\ref{varpietain}), plays a
role similar to the effective damping $\eta$ (\ref{eta}) for the
collective motion.
Moreover, $\eta_{\rm in}$  determines the correction to the liquid
drop mass parameter $M$ in (\ref{M0}) for the inertia $M(0)$
in "the zero frequency limit". Due to a smallness of the parameter
$kC$ (\ref{smallpar}), $\eta_{\rm in}$ is much smaller than $\eta$
(\ref{eta0}) for large particle numbers $A \approx 200-230$, as
seen from Fig.\ \ref{fig1},
\bea\l{etaintr}
 &&\eta_{\rm in} = \frac{\gamma
}{ 2 \sqrt{M |C_{\rm in}|}} \approx \eta^2 kC ~~~ \nonumber\\
&\approx& \frac{3 (L-1) \hbar
{\it \Gamma}_0 \eps_{{}_{\! {\rm F}}} }{5 \pi^2 p_{{}_{\! {\rm F}}} r_0}~ \sqrt{
\frac{6 l \mathcal{C}
\mathcal{G}_1 }{\pi \mathcal{G}_0 b_{{}_{\! S}} r_0^5~A}} ~\frac{1 + \pi^2
T^2 / c^2 }{T^2},~~~
\eea
 see (\ref{friction0}), (\ref{mass0}), (\ref{stiffness0}),
(\ref{Cintr}), and (\ref{smallpar}).

For such heavy nuclei ($A \approx 200-230$) and enough large
temperatures, $T \simg 5 ~\mbox{MeV} $, one has formally the
"underdamped" pole structure (\ref{underdamp}) ($\eta_{\rm in} < 1$)
for the parameters selected above. Using the expansion of the
poles $\om_{\pm}^{\rm in}$ (\ref{underdamp}) of the intrinsic
response function in powers of small $\eta_{\rm in}^2$
(\ref{etaintr}) up to terms of the order of $\eta_{\rm in}^4$,  one
writes
\bea\l{underdamp1}
&&\om_{\rm in}^{\pm} = \om_{\rm in}\left[\pm
\left(1-\frac{1}{2} \eta_{\rm in}^2\right) -i \eta_{\rm in}\right] \nonumber\\
&\approx& \pm \varpi_{{}_{\! {\rm LD}}}/\sqrt{kC}-i\Gamma_{+}/4 \quad{\rm
for}\quad \eta_{\rm in}^2 \ll 1;
\eea
 see (\ref{varpietain}), (\ref{overdamp1}) (for $\Gamma_{+}$ on the
very r.h.s.), and (\ref{Cintr}) (for $kC$ there) in the
derivation of the second equation. The "underdamped" poles
$\om_{\rm in}^{\pm}$ approach the real axis on a large distance from
the imaginary one as compared to the liquid drop frequency
$\varpi_{{}_{\! {\rm LD}}}=\sqrt{C/M}$, $|\om_{\rm in}^{\pm}| \gg
\om_{{}_{\! {\rm LD}}}$. They
have a small "width" $2\om_{\rm in}\eta_{\rm in}=\gamma/M\propto 1/T^2$
for our choice of large temperatures ($T \simg 5 {\rm MeV}$); see
(\ref{varpietain}), (\ref{friction0}), (\ref{mass0}), and
(\ref{wdthGT}). By this reason, for the "underdamped" case of
small $\eta_{\rm in}^2$ and low frequencies
$\om \siml \om_{{}_{\! {\rm LD}}}$, the
intrinsic response function $\chi(\om)$ is a smooth function of
$\om$.

For smaller temperatures $T ~\siml~ 4~ \mbox{MeV}$ and for our
parameters used in (\ref{etaintr}), one has the "overdamped" poles
(\ref{overdamp}) of the intrinsic response function $\chi(\om)$,
$\eta_{\rm in} > 1$. For such temperatures, $\eta_{\rm in}$
(\ref{etaintr}) is enough large. We can use therefore the
expansion of the "widths" $\Gamma_{\pm}^{\rm in}$ of
(\ref{overdamp}) in a small parameter $(M
\om_{\rm in}/\gamma)^2=(4 \eta_{\rm in}^2)^{-1}$ (see Fig.\ \ref{fig1}),
\bea\l{overdampintr}
&&\Gamma_{\pm}^{\rm in}= 4\om_{\rm in}\eta_{\rm in}
\left\{{1-\left(4 \eta_{\rm in}^{2}\right)^{-1} \atop {\left(4
\eta_{\rm in}^2\right)^{-1}}}\right\} \nonumber\\
&\approx& 2~\left\{{\gamma/M
\atop {1/(k\gamma)}}\right\} \quad {\rm for} \quad
\left(4\eta_{\rm in}^2\right)^{-1} \ll 1,
\eea
see
(\ref{varpietain}), (\ref{Cintr}) and (\ref{smallpar}). The
"intrinsic width" $\Gamma_{+}^{\rm in}$ in the upper row of
(\ref{overdampintr}) and the "collective width" $\Gamma_{+}$
(\ref{Gammaplus}) [see (\ref{overdamp1}] are the same.
$\Gamma_{-}^{\rm in}$ in the low row has the temperature dependence
as for $\Gamma_{-}$ in (\ref{Gammaminus}) but a different
$A$-dependence, $\Gamma_{-}^{\rm in} \propto A^{1/3}$ [see
(\ref{kfld}) and (\ref{friction0})]. Moreover,
$\Gamma_{-}^{\rm in} \gg \Gamma_{-}$ because of smallness of the
parameter $kC$ (\ref{smallpar}). It becomes clear after dividing
and multiplying the last expression for the $\Gamma_{-}^{\rm in}$ in
(\ref{overdampintr}) by the factor $C$ and using
(\ref{stiffness0}) and (\ref{friction0}).

The "intrinsic width" $\Gamma_{+}^{\rm in}$, see
(\ref{overdampintr}), is mainly larger than $\Gamma_{-}^{\rm in}$.
They become comparable when increasing temperature, i.e.,
$\Gamma_{+}^{\rm in}\simg \Gamma_{-}^{\rm in}$. As $\Gamma_{-}^{\rm in}$
from (\ref{overdampintr}),
\bel{Gammaminintr}
\Gamma_{-}^{\rm in} =\frac{10 \pi \mathcal{G}_0 b_{{}_{\! S}}
r_0^5 ~A^{1/3}}{27 (L-1) \mathcal{C} ~\tau},
\ee
 is large compared to the
characteristic collisional frequency $1/\tau$
(for the same choice of the parameters) the both poles are far
away from the zero, see more discussions of the "intrinsic widths"
below in connection with the heat pole in the next section
\ref{heatcorrfun}. Therefore, the intrinsic response function
$\chi(\om)$ (\ref{oscresponse}) is a smooth function of $\om$ for
the "overdamped" case of enough large $\eta_{\rm in}$ used in
the derivations of (\ref{overdampintr}) as for the
"underdamped" one discussed above. Thus, we expect that the "zero
frequency limit" based on the expansion of the intrinsic response
function $\chi(\om)$ is a good approximation for low frequencies
larger the critical value $\om_{\rm crit}$ within the
hydrodynamic condition (\ref{hydpolcond}) for all considered
temperatures.
It means that the definition of
the transport coefficients in this limit
(\ref{C0})-(\ref{M0}) is justified within the hydrodynamic
approximation (\ref{hydpolcond}).

The correction to the liquid drop mass parameter in the inertia
$M(0)$ (\ref{M0}) is always positive. This correction is the
decreasing function of the temperature and particle number which
can be presented approximately as
\bea\l{masscorr}
&&(M(0)-M)/M =
k\gamma^2/M =4\eta_{\rm in}^2 \nonumber\\
&\propto& ~\left(1 + \pi^2 T^2 /
c^2\right)^2 / \left(A~ T^4\right),
\eea
 see (\ref{etaintr}).
For smaller temperatures when the expansion in
(\ref{overdampintr}) is justified, this correction is equal
approximately to the ratio of the "intrinsic widths"
$\Gamma_{+}^{\rm in}/\Gamma_{-}^{\rm in}$ taken from (\ref{overdampintr}).
The relative mass correction
(\ref{masscorr}) and the "intrinsic width" ratio
$\Gamma_{+}^{\rm in}/\Gamma_{-}^{\rm in}$, see (\ref{overdampintr}),
decreases with temperature $T$
mainly as
$1/T^4$ if $T$ is not too big, as shown in Fig.\ \ref{fig1}.
The dimensionless inertia correction (\ref{masscorr}) is proportional
approximately to $1/A$. The zero frequency mass
$M(0)$ exceeds much the liquid drop inertia and turns asymptotically to
the latter for high temperatures, see Fig.\ \ref{fig1}.
Note, for enough large temperatures $T \simg 5 {\rm MeV}$ and
particle numbers $A \sim 200$ when $\eta_{\rm in}^4$ terms can be
neglected in accordance with  (\ref{etaintr}), all zero
frequency transport coefficients $C(0)$, $\gamma(0)$ and $M(0)$
[see (\ref{C0}), (\ref{gamma0}) and (\ref{M0})] approach the
corresponding liquid drop parameters.

It would be interesting now to get the "overdamped" correlation
function $\psi^{\prime\prime}(\om)$ determined by the imaginary
part of the corresponding response function
$\chi^{\prime\prime}(\om)$ (\ref{oscrespim}) through the
fluctuation-dissipation theorem, see (\ref{fludiptheor}) and
 (\ref{corrfun}). In the semiclassical approximation
(\ref{corrfun}) and (\ref{oscrespim})
for the first sound mode, one writes
\bea\l{corroscim}
 &&\frac{1}{T}\psi^{s \; \prime\prime}(\om)=\frac{2}{\om}\chi^{\prime\prime}(\om)
= \frac{2 }{4 M
k^2\varpi_{\rm in}\sqrt{\eta_{\rm in}^2+1}} \nonumber\\
&\times&\left(\frac{1
}{\om^2+(\Gamma_{-}^{\rm in})^2/4}- \frac{1
}{\om^2+(\Gamma_{+}^{\rm in})^2/4}\right).
\eea
 Using the approximations
as in (\ref{overdampintr}) and (\ref{Cintr}), one gets from
(\ref{corroscim})
\bea\l{corroscim1}
&&\frac{1}{T}\psi^{s
\;\prime\prime}(\om)= \frac{1 }{k} \left(\frac{\Gamma_{-}^{\rm in}
}{\om^2+(\Gamma_{-}^{\rm in})^2/4} \right.\nonumber\\
&-& \left. \frac{1 }{4 \eta_{\rm in}^2}\;
\frac{\Gamma_{+}^{\rm in} }{\om^2+(\Gamma_{+}^{\rm in})^2/4}\right)
\approx \frac{1}{k}\; \frac{\Gamma_{-}^{\rm in}
}{\om^2+(\Gamma_{-}^{\rm in})^2/4}.
\eea
 The second Lorentzian in the
middle is negligibly small compared to the first one because
\bel{gammapmincond}
\Gamma_{+}^{\rm in}\simg \Gamma_{-}^{\rm in} \gg
1/\tau \gg \om,
\ee
 and $4\eta_{\rm in}^2$ is large in these
derivations, see the discussions in between
(\ref{overdampintr}) and (\ref{Gammaminintr}). It seems that
we are left with the Lorentzian term of this correlation function
on very right of (\ref{corroscim1}) which looks as the
Landau--Placzek heat-pole correlation function
(\ref{corrfunhphof}) and equation (4.3.30) of \cite{hofmann} with
obvious constants $\psi^{(0)}$ and ${\it \Gamma}_T$. However, we can
not refer the found correlation function
$\psi^{s\;\prime\prime}(\om)$ (\ref{corroscim1}) to the heat pole
one. The "width" $\Gamma_{-}^{\rm in}$ of (\ref{overdampintr}) in
(\ref{corroscim1}) is finite and large compared to  the
characteristic collision frequency $1/\tau$ which, in turn, is
much larger considered frequencies $\om$, as shown above, see
(\ref{gammapmincond}). The limit $\Gamma_{-}^{\rm in} \to 0$ for a
fixed finite $\omega$  and the corresponding
$\delta(\om)$-function which would show the relation to the heat
pole correlation function do not make sense within the
approximation (\ref{hydpolcond}) used in (\ref{corroscim1}).
In particular, the response (\ref{oscresponse}) and the correlation
(\ref{corroscim1}) functions were derived for
enough large frequencies $\om \gg \om_{\rm crit}$ due to the
condition (\ref{hydpolcond}). Note also that the inertia
parameter $M$ (\ref{M0}) is not zero, as it should be for the heat
pole.

\subsubsection{HYDRODYNAMIC CORRELATIONS AND HEAT POLE}
\label{heatcorrfun}

For lower frequencies $\om$, which are smaller the critical value
$\om_{\rm crit}$, we should take into account the last additional term
in the denominator of (\ref{respfunas}) for the response
function. For such small frequencies, this friction term being
proportional to $\ups$ (\ref{gamma0hp}) becomes dominating as
compared to the liquid-drop one $\gamma=\gamma_{{}_{\! {\rm LD}}}$. Within this
approximation, we shall derive the heat-pole response and
correlation functions, and relate $\ups$
(\ref{gamma0hp}) of (\ref{respfunas}) with the corresponding heat
pole friction. This subsection will be ended by
discussions of the nuclear ergodicity.

For smaller frequencies,
\bel{heatpolcond}
\om \tau \ll
\om_{\rm crit}\tau \ll 1,
\ee
 (see the second equation in
(\ref{hydpolcond}) for the critical frequency $\om_{\rm crit}$),
one can neglect the friction $i\gamma \om$ term in the
denominator of the asymptotic response function (\ref{respfunas})
as compared to the last one, $\gamma \om \ll \ups C/ 2\om$. The
mass term there is even smaller than the friction one for
frequencies $\om \siml \om_{\rm crit}$ for  the considered parameters
and will be neglected too, $M \om^2 \ll \gamma\om$. In this
approximation, from (\ref{respfunas}) one
obtains the heat pole response function $\chi(\om) \approx
\chi^{\rm hp}(\om)$, which is similar to
(\ref{chicollhp}), (\ref{chichpqom}) for the infinite nuclear
matter,
\bel{hprespfunas}
\chi ^{\rm hp}(\om)= \frac{\om }{k^2 C_{\rm in}
\left(\om +i \Gamma^{\rm hp}/2\right)} \approx - \frac{\om }{k
\left(\om +i \Gamma^{\rm hp}/2\right)},
\ee
 where
\bel{heatpole}
\Gamma^{\rm hp}= -C \ups/C_{\rm in} \approx kC\ups.
\ee
 In these derivations, we used the
specific properties of the {\it intrinsic} response functions
which we now are interested in for analysis of the correlation
functions and ergodicity conditions \cite{hofmann}. In
(\ref{hprespfunas}) and in all approximate equations below in this subsection,
we
applied also the expansion in small parameter $kC$
(\ref{smallpar}) as in (\ref{Cintr}).

The real and imaginary parts of the response function
$\chi ^{\rm hp}(\om)$ (\ref{hprespfunas}) are, respectively,
\bea\l{rehprespas}
\chi ^{\rm hp\;\prime}(\om)&=& \frac{\om^2 }{k^2 C_{\rm in} \left[\om^2
+(\Gamma^{\rm hp})^2/4\right]} \nonumber\\
&\approx& - \frac{\om^2 }{k \left[\om^2
+(\Gamma^{\rm hp})^2/4\right]},
\eea
\bea\l{imhprespas}
\chi^{\rm hp\;\prime\prime}(\om)&=& - \frac{\om \Gamma^{\rm hp} }{2 k^2 C_{\rm in}
\left[\om ^2+(\Gamma^{\rm hp})^2/4\right]} \nonumber\\
&\approx& \frac{\om
\Gamma^{\rm hp} }{2 k \left[\om ^2+(\Gamma^{\rm hp})^2/4\right]}
\eea
up to small $kC$ corrections, see (\ref{smallpar}).

We shall derive now the correlation function
$\psi^{{\rm hp}\,\prime\prime}(\om)$  applying the
fluctuation-dissipation theorem (\ref{corrfun}) to the "intrinsic" response
function $\chi ^{{\rm hp}\;\prime\prime}(\om)$ (\ref{imhprespas})
obtained in the asymptotic limit (\ref{heatpolcond}). From
(\ref{corrfun}) and (\ref{imhprespas}) one gets
\bea\l{corrfunashp}
\frac{1}{T}\psi^{{\rm hp}\;\prime\prime}(\om)&=& \frac{2}{\om}
\chi ^{{\rm hp} \;\prime\prime}(\om) = -
\frac{\Gamma^{\rm hp}}{k^2 C_{\rm in}\;
\left[\om ^2+(\Gamma^{\rm hp})^2/4\right]} \nonumber\\
&\approx&
\frac{\Gamma^{\rm hp}}{k \left[\om ^2+(\Gamma^{\rm hp})^2/4\right]}.
\eea
 This correlation
function looks as the Landau--Placzek peak for the infinite Fermi
liquid, see (\ref{corrfunhphof}),
\bel{psifunhp}
\psi^{{\rm hp}\;\prime\prime}(\om)= \psi_{\rm hp}^{(0)} \frac{\hbar
{\it \Gamma}_T^{\rm hp} }{(\hbar \om)^2 + ({\it
\Gamma}_T^{\rm hp})^2/4}.
\ee
 It is identical to the r.h.s. of equation
(4.3.30) in \cite{hofmann}, but with the specific parameters
$\psi^{(0)}=\psi_{\rm hp}^{(0)}$ and
${\it \Gamma}_T={\it \Gamma}_T^{\rm hp}$,
\bel{psi0GammaT}
\frac{1}{T} \psi_{\rm hp}^{(0)}=-\frac{1}{k^2 C_{\rm in}}~
\approx \frac{1}{k}, \quad {\it \Gamma}_T^{\rm hp} =\hbar
\Gamma^{\rm hp}\approx \hbar kC\ups.
\ee
 The "width" $\Gamma^{\rm hp}$ in
(\ref{psi0GammaT}) is much smaller than the characteristic
collision frequency $1/\tau$,
\bel{Gammahptau}
\Gamma^{\rm hp} = \frac{13
\pi^4 \mathcal{G}_1 \mathcal{C}~ T^4 }{20\eps_{\rm F}^4b_{{}_{\! S}}r_0^5 {\it
\Gamma}_0~A^{1/3}~\tau} \ll \frac{1}{\tau},
\ee
see (\ref{gamma0hp})
and (\ref{smallpar}). The relationship (\ref{Gammahptau}) for
$\Gamma^{\rm hp}$ is in contrast to the one (\ref{gammapmincond}) for
the "intrinsic overdamped widths" $\Gamma_{\pm}^{\rm in}$
(\ref{overdampintr}) which are much larger the collision
frequency $1/\tau$ for the same selected parameters at all
temperatures $T \siml 10$ MeV and particle numbers
$A=200-230$.

For the following discussion of the friction coefficients,
we compare now the
"width" $\Gamma^{\rm hp}$ (\ref{heatpole}), (\ref{Gammahptau}) with
$\Gamma_{-}^{\rm in}$ in (\ref{overdampintr})
[see (\ref{friction0}), (\ref{stiffness0}), (\ref{gamma0hp}),
(\ref{sfirst}), (\ref{Cintr}) and (\ref{kfld})],
\bel{zetaintr}
\frac{\Gamma^{\rm hp}}{\Gamma_{-}^{\rm in}} =
\frac{\gamma C \ups }{2C_{\rm in}^2} \approx
\frac{1}{2}\gamma C k^2 \ups.
\ee
 For all temperatures and particle numbers
which we discuss here, this ratio is small,
\bel{zetapar}
\frac{\Gamma^{\rm hp}}{\Gamma_{-}^{\rm in}}
 = \frac{351 \pi^2 (L-1) \mathcal{C}^2 }{800
b_{{}_{\! S}}^2 r_0^{10}~A^{2/3}} ~\tbar^4 \approx
\frac{15\tbar^4 }{A^{2/3}}.
\ee
In the second
equation of (\ref{zetapar}) we used the same values of the
parameters as in (\ref{smallpar}). Note that the "width" of the
Landau--Placzek peak $\Gamma^{(0)}$, $\Gamma^{(0)} \sim
\tau_q^2/\tau \ll 1/\tau$ for $\taubar \ll 1$, is similar to
$\Gamma^{\rm hp}$ and is unlike $\Gamma_{\pm}^{\rm in}$
(\ref{overdampintr}) in (\ref{corroscim1}) for the
hydrodynamical sound correlation function. In contrast to the
hydrodynamical sound case, see (\ref{corroscim1}), we can consider
(\ref{corrfunashp}) for the correlation function approximation
in the zero width limit $\Gamma^{\rm hp} \to 0$ (or in the zero
temperature limit $T \to 0$) taking any small but finite frequency
$\om$ under the condition (\ref{heatpolcond}). Therefore, for such
frequencies $\om$, the correlation function (\ref{corrfunashp}) can
be approximated by $\delta(\om)$-like function as in
(\ref{corrdeltafunlim}) for the correlation function of the
infinite Fermi liquid (\ref{corrfunhphof}). Because of a very
close analogy of equation for the correlation function
$\psi^{{\rm hp}\,\prime\prime}(\om)$ (\ref{psifunhp}) to the
Landau--Placzek peak for the infinite Fermi-liquids in the
hydrodynamic limit, see (\ref{corrfunhphof}), and to
equation (4.3.30) of \cite{hofmann}, we associate the pole
(\ref{heatpole}) and corresponding asymptotics of the response
 (\ref{hprespfunas})
and correlation (\ref{corrfunashp}) functions with the "heat pole". As in the case of the
infinite nuclear matter, this pole for the finite Fermi-liquid
drop is situated at zero frequency $\om=0$. Moreover, they  are
both called as the "heat poles" because they disappear in the
zero temperature limit $T \to 0$ in line of the discussions near
equation (4.3.30) of \cite{hofmann} and after. In the case of the
infinite matter, we can see this property from
(\ref{corrfunhp}) because
${\tt C}_{\mathcal{V}}/{\tt C}_{\mathcal{P}}
\to 1 $) [or due to (\ref{psi0}) for $\psi^{(0)}$ in
(\ref{corrfunhphof})]. For the finite Fermi-liquid drop, the
reason is that $\ups \to 0$ in the zero temperature limit $T \to 0
$, see (\ref{gamma0hp}), and the only hydrodynamical sound
condition (\ref{hydpolcond}) is then satisfied with the response
function (\ref{oscresponse}) and correlation function
(\ref{corroscim1}) where the heat pole is absent, see the
discussion after (\ref{corroscim1}).

To get more explicit expressions for $\psi^{(0)}$ and ${\it
\Gamma}_T$ of (\ref{psi0GammaT}) we use now (\ref{kfld}),
(\ref{gamma0hp}), (\ref{friction0}) and (\ref{stiffness0}) for the
coupling constant $k^{-1}$, parameter $\ups$, friction $\gamma$
and stiffness $C(0)$, respectively. With these expressions, one
obtains approximately from (\ref{psi0GammaT})
\bel{psi0GammaT1}
\frac{1}{T} \psi_{\rm hp}^{(0)} = \frac{\mathcal{G}_0\eps_{{}_{\! {\rm F}}} b_{{}_{\! S}}
r_0^5 A^{4/3} }{9 \mathcal{C}} \approx
2 A^{4/3},
\ee
 \be\l{GammaT2m}
{\it \Gamma}_{T}^{\rm hp} = \hbar \Gamma^{\rm hp}
=\frac{13 \pi^6 \mathcal{G}_1
\mathcal{C} }{20 \eps_{\rm F}^4 b_{{}_{\! S}} r_0^5 {\it \Gamma}_0 ~
A^{1/3}}~ \frac{T^6 }{\left(1+\pi^2T^2/c^2\right)}.
\ee
  In the derivation of
(\ref{GammaT2m}), we used (\ref{sfirst}) for the first
sound solution $s_0=s_0^{(1)}$  ($n=1$) in (\ref{gamma0hp})
for $\ups$ and (\ref{wdthGT}) for the relaxation time $\tau$.
For simplicity, we neglected small
temperature corrections in the viscosity coefficient $\nu^{(1)}$
(\ref{shearvisone}) and in the first sound velocity $s_0^{(1)}$
(\ref{sfirst}). Other approximations are the same as well in the
derivation of (\ref{smallpar}) for $kC$ used in
(\ref{GammaT2m}) through (\ref{heatpole}). The
temperature dependences of the "intrinsic overdamped width"
$\Gamma_{-}^{\rm in}$ (\ref{Gammaminintr}) and "heat pole one"
$\Gamma^{\rm hp}$ (\ref{GammaT2m}), (\ref{Gammahptau}) are different,
namely $\Gamma^{\rm hp} \propto \tbar^4/\tau(0,T)$ and $\Gamma_{-}^{\rm in}
\propto 1/\tau(0,T)$ where the temperature dependence of the
relaxation time $\tau(0,T)$ can be found in (\ref{wdthGT}). The
both "widths" are the growing function of temperature as in
\cite{hofmann} but with a different power. The dependence on
particle number $A$ completely differs for these compared poles
being the growing function of $A$ for the "width"
$\Gamma_{-}^{\rm in}$, $\Gamma_{-}^{\rm in} \propto A^{1/3}$, and
decreasing function of $A$ for the $\Gamma^{\rm hp}$, $\Gamma^{\rm hp}
\propto A^{-1/3}$.  As noted above, like for the Landau--Placzek
peak [see (\ref{corrfunhphof}), (\ref{GammaT}) and (\ref{psi0})],
the heat pole with the "width" $\Gamma^{\rm hp}$ (\ref{GammaT2m})
exists only in heated systems with a temperature $T \neq 0$.
However, in contrast to the result (\ref{GammaT}),
(\ref{gambarhpt}) for the ${\it \Gamma}_T$ of the heat pole in the
infinite Fermi liquid, the heat pole "width" ${\it \Gamma}_T$
(\ref{GammaT2m}) disappears with increasing particle number $A$,
i.e., ${\it \Gamma}_T \rightarrow 0$ for $A \rightarrow \infty$. It
allows us to emphasize also that this kind of the heat pole
appears only in a {\it finite} Fermi system.

The correlation function $\psi^{{\rm hp}\;\prime\prime}(\om)$
(\ref{psifunhp}) was obtained approximately
near the pole $-i\Gamma^{\rm hp}/2$, see (\ref{heatpole}). The
corresponding $QQ$- correlation function
$\psi_{QQ}^{{\rm hp}\;\prime\prime}(\om)=k^2\psi^{{\rm hp}\;\prime\prime}(\om)$
is identical to the oscillator correlation function
$\psi_{\rm osc}^{{\rm hp}\;\prime\prime}(\om)$ defined through the imaginary part of
$\chi_{\rm osc}(\om)$ from the second equation of
(\ref{oscresponse}) at the zero mass parameter $M$, $M=0$, see
\cite{hofmann},
\bea\l{oscresphp}
\frac{1}{T}\psi_{\rm osc}^{{\rm hp}\,\prime\prime}(\om)&=&
\frac{2}{\om}\chi_{\rm osc}^{{\rm hp}\;\prime\prime}(\om)=
\frac{2}{|C_{\rm in}|}~ \frac{
|C_{\rm in}|/\gamma ^{\rm hp} }{\om^2 + \left(C_{\rm in}/\gamma
^{\rm hp}\right)^2} \nonumber\\
&\approx& 2 k ~ \frac{ 1/\left(k\gamma ^{\rm hp}\right)
}{\om^2+ 1/\left(k\gamma ^{\rm hp}\right)^2},
\eea
 see again
(\ref{Cintr}) for the last approximation. The
response (\ref{hprespfunas}) and
correlation (\ref{corrfunashp}) functions are
identical to the corresponding oscillator ones (\ref{oscresphp})
with a friction coefficient $\gamma ^{\rm hp}$,
\bel{frictionhp}
\gamma^{\rm hp} =2|C_{\rm in}|/\Gamma^{\rm hp} \approx 2k^{-1}
\left(\Gamma^{\rm hp}\right)^{-1} \approx 2/(k^2C\ups).
\ee
 Here,
the same equation (\ref{Cintr}) was used, $\Gamma^{\rm hp}$ is given by
(\ref{heatpole}), (\ref{GammaT2m}), $k^{-1}$ is the coupling
constant (\ref{kfld}). [For $C$ and $\ups$ in
(\ref{frictionhp}), one has (\ref{stiffness0}) and
(\ref{gamma0hp}), respectively.] According to
(\ref{corroscim1}) and (\ref{corrfunashp}) for the correlation
functions $\psi ^{s\;\prime\prime}(\om)$ and
$\psi^{{\rm hp}\;\prime\prime}(\om)$, with the help of (\ref{heatpole}),
(\ref{frictionhp}) and (\ref{gamma0}), one gets
\bea\l{corrfriction}
(1/2T)\psi^{s\;\prime\prime}(0)&=&\gamma=\gamma(0),\nonumber\\
(1/2T)\psi^{{\rm hp}\,\prime\prime}(0)&=& 2/(k^2C\ups)=\gamma^{\rm hp},
\eea
 in
line of the last right equation in (3.1.85) of \cite{hofmann}.

For the friction $\gamma ^{\rm hp}$ (\ref{frictionhp}) related to the
"heat pole width" $\Gamma^{\rm hp}$ (\ref{heatpole}), one
approximately writes
\be\l{frictionhpm}
\gamma ^{\rm hp} =\frac{40 \mathcal{G}_0 {\it \Gamma}_0
\eps_{\rm F}^5b_{{}_{\! S}}^2 r_0^{10} A}{117 \pi^6
\mathcal{G}_1 \mathcal{C}^2 T^6 }\;\left(1+\frac{\pi^2T^2}{c^2}\right),
\ee
 see
(\ref{GammaT2m}) for $\Gamma^{\rm hp}$ and (\ref{kfld}) for
$k^{-1}$ in (\ref{frictionhp}). We neglected here small
temperature corrections in the adiabatic in-compressibility
modulus $K=K^{\varsigma}$ (\ref{incompradexp}). The heat pole
friction $\gamma ^{\rm hp}$ (\ref{frictionhpm}) is proportional to
$1/T^6$ for smaller temperatures and $1/T^4$ for larger ones (due
to the cut-off parameter $c$). This decreasing temperature
dependence is much more sharp compared to the liquid drop one
$\gamma$ (\ref{friction0}), (\ref{wdthGT}); $\gamma \propto
1/T^2$ for smaller temperatures, and $\gamma$ is a constant for
large ones.
Notice, according to (\ref{frictionhp}), the "width" ratio
$\Gamma^{\rm hp}/\Gamma_{-}^{\rm in}$ (\ref{zetaintr}),(\ref{zetapar})
has a clear physical meaning as the ratio of the hydrodynamic
friction coefficient $\gamma$ (\ref{friction0}) to the heat pole
one $\gamma^{\rm hp}$ (\ref{frictionhpm}),
\bel{zintr1}
\Gamma^{\rm hp}/\Gamma_{-}^{\rm in} \approx \gamma/\gamma ^{\rm hp} \approx
\gamma(0)/\gamma ^{\rm hp}.
\ee
 A smallness of this ratio shown above
claims that the heat pole friction $\gamma ^{\rm hp}$ is much larger
than the typical hydrodynamic one $\gamma$, see more discussions
concerning this comparison of different friction coefficients
below.

As seen from the inequalities (\ref{heatpolcond}) with the
definition of $\om_{\rm crit}$ from (\ref{hydpolcond}), the heat
pole appears only in the "sound" branch $n=1$ and does not exist
for the Landau--Placzek branch of the solutions of
(\ref{despeq}) for $s$. We realize it immediately noting that
the width parameter $\ups$ (\ref{gamma0hp}) is proportional to
$s_0$ which is finite for $n=1$ and zero for $n=0$ case, see
(\ref{shp}) and (\ref{sfirst}), respectively.

As shown in \cite{hofmann}, for enough small ${\it \Gamma}_T$, the
coefficient $\psi^{(0)}$ in front of the Lorentzian-like correlation
function, see (\ref{corrfunhphof}) and (\ref{psifunhp}),
is related to the difference of
susceptibilities,
\bel{psichi}
(1/T) \psi^{(0)}
= \chi^T-\chi(0)=\chi^T-\chi^{\rm ad} +\chi^{\rm ad}-\chi(0).
\ee
Neglecting a small difference $\chi^T-\chi^{\rm ad}$ according to
(\ref{chiTchiad1}), (\ref{chiTchiad}), see Appendix C
and \cite{hofmann} for details, one notes that the ergodicity
condition (\ref{ergodicity1}) means smallness of the $(1/T)
\psi^{(0)}$ compared to the stiffness $C$.

However, from (\ref{psi0GammaT}), (\ref{psi0GammaT1}) one gets
a large quantity $(1/T) \psi_{\rm hp}^{(0)}/C \approx 1/(kC) \gg 1$.
Note, in the derivations of
(\ref{corrfunashp}), (\ref{psi0GammaT}) we took first $\om
\rightarrow 0$ (small $\om\tau$) for the finite ${\it
\Gamma}_T^{\rm hp}$, see also (\ref{gamma0hp}) for $\ups$ in the
second equation of (\ref{psi0GammaT}), and then, considered
${\it \Gamma}_T^{\rm hp} \rightarrow 0$ (small temperature limit $T
\to 0$ ).
We emphasize that the limits $\om \rightarrow 0$ and ${\it
\Gamma}_T^{\rm hp} \rightarrow 0$ are not commutative, i. e., the
result of the correlation function calculations depends on the
order of executing of these two operations like for the infinite
Fermi-liquid matter \cite{forster}. This
is obvious if we take into account that the "heat pole" last term
in the denominator of (\ref{respfunas}) appears in the next
($\tbar^4$) order in $\tbar$  and is proportional to $1/(\om
\tau)$ in contrast to the other classical (sound) hydrodynamic
terms, i.e., this $\ups$-term turns into zero for ${\it \Gamma}_T
\to 0 $ ($T \to 0$).

The relation (\ref{psichi}) was derived in \cite{hofmann} using
the opposite sequence of the above mentioned limits, namely, first
${\it \Gamma}_T \rightarrow 0$ and then $\om \rightarrow 0$ in
line of the recommendations of Forster \cite{forster} [first ${\it
\Gamma}_T \propto q^2 \rightarrow 0$ [or $\tau_q \to 0$, see
(\ref{GammaT}), (\ref{shp})], and then, $\om \rightarrow 0$ ($s
\rightarrow 0$) for the infinite Fermi-liquid]. In this case
there is no contradiction with the ergodicity for the finite
Fermi-liquid drop. In the limit ${\it \Gamma}_T^{\rm hp} \rightarrow
0$ ($T \to 0$) for {\it a finite} value of $\om$, the condition
(\ref{hydpolcond}) is fulfilled instead of (\ref{heatpolcond}),
and the "heat pole" term proportional to $1/\om$ in the
denominator of the response function (\ref{respfunas}) disappears
within the ESA used in the FLDM, as noted above. It means
formally that one can neglect $\psi_{\rm hp}^{(0)} $ in (\ref{psifunhp}),
and we have small quantities on the both sides of
(\ref{psichi}) taking into account the ergodicity condition
(\ref{ergodicity1}) derived in Appendix C.
It is not
obvious that the relation (\ref{psichi}) can be also derived for
the opposite consequence of the above mentioned limit transitions
unlike the Forster recommendations, i.e., taking first limit $\om
\rightarrow 0$ for a finite ${\it \Gamma}_T$, and then, considering
the limit ${\it \Gamma}_T \rightarrow 0$. In particular,
(\ref{corroscim1}) for the overdamped correlation function was
obtained for the last choice of the limit sequences.
Equation (\ref{corroscim1}) does have also the Lorentzian-like shape
but it is not related to the "heat pole" because the coefficient
in front of the Lorentzian is {\it not} equal to
$\chi^T-\chi(0)$.
This equation was derived only for
large $\Gamma_{-}^{\rm in}$ compared to the $1/\tau$, see
(\ref{gammapmincond}),
and is true {\it only} under these conditions and within
inequalities (\ref{hydpolcond}). There is no a $\delta(\om)$
function-like peak in (\ref{corroscim1}) for all possible
variations of the parameters for which this equation was derived.
The overdamped shape of the correlation function like
(\ref{corrfunhphof}) does not mean yet that this function is
the "heat pole" one though the opposite statement is true. We
point out again that
 $(1/T)\psi^{(0)}$ (\ref{corroscim1}) is really
large compared to the stiffness $C$, $(1/T)\psi^{(0)}=1/k$, and the
ergodicity condition (\ref{ergodicity1}) is fulfilled rather than
the relation (\ref{psichi}) between $(1/T)\psi^{(0)}$ and
$\chi^T-\chi(0)$ within the hydrodynamic conditions
(\ref{hydpolcond}).

Following
the Forster's recommendations \cite{forster},
i.e., take first the limit of small ${\it \Gamma}_T$  (${\it
\Gamma}_T \rightarrow 0$) or small temperature ($T \rightarrow 0$),
one gets the typical hydrodynamic response function
(\ref{oscresponse}) without "heat pole" terms. The next limit $\om
\rightarrow 0$ ($\om\tau \rightarrow 0$) in (\ref{oscresponse})
leads to the finite value,
\bel{isorespk}
\chi(0)=\frac{1
}{k^2C_{\rm in}} \approx - \frac{1}{k} -C,
\ee
 up to the
relatively small corrections of higher order in parameter $kC$
(\ref{smallpar}). This is in line of Appendix C,
and the
ergodicity condition (\ref{ergodicity1}) is fulfilled for the
finite Fermi-liquid drop within the ESA. Note that we accounted
above for the $kC$ correction at the second order
in (\ref{isorespk}). In this way, we got the relation
(\ref{kstiffCchi0}) between the coupling constant $k^{-1}$,
isolated susceptibility $\chi(0)$ and stiffness $C$ provided that
the condition (\ref{hydpolcond}) is true, see also
(\ref{kstiffC0chi0}) with the stiffness $C(0)=C$ of the "zero
frequency limit". Note also that the "heat pole" response
function $\chi^{\rm hp}(\om)$ (\ref{hprespfunas}) has a sharp peak
near the zero frequency, and hence, is not smooth, i.e., "the zero
frequency limit" for the transport coefficients can not be
applied in the case (\ref{heatpolcond}).

Thus, all properties of the finite Fermi liquids within the ESA
concerning the ergodicity relation (\ref{ergodicity1}), as applied
to (\ref{psichi}), are quite similar to the ones for the
infinite nuclear matter [besides the expressions (\ref{Gammaminus}),
$\Gamma_{-} \propto b_{{}_{\! S}}/A^{1/3}$, and
 (\ref{Gammahptau}),  $\Gamma^{\rm hp} \propto 1/( b_{{}_{\! S}} A^{1/3})$,
themselves depending on  $b_{{}_{\! S}}$]. Our study of these properties is helpful
for understanding the microscopic shell-model approach
\cite{hofmann,hofivmag,hofbook}. We point out
that the strength function corresponding to the asymptotics
(\ref{respfunas}) is the curve with the two maxima which are
related to the "heat pole" and standard (sound) hydrodynamic
modes. However, for intermediate frequencies $\om $ of the order
of $\om_{\rm crit}$ in the low frequency region, see
(\ref{hydpolcond}) and (\ref{heatpolcond}), the asymptotic
response function (\ref{respfunas}) can not be presented exactly
in terms of a sum of the two  oscillator response functions like
(\ref{oscrespcoll}). For instance, in this case we have the
transition from the "heat pole" mode to the sound hydrodynamic
peak, and the response function (\ref{respfunas}) is more
complex. We have a similar problem when the hydrodynamic
condition $\om \tau \ll 1$ becomes not valid. However, as shown in
the next subsection, such problems can be overcome approximately
using an alternative definition for the transport coefficients
suggested in \cite{hofmann}.

For larger frequencies, i.e., for $\om \tau$ larger or of the order
of 1, but within the low frequencies $\om$ smaller than $\Omega$,
see (\ref{x}), the equation for the collective motion becomes more
complicate. It is not reduced generally speaking to the second
order differential equation with the constant coefficients as in
the zero frequency limit of the hydrodynamic approach
(\ref{hydpolcond}). As shown and applied in
\cite{hofivyam,hofmann} (see also \cite{magkohofsh} in
connection to the FLDM), the problem of the definition of transport
coefficients can be nevertheless overcome by  defining them
through a procedure of fitting an oscillator response function
(\ref{oscrespcoll})  to selected peaks of the collective
response function $\chi_{QQ}^{\rm coll}(\om)$ of
(\ref{chicollfldm}) with respect to the parameters $M$, $C$ and
$\gamma$. Here such a fitting procedure would also be adequate for
temperatures mentioned above, especially because our response
function (\ref{chicollfldm}) has several poles (\ref{rootomega}),
for instance, with $i = 0,1,2; n=1$ and $i=0;n=0$. Some of them
are the overdamped poles close to the imaginary axis in the
$\om$-complex plane. This procedure can be done analytically in
the zero frequency limit provided that the response function
(\ref{chicollfldm}) can be approximated
by the oscillator
response functions as in (\ref{oscrespcoll}) or by
$\chi_{\rm osc}^{\rm hp}(\om)$ in (\ref{oscresphp}). In this case, we
have analytical fitting of the collective response function
(\ref{chicollfldm}) by these oscillator response functions and
get the expressions for the transport coefficients
(\ref{C0}) -
(\ref{M0}) in the zero frequency limit
(\ref{hydpolcond}) [or (\ref{frictionhp}) for the heat pole
friction in a smaller frequency region (\ref{heatpolcond})]. For
larger frequencies, we need to carry out the fitting procedure
numerically.

We should also comment a little more the definition of the
transport coefficients in the zero frequency limit in connection
to the one through the fitting procedure to avoid some possible
misunderstanding. The transport coefficients in the zero frequency
limit can be related to the "intrinsic" response function and its
derivatives taken at $\om \rightarrow 0$ \cite{hofmann,hofbook};
see (\ref{C0}),
 (\ref{gamma0}), and (\ref{M0}). For
application of this method of the transport coefficient
calculations, we should be carefully in the case when we have
several peaks in the strength function but we need to get the
transport coefficient, for instance, for the second or more high
peaks. In these cases the zero frequency limit might be applied
also, but we have first to remove all lower peaks in the collective
response function and take then the corresponding "intrinsic"
response function and its derivatives without these lower peaks.
In practical applications, this limit for the transport
coefficients obtained in a such way is close to the same limit for
the oscillator response function which fits the selected peak. The
latter could be also the second or more high one.

We shall consider now the hydrodynamical approximation
$\om \tau \ll 1$ for the response function, see
(\ref{respfunas}), for the two cases:  The sound response
function (\ref{oscresponse}) for the sound condition
(\ref{hydpolcond}) and the heat-pole response function
(\ref{hprespfunas}) for the heat pole condition
(\ref{heatpolcond}). The corresponding correlation functions are
the sound correlation function (\ref{corroscim1}) and the
heat-pole correlation one (\ref{corrfunashp}). These two different
approximations are realized for different consequence of the
limit transitions, i.e., the approximate result depends on the
consequence of their applying. The heat pole case
(\ref{hydpolcond}) is realized when we take first the limit $\om
\to 0$ for a finite width ${\it \Gamma}_T$, and then, ${\it
\Gamma}_T \to 0$ (or zero temperature limit $T \to 0$). This leads
approximately to the $\delta(\om)$-like function for the
correlation function. In contrast to this, the sound pole case
(\ref{heatpolcond}) is realized when we take first ${\it
\Gamma}_T \to 0$ (or $T \to 0$) to remove the last heat pole term
proportional to $\ups$ in the hydrodynamical response
(\ref{respfunas}), and then, $\om \to 0$. We like to follow this
last consequence of the limit transition in line of the Forster
recommendations \cite{forster} when we have the response
(\ref{oscresponse}) and correlation (\ref{corroscim1}) functions
without heat pole. In this case the transport coefficients for
$\om \tau \ll 1$ are the standard hydrodynamical ones
(\ref{C0}), (\ref{gamma0}) and (\ref{M0}) related to the parameters
of the standard hydrodynamical model (\ref{stiffness0}),
(\ref{friction0}) and (\ref{mass0}), respectively. Exception
should be done for the
modified mass parameter in (\ref{M0}) which turns into the
irrotational flow inertia (\ref{mass0}) for high temperatures.

\subsection{Discussion of the results}
\label{discuss}

In this subsection, we discuss the
results of the FLDM calculations for the collective response
function and transport coefficients. We shall explain now in more details
the application of the general fitting procedure for the definition
of the transport coefficients. We discuss also the
stiffness and inertia parameters found within the FLDM.
This subsection will be ended by
the discussion of the friction versus temperature. One
of the important points of this discussion is the "heat pole"
friction and comparison with the quantum shell-model calculations
\cite{hofivyam,hofmann,hofbook}.

We show first the imaginary part of the response function
$\chi_{QQ}^{\rm coll}(\om)$ (\ref{chicollfldm}) (its strength)  for
different temperatures in Fig.\ \ref{fig2}.  The total collective
response function $\chi_{QQ}^{\rm coll}$ is presented in
Fig.\ \ref{fig2} as a sum of the two branches $n=0$ and $1$ of
eigen-frequencies $\om^{(n)}$, see (\ref{rootomega}), in the
imaginary part (strength) of the response function
(\ref{chicollfldm}). They are related to the two different
solutions of the dispersion equation (\ref{despeq}) for the sound
velocity $s^{(n)}$. These solutions are similar to the
Landau--Placzek (Raleigh) and the sound (Brillouin) ones in normal
liquids. The latter are approached exactly by $s^{(0)}$ and
$s^{(1)}$ solutions for sound velocity $s$ in the
hydrodynamic limit $\om\tau \rightarrow 0$, which are related to
the eigen-frequencies of the infinite-matter vibrations
$\om^{(0)}$ (\ref{shp}) and $\om^{(1)}$ (\ref{sfirst}),
respectively. The integral collision term is parametrized in
terms of the relaxation time $\tau(\om,T)$ (\ref{relaxtime}),
(\ref{widthG}) with $c=20~\mbox{MeV}$. We took the nucleus Pu-230
with particle numbers $A=230$ as an example of enough heavy
nucleus.

For the intermediate temperatures $4~\mbox{MeV} ~\siml~ T~ \siml~
6~\mbox{MeV}$ we have the three peak structure. More detailed
plots for smaller frequencies are shown in Fig.\ \ref{fig3} for the
temperature $T=6~\mbox{MeV}$ for which the first two peaks ("heat
pole" and usual hydrodynamic ones) are seen better in a normal
scale. In Fig.\ \ref{fig3}, we show also the separate contributions
of the two branches $n=0$ (dotted line) and $n=1$
(dashed one) for the eigen-frequencies $\om^{(n)}$
(\ref{rootomega}) calculated from the secular equation
(\ref{poleseq}) at each $s^{(n)}$ ($n=0,1$) as in
Fig.\ \ref{fig2}. We present also the imaginary part of the
asymptotic response function (\ref{respfunas}) obtained
analytically above in the hydrodynamic frequent-collision limit.
As seen from Fig.\ \ref{fig3}, we found from (\ref{chicollfldm})
the $n=1$ mode with the two ($i=0, 1$) peaks  and the $n=0$ mode
with one peak ($i=0$) for small frequencies $\om$ and small
parameter $\om\tau$ in agreement with asymptotics
(\ref{respfunas}). The heat pole contribution is shown separately
by the dotted curve.
Note that the two curves for $i=0 $ and $1$ at $n=1$
in Fig.\ \ref{fig3} coincide because they both
were calculated without the last $\Upsilon$ term in
(\ref{respfunas}). For the dotted curve,
one has $\Upsilon \propto
s_0^{(0)}=0$, and for the dashed one,
the last $\Upsilon$ term in
(\ref{respfunas}) is omitted  under the asymptotical sound
condition (\ref{hydpolcond}). Therefore, the upper asymptotical
data (thin solid) marked also by the condition (\ref{hydpolcond})
are in factor about two larger than the dotted,
or dashed, or asymptotical (\ref{respfunas}) ones.

The third peak in Fig.\ \ref{fig2} appears for intermediate
temperatures and larger frequencies. This peak is coming from the
third pole $i=2$ which belongs to the branch $n=1$ in
(\ref{rootomega}). This is the essentially Fermi-liquid
underdamped mode due to the Fermi-surface distortions related to
the shear modulus $\lambda$ given by (\ref{shearmod}). Such a peak is
moving from a large zero-sound-frequency region of the giant
resonances  to smaller frequencies with increasing temperature.
The second ($i=1$) peak in the $n=1$ branch and first ($i=0$)
peak in the $n=0$ one in the low frequency region ($\om\tau \ll
1$) are related to the overdamped motion described approximately
by the overdamped oscillator response function like
(\ref{oscrespcoll}) for the same cut-off parameter $c=20~\mbox{MeV}$.
For $c=\infty$ the overdamped motion turns into the underdamped
one for large temperatures $T \simg 7~\mbox{MeV}$. The next
(third) peak in a more high frequency region ($\om\tau ~\simg~ 1$)
corresponds to the underdamped mode for the both $c$ values. The
first lowest peak in Figs.\ \ref{fig2} and \ref{fig3}, which is not seen
in Fig.\ \ref{fig2} being too close to the ordinate axis and
studied separately in Fig.\ \ref{fig3}, is due to the overdamped
"heat pole" $i\ups/2$ in the collective response function, see
(\ref{gamma0hp}) for $\ups$. The most remarkable property of
this "heat pole" peak for smaller temperatures is that it has
mainly a very narrow width (\ref{gamma0hp}) which increases with
the temperature as $T^6$, see the comments concerning the heat
pole "width" after (\ref{gamma0hp}) and (\ref{GammaT2m}).
This is in
contrast to the temperature behavior of the width $\Gamma_{-}$
(\ref{Gammaminus}) like $T^2$ for the hydrodynamic sound peak at
large temperatures.
Fig.\ \ref{fig2}  shows the three  peaks only for the intermediate
temperatures $4~ \siml~ T ~\siml~ 6~\mbox{MeV}$ because for
smaller temperatures the third peak moves to the high frequency
region larger $\Omega$ corresponding to the giant resonances and
first peak is very close to the ordinate axis.

The transport coefficients for such two- or three resonance
structure were calculated by a fitting procedure of the oscillator
response functions to the selected peaks. We subtract first the
"heat pole" peak known analytically, see (\ref{hprespfunas}), from
the total response function (\ref{chicollfldm}). We are left then
with the two-humped curve and fit then it by the sum of the two
oscillator response functions as (\ref{oscrespcoll}). One of
them which fits the first (hydrodynamic) peak in the curve with
the remaining two maxima is the overdamped oscillator response
function ($\eta ~>~ 1$) and other one (more high in the low energy
region) corresponds to the underdamped motion ($\eta ~<~ 1$). In
this way, we get the two consequences of the transport coefficients
presented in Figs.\ \ref{fig4}-\ref{fig7}. In these Figures,
the heavy squares
are related to the second, hydrodynamic-sound peak of
Figs.\ \ref{fig2},\ref{fig3} for the mostly overdamped modes with
the effective friction $\eta~
>~1$. The open squares show the third Fermi-liquid peak (see
Fig.\ \ref{fig2}) related to the underdamped motion ($\eta~ < ~1$)
and Fermi-surface distortions, very specific for the
Fermi liquids, in contrast to the normal liquids.

For the temperatures smaller about $6~\mbox{MeV}$ the second peak
$i=1$ in the total response function is {\it overdamped} and is
coming from the two poles $(i=1,n=1)$ and $(i=0,n=0)$ which are
close to the standard hydrodynamic approach. The third peak, due
to the Fermi-surface distortions as noted above, can not be found
in principle in the hydrodynamic limit. The main difference
between the second and third peaks can be found in the comparison
of the stiffness coefficient $C$ with the liquid-drop value
$C_{\rm LD}$ obtained both from the fitting procedure mentioned above.
For the third ("Fermi liquid" in sense of the relation to the
Fermi surface distortions specific for the Fermi liquids, in
contrast to normal ones) peak the stiffness $C$ is much high
than the liquid drop value $C_{\rm LD}$ in contrast to the second
(typical hydrodynamical) one for which the stiffness $C$ is very
close to $C_{\rm LD}$ almost for all temperatures, see
Fig.\ \ref{fig4}. It means that the third peak is essentially of
different nature than the second one because exists only due to
the Fermi-surface distortions. A measure of these distortions is
the anisotropy (or shear modulus) coefficient $\lambda$, see
(\ref{shearmod}), which disappears in the hydrodynamic limit.

For enough large temperature (larger than or of the order of $7~
\mbox{MeV}$~) all three peaks are not distinguished in
Fig.\ \ref{fig2}. For such large temperatures the fitting procedure
is a little modified to select these three peaks which are close
to each other. For the finite $c=20~\mbox{MeV}$ and all large
temperatures presented in Fig.\ \ref{fig2} nearly $7-10~\mbox{MeV}$,
we have one wide peak which can be analyzed as the superposition
of the three peaks, namely the "heat-pole", usual overdamped
hydrodynamic and underdamped "Fermi-liquid" ones. Subtracting the
first "heat pole" peak [see (\ref{hprespfunas})] as for lower
temperatures, we fit then the remaining curve by the only one
overdamped oscillator function like (\ref{oscrespcoll}) for $\eta
> 1$. We subtract then again this overdamped fitted oscillator function
from the response function (\ref{chicollfldm}) without the heat
pole one (\ref{hprespfunas}) and fit the rest by the single
underdamped oscillator. The found parameters of the two last
oscillator response functions are used as initial values for the
iteration fitting procedure of the sum of the two oscillator
response functions of the same types to the response function
(\ref{chicollfldm}) (without the heat pole). The found transport
coefficients are presented in Figs.\ \ref{fig4}-\ref{fig7}. For
enough large temperature nearly $10~\mbox{MeV}$ in the case
$c=\infty$ the only one underdamped oscillator can be used for
fitting procedure of one peak [after an exclusion of the heat
pole from (\ref{chicollfldm})].

We show also the mass parameters found from the above described
fitting procedure for several selected peaks in Fig.\ \ref{fig5}.
For the third "Fermi-liquid" peaks the mass parameter $M$ is
close to the liquid drop values $M_{\rm LD}$ related to the
irrotational flow. The mass parameter of the second
"hydrodynamic" peak, due to the mixture of the identical
($i=1,n=1$) and ($i=0, n=0$) poles, is significantly smaller than
the liquid drop value $M_{\rm LD}$ but finite. For the first "heat
pole" ($i=0,n=1$) peak the mass parameter can be approximated
only by zero. As noted above,
the stiffness parameter for the
third peak is much larger than the one for other (hydrodynamic)
poles which is mainly close to the liquid drop value (see
Fig.\ \ref{fig4}).
As shown in Figures \ref{fig4} and \ref{fig5},
for enough large temperatures the temperature
dependences of the stiffness ($C$) and mass ($M$) parameters are
close to their zero frequency limit, see (\ref{C0}) for $C(0)$
and (\ref{M0}) for $M(0)$. For
smaller temperatures, the inertia $M(0)$ [Fig.\ \ref{fig5}]
becomes essentially larger than that found from the response function
(\ref{chicollfldm}). It is in contrast to the stiffness $C(0)$
which is identical to the liquid-drop quantity in the
semiclassical limit $\hbar \rightarrow 0$ when $C(0)$ does not contain
quantum shell corrections.

Figs. \ref{fig6} and \ref{fig7} show the results for the friction
coefficient $\gamma/\hbar$ versus the temperature for the
collective response function $\chi_{QQ}^{\rm coll}(\om) =k^2(T)
\chi_{FF}^{\rm coll}(\om)$ related to the $\chi_{FF}^{\rm coll}(\om)$
(\ref{chicollfldm}). We used here the same parameters as well
in Figs.\ \ref{fig2} and \ref{fig3} for the response function.
The solid line for the friction
$\gamma$ (\ref{friction0}) corresponds to the response function
(\ref{oscrespcoll}) in the hydrodynamic limit (\ref{hydpolcond}),
the same as for the zero frequency approach (\ref{gamma0}). The
heavy squares show the result of the fit of (\ref{chicollfldm})
to the oscillator response function (\ref{oscrespcoll}).
We presented also the "heat
pole" contribution to the friction obtained from the fitting
procedure by one "heat pole" (overdamped) oscillator response
function (\ref{hprespfunas}), see circles in Fig. \ref{fig7}. We
might compare the results of this fit to the friction
analytically found in terms of the heat pole asymptotics
(\ref{frictionhp}) valid for smaller temperatures and shown by
solid thin lines in Figs.\ \ref{fig6} and \ref{fig7}. They are in a
good agreement for smaller temperatures where the overdamped
"heat pole" with the "width" $\ups$ (\ref{gamma0hp}) is more
important. This "heat pole" friction is too big as compared to
other friction components related to the hydrodynamical-sound
(full squares) and "Fermi-liquid" poles in  the usual scale of
Fig.\ \ref{fig6}. Therefore, we use the logarithmic scale in
Fig.\ \ref{fig7}.

Our FLDM friction, except for the "heat pole" one, is similar to
the corresponding result of SM calculations
\cite{hofivyam,hofmann}, see Fig.\ \ref{fig8}. A large SM
friction coming from the diagonal matrix elements in
Fig.\ \ref{fig8} and standard hydrodynamic friction
(\ref{friction0}) as well as heavy squares shown in
Figs.\ \ref{fig6} and \ref{fig7} are obviously similar. All these curves
for temperatures $T \simg 2~\mbox{MeV}$ show the mainly
diminishing friction, $\gamma \propto \tau \propto 1/T^2$ roughly
like in hydrodynamics, see (\ref{friction0}). Some deflection of
the friction temperature dependence in Fig.\ \ref{fig6} for large
temperatures $T$ from usual hydrodynamic one $1/T^2$, i.e., a
constant asymptotics is related to a different temperature
behavior of the ${\it \Gamma}(0,T)$ (\ref{widthG}) for a finite
and infinite cut-off parameter $c$: This ${\it \Gamma}(0,T)$ goes
to a constant for large temperatures if $c$ is finite and to zero
for $c=\infty$, see the solid and dashed lines in Fig.\ \ref{fig6}.

It is noted also a similarity concerning the third
("Fermi-liquid") peak presented by the lower open squares with
mainly increasing friction in Figs.\ \ref{fig6}, \ref{fig7} and by
joint full squares in Fig.\ \ref{fig8}. For $c=20~\mbox{MeV}$ and
temperatures smaller about $10~\mbox{MeV}$ the friction  of this
mode increases, see Figs.\ \ref{fig6}, \ref{fig7}, in contrast to
the standard hydrodynamic behavior (for $c=\infty$ this friction
increases first up to about $6-7~\mbox{MeV}$, and then, decreases
at larger temperatures). In Fig.\ \ref{fig8} the lower curve with
growing dependence on the temperature for $c= 20~\mbox{MeV}$ was
obtained by excluding the contribution of the diagonal terms in
the response function within the quantum approach based on the
SM, see \cite{hofmann,hofbook} for the detailed explanations.
Within the conceptions of the FLDM and classical hydrodynamics of
the normal liquid drops the first "heat pole" friction obtained
for enough small frequencies (\ref{heatpolcond}) within the
hydrodynamic collision regime $\om \tau \ll 1$ at finite
temperature is the physical mode which can be excited when this
regime might be realized like the Landau--Placzek pole for normal
liquids. However, the hydrodynamic collision regime  being still
within a low frequency region (enough small collision frequency
$1/\tau$) is expected to be not achieved in fission experiments.
Therefore, the friction is related mainly to another Fermi-liquid mode
corresponding to the only third peak owing to the Fermi-surface
distortions. The friction of this mode is much smaller than the
hydrodynamic one for small temperatures, and they become
comparable for high ones. The Fermi-surface distortion friction
can be characterized by completely other, mainly growing
temperature behaviour, see the lower curve marked by open squares
in Figs.\ \ref{fig6} and \ref{fig7}. Concerning the SM calculations, it
seems that we should omit the diagonal matrix elements, see
\cite{hofmann}, because of similar arguments: The hydrodynamic
collision regime seems to be not realized for nuclear fission
processes. (These diagonal matrix elements might correspond to the
physical hydrodynamic mode if it is excited, say in another
systems like a normal liquid drop). The quantum shell-model
friction without contributions of diagonal matrix elements is
related probably to another non-hydrodynamic mode, such as the third peak
for a Fermi-liquid drop, and this might be the physical reason
for an exclusion of these matrix elements.

Note that in the SM response-function derivations the diagonal
matrix elements mentioned above do not contribute in the Forster's
sequence of the limit transitions discussed at the end of the
previous section, first ${\it \Gamma}_T \to 0$, for exclusion of
the diagonal matrix elements at finite $\omega$, and then, $\om \to
0$ limit. In this case, we have not contribution of the diagonal
matrix elements in the friction, and we are left with the low
friction curves with increasing temperature dependence shown in
Figs.\ \ref{fig6}-\ref{fig8}. For the opposite limit sequence if we
consider first the small frequency limit $\om \to 0$ for the
finite (large) ${\it \Gamma}_T$ we have the contribution of
diagonal matrix elements to the friction shown by the curves
decreasing with temperature which correspond to the hydrodynamic
limit here. As noted above, the exclusion of diagonal matrix
elements for this last case could be justified because the
physical condition of the hydrodynamic limit $\om \tau \ll 1$ is
not probably realized in fission processes. In that case, we
expect the increasing friction; which has essentially other,
non-hydrodynamic nature. We might interpret it within the FLDM as
related to the third peak, due to the Fermi-surface distortions.


\section{NEUTRON-PROTON CORRELATIONS AND IVGDR}
\l{npcorivgdr}

\subsection{Extensions to the asymmetric nuclei}
\l{extensiontoas}

The FLDM was successfully applied for studying the global properties
of the isoscalar multipole
giant resonances having nice agreement of their basic characteristics,
such as the energies and sum rules, with experimental data
for collective excitations of heavy nuclei
\cite{strutmagden0,kolmagpl}. For the collective excitation modes in
asymmetric neutron-proton nuclei, the FLDM was straightly extended
 in particular for calculations of the IVGDR structure
\cite{denisov,kolmagsh,BMV,BMR}.
In this case,
one has the two coupled (isoscalar and isovector) Landau--Vlasov equations
for the dynamical variations of distribution functions,
$\delta f^{}_{\pm}(\r,\p,t)$, in the nuclear phase-space volume \cite{kolmagsh},
\bea\l{LVeq}
\frac{\partial }{\partial t}
    \delta f^{}_{\pm}(\r,\p,t) &+&
    \frac{\p }{m_\pm^*}
   {\bf \nabla}_r  \left[ \delta f^{}_{\pm}(\r,\p,t) ~~\right.\nonumber\\
  &+&\left.
    \delta \left(\eps-\eps_{{}_{\! {\rm F}}}\right)
    \delta \eps_{\pm}
    +V_{\rm ext}^{\pm}\right]
    =\delta St^{}_{\pm}.~~
\eea
Here $m_\pm^*$ are the isoscalar (+) and isovector (-) effective masses,
$\eps=p^2/(2 m_\pm^*)$ , $\eps_{{}_{\! {\rm F}}}=
(p_{\rm F}^{\pm})^2/(2 m_\pm^*)$  is the Fermi energy.
The splitting between
the Fermi momenta $p_{\rm F}^{\pm}$ is originated by the difference of the
neutron and proton potential well depths, due to the Coulomb interaction
\cite{migdal,kolmagsh},
\bel{momentumdef}
p_{\rm F}^\pm =p_{{}_{\! {\rm F}}}(1\mp \Delta,\quad \Delta=2(1+\mathcal{F}_0')
\mathcal{I}/3,
\ee
 where
$\mathcal{F}_0'=3J/\eps_{{}_{\! {\rm F}}}-1$  is the isotropic isovector
Landau constant of the quasiparticle interaction (\ref{fasymint}),
$J$ is the volume symmetry energy constant \cite{myswann69}.
The asymmetry parameter
$~\mathcal{I}=(N-Z)/A~$ is assumed to be small
near the nuclear stability line, $N$ and $Z$ are the neutron and proton
numbers in the nucleus ($A=N+Z$).
In (\ref{LVeq}), for the dynamical variations of the
self-consistent quasiparticle
(mean-field) interaction
$\delta \eps_\pm (\r,\p,t)$, one has
\bel{interaction}
\delta \eps_{\sigma}=\pi^2\hbar^3
\sum_{\sigma'}\left[\frac{F_{0,\sigma \sigma'}
}{p_{\rm F}^{\sigma'} m_{\sigma'}^*}
~\delta \rho_{\sigma'} +
\frac{m F_{1,\sigma \sigma'} }{m_{\sigma'}^* p_{\rm F}^\sigma\left(
p_{\rm F}^{\sigma'}\right)^2}
~\p \cdot {\bf j}_{\sigma'}\right].
\ee
The sum is taken over the sign index $\sigma=\pm$.
The dynamical variations
of the quasiparticle interaction
$\delta \eps_\pm$  at the first order with respect to the equilibrium energy
$p^2/(2 m_\pm^*)$ is defined through those of the
particle density,
\bel{densitydef}
\delta \rho_\pm(\r,t)=
\int \frac{2{\rm d}\p }{(2 \pi\hbar)^3}\;\delta f_\pm(\r,\p,t)
\ee
[zero $\p$-moments
of the dynamical distribution functions $\delta f_{\sigma}(\r,\p,t)$
(\ref{planewave})], and the current density,
\bel{currentdef}
{\bf j}_\pm(\r,t)=\int \frac{2{\rm d}\p }{(2 \pi\hbar)^3}
~\frac{\p}{m}~ \delta f_\pm(\r,\p,t)
\ee
(their first $\p$-moments).
The Landau interaction constants $F_{l,\sigma \sigma'}$ in
(\ref{interaction}) are defined by
expansion of the scattering quasiparticle's interaction amplitude
$F_{\sigma \sigma'}(\p,\p')$
in the Legendre polynomial series,
\bel{fasymint}
F_{\sigma \sigma'}(\p,\p')=F_{0,\sigma \sigma'} + F_{1,\sigma \sigma'}
{\hat p} \cdot {\hat p}'+ ..., \quad {\hat p}=\p/p.
\ee
For the sake of simplicity,
we assume that $F_{l,\sigma \sigma'}$ is a symmetrical matrix
($l \leq 1)$ and
$F_{l,pp}-F_{l,nn}$
is of the second order in parameter $\Delta$ [see below (\ref{LVeq})], and
can be neglected in the linear approximation with respect to $\Delta$,
\bel{symmetry}
F_{l,pp}=F_{l,nn},\quad F_{l,pn}=F_{l,np}.
\ee
Thus, we arrive at usual simple definitions for the isoscalar
$F_0$ and $F_1$ and isovector $F_0'$ and $F_1'$ Landau interaction
constants \cite{migdal,kolmagsh},
\bea\l{landauconst}
F_{l}&=&(F_{l,pp}+F_{l,pn})/2,
\nonumber\\
 F_{l}^\prime&=&(F_{l,pp}-F_{l,pn})/2,
\qquad l=0,1.
\eea
These constants are related to the
Skyrme interaction constants
in the usual way \cite{liu}.
The isoscalar ($\mathcal{F}_0$)
and isovector ($\mathcal{F}_0'$) isotropic interaction constants are
associated with
the volume in-compressibility modulus $K$ and symmetry energy constant  $J$,
respectively.
The anisotropic interaction constants $\mathcal{F}_1$ and
$\mathcal{F}_1'$ correspond to
the effective masses by equations $m_{+}^*=m(1+\mathcal{F}_1/3)$ and
$m_{-}^*=m(1+\mathcal{F}_1'/3)$.
The periodic time-dependent external
field in (\ref{LVeq}) is given by  $V_{\rm ext} \propto \exp(-i \om t)$
as in (\ref{extfield}). The collision term $\delta St^{}_{\pm}$ is
taken in the simplest $\tau_\pm$-relaxation time approximation
(\ref{intcoll}).
For simplicity, we consider in this section the low temperature
limit $T \rightarrow 0$ neglecting the difference between the local and
global equilibrium for the quasistatic distribution function.

Solutions of these equations (\ref{LVeq}) associated with the dynamic
multipole particle-density variations,
$\delta \rho_\pm(\r,t) \propto Y_{L0}({\hat r})$ in
the spherical coordinates $r$, $\theta$ , $\varphi$,  can be found
in terms of a superposition of the plane waves (\ref{planewave})
over angles of the wave vector $\q$  as
\bea\l{planewaves}
&& \delta f_{\pm}=\delta\left(\eps-
(p_{\rm F}^{\pm})^2 /2m_{\pm}^*\right) ~~\nonumber\\
&\times&\int {\d}\Om_{\bf q} \mathcal{A}_{\pm} Y_{L0}\left({\hat q}\right)~
\exp\left[i\left({\bf q}{\r}-\om t\right)\right] ,  \quad {\hat q}=\q/q,~~
\eea
$\om=p_{\rm F}^{\pm}s q \sqrt{NZ/A^2}/m_\pm^*$,  $q=|\q|$.
The factor $~\sqrt{NZ/A^2}~$ ensures the conservation of the center-of-mass
 position for the odd vibration multipolarities $~L~$
\cite{eisgrei}), in particular, for the dipole modes ($L=1$).
The amplitudes of the Fermi surface
distortions $\mathcal{A}_\pm~$ are determined by (\ref{LVeq}).
For the simplest case of the zero anisotropic interaction  ($F_1=F_1'=0$)
in the collisionless limit $\omega \tau \to \infty$,
the dispersion equation for the sound velocity $s$
takes the form:
\bea\l{dispeq}
&&4F_0F_0'\left(F_0 Q_1(s)-1\right) \nonumber\\
&-&\frac14\Delta^2 F_0^2{F_0'}^2
\left(\frac{s^2}{s^2-1} + Q_1(s)\right)^2=0,
\eea
(We accounted for a small $\Delta$
and large $\omega \tau$ at the zero temperature.)
This equation has the two solutions $s=s_n$ related to the main peak $n=1$
and $2$ for its satellite, see (26) of \cite{kolmagsh}
for the finite $\omega \tau$ and nonzero $F_1$ and $F_1'$.
In the limit $\Delta \to 0$, the dispersion equations
given by (25) of \cite{kolmagsh}
with our definitions for $s_1$ and $s_2$ modes $n=1$ and $2$
are resulted in the two (isovector and isoscalar) equations
for the equations for the separated zero sounds,
respectively,
\bel{splitdispeq}
Q_1(s)= 1/F_0', \qquad \mbox{and}\qquad
Q_1(s)= 1/F_0.
\ee

        For the finite Fermi-liquid drop with a sharp ES
\cite{strutmagden,magstrut,magboundcond},
the macroscopic boundary conditions for the pressures
and those for the velocities  were derived in
\cite{kolmagsh,magsangzh,BMRV}. For small isovector vibrations near
spherical shape, the radial mean-velocity $u_{r}$   and
momentum-flux-tensor $\Pi_{rr}$ components, defined through
the moments of the distribution function
$\delta f_{-}$ as solutions of the kinetic equation (\ref{LVeq}) [see
(\ref{veloc}) and (\ref{momentflux})]
are given by (\ref{bound1}) and  (\ref{bound2}) with
$u_{r}=u_{r}^{+}-u_{r}^{-}$ and
$\Pi_{rr}=\Pi_{rr}^{+}-\Pi_{rr}^{-}$.
The r.h.s.s of these boundary conditions are the isovector
ES velocity  $u_{{}_{\! S}}=R \dot{Q}_S Y_{L0}({\hat r})$   and  capillary
pressure exceed
\bel{boundcondiv}
\delta P_S=2 Q_S b_S^{(-)} \rho_{{}_{\! 0}} A^{1/3}
Y_{10}(\hat{r})/3,
\ee
given through the isovector
surface energy constant $b_S^{(-)} \propto \alpha_{-}$ [see (\ref{sigma}) and
(\ref{bsplusminus})],  where $Q_S$ is
the dynamical isovector-dipole ($L=1$)
amplitude of the motion of the neutron drop
ES against the proton one (\ref{surface}),
keeping also the volume and the position of the center of mass conserved).
Note that another interpretation of the surface symmetry-energy
constant $b_{\rm S}^{(-)}$ in (\ref{boundcondiv}) is considered  in
\cite{denisov,abrIVGDR}.
This constant essentially differs from the isovector stiffness introduced
in \cite{myswann69} for the description of the neutron skin as a
collective variable, see more detailed discussions in \cite{BMV,BMRV}.

       The energy constant,  $~D=\hbar \om A^{1/3}~$,
and energy weighted sum
rules (EWSR),
\be\l{strength}
{\tt S}_{1}=\frac{\hbar^2}{\pi} \int \d \om\; \om\; \Im \chi^{\rm coll}(\om),
\ee
for the IVGDR can be found from the collective response function
$\chi^{\rm coll}(\om)$ .
The response function (\ref{chicollrho}) is determined by the
transition density (\ref{densvolsurf}) generalized
to the dynamic isoscalar and isovector components \cite{BMR}:
\bea\l{trandenscl}
\delta \rho_{\pm}({\r},t) &=&
\delta \rho_{\pm}^{\rm vol}({\r},t)\; w_{\pm}(\xi) \nonumber\\
&-&
\frac{1}{a}\frac{\d w_{\pm}(\xi)}{ \d \xi}\; \overline{\rho}\;
\left[\delta R_{\pm}- \delta \aleph_{L}^{\pm} \;
Y_{L0} ({\hat r})\right],
\eea
where
$\delta \aleph_{L}^{\pm}$ is defined by the mass center conservation
($\int \d \r \;\r\; \delta \rho_{\pm}=0$),
$w_\pm(\xi)$ is given by (\ref{ysolplus}) and (\ref{ysolminus}).
In Fig.\ \ref{fig9etf}, a strong SO dependence of the isovector
density $w_{-}(\xi)$
is compared with
that of the isoscalar one $w_{+}(\xi)$
(low index ``+'' is omitted here and below)
for the SLy7 force as a typical example
\cite{magsangzh,BMRV}.
As  shown in \cite{BMRV}, the isoscalar $w(\xi)$, and therefore, the isovector
$w_{-}(\xi)$ densities depend rather strongly on the
most of the Skyrme forces \cite{chaban,reinhard} near the ES.
In  Fig.\ \ref{fig10etf} (in logarithmic scale),
one observes notable differences in the isovector densities $w_{-}$ derived
from different Skyrme forces
within the edge diffuseness. In particular,
this is important for the calculations of the neutron skins of nuclei
\cite{BMRV}.

We emphasize that the dimensionless densities, $w(x)$
(\ref{ysolplus}) and $w_{-}(x)$
(\ref{ysolminus}), shown in Figs. \ref{fig9etf} and \ref{fig10etf} were
obtained in
the leading ES approximation ($a/R \ll 1$) as functions of the
specific combinations
of the Skyrme force parameters, such as $\beta$ and $ c_{sym}$ of
(\ref{defpar}). Therefore, they are the universal distributions
independent of the specific properties of the nucleus such as the neutron and
proton numbers, and the deformation and curvature of the nuclear ES;
see also \cite{strtyap,strutmagden,magsangzh}.
These distributions yield approximately the spatial coordinate dependence
of local densities in the normal-to-ES direction $\xi$.
With the correct asymptotical behavior outside of the ES layer for any
ES deformation, they satisfy the leptodermic
condition $a/R \ll 1$, in particular,
for the semi-infinite nuclear matter.

The universal functions $w(\xi)$ (\ref{ysolplus})  and $w_{-}(x)$
(\ref{ysolminus}) of the leading order in the ESA
can be used [explicitly analytically in the quadratic
approximation for $\epsilon(w)$] for the calculations of the surface
energy coefficients
$b_S^{(\pm)}$ (\ref{sigma}), the neutron skin  and isovector stiffness
(see \cite{BMRV}). As shown in Appendices B and C of \cite{BMRV},
only these particle-density distributions  $w_{\pm}(\xi)$
within the surface layer
are needed through their derivatives [the lower limit of the integration
over $\xi$ in (\ref{sigma}) can be approximately extended to
$-\infty$ because of no contributions from the internal volume region
in the evaluation of the main surface terms of the pressure and energy].
Therefore,  the surface symmetry-energy coefficient
$k_{{}_{\! S}}$ in (\ref{bsplusminus}) and (\ref{Jm}) (also the neutron skin
and the isovector stiffness \cite{BMRV})
can be approximated analytically in terms of the functions
of the definite critical combinations of the Skyrme parameters such as $\beta$,
$c_{sym}$, $a$ [see (\ref{defpar})],
and the parameters of the infinite nuclear matter
($b_{\mathcal{V}}, \rho_\infty, K$). Thus,  they are independent of the
specific properties
of the nucleus (for instance, the neutron and proton numbers), and
the curvature and deformation of the nuclear surface in the
considered ESA.

Solving the Landau--Vlasov equations (\ref{LVeq})  in terms of the
zero sound plane waves
(\ref{planewaves}) with using the dispersion equations (26)
in \cite{kolmagsh}  for the sound velocities $s_n$ and
macroscopic boundary conditions
(\ref{bound1}) and (\ref{bound2}) with (\ref{boundcondiv}) on the
nuclear  ES,  from (\ref{chicollrho}) and (\ref{trandenscl}) one obtains
\bea\l{respfun}
 &&\chi_{L}^{\rm coll}(\om)=\sum_n \frac{\mathcal{A}_{L}^{(n)}(q)
}{\mathcal{D}_{L}^{(n)}(\om-i\Gamma/2)},\quad{\rm with}\quad
\mathcal{D}_{L}^{(n)}(\om) \nonumber\\
&=&
j_1'(qR)+\frac{3 \eps_{{}_{\! {\rm F}}} qR}{2b_S^{(-)}A^{1/3}}\;
\left[c_n j_1''(qR)+d_nj_1(qR)\right].
\eea
Here,
$~c_1\approx 1-s_1^2+\mathcal{F}_0'~$,
$d_1\approx 1-s_1^2+\mathcal{F}_0'$   for the main ($n=1$)  IVGDR
peak. Small anisotropic $\mathcal{F}_1$ and $\mathcal{F}_1'$
corrections
 and more bulky expressions for  $s_2$  of the satellite ($n=2$) peak
of a smaller ($\propto I$)
strength were omitted (see (D11) in \cite{kolmagsh} for more precise
expressions). We present  here also the simplest expressions for the amplitudes,
$\mathcal{A}_1(q)\approx -\rho_\infty R^3 j_1(qR)/(m \om^2)$  and
$\mathcal{A}_1(q)\propto \Delta \propto I$
for the $n=1$ and $2$ modes [see a more complete equation
(60) in \cite{kolmagsh}]. The Bessel functions  $j_1(z)$
and its derivative  $j_1'$ were defined after (\ref{potensolut}) ($L=1$).
The poles of the response function
$\chi^{\rm coll}(\om)$
(\ref{respfun}) (roots $\om_n$ of
the equation $D^{(n)}(\om-i \Gamma/2)=0$ or $q_n$ )
determine the IVGDR energies $\hbar \om$ as their
real part (the IVGDR width $\Gamma$
is determined  by their imaginary part).
The residue $\mathcal{A}_n$   is important for the calculations
of the EWSR (\ref{strength})
at a small width of the IVGDR $\Gamma$.
Note that the expression like  (\ref{respfun})
for the only one  main peak (without the IVGDR structure)
in symmetrical nuclei ($N=Z$)
with using the phenomenological boundary conditions
which have the same form as (\ref{bound1}) and (\ref{bound2}),
where however the isovector neutron-skin stiffness was applied
instead of the surface symmetry-energy constant $b_{{}_{\! S}}^{(-)}$ in the
capillary pressure exceed (\ref{boundcondiv})
was obtained earlier  in \cite{denisov}.

\subsection{ Discussions of the asymmetry effects}
\l{discaseff}

      The isovector surface energy constants $k_{{}_{\! S}}$ (\ref{bsplusminus})
in the ESA using the simplest quadratic approximation
for $\epsilon(w)$ of
the energy density (\ref{enerden})
are shown in Table 1 for several critical Skyrme forces
\cite{chaban,reinhard}. These constants are rather sensitive to the choice of
the Skyrme forces. The modulus of  $k_{{}_{\! S}}$
for the Lyon Skyrme forces SLy4-7
\cite{chaban} is significantly larger than for other forces,
all of them much smaller than those related to
\cite{myswann69,myswnp80pr96,myswprc77,myswiat85}.
For T6 \cite{chaban}, one has $\mathcal{C}_{-}=0$,
and therefore,  $k_S=0$,
in contrast to all of other forces shown in Table 1. Notice that the
isovector gradient terms  which are important for the consistent
derivations within the ESA
\cite{BMRV} are not also included  ($\mathcal{C}_{-}=0$)  into the energy density in \cite{danielewicz1,danielewicz2}. For RATP \cite{chaban},
the isovector stiffness ($\propto -1/k_{{}_{\! S}}$),
corresponding inversed $k_{{}_{\! S}}$ but with the opposite sign
\cite{BMRV}, is even negative as $\mathcal{C}_{-}>0$
($k_{{}_{\! S}}>0$). The reason of significant differences in these  values
might be related to those of the critical  isovector
Skyrme parameter $\mathcal{C}_{-}$
  in the gradient terms of the energy density (\ref{enerden}).
Different experiments used for fitting this parameter were found to be
almost insensitive in determining uniquely its value, and hence, $k_S$
[or $b_S^{(-)}$, see (\ref{bsplusminus})],
as compared to the well-known isoscalar surface-energy constant $b_S^{(+)}$.
The isovector  surface-energy constant $k_{{}_{\! S}}$  (\ref{bsplusminus})
and the corresponding stiffness
depend much on the SO $\beta$ parameter through the constant
$\mathcal{J}_{-}$ (\ref{Jm}).

The IVGDR energy constants $D=\hbar \om^{(-)} A^{1/3}$
of the hydrodynamic model (HDM)
are roughly in good agreement with the well-known experimental value
$D_{\rm exp}\approx 80$ MeV for heavy nuclei within a precision better
or of the order of 10\%,
as shown in \cite{BMV,BMRV} (see also
\cite{denisov,kolmagsh,plujko}).
More precise $A^{-1/3}$ dependence of $D$
seems to be beyond the accuracy of these HDM calculations. This takes place
 even accounting more consistently for the ES motion because of several
other reasons (the macroscopic Fermi-surface distortions
\cite{denisov}, also including  structure of the IVGDR
\cite{kolmag,kolmagsh,BMR,abrIVGDR,abrdavpl,plujko},
curvature, Coulomb, quantum-shell,
and pairing \cite{belyaevzel} effects
towards the realistic
self-consistent calculations
based on the Skyrme HF approach
\cite{vretenar1,vretenar2,ponomarev,nester1,nester2}.
Larger values 30-80 MeV of the isovector stiffness
\cite{myswann69}
(smaller $k_{{}_{\! S}}$ ) were found in
\cite{myswnp80pr96,myswiat85,vinas2,brguehak}.
With smaller $|k_{{}_{\! S}}|$ (see Table 1, or larger the isovector
stiffness) the fundamental parameter of the LDM expansion in
\cite{myswann69,myswnp80pr96}
is really small for $A \simg 40$, and therefore, the results
obtained by using this expansion are justified \cite{BMRV}.

        Table 1 shows also the mean IVGDR energies $D$
obtained \cite{BMV,BMRV} within a more precised FLDM \cite{kolmagsh}.
The IVGDRs even for the spherical nuclei have a double-resonance structure,
the main peak $n=1$ which exhausts mainly the EWSR
for almost all Skyrme forces and the satellite one $n=2$ with
the significantly smaller EWSR contributions proportional to
the asymmetry parameter $I$, typical for heavy nuclei. The
last row shows the average
$D(A)$ weighted by their EWSR
distribution in rather good agreement with the experimental data
within the same accuracy about 10 \%, and in agreement with the results
of different other macroscopic IVGDR models
\cite{denisov,abrIVGDR,abrdavpl,plujko}.
Exclusion can be done (see Table 1) for  the
Skyrme forces SIII \cite{chaban} and SkL3 \cite{reinhard} where we
obtained a little larger IVGDR energies.
Note that the main characteristics of the
 IVGDR described by mean $D$
are almost insensitive to the isovector surface-energy constant
$k_{{}_{\! S}}$ \cite{BMRV,BMV}. Therefore, we suggested \cite{BMRV,BMR}
to study the IVGDR
two-peak (main and satellite) structure in order to fix the ESA value of
$k_{{}_{\! S}}$ \cite{BMRV} from comparison with the experimental data
\cite{adrich,wieland,kievPygmy} and theoretical results
\cite{vretenar1,vretenar2,ponomarev,nester1,nester2,ditoro}.


 \section{NUCLEAR COLLECTIVE ROTATIONS}
\l{semshellmi}

\subsection{General ingradients of the cranking model}
\l{cranmod}

Within the cranking model, the nuclear collective
rotation of the Fermi independent-particle system
associated with a many-body Hamiltonian,
$H^{\boldsymbol\om}=H+H_{\rm CF}^{\boldsymbol\om}$,
can be described, to a good approximation \cite{eisgrei},
in the restricted subspace of Slater determinants, by
the eigenvalue problem for
 a s.p.\ Hamiltonian, usually called the {\it Routhian}.
For this Routhian, in the body-fixed rotating frame \cite{bohrmot,mix,fraupash},
one has
\bel{raussian}
h^{\boldsymbol\om}=h + h_{\rm CF}^{\boldsymbol\om},\qquad
h_{\rm CF}^{\boldsymbol\om}=-\boldsymbol\om \cdot
\left(\boldsymbol\ell + {\bf s}\right),
\ee
where $h_{\rm CF}^{\boldsymbol\om}$ is the s.p.\ cranking field which is
approximately equal to the Coriolis interaction
(neglecting a smaller centrifugal term, $\propto \om^2$).
The Lagrange multiplier $\boldsymbol\om$
(rotation frequency of the body-fixed coordinate system) is defined
through the constraint on the nuclear angular momentum ${\bf I}$,
evaluated through the quantum average
$\langle \boldsymbol\ell +{\bf s} \rangle^{\boldsymbol\om}={\bf I} $, of the
total s.p.\ operator, $\boldsymbol\ell + {\bf s}$,
where $\boldsymbol\ell$
is the orbital angular momentum and ${\bf s}$ is the spin of the quasiparticle,
thus defining a function
$\boldsymbol\om=\boldsymbol\om({\bf I})$.
The quantum average of the total s.p.\ operator $\boldsymbol\ell
+ {\bf s}$ is obtained by evaluating expectation values of the
many-body Routhian
$H_{\rm CF}^{\boldsymbol\om}$ in the subspace of Slater determinants.
For the specific case of a rotation around the $x$ axis ($\om=\om_x$)
which is perpendicular to the symmetry $z$ axis of the axially-symmetric mean
field $V$, one has (dismissing for simplicity spin (spin-isospin) variables),
\bel{constraint0}
\langle \ell_x \rangle^{\om} \equiv
d_s \sum_i n_i^{\om} \int \d \r
\;\psi_i^{\om}\; \left(\r\right) \; \ell_x\;
\overline{\psi}_i^{\om}\left(\r\right)=I_x,
\ee
where $d_s$ as the spin (spin-isospin) degeneracy in the case of the
corresponding symmetry of the mean potential $V$.
The occupation numbers $n_i^{\om}$
for the Fermi system of independent nucleons are given by
\bel{ocupnumbi}
n_i^{\om}\equiv n\left(\eps_i^{\om}\right)
=\{1+ \exp\left[(\eps_i^{\om} - \mu^{\om})/T\right]\}^{-1}.
\ee
In (\ref{constraint0}),  $\psi_i^{\om}(\r)$ are the eigenfunctions
and
$\overline{\psi}_i^{\om}(\r)$ their complex conjugate,
$\eps_i^{\om}$
the eigenvalues of the Routhian $h^{\om}$ (\ref{raussian}),
$\mu^{\om}$ is the chemical potential. For relatively small frequencies
$\om$ and temperatures $T$, $\mu^{\om}$ is to a good approximation
equal to the Fermi energy,
$\mu^{\om} \approx \eps_{{}_{\! {\rm F}}} =\hbar^2 k_{\rm F}^2/2 m^*$,
where $k_{{}_{\! {\rm F}}}$ is the Fermi momentum in units of $\hbar$.
From (\ref{constraint0}), the rotation frequency $\om$ can be
specifically expressed in terms of a given angular momentum of nucleus
$I_x$, $\om=\om\left(I_x \right)$.
Within the same approach,
one approximately has for the particle number
\bel{partconspert}
A= d_s\sum_i n_i^{\om}
\int \d \r\; \psi_i^{\om}(\r)\;\overline{\psi}_i^{\om}(\r)
\approx d_s \int \d \eps \; n(\eps),
\ee
which determines the chemical potential $\mu^{\om}$ for a given number of
nucleons $A$. As we introduce the continuous
parameter $\om$ and ignore the uncertainty
relation between the angular
 momentum and angles of the body-fixed coordinate system,
the cranking model is semiclassical
in nature \cite{ringschuck}.
Thus, we may consider the collective MI $\Theta_x$ (for a rotation
around the $x$ axis, and omitting, to simplify the notation, spin and
isospin variables) as a response of the quantum average
$\delta \langle \ell_x \rangle^{\om}$ (\ref{constraint0}), to the
external cranking field $h_{\rm CF}^{\om}$ in (\ref{raussian}).
Similarly to the magnetic or isolated
susceptibilities
\cite{richter,fraukolmagsan,magvvhof,yaf},
one can write
\bel{response}
\delta \langle \ell_x \rangle^{\om}=
\Theta_x(\om) \delta\om,
\ee
where
\bea\l{thetaxdef}
&&\Theta_{x}(\om)=
\partial \langle \ell_x\rangle^\om/\partial \om=
\partial^2 E(\om)/\partial \om^2,
\quad\;\;\;
\nonumber\\
&& E(\om)=\langle h \rangle^\om
\equiv d_s \sum_i n_i^{\om}\int \d \r
\;\psi_i^{\om}\; \left(\r\right) \;h \;
\overline{\psi}_i^{\;\om}\left(\r\right).\;\;\;
\eea
Traditionally \cite{mix,dfcpuprc2004,mskbPRC2010},
another parallel (alignment)
rotation with respect to the
symmetry $z$ axis can be also considered as presented
in Appendix A of \cite{mskbPRC2010}.

As was shown in
\cite{inglis,bohrmotpr,belyaevzel,valat,bohrmot,fraupash,mix}, one can treat
the term
$-\boldsymbol\om \cdot \boldsymbol\ell = -\om \, \ell^{}_x$ as
a perturbation for a nuclear rotation around the $x$ axis.
With the constraint (\ref{constraint0})
and the MI (\ref{thetaxdef}) treated in second order perturbation theory,
one obtains the well known Inglis cranking formula.
Instead of carrying out the rather involved calculations
presented above, one could, to obtain the yrast line energies $E(I_x)$
for small enough temperatures $T$ and frequencies $\om$, approximate
the angular frequency by $\om=I_x/\Theta_x$ and write the energy in the form
\bel{yrast}
E(I_x)= E(0) + \frac{I^{2}_x}{2 \Theta_x}.
\ee
As usually done, the rotation term above needs to be quantized
through $I^{2}_x \rightarrow I_x (I_x+1)$ in order to study the
rotation bands.

\subsection{Self-consistent ETF description of nuclear rotations}
\l{etfmi}

Following
reference \cite{bartelnp}, a microscopic description of rotating nuclei
was obtained in the Skyrme Hartree--Fock formalism, within the Extended
Thomas--Fermi density-functional theory up to order $\hbar^2$.
Within a variational space restricted to Slater determinant,
 the minimization of the
expectation value of
the nuclear Hamiltonian
lead to the s.p. Routhian $h^{\boldsymbol\omega}_{\isospin}$ (\ref{raussian})
that is determined by a one-body potential $V_{\isospin}({\bf r})$,
a spin-orbit field ${\bf W}_{\isospin}({\bf r})$ and an effective mass
form factor $f^{\rm eff}_{\isospin}({\bf r})=m/m^*_{\isospin}$ (see
also \cite{brguehak}). In addition,
in the case when the time
reversal symmetry is broken, a cranking field form factor
$\boldsymbol\alpha_{\isospin}({\bf r})$ and a spin field form factor
${\bf S}_{\isospin}({\bf r})$
also appear. In this subsection the (roman) subscript $\isospin$
refers to the nucleon isospin
($\isospin = \{n,p\}$) and should not be confused
with the wave number $q$ in other sections.
All these fields can be written
as functions of local densities and their derivatives, like the
neutron-proton particle
densities $\rho_{\isospin}({\bf r})$, the kinetic energy densities
$\tau_{\isospin}({\bf r})$,
the spin densities (also referred to
as spin-orbit densities)
${\bf J}_{\isospin}({\bf r})$, the current densities
${\bf j}_{\isospin}({\bf r})$, and
the spin-vector densities
$\boldsymbol\rho_{\isospin}({\bf r})$.
Note that in the present subsection, $\tau_{\isospin}({\bf r})$
stands for the kinetic energy
density which should not be confused with the relaxation time in
previous sections (here, however, with a different subscript ${\rm q}$
as compared to $q$ in 
sections \ref{kinapp},\ref{respfuntheor},\ref{npcorivgdr} and Appendices
A,B). In principle, two
additional densities appear, a spin-vector kinetic energy density
$\boldsymbol\tau_{\isospin}({\bf r})$ and a tensor coupling
$J^{}_{\alpha \beta}({\bf r})$
between spin and gradient vectors, which have, however, been neglected
since their contribution should be small, as suggested by \cite{BFH87}.

The cranking-field form factor
$\boldsymbol\alpha_{\isospin}({\bf r})$ contains two
contributions.
One of them is coming from the orbital part of the constraint,
$-\boldsymbol\om \, \boldsymbol\ell$,
which has been shown in \cite{BV78} to correspond to the Inglis cranking
formula \cite{inglis}. The other,
a Thouless--Valatin self-consistency
contribution \cite{TV62} has its origin in the self-consistent response
of the mean field to the time-odd part of the density matrix generated by
the cranking term of the Hamiltonian.
The aim is now to find functional relations for the local densities
$\tau_{\isospin}(\r)$, ${\bf J}_{\isospin}(\r)$,
${\bf j}_{\isospin}(\r)$ and $\boldsymbol\rho_{\isospin}(\r)$ in
terms of the particle densities $\rho_{\isospin}({\bf r})$,
in contrast to those given by
Grammaticos and Voros \cite{GV79} in terms of the form factors
$V_{\isospin}$,
$f^{\rm eff}_{\isospin}$, ${\bf W}_{\isospin}$,
$\boldsymbol\alpha_{\isospin}$ and
${\bf S}_{\isospin}$.
Taking advantage of the fact that, at the leading Thomas--Fermi order,
the cranking field form factor is given by \cite{bartelnp}
\bel{ETF02}
  \boldsymbol\alpha^{({\rm TF})}_{\isospin}
                        = f^{\rm eff}_{\isospin}\left( {\bf r}
    \times \boldsymbol\omega \right),
\ee
one simply obtains the rigid-body value for the Thomas--Fermi current density
\bel{ETF03}
   {\bf j}^{({\rm TF})}_{\isospin} = \displaystyle \frac{m}{\hbar}
   \left( \boldsymbol\om \times {\bf r} \right) \, \rho_{\isospin}.
\ee
This result is not that trivial, since it is only through the effect
of the Thouless--Valatin self-consistency terms that such a simple
result is obtained. Notice also that (\ref{ETF03}) corresponds to a
generalization to the case $f^{\rm eff}_{\isospin} \neq 1$ of a result
already found by Bloch \cite{Bl54}.
Equation  (\ref{ETF03})
can be also considered as an extension of
the Landau quasiparticle (generalized TF) theory
\cite{landau,abrikha} presented in
Secs. \ref{etfmi}, \ref{cranmod} to the case of rotating Fermi-liquid systems,
cf. (\ref{ETF03}) with (\ref{currentdef})
for the current density as an average of the \textit{particle}
velocity, $\p_{\rm rot}/m=\boldsymbol\om \times \r$,
rotating with the frequency
$\boldsymbol\om$.
In particular, the re-normalization of the cranking field form factor
$\boldsymbol\alpha^{({\rm TF})}_{\isospin}=
f_{\rm q}^{\rm eff} \boldsymbol\alpha_{\rm o}$ with
$\boldsymbol\alpha_{\rm o} =
\left( {\bf r} \times \boldsymbol\omega \right)$,
by (\ref{ETF02}) can be also explained
as related to the effective mass corrections,
$f_{\rm q}^{\rm eff} \neq 1$, obtained
by Landau \cite{landau} with using
both the Galileo principle and the Thouless--Valatin self-consistency
corrections to a particle mass $m$ due to the quasiparticles'
(self-consistent) interaction
through a mean field. They lead in \cite{bartelnp}
to  the self-consistent TF angular momentum
of the \textit{quasiparticle}
$\boldsymbol\ell_{\isospin} =f_{\isospin}^{\rm eff} \boldsymbol\ell_{\rm o}$ with the
classical angular momentum $\boldsymbol\ell_{\rm o}= \r \times \p $
of the particle,
so that $-\boldsymbol\om\cdot \boldsymbol\ell_{\isospin}=
\boldsymbol\alpha_{\isospin} \cdot \p$. This effect is similar to that
for the kinetic energies of the \textit{quasiparticles},
$\eps_{\isospin}=p^2/(2 m^*_{\isospin})=f_{\isospin}^{\rm eff}\eps_{\rm o}$
where $\eps_{\rm o}=p^2/(2m)$, see after (\ref{LVeq}).
With this transparent connection to the Landau quasiparticle
theory, it is clear that
there is no contradictions with the TF limit
of the current densities (\ref{currentdef}), $\hbar \rightarrow 0$,
accounting for the particle
densities (\ref{densitydef}), as well as
with the definitions in subsections \ref{extensiontoas} and
\ref{semshell}, because  $\hbar$ in (\ref{ETF03})
 appears formally due to a traditional
 use of the dimensionless units for the angular momenta in the
quantum-mechanical picture to compare with experimental
nuclear data. Another reason is related to
a consistent treatment of the essentially quantum
spin degrees of freedom, beyond the Landau quasiparticle approach
to the description of Fermi liquids,
which have no \textit{straight} classical limit,
in contrast to the orbital angular momentum
$\boldsymbol\ell$.
The convergence in the TF limit $\hbar \rightarrow 0$ can be realized
for smooth already quantities after the statistical (macroscopic)
averaging over many
s.p.\ (more generally speaking, many-body) quantum states
to remove the fluctuating (shell) effects  which
appear in the denominators of the exponents within the POT (see 
Sec.\ \ref{semshell} for
more detailed discussions). Finally, the spin paramagnetic  effect
 can be considered as a macroscopic one in the MI
like the orbital diamagnetic contribution.
For instance, the spin-vector density does not
have a \textit{straight} classical analogue,
such as the orbital angular momentum,
and is considered as the object of leading order  $\hbar$.

Starting from these results and taking advantage of the fact that in
the functional ETF expressions up to the order
$\hbar^2$, it is sufficient to replace quantities, such as
the cranking field form factor $\boldsymbol\alpha_{\isospin}$, by
their Thomas--Fermi expressions
(after the statistical averaging mentioned above).
In order to obtain a
semiclassical expression,
that is correct to that order in $\hbar$, one obtains for the spin-vector
densities $\boldsymbol\rho_n$ and $\boldsymbol\rho_p$, which are of order
$\hbar$ in the considered ETF expansion, a system of linear
equations. They  can be easily resolved \cite{bartelnp}.
One also notices from this system of
equations that the spin-vector densities are proportional to the
angular velocity $\omega$.
Exploiting the well known analogy of the microscopic Routhian problem with
electromagnetism, one may then define {\it spin susceptibilities}
$\chi_{\isospin}$,
\bel{ETF04}
  \boldsymbol\rho_{\isospin} = \hbar \,
\chi_{\isospin} \, \boldsymbol\omega \; .
\ee

The key question now is to assess the sign of
these susceptibilities and to
decide whether or not the corresponding alignment is of a
``Pauli paramagnetic''
character. The study of \cite{bartelnp} shows that this is the case, i.e., that
the spin polarization is, indeed, of paramagnetic character, thus confirming
the conclusions of the work performed
by Dabrowski \cite{Da75} in a simple model of non-interacting
nucleons.

Since the cranking field factor
$\boldsymbol\alpha_{\isospin}$ is, appart from that of the
constraining field
$\boldsymbol\alpha_{\rm o}$ determined
only by the current densities ${\bf j}_{\isospin}$
and the spin-vector densities
$\boldsymbol\rho_{\isospin}$, one can then write down \cite{bartelnp}
the contributions to the
current densities ${\bf j}_{\isospin}$
going beyond the Thomas--Fermi approach.
The semiclassical corrections
of order $\hbar^2$ can be split into
contributions
$(\delta {\bf j}_{\isospin})_{\ell}$ and
$(\delta {\bf j}_{\isospin})_{s}$ coming
respectively from the orbital motion and the spin degree of freedom. It is
found \cite{bartelnp} that the orbital correction
$(\delta {\bf j}_{\isospin})_{\ell}$
corresponds to a surface-peaked {\it counter-rotation} with respect to the
rigid-body current proportional
to $\left( \boldsymbol\omega \times {\bf r} \right)$,
thus recovering the Landau diamagnetism characteristic of a finite Fermi gas.
With the expressions of the current densities ${\bf j}_{\isospin}$
and the spin-vector
densities $\boldsymbol\rho_{\isospin}$ up to order $\hbar^2$,
one can write down the
corresponding ETF expressions for the kinetic energy density
$\tau_{\isospin}(\r)$ and
spin-orbit density ${\bf J}_{\isospin}(\r)$.

Having now at hand the ETF functional expressions up to order $\hbar^2$ of all
the densities entering our problem, one is able to write down the energy of the
nucleus in the laboratory frame as a functional of these local densities,
\bel{endentau}
E = \int \d \r \; \rho \; \mathcal{E}[\rho_{\isospin},
\tau_{\isospin}, {\bf J}_{\isospin}, {\bf j}_{\isospin},
\boldsymbol\rho_{\isospin}],
\ee
where $\rho=\rho_n+\rho_p$ as in Appendix D,
$\rho \approx \rho_\infty w_{+}$.
Upon some integration by parts, one finds that
$\mathcal{E}$ can be written as a sum of
 the energy density per particle of
the non-rotating system
$\mathcal{E}(0)$ and its rotational part,
in line of (\ref{yrast}).
Within the ETF approach, one has from (\ref{endentau})
\bel{endentauSCM}
E_{\rm ETF}= \int \d\r \rho \mathcal{E}(0) +
\frac12 \Theta_{\rm ETF}^{({\rm dyn})}\,\om^2,
\ee
where $\Theta_{\rm TF}^{({\rm dyn})}$ is the ETF dynamical moment of
inertia for the
nuclear rotation with
the frequency $\boldsymbol\om$. This MI is given in the form:
\bea\l{ETF05}
  \Theta_{\rm ETF}^{({\rm dyn})} &=& m \sum_{\isospin} 
\int \d \r \; \left\{ r_\perp^2 \,
\rho_{\isospin}
  - \left(3 \pi^2 \right)^{-2/3} f^{\rm eff}_{\isospin} \; \rho^{1/3}_{\isospin} 
\right.\nonumber\\
&+&  \left.
       \left[\frac{\hbar^2}{2m} + W^{}_0 \left(\rho + \rho_{\isospin}\right)
\right] \, \chi_{\isospin}   \right\},
\eea
where
$ r_\perp$ is the distance of a given point to the rotation axis and
$W^{}_0$ is the Skyrme-force strength parameter of the spin-orbit
interaction \cite{brguehak}.

One notices that the Thomas--Fermi term which comes from the orbital motion
turns out to be the rigid-body moment of inertia. Semiclassical corrections
of order $\hbar^2$ come from both the orbital motion
($\Theta_{\rm orb.}^{({\rm dyn})}$) and from the spin degrees of freedom
($\Theta_{\rm spin}^{({\rm dyn})}$). The contribution $\Theta_{\rm orb.}^{({\rm dyn})}$
is
negative corresponding to a surface-peaked counter rotation in the rotating
frame. Such a behavior is to be expected
for a N-particle system bound by
attractive short-range forces (see \cite{BJ76}).
The spin contribution
$\Theta_{\rm spin}^{({\rm dyn})}$ turns out to be of the {\it paramagnetic} type,
thus leading to a positive contribution which corresponds to an alignment of
the nuclear spins along the rotation axis. It can
also be shown (see \cite{BBQ93})
that the ETF kinematic moment of inertia,
\bel{mikin}
  \Theta_{\rm ETF}^{({\rm kin})} = \displaystyle
\frac{\langle{\boldsymbol\ell}+{\bf s}\rangle^{\om}}{\omega},
\ee
is identical
to the ETF dynamical moment of
inertia presented above.

It is now interesting to study the importance of the Thouless--Valatin
self-consistency terms.  This has accomplished by calculating
the moment of inertia
in the Thomas--Fermi approximation but omitting, this time, the
Thouless--Valatin terms. One then finds \cite{bartelnp} the following
expressions for the dynamical moment of inertia, in what is simply the
Inglis cranking (IC) limit
\bea\l{ETF06}
  \Theta_{\rm IC}^{({\rm dyn})} &=& m \sum_{\isospin} \int \d \r \;
\left[ \frac{\rho_{\isospin}}{\left(f^{\rm eff}_{\isospin}\right)^2} \right.
\nonumber\\
  &+& \left.\frac{m B_3}{\hbar^2} \, \rho_{\isospin} \, \rho_{\bar {\isospin}}
          \left(\frac{1}{f^{\rm eff}_{\isospin}} -
\frac{1}{f^{\rm eff}_{\bar {\isospin}}} \right)^{\!\!2} \,
                                        \right] r_\perp^2,
\eea
where ${\bar \isospin}$ is the {\it other}
charge state (${\bar \isospin} \!\!=\!\! p$ when
$\isospin \!\!=\!\! n$ and vice-versa)
and $B_3$ is defined through the Skyrme force parameters
$t^{}_1, t^{}_2, x^{}_1$ and $x^{}_2$ (see \cite{bartelnp}).
Apart from the corrective term in $\rho_{\isospin} \, \rho_{\bar {\isospin}}$,
one notices
that the first term in the expression above, which is the leading term,
 yields, at least for a standard HF-Skyrme
    force where $f^{\rm eff}_{\isospin} \geq 1$, to a smaller moment of
    inertia than the corresponding term in (\ref{ETF05})
containing the Thouless--Valatin corrections.
It is also worth noting that in
this approximate case, the kinematic moment of inertia is given by
\bel{ETF07}
  \Theta_{\rm IC}^{({\rm kin})}
        = m \sum_{\isospin} \int \d \r \;
\frac{\rho_{\isospin}}{f^{\rm eff}_{\isospin}} \;
r_\perp^2,
\ee
which turns out to be quite different from the above given
dynamical moment
of inertia, (\ref{ETF06}),
obtained in the same limit (Thomas--Fermi limit, omitting the
Thouless--Valatin self-consistency terms).

To investigate the importance of the different contributions to the total
moment of inertia, we have performed
self-consistent ETF calculations up to
order $\hbar^4$ for 31 non-rotating nuclei,
imposing a spherical symmetry,
and
using
the SkM$^*$ Skyrme effective nucleon-nucleon interaction \cite{BQB82}. Such
calculations yield variational semiclassical density profiles for neutrons
and protons \cite{brguehak} which are then used to calculate the above given
moments of inertia. The nuclei included in our calculations are $^{16}$O,
$^{56}$Ni, $^{90}$Zr, $^{140}$Ce, $^{240}$Pu and three isotopic chains for
Ca ($A \!=\! 36-50$), Sn ($A \!=\! 100-132$) and Pb ($A \!=\! 186-216$).
The results of these calculations
are displayed in figure \ref{fig11} taken from
\cite{bartelnp}.

One immediately notices the absence of any significant isovector dependence.
The good reproduction of the total ETF moment of inertia obtained by the
Thomas--Fermi (rigid-body) value is also quite striking. One finds that the
orbital and spin semiclassical corrections are not small individually but
cancel each other to a large extent. To illustrate this fact the ETF moments
obtained by omitting only the spin contribution are also shown on the figure.
One thus obtains a reduction of the Thomas--Fermi result that is about 6\% in
$^{240}$Pu but as large as 43\% in $^{16}$O.

The Inglis cranking approach performed at the Thomas--Fermi level underestimates
the kinematic moment of inertia by as much as 25\% and the dynamical moment of
inertia by about 50\% in heavy nuclei, demonstrating in this way the importance
of the Thouless--Valatin self-consistency terms.

In \cite{bartelnp}, a crude estimate of the semiclassical corrections due to orbital and spin degrees of freedom has been made by considering the nucleus as a piece of
symmetric nuclear matter (no isovector dependence as already indicated by the
self-consistent results shown in figure \ref{fig11} above).
It turns out that these
semiclassical corrections have an identical $A$ dependence
($A_{}^{-2/3}$ relative to the leading order Thomas--Fermi,
i.e.\ rigid-body, term)
\bel{ETF08}
  \Theta_{\rm  ETF} = \Theta^{\rm (RB)}
     \left[ 1 + \left( \eta_{\ell} +
\eta_s \right) A^{-2/3} \right] \; .
\ee
A fit of the parameters $\eta_{\ell}$ and $\eta_s$ to
the numerical
results displayed
in Fig.\ \ref{fig11} yields $\eta_{\ell} = -1.94$ and $\eta_s = 2.63$
giving a total (orbital
+ spin) corrective term of $0.69 \, A^{-2/3}$. For a typical rare-earth
nucleus (A = 170) all this would correspond to a total corrective term equal to
2.2\% of the rigid-body value, resulting from a -6.3\% correction
for the orbital
motion and a 8.5\% correction for the spin degree of freedom.

Whereas in the calculations that lead to figure \ref{fig11} above,
spherical symmetry
was imposed, fully variational calculations have been performed in
\cite{bartelpl}, imposing however the nuclear shapes to be of spheroidal form.
In this way, the
nuclear rotation clearly impacts on the specific form of the
matter densities $\rho^{}_n$ and $\rho^{}_p$ which, in
turn, in the framework
of the ETF approach determine all the other local densities, as explained
above.

Trying to keep contact with usual shape parametrizations,
by the standard quadrupole
parameters $\beta$ and $\gamma$ equating the semi-axis lengths of the spheroids
with the lengths of a standard quadrupole drop.

As a result,
figure \ref{fig12} shows the evolution of the equilibrium
solutions (the ones that minimize the energy for given angular momentum I)
as a function of I. One clearly observes that at low
values of
the angular momentum
(I in the range between 0 and 50 $\hbar$) the nuclear drop takes on
an oblate
shape, corresponding to increasing values of the quadrupole parameter $\beta$
with increasing I values, but keeping the non-axiality parameter fixed at
$\gamma = 60^{\circ}$.  For larger values of the total angular momentum
(I beyond 55 $\hbar$), one observes a transition into triaxial shapes, where
the nucleus evolves rapidly to more and more elongated shapes. For even
higher values of I (I beyond 70 $\hbar$) the nucleus approaches
the fission
instability. These results are in excellent qualitative agreement with those
obtained by Cohen, Plasil and Swiatecki \cite{CPS74} in a rotating LDM.

It is
amusing to observe
here
a {\it backbending} phenomena at the semiclassical
level when one is plotting,
as usual, the moment of inertia $\Theta_{\rm ETF}$
vs the rotational angular momentum,
see Fig.\ \ref{fig13}.
One should, however,
insist on the fact that this {\it backbending} has strictly nothing to do with
the breaking of a Cooper pair. The rapid increase
of the moment of inertia
at about I$ \,= 60 \hbar$
with a practically constant (or even slightly
decreasing)
rotational frequency $\om$ comes simply
from the fact that at such a value of I
(between I $\approx$ 60 and I $\approx$ 70) the nucleus elongates
substantially
increasing in this way its deformation and at the same time its
moment of inertia.

It is therefore interesting to notice that the semiclassical ETF
approach leads to a moment of inertia that is very well approximated
by its Thomas--Fermi, i.e. rigid-body value. Thouless--Valatin terms which
arise from the self-consistent response of the mean field to the time-odd
part of the density matrix generated by the cranking piece of the
Hamiltonian are naturally taken care of in this approach. Semiclassical
corrections of order $\hbar^2$ coming from the orbital motion and the spin
degree of freedom are not small individually, but compensate each other
to a large extent. One has, however, to keep in mind  that the
shell and pairing effects, that go beyond the ETF approach, are not included
in this description. These effects are not only both present, but influence
each other to a large extent, especially for collective high-spin
rotations of strongly deformed nuclei, as shown
in \cite{belyaevhighspin,pomorbartelPRC2011,sfraurev}.

\subsection{MI shell structure and periodic orbits}
\l{semshell}

We shall outlook first the basic points of the POT for the semiclassical
level-density and free-energy shell corrections
\cite{strut,fuhi,migdalrev}. We apply then the POT
for the derivation of the MI through the rigid-body MI
(with the shell corrections, see Appendix E)
in the NLLLA
related to the equilibrium collective rotation
with a given frequency $\omega$ \cite{mskbPRC2010}. For simplicity,
we shall discard the spin and isospin degrees of freedom, in particular,
the spin-orbit and asymmetry interaction.

Notice also that from the results presented in
Figs. \ref{fig11} and \ref{fig13} (with the help of
Fig. \ref{fig12}), one may conclude that the main contribution to
the moment of inertia
of the strongly deformed heavy nuclei can be found within the
ETF approach to the rotational problems as a smooth
rigid body MI.

\medskip

\subsubsection{
GREEN'S FUNCTION TRAJECTORY EXPANSION}
\label{greenfun}

For the derivations of shell effects \cite{strut} within the POT
\cite{gutz,strutmag,bt,creglitl,sclbook,migdalrev},
it turns out to be helpful
to use the coordinate representation of the MI
through the Green's functions $G\left(\r_1,\r_2;\eps \right)$
\cite{magvvhof,yaf,gzhmagsit,mskbg,mskbPRC2010},
\bea\l{thetaxdefG}
\Theta_{x}&=&\frac{2 d_s}{\pi}\;\int^{\infty}_0 \d\eps\;
n({\eps}) \int \d \r_1 \int \d \r_2 \; \ell_{x}(\r_1)\; \ell_{x}(\r_2)
\nonumber\\
&\times&
 \Re \left[ G\left(\r_1,\r_2;\eps \right) \right]
\Im \left[ G\left(\r_1,\r_2;\eps\right) \right].
\eea
The Fermi occupation numbers $n(\eps)$ (\ref{ocupnumbi})
are approximately considered at $\om=0$ ($\eps=\eps_i$).
In (\ref{thetaxdefG}), $\ell_x(\r_1)$ and $\ell_x(\r_2)$
are the s.p.\ angular-momentum projections onto the perpendicular rotation
$x$ axis at the spatial points $\r_1$ and $\r_2$, respectively.
With the usual energy-spectral representation for the one-body
Green's function $G$ in the mean-field approximation, one finds
the standard cranking model expression, which however includes
the diagonal matrix elements of the operator $\ell_x$.
In this sense, equation (\ref{thetaxdefG}) looks more general beyond the
standard perturbation approximation,
see \cite{mskbPRC2010}.
Moreover, the quantum criterion of the application
of this standard cranking model approximation, which is a smallness
of the cranking field perturbation $h^\om_{\rm CF}$ in
(\ref{raussian}) as
compared to the distance between
the neighboring states of the non-perturbative spectrum, becomes weaker
in the semiclassical approach, see more comments below in relation to
\cite{belyaevfirst,belyaevbif}.

For the MI calculations by (\ref{thetaxdefG}),
through the Green's function $G$, one may use the
semiclassical Gutzwiller trajectory expansion \cite{gutz} extended to
continuous symmetry \cite{strutmag,smod,magosc,creglitl,magkolstr,sclbook}
and symmetry breaking \cite{spheroidptp,maf,sclbook,migdalrev} problems,
\bel{GRdefsem}
G\left(\r_{1},\r_{2};\eps\right)=
\sum_{\rm CT} G_{\rm CT}\left(\r_{1},\r_{2};\eps\right),
\ee
where
\bea\l{Gct}
&&G_{\rm CT}\left(\r_{1},\r_{2};\eps\right)=
\mathcal{A}_{\rm CT}\left(\r_{1},\r_{2};\eps\right) \nonumber\\
&\times&
\exp\left[\frac{i}{\hbar}{\rm S}_{\rm CT}\left(\r_{1},\r_{2};\eps\right)
 - \frac{i \pi}{2}\sigma_{{}_{\! {\rm CT}}}-
i\phi_{\rm d}\right].
\eea
The sum runs over all isolated
 classical
trajectories (CTs) or their families inside the
potential well $V(\r)$ which, for a given energy $\eps$, connect the two
spatial points $\r_1$ and $\r_2$.
Here ${\rm S}_{\rm CT}$ is the classical action along such a CT,
and $\sigma_{{}_{\! {\rm CT}}}$ denotes the phase associated with the Maslov index
through the number of caustic and turning points along the path CT,
$\phi_{\rm d}$
is the constant phase depending on the dimension
of the problem \cite{sclbook,strutmag,maf,migdalrev}.
The amplitudes $\mathcal{A}_{\rm CT}$ of the Green's function depend on the
classical stability factors and trajectory degeneracy, due to the symmetries
of that potential \cite{sclbook,strutmag,smod,spheroidptp,maf}.
For the case of the isolated CTs
\cite{gutz,sclbook},
 one has the explicit semiclassical expression for the amplitudes
through the stability characteristics of classical dynamics,
\bel{Agutz}
\mathcal{A}_{\rm CT}(\r_{1},\r_{2};\eps)
=-\frac{1}{2\pi\hbar^2}\,\sqrt{\Big|\mathcal{J}_{\rm CT}({\bf
p}_{1},t_{{}_{\! {\rm CT}}};\r_{2},\eps)\Big|}.
\ee
Here, $\mathcal{J}_{\rm CT}(\p_{1},t_{{}_{\! {\rm CT}}};\r_{2},\eps)$
is the Jacobian for the transformation between the two sets of variables
$\p_{1},t_{{}_{\! {\rm CT}}}$ and $\r_{2},\eps$; $\p_1$ and $t_{{}_{\! CT}}$ are the
initial momentum and time of motion of the particle along a
CT, $t_{{}_{\! {\rm CT}}}=\partial S_{\rm CT}/\partial\eps$ ,
$\r_2$ and $\eps$ are its final coordinate and energy.
In more general case, if the mean field Hamiltonian $h$ obeys a
higher symmetry like that of spherical or harmonic-oscillator
potentials with rational ratios of frequencies,
one has to use other expressions for the amplitude
$\mathcal{A}_{\rm CT}(\r_{1},\r_{2};\eps)$ for close trajectories
of a finite action (with reflection from the potential boundary),
taking into account such symmetries.
They account for an enhancement in $\hbar$ owing to their classical
degeneracy (see \cite{strutmag,sclbook,maf,migdalrev} and the discussion
in subsection  below). In the case of
the bifurcation of POs, generated by a symmetry-breaking, one
may use the ISPM
\cite{spheroidptp,maf}, especially
for superdeformed shapes of the potential. Some examples
of the specific amplitudes for the degenerate families of
closed POs in the harmonic oscillator (HO) potential are given in 
Appendix E of
\cite{mskbPRC2010}.
Note that (\ref{Agutz}) can be applied for any potential wells for the
contributions of closed and
non-closed trajectories which can be considered as isolated
(no PO families) ones for the given end points $\r_1$ and $\r_2$.

Among all of CTs in (\ref{GRdefsem}),
we may single out ${\rm CT}_{0} $
which connects directly ${\bf r}_1 $ and ${\bf r}_2 $ without
intermediate turning points, see Fig.\ \ref{fig14}.
It is associated with the component
$G_{{\rm CT}_0}$ of the sum (\ref{GRdefsem}) for the semiclassical
Green's function.
Therefore, for the Green's function $G(\r_1,\r_2;\eps)$ (\ref{GRdefsem}),
one has then a separation,
\bel{Ggsplit}
G = G_{CT_{0}} + G_{1} \approx  G_0 + G_1.
\ee
In the NLLLA \cite{gzhmagsit,gzhmagsit,mskbPRC2010},
 \be\l{nllla}
s_{{}_{\! 12}}  \ll \hbar/p_{{}_{\! {\rm F}}},
\ee
the first term
$G_{CT_{0}}$ of the splitting in the middle of (\ref{Ggsplit}) is given
by
\be\l{G0}
G_{CT_{0}} \approx G_0(s_{{}_{\! 12}},p)=
-\frac{m}{2 \pi \hbar^2  s_{{}_{\! 12}}} \;
\exp\left[\frac{i}{\hbar} s_{{}_{\! 12}} \, p\left(\r\right) \right],
\ee
where $p\left(\r\right) = \sqrt{2 m [\eps - V(\r)]\,}~$, $~V(\r)~$
is a mean nuclear potential,
\bel{srvar}
s_{{}_{\! 12}}=\left|\r_2-\r_1\right|,\qquad \r=\left(\r_1+\r_2\right)/2,
\ee
$p=|\p|$, $\p=(\p_1+\p_2)/2$.
The second term $G_1$ in (\ref{Ggsplit})
is the fluctuating part of the Green's function
(\ref{GRdefsem}) determined  by all other trajectories
${\rm CT}_1 \neq {\rm CT}_{0}$ in the sum
(\ref{GRdefsem}) with reflection points at the potential surface
(see one of such trajectories ${\rm CT}_1$ in Fig.\ \ref{fig14}),
\bel{Gosc}
G_{1}(\r_{1},\r_{2};\eps)=
\sum_{{\rm CT}_1}
G_{{\rm CT}_1}\left(\r_{1},\r_{2};\eps\right),
\ee
where $G_{{\rm CT}_1}$ is the Green's function component (\ref{Gct}) taken at
the ${\rm CT} \neq {\rm CT}_0$, i.e., ${\rm CT}_1$.

\medskip

\subsubsection{
LEVEL-DENSITY AND ENERGY SHELL CORRECTIONS}
\l{general}

The
level density, $g(\eps)=\sum_i\delta(\eps-\eps_i)$, where
$\eps_i$ is the quantum spectrum, is identically
expressed in terms of the Green's
function $G$ as
\bel{totdensgen}
g(\eps)=- \frac{1}{\pi}\;\Im
\int \d {\bf r}\; \left[G({\r}_1,{\r}_2;\eps)
\right]_{\r_1 \rightarrow \r_2 \rightarrow \r}.
\ee
According to (\ref{Ggsplit}), this level density
can be presented semiclassically
as a sum of the smooth and oscillating components
\cite{gutz,strutmag,sclbook,migdalrev},
\bel{totdensscl}
g_{\rm scl}(\eps)=g_{{}_{\! {\rm ETF}}}(\eps) + \delta g_{\rm scl}(\eps),
\ee
where
$g_{{}_{\! {\rm ETF}}}(\eps)$ is given by the ETF approach related to the
component $G_0$ in (\ref{Ggsplit}) in the NLLLA (\ref{nllla})
$\r_1 \rightarrow \r_2 \rightarrow \r$
\cite{brguehak,sclbook,gzhmagfed,migdalrev}. The local part of
$g_{{}_{\! {\rm ETF}}}(\eps)$ is the main simplest Thomas--Fermi (TF) level density
$g_{{}_{\! {\rm TF}}}(\eps)$ \cite{sclbook}.
The second oscillating term $\delta g_{\rm scl}(\eps)$
of the level density (\ref{totdensscl}) corresponds to the fluctuating
$G_1$ in the sum
(\ref{Ggsplit}) for the Green's function $G$ near
the Fermi surface. The stationary phase conditions for the
(standard or improved)
SPM evaluation
of the integral taken from $G_1$ over the spatial coordinates $\r$
are the PO equations.
As the result, one arrives at the sum over PO sum for this
oscillating level density
\cite{gutz,strutmag,bt,sclbook},
\bea\l{dlevdenscl}
&&\delta g_{\rm scl}(\eps)
= \Re \sum_{\rm PO} \delta g_{{}_{\! {\rm PO}}}(\eps)\quad \mbox{with}\qquad
\nonumber\\
&&\delta g_{{}_{\! {\rm PO}}}(\eps)= \mathcal{B}_{\rm PO}\;
\exp\left[\frac{i}{\hbar}\; S_{\rm PO}(\eps) -
i \frac{\pi}{2}\;\sigma_{{}_{\! {\rm PO}}} -i \phi_{d}\right],\qquad
\eea
where $B_{\rm PO}$ is an amplitude of the oscillating PO terms, see
\cite{gutz,strutmag,bt,creglitl,sclbook,migdalrev,spheroidptp}.
The above sum runs over the isolated POs and, in the case of
degeneracies owing to the symmetries of
  a given potential well, over all families of POs.
  $\mathcal{B}_{\rm PO}$ is the oscillation amplitude depending on the
  stability factors, $S_{\rm PO}(\eps)$ the action integral along a given
  PO, and $\sigma_{{}_{\! {\rm PO}}}$ is
the Maslov phase associated with the turning and
  caustic points along the PO, see
\cite{sclbook,maf,migdalrev} for the detailed explanations.

The semiclassical free-energy shell corrections,
$\delta F_{\rm scl}$ at finite temperature
($T \siml \hbar \Om \ll \eps_{{}_{\! {\rm {\rm F}}}}$),
can be expressed through the PO components of the energy shell corrections
$\delta U_{\rm scl}$ \cite{strutmag,sclbook,migdalrev}
(see Appendix E.1),
\bea\l{descl}
\delta U_{\rm scl} &=& \Re \sum_{\rm PO} \delta U_{\rm PO},
\nonumber\\
\delta U_{\rm PO}&=&d_s \; \frac{\hbar^2}{t^{2}_{\rm PO}} \;
\delta g_{{}_{\! {\rm PO}}}(\mu),
\eea
with the exponentially decreasing temperature-dependent factor
\cite{strutmag,kolmagstr,richter,fraukolmagsan,sclbook,mskbPRC2010},
\bea\l{fpotau}
\delta F_{\rm scl}&=&\Re \sum_{\rm PO} \delta F_{\rm PO} \nonumber\\
&=&
\Re \sum_{\rm PO}\frac{\pi t_{{}_{\! {\rm PO}}}T/\hbar}{
\sinh\left(\pi t_{{}_{\! {\rm PO}}}T/\hbar\right)}\;\delta U_{\rm PO}.
\eea
Finally through (\ref{descl}), the shell corrections
$\delta F_{\rm scl}$ and $\delta U_{\rm scl}$ are determined by
the PO level-density shell-correction components
$\delta g_{{}_{\! {\rm PO}}}(\eps)$ of (\ref{dlevdenscl}) at the chemical potential,
$\eps=\mu \approx \eps_{{}_{\! {\rm F}}}$.
In (\ref{descl}), one has the additional factor,
$\propto 1/t^2_{\rm PO}$,
which yields the convergence of the PO sum (without averaging of
$\delta g(\eps)$
over the s.p.\ spectrum), $t_{{}_{\! {\rm PO}}}$ is the time of motion along
the PO (PO period). Another (exponential) convergence of
$\delta F_{\rm scl}$
(\ref{fpotau}) with increasing the period $t_{{}_{\! {\rm PO}}}$ and temperature
$T$ is giving by the temperature-dependent factor in front of
$\delta U_{\rm PO}$.

\subsubsection{FROM CRANKING MODEL TO THE RIGID BODY ROTATION}
\l{relperprigrot}

Substituting (\ref{Ggsplit}) into (\ref{thetaxdefG}), one has a sum
of several terms,
\bel{thetaxsum}
\Theta_{x\; {\rm scl}} \approx \Theta_x^{00} + \Theta_x^{01} +
\Theta_x^{10} + \Theta_x^{11},
\ee
where
\bea\l{thetaxnnp}
\Theta_x^{n n'}&=&\frac{2 d_s}{\pi}\int \d \eps \;n(\eps)
\int \d \r_1 \int \d \r_2\;
\ell_x\left(\r_1\right)\;\ell_x\left(\r_2\right)
\nonumber\\
&\times&
\Re \left[G_{n}\left(\r_1,\r_2;\eps\right)\right]\;
\Im \left[G_{n'}\left(\r_1,\r_2;\eps\right)\right].
\eea
Indexes $n $ and $n'$ run independently
the two integers 0 and 1. As shown in Appendix E.2a,
the main smooth part of the semiclassical
MI $\Theta_{x\; {\rm scl}}$ (\ref{thetaxsum})
is associated with the TF (ETF)
rigid-body component through the first term $\Theta_x^{00}$
averaged over the phase-space variables; see section \ref{etfmi}, also
\cite{bartelpl,gzhmagsit,mskbPRC2010}, and previous publications
\cite{bloch,amadobruekner,rockmore,jenningsbhbr}.
The statistical averaging over phase space coordinates removes
the non-local long-length correlations.
The $\hbar$ corrections of the smooth ETF approach to the TF
approximation were obtained in
\cite{jenningsbhbr,bartelnp,bartelpl},
see Sect. \ref{etfmi} for the review of these
works.

 Using the transformation of the coordinates
$\r_1$ and $\r_2$ to the center-of-mass and relative ones
$\r$ and $\s_{{}_{\! 12}}$,
\bel{newcoord}
\r=(\r_1+\r_2)/2 \qquad {\rm and}\qquad \s_{{}_{\! 12}}=\r_2-\r_1,
\ee
in (\ref{thetaxnnp}), respectively, one simplifies much the calculations of
the oscillating terms,
$~
\Theta_x^{01} +
\Theta_x^{10} + \Theta_x^{11}~$.
In this way, one finds that the shell
component $\delta \Theta_x^{01}$ of $\Theta_x^{01}$
[see (\ref{thetaxnnp}) at $n=0$
and $n'=1$]
is dominating in the MI shell correction
$\delta \Theta_{x\; {\rm scl}}$ within the NLLLA (\ref{nllla}),
see Appendix E.2b.
Indeed, in this approximation, substituting the
components, $G_0$ and $G_1$, of the Green's function (\ref{Ggsplit})
[see (\ref{G0}) for $G_0$] into
(\ref{thetaxnnp}) for $\Theta_x^{01}$,
and  using the averaging over the phase-space
variables in the fluctuating (shell) part $\delta
\Theta_{x}$ of $\Theta_{x}$,
one results in the relationship for the
corresponding shell corrections (see Appendix E.2b):
\bel{dmilocapproach}
\delta \Theta_{x\; {\rm scl}} \approx
\delta \Theta_{x}^{01} \approx \delta \Theta_{x}^{\rm (RB)}.
\ee
Here, $\delta \Theta_x^{\rm (RB)}$ is
the shell correction to the rigid-body
MI $\Theta_x^{\rm (RB)}$,
 which is related to the semiclassical particle-density
$\rho(\r)$ through
\bel{rigbodmomgen}
\Theta_x^{\rm (RB)} =
m \int \d \r\;r_{\perp x}^2\; \rho\left(\r\right),
\ee
with
\bel{rperpcoord}
r_{\perp x}^2=y^2+z^2.
\ee
The particle density $\rho(\r)$, and therefore, the MI
(\ref{rigbodmomgen}),  can be expressed in terms of the Green's
function $G$,
\bel{denpartgen}
\rho(\r) = -\frac{d_s}{\pi}\; \Im \int \d \eps\; n\left(\eps\right)\;
\left[G\left(\r_1,\r_2;\eps\right)\right]_{\r_1 \rightarrow \r_2 \rightarrow \r}.
\ee
With the splitting of the Green's function (\ref{Ggsplit}),
one obtains the semiclassical sum of the smooth and oscillating (shell)
components \cite{strutmagvvizv1986,brackrocciaIJMPE2010}:
\bel{denpartscl}
\rho(\r)\approx  \rho_{\rm scl}(\r)=\rho_{{}_{\! {\rm ETF}}}(\r) +
\delta \rho_{\rm scl}(\r).
\ee
The integration over $\eps$ in (\ref{denpartgen}) is performed
over the whole s.p.\ energy spectrum.
For the Green's function $G$, we applied the semiclassical expansion
(\ref{GRdefsem}) in terms of the sum (\ref{Ggsplit}) of CTs in the
last equation for the semiclassical particle density $\rho_{\rm scl}(\r)$.
The first term in (\ref{denpartscl}) is the (extended) Thomas--Fermi component
(see Appendix E.2a).
Substituting the particle density splitting (\ref{denpartscl}) into
(\ref{rigbodmomgen}), one has the corresponding semiclassical
expression
of the rigid-body MI,
\bel{rigmomsplit}
\Theta_x^{\rm (RB)} \approx \Theta_{x\; {\rm scl}}^{\rm (RB)} =
\Theta_{x\; {\rm ETF}}^{\rm (RB)} +
\delta \Theta_{x\; {\rm scl}}^{\rm (RB)}.
\ee
We introduced the shell corrections $\delta \rho$
(see \cite{brackrocciaIJMPE2010}) to the particle
density $\rho$ and $\delta \Theta_{x {\rm scl}}^{\rm (RB)}$ to the rigid-body MI
$\Theta_x^{\rm (RB)}$, and their semiclassical counterparts,
\bel{drigbodmomgen}
\delta \Theta_{x}^{\rm (RB)} \approx
\delta \Theta_{x \; {\rm scl}}^{\rm (RB)} =
m \int \d \r\;r_{\perp x}^2\; \delta \rho_{\rm scl}\left(\r\right),
\ee
where
\bel{ddenpart}
\delta \rho_{\rm scl}\left(\r\right)=-\frac{d_s}{\pi}\;
\Im \sum_{{\rm CCT}_1}\int \d \eps\; n\left(\eps\right)\;
G_{{\rm CCT}_1}\left(\r_1,\r_2;\eps\right),
\ee
where $G_{{\rm CCT}_1}$ is given by (\ref{Gct}) with ${\rm CT}$ being the
closed ${\rm CT}_1$, i.e.,
${\rm CCT}_1$ ($\r_1\rightarrow \r_2 \rightarrow \r$).
With the smooth (extended) TF MI component (\ref{TFrig}),
see also the section \ref{etfmi}, the equation (\ref{dmilocapproach})
yields semiclassically
\bel{milocapproach}
\Theta_{x\; {\rm scl}} \approx \Theta_{x\; {\rm scl}}^{\rm (RB)},
\ee
that is in agreement with the adiabatic picture
of the statistically equilibrium rotation \cite{mskbPRC2010}.
Note that the non-adiabatic MI at arbitrary rotation
frequencies for the HO mean field by Zelevinsky \cite{zelev} was extended
to the finite
temperatures in \cite{mskbPRC2010}.

We emphasize that due to
an averaging over the phase space variables, one survives with
the NLLLA.
Note also that the classical angular-momentum projection (\ref{l2})  in the
rotating body-fixed coordinate system is caused by the global rotation
with a given frequency $\omega$ rather than by
the motion of particles along the trajectories  inside the nucleus
with respect to this system,
considered usually in the cranking model.
According to the time-reversible symmetry of the Routhian,
the particles are, indeed, moving in the non-rotating coordinate system
along these trajectories in both opposite directions. 
Their contributions to the total angular momentum of the nucleus turns out to
be zero. Performing then
the integration over $\s$ in (\ref{dthetax01new})
 in the spherical coordinate system,
one obtains the rigid-body shell correction
$\delta \Theta_x^{\rm (RB)}$ in the NLLLA
as explained in Appendix E.2.
Note that
the cranking model for the nuclear rotation
implies that the correlation (non-local) corrections to
(\ref{dthetax01new}) and (\ref{rigbodmomgen}) should be
small enough with respect to the main rigid-body shell component
$\delta \Theta_x^{\rm (RB)}$ to be neglected within the adiabatic picture
of separation of the global rotation of the Fermi system
from its vibration and then,
both from the internal motion of particles.
Other contributions,
except for a smooth rigid-body part coming from $\Theta_x^{00}$, like
$\Theta^{10}_{x}$ and $\Theta^{11}_{x}$, as
referred to the fluctuation (non-local) correction
to the rigid body MI are found semiclassically to be negligibly small
in the NLLLA
due to the averaging
over phase-space variables, see Appendix E.2b.
In particular, for the HO Hamiltonian, it was shown that there is almost
no contribution of the
$\delta \Theta_x^{11}$ at leading order in $\hbar$ in \cite{mskbPRC2010}.
Thus, with the semiclassical precision, from the
adiabatic cranking model expression (\ref{thetaxdefG}) we come to the
MI of the statistically equilibrium rotation (\ref{milocapproach}),
which must be the rigid-body MI, according to
the general theorem of the statistical physics. This is in agreement
with the ETF approach of section \ref{etfmi}.
Our semiclassical derivations, valid for the rotation frequencies
$\hbar \om \ll \hbar \Om$, are beyond the quantum criterion
of the application of the standard
2nd order perturbation approach within the cranking model where
$\hbar \om$ is small as compared to the distance between the
neighboring levels of quantum spectra. We point out that this weakness
of the perturbation theory criterion is similar to that with the statistical
averaging in the heated Fermi systems and with accounting for the pairing
correlations \cite{belyaevhighspin,belyaevbif}, where the role of the
distance between the quantum neighboring energy levels plays the
temperature and the pairing gap, as distance between gross shells
$\hbar \Omega$ (\ref{hom}) in the POT \cite{strutmag}, respectively.

\medskip

\subsubsection{
SHELL CORRECTIONS TO THE RIGID-BODY MI}
\l{SCperprigrot}

Using (\ref{ddenpart}) for calculations of
the MI rigid-body shell correction
$\delta \Theta_{x\; {\rm scl}}^{\rm (RB)}$ (\ref{drigbodmomgen}),
one may exchange the order of integrations over the coordinate $\r$
and energy $\eps$.
By making use also of the semiclassical
trajectory expansion (\ref{GRdefsem}) for the oscillating Green's
function component
$G_1\left(\r_1,\r_2;\eps\right)$ of the sum (\ref{Ggsplit}), one finds
\bea\l{dTxrigSCL}
&&\delta \Theta_{x\; {\rm scl}}^{\rm (RB)} =
-\frac{m d_s}{\pi}\;
\Im \sum_{{\rm CCT}_1} \int \d \eps\;
n(\eps) \qquad \nonumber\\
&\times& \int \d \r\;\left\{r_{\perp x }^2 \mathcal{A}\left(\r,\r;\eps\right)
\right.\qquad\nonumber\\
&\times& \left.
\exp\left[\frac{i}{\hbar}\;
 S\left(\r_1,\r_2;\eps\right) -
\frac{i \pi}{2}\sigma - i\phi_{d}\right]
\right\}_{{\rm CCT}_1}.\qquad
\eea
As usually, with the  semiclassical precision, we evaluate
the spatial integral by the SPM extended to continuous symmetries
\cite{strutmag,sclbook,migdalrev} and the
bifurcation phenomena (ISPM)
\cite{ellipseptp,spheroidptp,maf,migdalrev,magvlasar}.
The SPM (ISPM) condition writes
\bea\l{spmcond}
&&\left[\frac{\partial S\left(\r_1,\r_2;\eps\right)}{\partial \r_1}
+ \frac{\partial S\left(\r_1,\r_2;\eps\right)}{\partial \r_2}
 \right]^{\ast}_{{\rm CCT}_1}
\nonumber\\
 &\equiv&
\left(-\p_1+\p_2\right)^{\ast}_{{\rm CCT}_1},
\eea
where the asterisk means the SPM value of the spatial coordinates
and momenta,
$~\r_j=\r_j^{\ast}~$ and $~\p_j=\p_j^{\ast}~$ ($~j=1,2$)
at the closed ${\rm CT}_1$s
in the phase space,
$\r^{\ast}_1=\r^{\ast}_2$ and $\p^{\ast}_1=\p^{\ast}_2$. Thus, with the
standard relations for the canonical variables by using the action as a
generating function, one arrives
at the PO condition  on right of (\ref{spmcond}).
Within  the simplest ISPM
\cite{spheroidptp,maf,migdalrev,magvlasar},
the other smooth factors $r_{\perp x}^2$ and
$\mathcal{A}_{{\rm CT}_1}\left(\r,\r,\eps\right)$
of the integrand in (\ref{dTxrigSCL})
 can be taken off the
integral over $\r$ at these stationary points. Assuming that the
quantum averages
$\langle (y^2+z^2)^2 \rangle/\eps$ are smooth enough functions of $\eps$
as compared to other factors, for instance, $\delta n$,
one may take them approximately also off the integral
over $\eps$ at the chemical potential, $\eps=\mu$. For example,
for the HO potential (see \cite{mskbPRC2010}),
they are simply exact constants.
Therefore, the main contribution into the integral
in (\ref{dTxrigSCL}) is coming from
the PO stationary-phase points, determined by (\ref{spmcond}),
 as for calculations of the
level-density shell corrections $\delta g_{\rm scl}$ (\ref{dlevdenscl})
\cite{sclbook,strutmag,migdalrev,mskbPRC2010}.
The SPM condition (\ref{spmcond}) is
 identity for any stationary point of the classically accessible
spatial region for a particle motion filled by PO families
in the case of their high
degeneracy
$\mathcal{K}\geq 3$.
For instance, it is the case for the contribution  of
the three dimensional (3D) orbits in the axially symmetric HO-potential well
with commensurable frequencies, $\om_x=\om_y=\om_\perp$ and $\om_z$
\cite{magosc,mskbPRC2010}.
The stationary points occupy
some spatial subspace for a smaller degeneracy $\mathcal{K}$.
In the latter case of the equatorial orbits (EQs) ($\mathcal{K}=2$)
in this HO potential well,
the SPM condition is
identity in the equatorial plane $z=0$.
Following
similar derivations of  the oscillating component $\delta g_{\rm scl}$
(\ref{dlevdenscl})
of the level density $g_{\rm scl}(\eps)$ (\ref{totdensscl})
and free-energy shell correction $\delta F_{\rm scl}$ (\ref{fpotau}),
one expands the
smooth amplitudes and action phases of the
MI shell corrections
$\delta \Theta_{\kappa {\rm scl}}^{\rm rig}$ (\ref{dTxrigSCL}) up to the
first nonzero terms
(see Appendix C of \cite{mskbPRC2010} and Appendix E.2
here).
Finally, from (\ref{dTxrigSCL}), one obtains \cite{mskbPRC2010}
\bel{dTxrigSCLgen}
\delta \Theta_{x {\rm scl}}^{\rm (RB)}=\frac{m}{\mu}\;\Re \sum_{\rm PO}
\langle r_{\perp x}^2 \rangle_{{}_{\! {\rm PO},\mu}}\;\delta F_{\rm PO},
\ee
where $\langle r_{\perp x}^2 \rangle_{{\rm PO},\mu}$
is the average given by
\bel{avsqcoor}
\langle r_{\perp x}^2 \rangle_{{\rm PO},\eps} =
\frac{\int \d \r\;\mathcal{A}_{\rm PO}\left(\r,\r;\eps\right)\;r_{\perp x}^2
 }{\int \d \r\;\mathcal{A}_{\rm PO}\left(\r,\r;\eps\right)}
\ee
at $\eps=\mu$, $\mathcal{A}_{\rm PO}(\r,\r;\eps)$ are the
Green's function amplitudes
for a closed ${\rm CT}_1$ in the phase space, i.e., PO.
Integration over $\r$ is performed  over the classically
accessible region of the spatial coordinates.
Semiclassical expression (\ref{dTxrigSCLgen})
is general for any potential well.
Shorter POs are
dominating in the PO sum (\ref{dTxrigSCLgen})
\cite{sclbook,strutmag,kolmagstr,mskbPRC2010,migdalrev},
see
(\ref{fpotau}), (\ref{descl}). Therefore,
according to (\ref{fpotau}) for $\delta F_{\rm scl}$,
we obtain approximately the relation
\bel{dTxrigSCLgen1}
\delta \Theta_{x {\rm scl}}^{\rm (RB)}\approx
\frac{m}{\mu}\;
\langle r_{\perp x}^{2}\rangle_{\mu}\;
\delta F_{\rm scl},
\ee
where $\langle r_{\perp x}^2 \rangle_{\mu}$ is an
average value of the quantity (\ref{avsqcoor}),
independent of the specific PO, at $\eps=\mu$ over short
dominating POs.

For the axially symmetric
 HO potential well with the commensurable frequencies $\om_\perp$
and $\om_z$, as the simplest example,
the integration in (\ref{avsqcoor}) over $\r$
for the 3D contribution means over the 3D volume occupied by
the 3D families of orbits.
For the EQ component
the integral is taken
over the 2D spatial region filled by the EQ families
in the equatorial ($z=0$) plane \cite{mskbPRC2010}.
In the incommensurable-frequency
case (irrational $\om_\perp/\om_z$), one has the only EQ-orbit contributions.
The average (\ref{avsqcoor}) can be easily calculated
by using the Green's  function amplitudes $\mathcal{A}_{\rm PO}$ 
for 3D and
for EQ orbits, which are given in \cite{magosc,mskbPRC2010}.
Finally, for the considered HO potential, one may arrive at
\bel{dthetaxdehoscl}
\delta \Theta_{x\; {\rm scl}} \approx \delta \Theta_{x {\rm scl}}^{\rm (RB)}=\frac{1 +
\eta_{\rm HO}^2}{3 \om_\perp^2}\;\delta F_{\rm scl},
\ee
where $\delta F_{\rm scl}$ is the semiclassical PO sum
(\ref{fpotau}), (\ref{descl})
for the semiclassical free-energy shell-corrections,
$\eta_{{}_{\!{\rm HO}}}=\om_\perp/\om_z$ is the deformation parameter.
For the parallel (alignment) rotations around the symmetry axis,
 one finds similar relations of the MI through the rigid-body MI to
the free-energy shell corrections.
Moreover, one has such relations
for the smooth TF parts, in particular for the HO case, see Appendices E.2.1
here and D1 in
\cite{mskbPRC2010}.
Thus, for the total
moment $\Theta_x$ [see (\ref{thetaxsum})], one may prove semiclassically
within the POT, up to the same $\hbar$ corrections in a
smooth TF part, that
the shell MI and free-energy shell corrections are approximately
proportional, in particular exactly
 for that
HO Hamiltonian \cite{mskbPRC2010}:
\bel{thetaxhoscl}
\Theta_{x\; {\rm scl}}=\frac{1 + \eta^2}{3 \om_\perp^2}\;F_{\rm scl},\qquad
F_{\rm scl}=F_{{\rm ETF}}+\delta F_{\rm scl}.
\ee
We emphasize that the POT expressions (\ref{dthetaxdehoscl})
for $\delta \Theta_{x\; {\rm scl}}$ and
(\ref{thetaxhoscl})
of $\Theta_{x\; {\rm scl}}$  were derived
without a direct use of the statistically equilibrium rotation
condition \cite{bohrmot,mskbPRC2010}.

Substituting the semiclassical PO expansion
(\ref{descl}) for the free-energy shell correction
$\delta F_{{\rm scl}}$ (\ref{fpotau}) (after \cite{spheroidptp})
for 3D orbit families and
for EQ POs into (\ref{dthetaxdehoscl}),
one arrives  finally at the explicit POT expressions for the MI
shell corrections $\delta \Theta_x$
in terms of the characteristics of the classical
POs. For the mean field with the spheroidal shapes and sharp edges
(spheroid cavity), these derivations can be performed similarly
as for the HO Hamiltonian in \cite{mskbPRC2010}
but with accounting
for the specific PO degeneracies.
Note that
the parallel, $\delta \Theta_z$,
and perpendicular, $\delta \Theta_x$,
MI shell components  are expressed through the 3D and EQ POs
through the free-energy shell correction which contains generally
speaking both them for the deformations larger the bifurcation ones.
The dominating
 contributions
of one of these families or coexistence of both together
depend on the surface deformation parameter (semi-axis ratio
of spheroid). For the critical deformations and on right of them, one observes
the significant enhancement of the MI shell corrections through the PO
level-density amplitudes $\mathcal{B}_{\rm POT}$
[see (\ref{dlevdenscl})] of the free-energy shell corrections
(\ref{fpotau}), (\ref{descl}).

\subsubsection{COMPARISON OF SHELL STRUCTURE CORRECTIONS WITH\\ QUANTUM RESULTS}
\l{qmsclcomparison}

Fig.\ \ref{fig15} shows the
semiclassical free-energy shell correction $\delta F_{\rm scl}$,
[(\ref{fpotau}), (\ref{descl}), see also \cite{magosc,mskbPRC2010}]
vs the particle-number variable, $A^{1/3}$,
at a small temperature of $T=0.1\; \hbar \omega_{{}_{\! 0}}$ for different critical
symmetry-breaking and bifurcation deformations
$\eta_{{}_{\! HO}}=1$, $6/5$, and $2$ of the HO potential
\cite{sclbook,mskbPRC2010} with
the corresponding
quantum SCM results for the same
deformations.
This comparison also shows practically a perfect agreement
between the
semiclassical,
(\ref{fpotau}) and (\ref{descl}),
and quantum results.
For the spherical case ($\eta_{{}_{\! HO}}=1$),
one has only contributions of the families
of 3D orbits with the highest degeneracy $\mathcal{K}=4$.
At the bifurcation points $\eta_{{}_{\! HO}}=6/5$ and $2$ the relatively simple
families of these 3D POs appear
along with EQ orbits of smaller degeneracy.
For $\eta_{{}_{\! HO}}=6/5$, one mainly has the contributions from EQ POs
because the 3D
orbits are generally too long in this case.
For the bifurcation point $\eta_{{}_{\! HO}}=2$,
 one finds an interference of the two
comparably large contributions of EQ and 3D
orbits
with  essentially the different time periods $t_{{}_{\! EQ}}$ and
$t_{{}_{\! 3D}}$, respectively.

The quantum (QM) and semiclassical
(SCL) shell corrections to
the MI $\delta \Theta^{}_x$ of (\ref{dthetaxdehoscl})
are compared in Fig.\ \ref{fig16}.
An excellent agreement is observed
between the semiclassical and quantum results as for the free-energy shell
corrections $\delta F$.
 It is not really astonishing
because of the proportionality of the $\delta \Theta^{}_x$ to
$\delta F$ [see (\ref{dthetaxdehoscl})].
One finds in particular the same clear interference of contributions of
3D and EQ POs in the shell corrections to the MI at $\eta_{{}_{\! HO}}=2$.
The exponential decrease of shell oscillations with increasing temperature,
due to the temperature factor
in front of the PO energy-shell correction components $\delta E_{\rm PO}$
in (\ref{fpotau})
is clearly seen in Fig.\ \ref{fig16}. As the MI and free-energy
shell corrections are basically proportional [see (\ref{dTxrigSCLgen})]
for any mean potential well, we may emphasize
the amplitude enhancement of the MI near the bifurcation
deformations due to that
for the energy-shell corrections found in
\cite{ellipseptp,spheroidptp,maf,migdalrev,magvlasar}.
The
critical temperature for a disappearance of shell effects in the MI
is found for prolate deformations ($\eta>1$)
and particle numbers
$A \sim 100-200$, approximately at $T^{}_{cr}=
\hbar \omega^{}_{EQ}/\pi \sim \hbar \omega^{}_{0}/\pi \approx 2-3$ MeV
just as for $\delta F$, see \cite{sclbook,strutmag,mskbPRC2010}. This effect
is also general for any potentials.
The particle-number dependence of the shell corrections
 $\delta \Theta^{}_z$  to the total MI $\Theta^{}_z$
(alignment)
is not shown because it is similar to that of $\delta \Theta^{}_x$
through their approximate relations,
$\delta \Theta^{}_z \propto \delta \Theta^{}_x \propto \delta F$.


\section{Conclusions}
\label{concl}

We derived the dynamical equations of motion, such as the conservation
of the particle number, momentum and energy as well as the
general transport equation for the entropy for \textit{low}
frequency excitations in nuclear matter within the Landau
quasiparticle theory of \textit{heated} Fermi-liquids.
Our approach is based essentially on the
Landau--Vlasov equation for the distribution function, and it
includes all its moments in phase space, in contrast to several
truncated versions of fluid dynamics
similar to the hydrodynamic description
in terms of a few first moments. From the dynamics of the
Landau--Vlasov equation for the distribution function, linearized
near the \textit{local} equilibrium, we obtained the momentum flux
tensor and heat current in terms of the shear modulus, viscosity,
in-compressibility and thermal conductivity coefficients as for
very viscose liquids called sometimes \textit{amorphous solids}.
We obtain the dependence of these coefficients on the
         temperature, the frequency and the Landau interaction parameters.
We derived the \textit{temperature
expansions} of the density-density and
temperature-density response functions for nuclear matter and got
their \textit{specific expressions for small temperatures} as
compared to the chemical potential. The \textit{hydrodynamic limit}
of normal liquids for these response functions \textit{within the
perturbation theory} was obtained from the Landau--Vlasov equation for
both distribution function and sound
     velocity, as for an eigenvalue problem.
In this way we found the Landau--Placzek and first sound peaks in
the corresponding strength functions as the hydrodynamic limit of
the Fermi-liquid theory for heated Fermi-systems. The former (heat
pole) peak was obtained only because of the use of the local
equilibrium in the Landau--Vlasov linearized dynamics instead of
the global static Fermi-distribution of the giant multipole-resonance
physics. This is very important for the dispersion equation and
its wave velocity solutions.

We got the \textit{isolated, isothermal and adiabatic
susceptibilities for the Fermi-liquids} and showed that they
satisfy the \textit{ergodicity} condition of equivalence of the
isolated and adiabatic susceptibilities as well as the general
Kubo inequality relations. We found the \textit{correlation function}
using the fluctuation-dissipation theorem and discussed its
relation to the susceptibilities and Landau--Placzek "heat pole" in
the hydrodynamic limit.

We applied the theory of heated Fermi-liquids to the Fermi-liquid
drop model of finite nuclei within the Landau--Vlasov dynamics in the
nuclear interior and macroscopical boundary conditions in the
effective sharp surface approximation. Solutions of this
problem in terms of the response functions and transport
coefficients were obtained. We considered the hydrodynamic limit
of these solutions and found the ``heat pole'' correlation function
for frequencies smaller than some critical frequency. The latter
was realized only because of using the local equilibrium for the
distribution function. The isolated, isothermal and adiabatic
susceptibilities for finite nuclei within the FLDM in the ESA
were derived. We showed that the ergodicity condition is satisfied
also for finite Fermi-systems as for infinite nuclear
matter in the same ESA.

We found a three-peak structure of the collective strength
function: the "heat", standard hydrodynamic and essentially
Fermi-liquid peaks. The conditions for the existence of such modes
were analyzed and the temperature dependence of their transport
coefficients such as friction, stiffness and inertia were
 obtained
in particular, in the hydrodynamic limit.
We arrived at the increasing temperature dependence of the
friction coefficient for the specific Fermi-liquid mode which
exist due to the Fermi-surface distortions. At enough large
temperatures, we showed a nice agreement with the results for the
friction which were obtained earlier within the microscopic
shell-model approach of \cite{hofbook}. The correlation functions
found in the FLDM and quantum shell models were discussed in
relation to the susceptibilities and ergodicity properties of
finite nuclei.

The expression for the surface symmetry-energy constant $k_{{}_{\! S}}$
was derived from simple isovector solutions
of the particle density and energies in the leading ES
approximation. We used them for the calculations
of the energies, sum rules of the  IVGDR strength and the transition
densities
within the HDM and FLDM  \cite{kolmagsh} for several Skyrme-force parameters.
The surface symmetry-energy constant
depends much on the fundamental
well-known parameters of the Skyrme forces, mainly through the
coefficient in the  density gradient terms of the isovector part of the
energy density.
The value of this isovector constant
 is rather sensitive  also
on the SO interaction.
The IVGDR strength is split into the two main and satellite peaks.
The mean energies and EWSRs within both HDM and FLDM are in fairly
good agreement with the
experimental data.

Semiclassical functional expressions were derived
 in the framework of the Extended Thomas--Fermi approach. We used
 these analytical expressions to obtain a self-consistent description
 of rotating nuclei where the rotation velocity impacts on the structure
 of the nucleus. It has been shown that such a treatment leads, indeed,
 to the Jacobi phase transition to triaxial shapes as already predicted
 in \cite{CPS74} within the rotating LDM. We emphasize
 that the rigid-body moment of inertia gives a quite accurate
 approximation for the full ETF value. Being aware of the mutual influence
 between rotation and pairing correlations
 \cite{belyaevhighspin,pomorbartelPRC2011,sfraurev}, it would be
 especially  interesting to work on an approach that is able to
determine  the nuclear structure depending on its angular velocity,
as we have  done here in the ETF approach, but taking pairing correlations
and their rotational quenching into account.

We derived also the shell corrections
of the MI
in terms of free-energy shell corrections
within the nonperturbative extended POT through those of the rigid-body MI of
the equilibrium rotations,
which is exact for the HO potential.
For the HO, we extended
to the finite temperature case the Zelevinsky
derivation of the non-adiabatic
MI at any rotation frequency.
For the deformed HO potential, one finds a perfect agreement
between the semiclassical POT and quantum
results for the free-energy
and the MI shell corrections
at several critical deformations and temperatures.
For larger temperatures, we show that the short EQ orbits are
mostly dominant.
For small temperatures, one observes a remarkable interference of the
short 3D and EQ orbits in the superdeformed region.
An exponential decrease of all shell corrections
with increasing temperature is observed, as expected. We point out
also the amplitude enhancement of the MI shell corrections due to
the bifurcation catastrophe phenomenon.

As further perspectives, it would be worth to apply our results to
calculations of the IVGDR structure within the Fermi-liquid droplet
model to determine the value of the fundamental surface symmetry-energy
constant from comparison with experimental data for the pygmy resonance
\cite{adrich,wieland} and theoretical calculations
\cite{vretenar1,vretenar2,ponomarev,nester1,nester2,BMR}.
For further extensions to
the description of  the isovector low-lying collective states, one has first to
use the POT for including semiclassically the shell effects
\cite{strutmag,sclbook,gzhmagfed,blmagyas,BM}.
It would be also worth to apply this semiclassical theory
to the shell corrections of the MI for the spheroid cavity
and for the inertia parameter of the
low-lying collective excitations in nuclear dynamics involving
magic nuclei
\cite{dfcpuprc2004,magvvhof,yaf,gzhmagfed}. One of the most attractive
subject of the semiclassical periodic orbit theory, in line of the main
works of S.T. Belyaev
\cite{belyaevfirst,belyaevzel,belzel,belsmitolfay87},
is its extension
to the {\it pairing} correlations
\cite{brackquenNPA1981,brackrocciaIJMPE2010},
and their influence on the collective
vibrational and rotational excitations in heavy deformed neutron-rich
nuclei
\cite{belyaevhighspin,pomorbartelPRC2011,sfraurev} (see also
\cite{abrpairing} for the semiclassical phase-space dynamical approach
to the Hartree--Fock--Bogoliubov theory).

\bigskip

\section*{Acknowledgement}

\label{aknow}

\medskip

Authors gratefully acknowledge 
H.\ Hofmann for many suggestions and fruitful discussions, also 
 S.\ Aberg, V.\ I.\ Abrosimov, J.\ Blocki, R.\ K.\ Bhaduri, M.\ Brack, 
V.\ Yu.\ Denisov, S.\ N.\ Fedotkin,
H.\ Heisenberg, F.\ A.\ Ivanyuk,
V.\ M.\ Kolomietz, M.\ Kowal, J.\ Meyer, V.\ O.\ Nesterenko, V.\ V.\ Pashkevich,
M.\ Pearson, V.\ A.\ Plujko, P.\ Ring, V.\ G.\ Zelevinsky, 
A.\ I.\ Sanzhur, S.\ Siem, J.\ Skalski,
and X.\ Vinas for many useful discussions.  One of us (A.G.M.) is
also very gratitude for a nice hospitality during his working visits of the
Technical Munich University in Garching and the University 
of Regensburg in Germany, the Interdisciplinary 
Hubert Curien Institute  of the Louis Pasteur University 
in Strassburg of France,  and 
National Centre for Nuclear Research in Otwock-Swierk of Poland.


\vspace{1cm}

\noindent
\begin{appendix}
\setcounter{equation}{0}
\renewcommand{\thesubsection}{\Alph{section}.\arabic{subsection}}
\renewcommand{\theequation}{\mbox{\Alph{section}.\arabic{equation}}}
\noindent
\section{Elements of Landau theory for equilibrized systems}
\label{app1}
\subsection{Thermodynamic relations}
\label{app1thermrel} Let us begin recalling the fundamental
equations $TdS = dE-\mu dN +\press \d \mathcal{V}$ and $-S\d T = \d F-\mu
\d N +\press \d \mathcal{V}$, which are related to each other by the
Legendre transformation $F=E-TS$. They imply the following
relations for the chemical potential $\mu$ and pressure
$\press$: 
\bel{chempot} 
\mu =-T\left(\frac{\partial S }{\partial
N}\right)_{E,{\mathcal{V}}}= \left(\frac{\partial E }{\partial
N}\right)_{S,{\mathcal{V}}}= \left(\frac{\partial F }{\partial
N}\right)_{T,{\mathcal{V}}}, 
\ee 
%
\bel{pressure} \press =
-\left(\frac{\partial E }{\partial \mathcal{V}}\right)_{S,N}=
-\left(\frac{\partial F }{\partial \mathcal{V}}\right)_{T,N}. 
\ee 
For
homogeneous systems the {\it intensive} quantities depend only on
two independent variables. For instance, the entropy per particle
$S/N=\varsigma(E/N,\mathcal{V}/N)$ only depends on the energy and
volume per particle, $E/N $ and $\mathcal{V}/N$ respectively. For
such systems, the adiabadicity condition may simply be expressed as
$\varsigma = {\rm const}$. 
Commonly in
nuclear physics, one uses  the particle density $\rho=N/\mathcal{V}$,
in which case the chemical potential can be expressed as
\bel{chempothom} 
\mu = \left({\partial \phi \over
\partial \rho}\right)_T 
\ee
 with $\phi = F/\mathcal{V}$ being the
free internal energy per unit volume.

For differential quantities there exist various variants of the  
Gibbs-Duheim relation 
\bel{giduvar} 
\d \phi = -\varsigma \rho \d T + \mu \d \rho \qquad \mbox{or} 
\ee
%
\bea\l{gibbsduh} 
&&\d\press = \varsigma \rho \d T + \rho \d \mu~
\quad{\rm implying} \quad \left({\partial \varsigma \over
\partial \mu}\right)_T \nonumber\\
&=& \frac{1}{ \rho} \left[\left(\frac{\partial \rho
}{\partial T}\right)_\mu- \varsigma \left(\frac{\partial \rho
}{\partial \mu}\right)_T \right], 
\eea
  as follows from
Legendre transformations. Thus for the derivatives of the pressure
$\mathcal{P}$, considered as functions of $T$ and $\rho$, one gets from
(\ref{gibbsduh}):
\bea\l{dpdmt}
 &&\left(\frac{\partial \press }{\partial \rho} \right)_T =
    \rho \left(\frac{\partial \mu }{\partial \rho}\right)_T,
    \qquad
    \left(\frac{\partial \press }{\partial T} \right)_\rho =\nonumber\\
  &=&\rho\left(\varsigma-
    \left(\frac{\partial \rho } {\partial T}\right)_{\mu}
    \left(\frac{\partial \mu }{\partial \rho}\right)_T\right).
\eea
 In deriving such relations, it is useful to employ special
properties of the Jacobian,  which allows
one to perform transformations between different variables (see
e.g.,
\cite{brenig}).
These relations will be used below to get the specific heats as
well as the isothermal and adiabatic compressibilities, together
with the corresponding susceptibilities. At first, we shall look at
{\it in}-compressibilities  defined by the
derivative of the pressure over the particle density
 (multiplied by a factor of 9). At {\it
constant entropy per particle} $\varsigma$, the {\it adiabatic}
in-compressibility $K^{\varsigma}$ writes 
\bel{incompraddef}
K^{\varsigma} \equiv 9 \left(\frac{\partial \press }{\partial
\rho}\right)_\varsigma = 9 \left(\frac{\partial \press }{\partial
\mu}\right)_\varsigma \left(\frac{\partial \mu }{\partial
\rho}\right)_\varsigma. 
\ee 
To get the corresponding quantity
at constant temperature $K^{T}$, one only needs to replace
$\varsigma$ by $T$. According to (\ref{dpdmt}) and
(\ref{chempothom}), one obtains 
\bel{incomprTdef} 
K^{T} \equiv 9
\left(\frac{\partial \press }{\partial \rho}\right)_T = 9 \rho
\left(\frac{\partial \mu }{\partial \rho}\right)_T = 9 \rho
    \left(\frac{\partial^2 \phi }{\partial \rho^2}\right)_T.
\ee

Next we turn to the specific heats at constant volume and constant
pressure. If measured per particle, they can be defined in terms
of the entropy per particle $\varsigma$  as 
\bel{specifheatpdef}
{\tt C}_{\mathcal{V}} =
    T \left(\frac{\partial \varsigma }{\partial T} \right)_{\mathcal{V}}=
    T \left(\frac{\partial \varsigma }{\partial T} \right)_\rho,
\quad {\tt C}_\press =
    T \left(\frac{\partial \varsigma }{\partial T} \right)_\press.
\ee 
They obey the following, well known relation to the
in-compressibilities \cite{forster,brenig} 
\bel{cvcpkSkT} 
\left(\frac{{\tt C}_\press 
}{{\tt C}_{\mathcal{V}}}\right)= 
\frac{\left(\partial \press /\partial
\rho\right)_\varsigma }{\left(\partial \press /\partial \rho \right)_T} = 
\frac{K^{\varsigma} }{K^{T}}. 
\ee

For the variation of the entropy $\varsigma$ per particle,  one
finds %
\bel{entropypart}
 \d \varsigma = - \frac{1}{\rho}
\left[\varsigma + \left(\frac{\partial \mu }{\partial
T}\right)_\rho\right] \d \rho + \frac{{\tt C}_{\mathcal{V}}}{T} \d T,
\ee 
after using (\ref{giduvar}) and the  specific heat ${\tt
C}_{\mathcal{V}}$ of (\ref{specifheatpdef}). To get the first term we
applied 
\bel{derphi} -\left(\frac{\partial (\varsigma
\rho) }{\partial \rho}\right)_T= \left(\frac{\partial \mu 
}{\partial T}\right)_\rho \equiv \frac{\partial^2 \phi }{\partial
\rho \partial T}, 
\ee 
which is a consequence of (\ref{giduvar}).

\subsection{Landau theory proper}
\label{landtheorprop}

In the following, we will repeat some important relations discussed
in \cite{heipethrev} without arguing much about their proofs.
These relations will be needed to derive some specific
thermodynamic properties for quantities, as the entropy or the
specific heats.
A basic element in Landau theory is the microscopic
expression for the entropy per particle,  
\bea\l{entropydef} 
\varsigma &=&
    - \frac{1}{\rho} \int 
\frac{2 d \p }{\left(2 \pi \hbar \right)^3}
    \left[f_\p \ln f_\p + \left(1-f_\p\right)
    \ln \left(1-f_\p\right) \right]\nonumber\\
   &=& \left\langle \frac{p^2 }{3 m^*}
    \left(\frac{\eps_\p-\mu }{T}\right) \right\rangle /
\left\langle \frac{p^2 }{3 m^*} \right\rangle. 
\eea
 in terms of
the Fermi distribution  $f_{\p}$ [c.f. (\ref{fgeq})]. The (static)
quasiparticle density $\rho$ in (\ref{entropydef}) may be
expressed as 
\bel{densstat} 
\rho  =\frac{N}{\mathcal{V}} =  \frac{p_{\rm F}^3
}{3 \pi^2 \hbar^3}=
     \int \frac{2 \d \p }{\left(2 \pi \hbar \right)^3} f_\p
 = \mathcal{N}(T) \left\langle \frac{p^2 }{3 m^*} \right\rangle,
\ee 
with the density of states $\mathcal{N}(T)$ 
(\ref{enerdensnt}). The additional factor 2 in the integration
measure accounts for the spin degeneracy. The expressions on the
right in both (\ref{entropydef}) and (\ref{densstat}) are obtained
after integrating by parts. The brackets $<\cdots>$ denote some
kind of average, which if written for any quantity $A(\r,\p,t)$ is
defined as 
\bel{averag} 
\langle A(\r,\p,t) \rangle
    =\frac{1}{\mathcal{N}(T)} \int \frac{2 \d \p^\prime
    }{(2 \pi \hbar)^3}
    \left(-\frac{\partial f_{\p^\prime}
    }{\partial \eps_{\p^\prime}}\right)
    A(\r,\p^\prime,t).
\ee
 In addition to the $\mathcal{N}(T)$, one needs 
\bel{mcapt} 
\mathcal{M}(T) = \mathcal{N}(T) \left\langle
\frac{\eps_\p-\mu }{T}
    \right\rangle.
\ee
 From (\ref{densstat}) one may derive (see (2.9) and
(2.11) of \cite{heipethrev}) 
\bel{drhomt} 
\d \rho =
    \frac{\mathcal{N}(T) }{\mathcal{G}_0}~ \d \mu +
\frac{\mathcal{M}(T)}{\mathcal{G}_0}~\d T,
\ee
[see also (\ref{effmass}) for $\mathcal{G}_0$] which allows
one to express the isothermal in-compressibility $K^{T}$ 
(\ref{incomprTdef}) by (\ref{isotherk}).
For the variation of the pressure with temperature, one gets
from (\ref{dpdmt}) and (\ref{drhomt}) 
\bel{dpdtr}
    \left(\frac{\partial \press }{\partial T} \right)_{\rho} =
    \left(\varsigma- \frac{\mathcal{M}(T)}{\mathcal{N}(T)}\right) \rho =
    \frac{2}{3} \rho {\tt C}_{\mathcal{V}}.
\ee 
For the proof of the second equation, we refer to (3.35) of
\cite{heipethrev} (mind however a difference in the notations for
the specific heat: Our $\rho {\tt C}_{\mathcal{V}}$ is identical to the
$ {\tt C}_{\mathcal{V}}$ of \cite{heipethrev}). For our 
$ {\tt C}_{\mathcal{V}}$, one may derive the formula (see (3.34) of
\cite{heipethrev}) 
\bel{specheatveq} 
{\tt C}_{\mathcal{V}} = 
\frac{T \mathcal{N}(T)}{\rho} \left\langle 
\left[\frac{\eps_\p-\mu }{T}
   -\frac{\mathcal{M}(T)}{\mathcal{N}(T)} \right]^2 \right\rangle.
\ee 
Collecting (\ref{incomprTdef}), (\ref{isotherk}) and
(\ref{dpdtr}), one can write the variation of the pressure 
as
\bel{dpress} 
{\rm d}\press = \rho 
\left(\frac{\mathcal{G}_0}{\mathcal{N}(T)} 
{\rm d} \rho + \frac{2}{3} {\tt C}_{\mathcal{V}} {\rm d} T\right). 
\ee

Thermodynamic quantities such as
in-compressibilities and susceptibilities are calculated under
different conditions as fixed temperature or entropy. As one
knows (see, e.g., 
\cite{forster}), these
(in)-compressibilities may be associated to different sound
velocities. To make use of the adiabaticity condition mentioned
earlier, we need the derivatives of the entropy per particle
$\varsigma(\rho,T)$. The ones arising in (\ref{entropypart}) can
be simplified by exploiting the specific Fermi-liquid expressions
given in (\ref{drhomt}) and the second relation of (\ref{dpdtr})
between the entropy per particle $\varsigma$ and the specific heat
${\tt C}_{\mathcal{V}}$,
\bel{dspdmt}
  \d \varsigma =
   - \frac{2}{3 \rho} {\tt C}_{\mathcal{V}} \d \rho +
    \frac{{\tt C}_{\mathcal{V}}}{T} \d T.
\ee 
Next we turn to the  adiabatic in-compressibility
$K^{\varsigma}$ (\ref{incompraddef}).  It may be expressed by
the isothermal one $K^{T}$ given in (\ref{isotherk}), 
see (\ref{Kadiabat}).
To derive this relation, 
the Jacobian transformation from $(\rho,\varsigma)$ to $(\rho,T)$ for
the derivatives of the pressure in (\ref{incompraddef}) has been applied
[mind also
(\ref{dpdtr}), (\ref{incomprTdef}) and (\ref{dspdmt})]. Finally,
for the ratio of the specific heats, we find from
(\ref{cvcpkSkT}), (\ref{isotherk}) and (\ref{Kadiabat})
\bel{cpcvkakt} 
\left(\frac{{\tt C}_\press }{{\tt C}_{\mathcal{V}}}\right)= 
1 + \frac{4 T {\tt C}_{\mathcal{V}} \mathcal{N}(T)}{9 \rho
\mathcal{G}_0}. 
\ee

\subsection{Low temperature expansion}
\label{lowtempexp}

In this subsection, we address the temperature dependence of the
quantities introduced above. It may be derived as discussed in
\cite{heipethrev} and conveniently
expressed by expansions in terms of $\tbar = T/\eps_{{}_{\! {\rm F}}}$; with
$\eps_{{}_{\! {\rm F}}}$ being the Fermi energy at zero temperature, 
$\eps_{{}_{\! {\rm F}}}=p_{\rm F}^2
/ (2m^*)=(3\pi^2\hbar^3 \rho)^{2/3}/(2m^*)$.  For some of the
quantities discussed below we shall include terms of third order
in $\tbar = T/\eps_{{}_{\! {\rm F}}}$, which are not considered in \cite{heipethrev}.

From (\ref{densstat}) one gets for the particle density
$\rho(\mu,T)$ 
\bel{densexp} 
\rho =
    \frac{\left(2 m^* \mu\right)^{3/2} }{3 \pi^2 \hbar^3}
    \left(1 + \frac{\tp }{8} \right)
\ee 
as function of the chemical potential $\mu$ and the
temperature  $T$.  For the chemical potential $\mu$, one obtains
\bel{chemicpot} \mu = \eps_{{}_{\! {\rm F}}} \left(1-\frac{\tp }{12}
\right),
\ee 
which is typical for a system of independent fermions.
 At this stage it may be worth while to mention
that the formulas presented here remain largely unchanged in case
of the presence of a density dependent potential $V(\rho)$. As
long as such a potential does not depend on the momentum, we may
just change our s.p.\ energy $\eps_\p^{\rm g.e.}$ to $
p^2/(2 m^*) + V(\rho)$, and the chemical potential $\mu$ to the
$\mu'=\mu-V(\rho)$ of \cite{heipethrev}.

For the density of states $\mathcal{N}(T)$ of the quasiparticles, one
finds from (\ref{enerdensnt}) 
\bel{capnexp} 
\mathcal{N}(T) = 
\mathcal{N}(0)\left(1-\frac{\tp }{12} \right), 
\ee 
where $\mathcal{N}(0)$ is given by (\ref{nzero}).
Similarly, for 
$\mathcal{M}(T)$ defined in (\ref{mcapt}),  one gets 
\bel{mcapexp} 
\mathcal{M}(T) = 
\frac{\pi^2}{6}\;\mathcal{N}(0)\; \tbar
    \left(1 + \frac{13\tp }{60}\right).
\ee 
As different to \cite{heipethrev}, we include a temperature
correction here, which is of interest for some of the quantities
described in the text. The specific heat 
${\tt C}_{\mathcal{V}}$ (\ref{specheatveq}) per particle for the constant volume
becomes 
\bel{specifheatv} 
{\tt C}_{\mathcal{V}}=
   \frac{\pi^2{\bar T}}{2}
    \left(1-\frac{3 \tp }{10} \right).
\ee 
For the isothermal in-compressibility $K^{T}$, one
gets from (\ref{isotherk}) and (\ref{capnexp}) 
\bel{isotherkexp}
K^{T} = 6 \eps_{{}_{\! {\rm F}}} \mathcal{G}_0
    \left(1 + \frac{\tp }{12} \right).
\ee 
Likewise, for the in-compressibility modulus $K^{\varsigma}$
(\ref{Kadiabat}) at constant entropy $\varsigma$ per particle, one
obtains 
\bel{incompradexp} 
K^{\varsigma} =
    6 \eps_{{}_{\! {\rm F}}} \mathcal{G}_0 \left[1 + \frac{\tp }{12}
    \left(1 + \frac{4 }{\mathcal{G}_0}\right) \right]\;.
\ee 
Using (\ref{incompradexp}), the adiabatic sound velocity
$v^{(\varsigma)}$ (cf. 
\cite{forster,brenig}) 
can be expressed as
\bel{speedad} 
v^{(\varsigma)} =
    \sqrt{\frac{K^{\varsigma} }{9 m}} = v_{{}_{\! {\rm F}}} s^{\varsigma},
\ee 
where 
\bel{velocad} 
s^{\varsigma} = \sqrt{{\mathcal{G}_0 
\mathcal{G}_1 \over 3}
    \left[1 + \frac{\tp }{12} \left(1 + 
\frac{4 }{\mathcal{G}_0}\right)\right]}.
\ee 
The ratio of the specific heats (\ref{cvcpkSkT}) may be
calculated using either the expansions of the in-compressibilities
(\ref{incompradexp}) and (\ref{isotherkexp}) or
(\ref{cpcvkakt}) together with (\ref{capnexp}) and
(\ref{specifheatv}). Finally, one gets 
\bel{cpcvexp} 
\left(\frac{{\tt C}_\press
}{{\tt C}_{\mathcal{V}}}\right) =
    1 + \frac{\tp }{3 \mathcal{G}_0}.
\ee 
Thus, from (\ref{specifheatv}) and (\ref{cpcvexp}),
\bel{specifheatp} C_\press =
   \frac{\pi^2 {\bar T}}{2}
    \left[1-\frac{3 \tp }{10}
    \left(1-\frac{10 }{9 \mathcal{G}_0} \right)\right]
\ee 
is the specific heat at the fixed pressure.

\subsection{Thermodynamic relations for a finite Fermi-liquid drop}
\label{thermgibbs}

In this subsection, we apply the formulas derived above to extend
the derivations of the boundary conditions in
\cite{strutmagbr,magstrut,magboundcond} to the case of equilibrium
at  a finite $T$. Like in these papers, the finite Fermi-liquid drop
is treated in the effective sharp surface approximation, see
subsection \ref{boundary} and Appendix D.
Applying to the standard thermodynamic relations $~{\rm d}E=T{\rm d}S
-\press {\rm d}\mathcal{V}-\press_Q{\rm d}Q~$ and $~{\rm d}G=-S{\rm
d}T + \mathcal{V} {\rm d}\press-\press_Q{\rm d}Q~$, we include the
change of the collective variable $Q$
(see, e.g., \cite{hofmann,hofbook}). $G$ is the Gibbs free
energy $G=F+\press \mathcal{V}=E+TS+\press \mathcal{V}$, defined
similarly to the free energy $F$ with only the volume 
$\mathcal{V}$
replaced by the pressure $\press$.
For the FLDM 
it is more convenient to use $G$
rather than $F$, simply because in general volume may not be
conserved but the pressure has to be fixed by the boundary
condition (\ref{bound2}). The Gibbs free energy is used for
deriving these boundary conditions as well as for the
calculations of the coupling constants and susceptibilities
associated to the operator ${\hat F}(\r)$ (\ref{foperl}).

For the following derivations, we need the relations for the
thermodynamical potentials per particle. The Gibbs free energy per
particle $G/N$ which is identical to the chemical potential
$\mu$  
is related to the corresponding
free energy $F/N$ by the relation $G/N \equiv
\mu=F/N+\press/\rho$~. For a finite Fermi-liquid drop where the
particle density $\rho$ is function of the coordinates (smooth
inside and sharp decreasing in the surface region) they are
written as in \cite{strutmagbr,magstrut,magboundcond} through the
variational derivatives $\delta g/\delta \rho$ and $\delta
\phi/\delta \rho$ with the thermodynamical potential densities
$g$ and $\phi$ per unit of volume, respectively, and this
relation reads now 
\bel{gibbspart} 
\frac{\delta g }{\delta \rho}
\equiv \mu= \frac{\delta \phi }{\delta \rho} +
\frac{\press }{\rho}.
\ee
 These densities depend on the coordinates through 
$\rho$ and its gradients. Their calculation is carried out
from the variations of the corresponding total integral
quantities $G$ and $F$ with the following integration by parts,
see \cite{strutmagbr,magstrut,magboundcond} for details. Taking
into account also that the particles in the Fermi-liquid drop
move in a mean field $V$ with the coordinate dependence similar to
the density $\rho$, one gets from (\ref{gibbspart})
\bea\l{thermrelmu} 
{\rm d} \frac{\delta g }{\delta \rho} &\equiv&
{\rm d} \mu= -\varsigma {\rm d} T + \frac{1}{\rho} {\rm d}\press
+{\rm d} V\quad{\rm with}\nonumber\\ {\rm d}
V&=&-\left(\press_Q/N\right) {\rm d} Q. 
\eea
From (\ref{thermrelmu}) one has 
\bel{gradrelmu} 
{\bf \nabla} \mu
=-\varsigma {\bf \nabla} T + \frac{1}{\rho} {\bf \nabla}\press
+{\bf \nabla} V. 
\ee

For the derivation of the boundary condition (\ref{bound2}), we
used (\ref{gradrelmu}) for the transformations of
(\ref{momenteq}) instead of (17) of \cite{magstrut}. The
one-to-one correspondence of this derivation with that explained in
\cite{strutmagbr,magstrut,magboundcond} becomes obvious if we note
that equation (17) 
was found from 
\bel{gradrelener} 
{\bf \nabla}
\eps =T{\bf \nabla} \varsigma + \frac{1}{\rho} {\bf
\nabla}\press +{\bf \nabla} V, 
\ee
 for the adiabatic condition of
a constant entropy per particle ($\eps$ here is the same as
$\delta \eps / {\delta \rho}$ in the notation of
\cite{magstrut}). The variational derivative $\delta g / {\delta
\rho}$ (\ref{gibbspart}) (or the chemical potential $\mu$)
appears now in the following key equation for the derivation of
the surface condition (\ref{bound2}): 
\bel{keybound}
\rho_\infty\left(\frac{\delta g }{\delta
\rho}\right)_{S}^{\rm vol}= -b_{\mathcal{V}}\rho_\infty +2 \alpha \mathcal{H},
\ee 
where $b_{\mathcal{V}}$ is the nucleon binding energy in the infinite
nuclear matter, $\mathcal{H}$ the mean curvature of the nuclear
surface, $\mathcal{H}=1/R_0$ for the spherical shape at equilibrium.
Index "vol" means that the Gibbs free energy per particle is
considered as that found in the nuclear 
 interior. Hence, is a
smooth quantity taken at the nuclear surface as the quantities
in the l.h.s. of the boundary conditions (\ref{bound1}) and
(\ref{bound2}) within the precision of the ESA.

The temperature $T$ and chemical potential $\mu$ in
(\ref{thermrelmu}) and (\ref{keybound}) are constants as
function of the coordinates $\r$ within our Fermi-liquid-drop
interior at equilibrium. 
With these properties, one gets 
\bel{derVderP}
{\bf \nabla} V =-\frac{1 }{\rho} {\bf \nabla}\press =
\frac{K }{9
\rho_{{}_{\! 0}}} {\bf \nabla} \rho.  
\ee 
In the second equation, we applied
(\ref{dpress}) which shows that the expression in the middle
of (\ref{derVderP}) is proportional to the gradient of the
particle density with some smooth coefficient related to the
in-compressibility  $K$. The relation (\ref{derVderP}) will be
used in the Appendix C  
for the calculation of several coupling
constants and susceptibilities for the constant temperature and
entropy, as well as for the static limit $\om \to 0$, with the
corresponding in-compressibility modulus and particle density in
the last equation (\ref{derVderP}).

For the derivations of the susceptibilities in Appendix C 
and ratio of the surface energy constants (\ref{bstbs}), we need  
here also the following thermodynamic relation:
\bea\l{ctcaddif}
\left(\frac{\partial^2 G}{\partial Q^2}\right)_T &-& 
\left(\frac{\partial^2 E}{\partial Q^2}\right)_S=
\left[\left(\left(\frac{\partial^2 G}{\partial T^2}\right)_Q\right)^{-1}
\!\!\left(\frac{\partial^2 G}{\partial T\partial Q}\right)^2\right]_{Q=0}
\nonumber\\ 
&=&-\left[\left(\left(\frac{\partial S}{\partial
T}\right)_Q\right)^{-1} \left(\frac{\partial S}{\partial
Q}\right)^2\right]_{Q=0}. 
\eea

We obtained this relation as explained in Appendix A1 in
\cite{hofmann} with the only one change of the free energy $F$ to
the Gibbs free energy $G$.  
The derivatives in
these equations should be considered for the constant pressure
instead of the volume of the Fermi-liquid drop.


\setcounter{equation}{0}
\section{Stress tensor and heat current}
\label{app2}
We shall derive
the specific expressions for the shear modulus and viscosity
in the stress tensor
$\sigma_{\alpha\beta}$ (\ref{presstens}) representing it in the form
(\ref{prestensone}) with (\ref{presslamb}) and (\ref{pressnu})
in subsection B.2. 
We are going also to obtain the expression for
the thermal conductivity in the heat current in subsection B.1. 
The next subsection B.3  
is devoted to the long wave approximation
for the above mentioned coefficients.  
In the latter subsection,
we derive some basic formulas for this approximation which are
used for the response function in whole section \ref{longwavlim}, 
beside
the above mentioned coefficients, in particular equations
for poles of the response function.

\subsection{Stress tensor, shear modulus and viscosity}
\label{app2stress}

For the calculation of the stress tensor $\sigma_{\alpha\beta}$
(\ref{presstens}), we shall show first that it really has
the form given in (\ref{prestensone}), (\ref{presslamb}), (\ref{pressnu})
with some coefficients $\lambda$ and $\nu$ 
in Appendix B.1a, 
and then, find their specific expressions in B.1b. 

\subsubsection{STRESS TENSOR FOR  FERMI LIQUIDS}
\label{stresscontfliq}

First, after a short calculation of the r.h.s. of (\ref{presslamb})
and (\ref{pressnu}), one simply gets 
\bel{prestensone1}
{\tilde \sigma}_{\alpha\beta}= -\left(\frac{\lambda 
}{\om} -i \nu\right)~ \left(q_\beta {\tilde u}_\alpha + q_\alpha
{\tilde u}_\beta -\frac{2}{3}\q{\tilde \u}
\delta_{\alpha\beta}\right) .
\ee
To simplify more these expressions we
note now, that the Fourier components
${\tilde \sigma}_{\alpha\beta}$ (\ref{prestensone1}) of the stress tensor
$\sigma_{\alpha\beta}$ (\ref{presstens}) is a symmetric
tensor with the two independent components ${\tilde
\sigma}_{zz}$ and ${\tilde \sigma}_{xz}$ in the Cartesian
coordinate system $(x,y,z)$ with the axis $z$ directed to the wave
vector ${\bf q}$ because of axial symmetry. The tensor (\ref{prestensone1})
has also zero trace. Hence, from the set of
equations (\ref{prestensone1}) only two independent ones 
survive, namely, 
\bel{prestensone2}  
{\tilde \sigma}_{zz}=
-\frac{4}{3} \left(\frac{\lambda }{\om} - i \nu\right)~ q
{\tilde u}_z,\quad {\tilde \sigma}_{xz}= 
-\left(\frac{\lambda }{\om} - i \nu\right)~q{\tilde u}_x, 
\ee
 with
\bel{prestensone2b} 
{\tilde \sigma}_{xx}= {\tilde
\sigma}_{yy}= -\frac{1}{2} {\tilde \sigma}_{zz}, \quad
 {\tilde \sigma}_{yz}= {\tilde \sigma}_{xz}
 \quad {\rm and} \quad  {\tilde \sigma}_{xy}=0.
\ee

On the other hand, the stress tensor $\sigma_{\alpha \beta}$ (\ref{presstens})
in the l.h.s. of (\ref{prestensone2})
is determined by the distribution function 
$\delta {\tilde f}_{\rm l.e.}(\q,\p,\om)$ in the
plane-wave representation, see
(3.10) from \cite{heipethrev},
\bea\l{basiceq}
&&\delta {\tilde f}_{\rm l.e.}(\q,\p,\om)= 
\left(\frac{\partial f_{\p} 
}{\partial \eps_{\p}}\right)_{\rm g.e.} 
\left\{\frac{\om }{\mathcal{D}_{\p}}
\left[\delta {\tilde \mu} +\frac{m }{m^*} \p{\tilde
\u}\right.\right.\nonumber\\
&+&\left.\left. \left(\frac{\eps_{\p}-\mu }{T}\right)_{\rm g.e.} 
\delta {\tilde T}
- \frac{\mathcal{F}_0 }{\mathcal{N}(T)} \delta {\tilde
\rho}\right]\right.\nonumber\\ 
&-&\left.
\frac{\q \v_{\p} }{\mathcal{D}_{\p}} 
\left[\delta {\tilde \mu}+\p{\tilde \u}+
\left(\frac{\eps_{\p}-\mu }{T}\right)_{\rm g.e.} \delta {\tilde
T}\right]\right\}.
\eea
This expression can be easy derived from (\ref{landvlas})
after not too lengthy and simple transformations 
\cite{heipethrev}, besides of the adaptation to our notations.
We substitute then the distribution function
$\delta {\tilde f}_{\rm l.e.}(\q,\p,\om)$ given by (\ref{basiceq})
to the l.h.s. of (\ref{prestensone2}) through
(\ref{presstens}) in the considered representation.
In this way, we easy realize that the stress tensor (\ref{presstens})
has the above mentioned symmetry properties, and
its components ${\tilde \sigma}_{zz}$, and $
{\tilde \sigma}_{xz}$ of l.h.s. of (\ref{prestensone2})
with some shear modulus $\lambda$ and viscosity $\nu$
are indeed proportional to $q {\tilde
u}_z$ and $q {\tilde u}_x$, respectively.
As result, these stress tensor components can be represented for convenience
in terms of the
two dimensionless quantity $\chizz$ and $\chixz$ independent of
the mean velocity $\u$, 
\bel{dpresszz}
 {\tilde \sigma}_{zz}=
    -\frac{\rho_{{}_{\! 0}} \eps_{{}_{\! {\rm F}}} }{v_{{}_{\! {\rm F}}}} \chizz {\tilde u}_z , \qquad
     {\tilde \sigma}_{xz}=
    -\frac{\rho_{{}_{\! 0}} \eps_{{}_{\! {\rm F}}} }{v_{{}_{\! {\rm F}}}} \chixz {\tilde u}_x,
\ee 
where 
\bel{chizzfun} 
\chizz = \mathcal{J}_1 \frac{\rho_{{}_{\! 0}} }{\mu}
    \frac{\delta {\tilde T} }{\delta {\tilde \rho}} + \mathcal{J}_2,
\ee
 \bel{J1def} 
\mathcal{J}_1=\frac{2i \mathcal{N}(T) }{s \tau \eps_{{}_{\! {\rm F}}}
\mathcal{N}(0)} \left\langle P_2\left({\hat p}_z\right)~
\frac{\eps_\p 
}{\mathcal{D}_\p} \left(\frac{\eps_\p-\mu }{T}
-\frac{\mathcal{M}(T)}{\mathcal{N}(T)}\right)\right\rangle_{\rm g.e.}, 
\ee
\bea\l{J2def}
\mathcal{J}_2&=&-\frac{4 \mathcal{N}^2(T) }{3 s 
\eps_{{}_{\! {\rm F}}} \mathcal{N}^2(0)} 
\left\langle \eps_\p P_2\left({\hat p}_z\right) 
\left[\frac{\om }{\mathcal{D}_\p}~\left(
\frac{\mathcal{N}(0) }{\mathcal{N}(T)} + 
\frac{3 p \om }{2 \mathcal{G}_1 q \eps_{{}_{\! {\rm F}}}}~ {\hat p}_z \right) 
\right.\right. \nonumber\\
&-&\left.\left. \frac{\q \v_\p }{\mathcal{D}_\p}~ 
\left(\frac{\mathcal{G}_0 
\mathcal{N}(0) }{\mathcal{N}(T)} + \frac{3 p\om 
}{2q\eps_{{}_{\! {\rm F}}}}~{\hat p}_z\right)\right] \right\rangle_{\rm g.e.},
\eea
$P_2(x)$ is the Legendre polynomial,  ${\hat p}_\alpha$ is the
$\alpha$ component of the unit vector ${\hat p}$ defined in
(\ref{intampfpp}). Other quantities were defined in Secs.\
\ref{eqmotion}, \ref{respfunsec} and Appendix D,  
see also (\ref{dfgeqrpt}),
(\ref{enerdensnt}), (\ref{mcapt}), (\ref{nzero}),
(\ref{domindp}) and (\ref{som}). $\chixz$ in
(\ref{dpresszz}) is given by 
\bea\l{chixzdef}
&&\chixz=-\frac{3 \om \mathcal{N}(T) }{s p_{\rm F}^3 \mathcal{N}(0)} 
\left\langle p^3
\frac{{\hat p}_x^2{\hat p}_z }{\mathcal{D}_\p} 
\left(\frac{s }{\mathcal{G}_1 } -
\frac{p }{ p_{{}_{\! {\rm F}}}} {\hat p}_z\right)\right\rangle\quad
\nonumber\\
&=&-\frac{3 \om \mathcal{N}(T) }{2 s p_{\rm F}^3 \mathcal{N}(0)} 
\left\langle p^3
\frac{(1-{\hat p}_z^2){\hat p}_z }{\mathcal{D}_\p} \left(
\frac{s }{\mathcal{G}_1 }-\frac{p}{p_{{}_{\! {\rm F}}}} {\hat p}_z\right)
\right\rangle. \quad
\eea 
In the second equation, we used the invariance of the average
in the first equation with respect to the replace ${\hat p}_x
\rightarrow {\hat p}_y$, due to the axial symmetry and the equation
$\sum_\alpha {\hat p}_\alpha^2=1$ for the unit vector ${\hat p}$. We
applied also the thermodynamic relation (\ref{dmu}) for 
$\delta {\tilde \mu}$ in the distribution function (\ref{basiceq})
in these derivations.

\subsubsection{THE SHEAR MODULUS AND VISCOSITY}
\label{calclambnu}

The shear modulus $\lambda$ and viscosity $\nu$ can be now found
from the comparison of (\ref{prestensone2}) for continuous matter
and explicit expressions
(\ref{dpresszz}) obtained above from the Fermi-liquid distribution function
$\delta {\tilde f}_{\rm l.e.}(\q,\p,\om)$ (\ref{basiceq}) for the same
stress tensor components ${\tilde \sigma}_{zz}$ and ${\tilde \sigma}_{xz}$.
Indeed, substituting (\ref{dpresszz}) to the l.h.s. of
(\ref{prestensone2}) and canceling the velocity field
components from their both sides,  one finds 
\bel{lamviseq} 
\mathcal{J}_1~\frac{\rho_{{}_{\! 0}}}{\mu_{\rm g.e.}}~ 
\frac{\delta {\tilde T} }{\delta
{\tilde \rho}} +\mathcal{J}_2
    =\frac{4}{3} \chi _{xz},\quad
\lambda-iv_{{}_{\! {\rm F}}} qs\nu=\rho_{{}_{\! 0}}\eps_{{}_{\! {\rm F}}} s~\chixz\;. 
\ee
From the first equation one has the ratio 
\bel{dtdrho}
\frac{\delta {\tilde T} }{\delta {\tilde \rho}} = \frac{\mu_{\rm g.e.}
}{\rho_{{}_{\! 0}} \mathcal{J}_1}~ \left(\frac{4}{3} \chi _{xz}-
\mathcal{J}_2\right). 
\ee
Separating real and imaginary parts in the second
equation, one obtains the shear modulus $\lambda$ and viscosity
$\nu$: 
\bel{shearmod} 
\lambda=
      \frac{\mid s \mid ^2 \chixz ^\prime }{s ^\prime}\rho_{{}_{\! 0}}
\eps_{{}_{\! {\rm F}}}q
\ee
 and 
\bel{viscos} \nu =
    -\frac{s ^{\prime\prime} \chixz ^\prime +
    s ^\prime \chixz ^{\prime\prime} }{s^\prime }~
    \frac{\rho_{{}_{\! 0}} \eps_{{}_{\! {\rm F}}} }{v_{{}_{\! {\rm F}}}}.
\ee 
With these constants $\lambda$ and $\nu$,  the equations
(\ref{prestensone}), (\ref{presslamb}), (\ref{pressnu}) and  
(\ref{presstens}) are identities.

The aim of the following derivations of the shear modulus and the viscosity
is to simplify $\mathcal{J}_1$
(\ref{J1def}), $\mathcal{J}_2$ (\ref{J2def}) and $\chixz$
(\ref{chixzdef}). For this aim, we make use of
transformations of the averages like $\langle p^k{\hat p}_z^l
\eps_\p^m (\q\v_\p)^n/\mathcal{D}_\p \rangle_{\rm g.e.}$ with
some integer numbers $0 \leq k \leq 4$, $0 \leq l \leq 4$, $m=0,1$
and $n=0,1$ in terms of more simpler functions $\chin$ 
($n=0,1,2$) introduced
in \cite{heipethrev} for the response functions, see
(\ref{chinfun}). For these functions, one has simple temperature and
hydrodynamic expansions presented below at the end subsection
of this Appendix B. Using
such enough lengthy and simple algebraic derivations, one
finally gets
\bea\l{J1}
\mathcal{J}_1&=&
    \frac{1}{s (1-i s \taubar)}
    \left[\left(1+ \tbar \frac{\mathcal{M}(T) }{\mathcal{N}(T)}
    +\frac{3}{\taubar ^2}
    (1-i s \taubar)^2\right)\frac{
\chione }{\mathcal{N}(0)}\right.\nonumber\\
    &-&\left.{\bar T}\left(\frac{\pi^2 }{3} 
\cvbar-\frac{\chitwo }{\mathcal{N}(0)}
    \right)\right],
\eea
\bea\l{J2}
   && \mathcal{J}_2=
    \frac{2i \mathcal{N}^2(T) }{3s \taubar (1-is \taubar) \mathcal{N}^2(0)}
    \left[3s\left(1-i s \taubar\right)^2\right.\nonumber\\
&+& \left.{3i \taubar \over {\mathcal{G}_1}}
    \left(1-i s \taubar\right)
\left(s^2-\frac{\mathcal{G}_0
    \mathcal{G}_1 \mathcal{N}(0) }{3 \mathcal{N}(T) }\right)
    +\frac{s \mathcal{N}(0)\taubar ^2 }{\mathcal{N}(T)}\right]\nonumber\\
    &\times& \left[ \left(\frac{3 \mathcal{N}(0)
    }{ \mathcal{N}(T)\taubar ^2}(1-i s \taubar)^2 
    + 1 + {\bar T}\frac{\mathcal{M}(T) }{\mathcal{N}(T)} \right)
    \frac{\chizero }{\mathcal{N}(T)} \right.\nonumber\\
    &-&\left. 1
    -{\bar T}\frac{\mathcal{M}(T) }{ \mathcal{N}(T)}
    +{\bar T}\frac{\chione }{\mathcal{N}(T)} \right],
\eea
\bea\l{chixz}
&&\chixz = -\frac{3i}{\taubar}
    \left(1-\frac{i \mathcal{F}_1 s \taubar }{3 \mathcal{G}_1}\right)
    \left[\left(\frac{(1-i s \taubar)^2\mathcal{N}(0)
    }{ \taubar ^2 \mathcal{N}(T) } +1\right.\right.\nonumber\\
&+&\left.\left.{\bar T}
    \frac{\mathcal{M}(T) }{\mathcal{N}(T)} \right)
    \frac{\chizero }{\mathcal{N}(0)} 
    +  \tbar \frac{\chione }{\mathcal{N}(0)}
    -\frac{1}{3}\left(1+{\bar T}
\frac{\mathcal{M}(T) }{\mathcal{N}(T)}\right)
    \frac{\mathcal{N}(T) }{\mathcal{N}(0)} \right]\nonumber\\
    &=& \chixz ^\prime + i\chixz ^{\prime\prime}.
\eea 
Note that the shear
modulus $\lambda$ (\ref{shearmod}) and viscosity
$\nu$ (\ref{viscos})
depend on the sound velocity $s$, and hence,
on the solution of the Landau-Vlasov
equation (\ref{landvlas}) for $s$ [(\ref{J1}),(\ref{J2})
and (\ref{chixz})].

\subsection{Heat current}
\label{app2heat}

For the following derivations of
the thermal conductivity $\kappa$ in Fermi liquids, we need to
derive the equation for the temperature $T$ from the general
transport equation (\ref{entropyeq}). The latter equation
(\ref{entropyeq}) in the linear approximation with respect to the
dynamical variations $\delta f$ in terms of the moments, such as the
velocity field $\u$ (\ref{veloc}), particle density $\delta \rho$,
entropy density per particle $\delta \varsigma$ and so on, writes
\bel{entropeqlin} 
\rho T {\partial \varsigma \over
{\partial t}}
    +{\bf \nabla} \cdot {\bf j}_T = 0,
\ee 
where ${\bf j}_T$ is the heat current given in terms of the
thermal conductivity $\kappa$ and temperature gradient by
(\ref{currheat}). By making use of the thermodynamic relation
for the entropy $\varsigma$ per particle, 
\bel{thermeqspt} 
\d \varsigma=
    \left(\frac{\partial \varsigma }{\partial \press}\right)_T
    \d \press
    +\left(\frac{\partial \varsigma }{\partial T}\right)_\press
    \d T,
\ee  
and  the well known arguments
to get the
thermal conductivity equation, we consider the process with the
{\it constant pressure} rather than the constant of particle density.
(We again omitted the symbol variation $\delta$ as in Sec.\ \ref{conserveqs}).
With the help of (\ref{thermeqspt}), one then results in the
Fourier thermal conductivity equation 
\bel{fouriereq}
   \rho {\tt C}_\press \frac{\partial T }{\partial t}-
   \kappa \triangle T = 0,
\ee 
where ${\tt C}_\press$ is the specific heat for the
constant pressure per particle, see (\ref{specifheatp}).
(Equation (\ref{currheat}) was also used in (\ref{fouriereq}) for the
heat current ${\bf j}_T$). Solving equation (\ref{fouriereq}) for the
temperature $T(\r,t)=T_{\rm g.e.}+\delta T$ in terms of the plane
waves for the dynamical part of the temperature $\delta T(\r,t)$
as in (\ref{planewave}) and using the relations
(\ref{som}), one gets 
\bel{kappadef} 
\kappa=i \rho {\tt C}_{\mathcal{P}} v_{{}_{\! {\rm F}}}s/q. 
\ee 
Notice, the thermal conductivity $\kappa$
(\ref{kappadef}) depends on the sound velocity $s$ as the shear
modulus $\lambda$ (\ref{shearmod}) and viscosity
$\nu$ (\ref{viscos}), and therefore, on the solution of the Landau-Vlasov
equation (\ref{landvlas}) for $s$.

\subsection{Long wave-length limit}
\label{app2exp}

As shown in section \ref{respfunsec} and subsections 
B.1a and B.1b, 
many physical quantities, such as the response
functions, see (\ref{ddresp}), the shear modulus (\ref{shearmod})
and viscosity (\ref{viscos}) can be expressed in terms of the
same helpful functions $\chin$ (\ref{chinfun}). By this reason,
it is easy to get their 
LWL limit 
by expanding the only $\chin$ in small parameter $\tau_q$.

For small $\taubar$, one can use asymptotic expansions 
for the Legendre function of second kind
$Q_1(\zeta)$ 
and its derivatives
which
enter $\chin$ with its derivatives, according to (\ref{chitemzero}),
(\ref{chitemone}) and (\ref{chitemtwo}). This approximation is
valid for large arguments $\zeta$. Substituting these expansions
into the functions $\chin$ (\ref{chinfun}) and $\wp$
(\ref{alphas}), one gets to fourth order in $\taubar$:
\bea\l{chiexpzero} 
\chizero &=&
    \frac{\taubar^2 }{3} \left[1 + 2 i s_0 \taubar
    -\left(2 s_1 + 3 s_0^2 +\frac{3}{5}\right) \taubar^2\right.
\nonumber\\
    &-& \left. \frac{\tp }{4} \taubar^2 \right] \mathcal{N}(0),
\eea 
\bel{chiexpone} 
\chione =
    \frac{\pi^2 \tbar \taubar^2 }{9}
    \left[1+2 i s_0 \taubar
    -\left(2 s_1 + 3 s_0^2 +\frac{6}{5}\right) \taubar^2\right] 
\mathcal{N}(0),
\ee 
%
%
\bea\l{chiexptwo}
&&\chitwo =
    \frac{\pi^2 \taubar^2 }{9}
    \left[1+2 i s_0 \taubar
    -\left(2 s_1 + 3 s_0^2 +\frac{3}{5}\right) \taubar^2 
    -\frac{\tp }{60}\right.\nonumber\\
    &\times&\left.\left(1+2 i s_0 \taubar
    -\left(2 s_1 + 3 s_0^2 - 60 \right) \taubar^2 \right) \right] 
\mathcal{N}(0).
\eea
With these expressions, 
one obtains the collective response
function $\chi _{DD}^{\rm coll}$ of (\ref{ddresp}), (\ref{despfunc}) 
through
\bea\l{ampsexp}
&&\aleph(s) \equiv \aleph(\taubar,s_0,s_1) = \frac{\taup }{27}
   \left\{-3 i s_0 + \left(1 + 6 s_0^2 +3 s_1\right) \taubar
\right.\nonumber\\
   &+& \left.\frac{\tp }{120}
   \left[93 i -\left(2+186 s_0^2 + 93 s_1\right)
   \taubar \right]\right\} \mathcal{N}^2(0),
\eea
\bea\l{dszexpzero}
&&D_0(s) \equiv D_0(\taubar,s_0,s_1) =
    \frac{\taup }{9}
    \left\{-i s_0 \left(1 - 3 s_0^2\right)\right.\nonumber\\
   &+& \left. \frac{1}{15}
    \left[5 + 3 s_0^2 \left(1-30 s_0^2\right)+
    15 s_1\left(1 -9 s_0^2 \right)\right] \taubar ~~~\right.\nonumber\\
    &+& \left. \frac{\tp }{120}
    \left[-i s_0 \left(19+93 s_0^2\right)
+\frac{1}{15}
    \left(-160 +18 s_0^2\right.\right.\right.\nonumber\\
&\times&\left.\left.\left.\left(54  +155 s_0^2\right) 
+ 
    5 s_1 \left(57 +837 s_0^2 \right)\right)\taubar
    \right]\right\} \mathcal{N}(0)~~~
\eea
[also for the temperature-density response function (\ref{dtresp})].
These two quantities determine the expansion of the function $D(s)
\equiv D(\taubar,s_0,s_1)$ (\ref{despfunc}) in powers of $\taubar$,
and  then approximately, the excitation modes given by the
dispersion relation (\ref{despeq}). Indeed, equaling zero the
coefficients which appear in front of the each power of $\taubar $
in this expansion of $D(\taubar,s_0,s_1)$, we get  
equations for the unknown quantities $s_0$ and $s_1$ of
(\ref{somexp}), 
\bea\l{eqzero} 
&&\frac{i s_0 }{\mathcal{G}_1}
\left[s_0^2 -\frac{\mathcal{G}_0 \mathcal{G}_1 }{3}
 + \frac{\tp }{120}
 \left(-40 \mathcal{G}_1 + 21 \mathcal{G}_0 \mathcal{G}_1
\right.\right.\nonumber\\ 
&-&\left.\left. 63 s_0^2
 - 30 \mathcal{G}_1 s_0^2\right)\right] = 0,
\eea
\bea\l{eqone}
&&\frac{1}{45 \mathcal{G}_1}\left\{5 \mathcal{G}_0 \mathcal{G}_1 - 3 s_0^2
\left[5 +
        2 \mathcal{G}_1 \left(2 - 5 \mathcal{G}_0 \right) + 30 s_0^2 \right] +
        15 s_1 \right.\nonumber\\
&\times&\left.
\left(\mathcal{G}_0 \mathcal{G}_1 -
        9 s_0^2\right) 
  +\frac{\tp }{120}
   \left[40 \mathcal{G}_1\left( \mathcal{G}_0 -5\right)
   - 6s_0^2 \left(60 - 287 \mathcal{G}_1 \right.\right.\right.\nonumber\\
&+& \left.\left.\left.
   105 \mathcal{G}_0 \mathcal{G}_1 +
           15 s_0^2 \left(21 + 10 \mathcal{G}_1 \right)\right)
        + 15 s_1 \right.\right.\nonumber\\
&\times&\left.\left.
\left(40 \mathcal{G}_1 - 
21 \mathcal{G}_0 \mathcal{G}_1 +
           9 s_0^2\left(21 + 10 \mathcal{G}_1 \right) \right)\right]\right\} 
    = 0.
\eea
Solving these equations, one obtains the position of the poles as
given in (\ref{shp}) and (\ref{sfirst}).

The shear modulus ($\lambda$) and viscosity ($\nu$) coefficients
enter the response function $\chi_{FF}^{\rm coll}$ (\ref{chicollfldm})
and (\ref{glx}) in terms of the sum $(\lambda - i \nu \om)/\rho_{{}_{\! 0}}
\eps_{{}_{\! {\rm F}}}$. The  
LWL expansion of this sum can be
obtained with help of (\ref{chiexpzero}), (\ref{chiexpone})
and (\ref{somexp}) and expansions of all static quantities in
$\tbar$ there, see Appendix A, but with taking into account fourth
order terms,
\bea\l{chixzexp}
&&(\lambda - i \nu \om)/\rho_{{}_{\! 0}} \eps_{\rm F}= s \chi_{xz}(\tbar,\om\tau)=
-i \frac{2}{5}\left(1 + \frac{5 \tp }{12} \right) \om\tau +
\nonumber\\
&&  \frac{\pi^4 \tbar^4 }{2160}
  \left[13 \left(\mathcal{G}_0 - \frac{4}{5} \mathcal{G}_1
  - \mathcal{G}_0 \mathcal{G}_1\right) -
     \frac{13 i  \mathcal{G}_0 \mathcal{G}_1 }{\om\tau} + 
\right.\nonumber\\
   &&  \left. \frac{i \om\tau }{25 \mathcal{G}_0}
     \left(780 + 2815 \mathcal{G}_0 + 52 \mathcal{G}_1
     + 260 \mathcal{G}_0 \mathcal{G}_1\right)\right].
\eea
Separating the real and imaginary parts in these equations, one
gets the LWL approximation of the both {\it real}
coefficients $\lambda$ and $\nu$.  
The terms linear in
$\om \tau$ determine the hydrodynamic viscosity $\nu^{(1)}$
(\ref{shearvisone}), and the terms proportional to $1/\om\tau$ are
related to $\nu^{(2)}$ (\ref{shearvishp}),  see 
discussions of the "heat pole" for the FLDM transport coefficients 
in Sec.\ \ref{transprop}.

The  LWL approximation for
the thermal conductivity $\kappa$ is determined by 
equation (\ref{kappadef})
and solutions (\ref{shp}) for the heat pole 
and (\ref{sfirst0}) for the sound velocity
$s$ of the
dispersion equations (\ref{eqzero}) and
(\ref{eqone}).

The explicit final expressions for the viscosity $\nu$ and the
thermal conductivity $\kappa$ are presented and discussed in
subsection 
II D in the LWL limit in
connection with the first sound and overdamped (heat pole) modes,
see (\ref{shearvisonehp}) and (\ref{kappaexp}). As seen
immediately from (\ref{chixzexp}), the linear terms in
$\taubar$ for the shear modulus $\lambda$ appear as high
temperature corrections proportional to ${\bar T}^4$. They are
regular in $\om\tau$, and therefore, are totally immaterial, see
more discussions in the subsection mentioned above.
In the
linear approximation in $\taubar$ of the  
LWL limit,
it is easy to check that the
derivative $\delta {\tilde T} / \delta {\tilde \rho}$ 
(\ref{dtdrho}) is the same as obtained in terms of the response
functions in (\ref{dtdrhoexp}), and therefore, the in-compressibility
$K_{\rm tot}$ (\ref{incomprtot}) turns into the adiabatic one.


\setcounter{equation}{0}
\section{Coupling constants and
susceptibilities} \label{app4}

Let us consider the change of average $\langle {\hat F}(\r)
\rangle$ of the operator ${\hat F}(\r)$ (\ref{foper}) due to a
quasistatic variation of the particle density $\rho_{\rm qs}(\r,Q,T)$,
\bel{quasavf} 
\delta \langle {\hat F}(\r) \rangle_{\rm qs}^{\tt X} =
\int~{\rm d} \r ~{\hat F}^{\tt X}(\r)~ \delta \rho_{\rm qs}^{\tt X}
=-\chi_{FF}^{\tt X}\; \delta Q\;, 
\ee
where 
\bel{quasdens} 
\delta \rho_{\rm qs}^{\tt X}
=\left[\left(\frac{\partial \rho_{\rm qs} }{\partial Q}\right)_T
+\left(\frac{\partial \rho_{\rm qs} }{\partial T}\right)_Q 
\frac{\delta T}{\delta Q}\right]^{\tt X} \delta Q \;.
\ee 
The index "${\tt X}$" shows one of the conditions of the constant temperature 
(${\tt X}=T$), entropy 
(${\tt X}$ is "ad") or static limit $\om \to 0$ (${\tt X}$ is "$\om=0$").
We shall follow the notations of \cite{hofmann,hofbook} omitting the
index $\om=0$ for the coupling constant ($k_{\om=0} \equiv k$),
surface energy constant ($b_{{}_{\! S}}^{\om=0} \equiv b_{{}_{\! S}}$), and
in-compressibility [$K^{\om=0} \equiv K=K_{\rm tot}$ for $\om=0$, see
(\ref{incomprtot})]. We write it as the zero argument for the
isolated susceptibility, $\chi^{\om=0} \equiv \chi(0)$, and
stiffness coefficient, $C^{\om=0}=C(0)$. $f^{\tt X}$ and
$\delta f^{\tt X}$ denote the quantity $f$ and its variation
provided that the  
${\tt X}$ condition is
carried out. The index "qs" stands for the quasistatic quantities
as in \cite{hofmann} and will be omitted within this Appendix. 
Note that the
operator ${\hat F}(\r)$ (\ref{foperl}) depends in the FLDM on 
${\tt X}$ through the derivatives of the particle density, and by
this reason, the upper index ${\tt X}$ appears in 
${\hat F}^{\tt X}(\r)$ of (\ref{quasavf}).
From the first of (\ref{quasavf}) with (\ref{foperl}) and
(\ref{quasdens}), one gets the self-consistency condition
(\ref{selfconsist}),
\bel{Favfld} 
\delta \left\langle {\hat F}(\r)
\right\rangle^{\tt X} =k_{\tt X}^{-1} \delta Q,
\ee
 with the following expression for the
coupling constant, 
\bea\l{couplfld} 
k_{\tt X}^{-1}
&=& - R_0 \int~{\rm d} \r \left\{\frac{\partial V }{\partial r}
Y_{L0}({\hat r}) \left[\left(\frac{\partial \rho }{\partial
Q}\right)_T \right.\right.\nonumber\\
&+&\left.\left.\left(\frac{\partial \rho }{\partial T}\right)_Q
\frac{\delta T }{\delta Q}\right]\right\}_{Q=0}^{\tt X}\mathcal{O}_1, 
\eea 
within the ESA parameter of smalleness
$a/R \sim A^{-1/3}\ll 1$,
$\mathcal{O}_1=1+\mathcal{O}\left(a/R\right)=
1+\mathcal{O}\left(A^{-1/3}\right)$.
For the susceptibilities $\chi_{FF}^{\tt X}$ defined by (\ref{quasavf}) 
up to small corrections of the
order of $A^{-1/3}$ in the same approximation,  from
(\ref{couplfld}) for $k_{\tt X}^{-1}$ one gets 
\bel{chix} 
\chi^{\tt X}=-k_{\tt X}^{-1}\mathcal{O}_1.
\ee
We omit also the low indexes "FF"
for the susceptibilities. Note that we have not identities of
$-k_{\tt X}^{-1}$ to $\chi^{\tt X}$ because we neglected earlier
high order $A^{-1/3}$ corrections in the derivation of the
operator ${\hat F}$ (\ref{foperl}), in particular, in the FLDM
approximation (\ref{denseq}) for the quasistatic particle density
$\rho_{\rm qs}$.  
The equation (\ref{chix}) is in
agreement with (\ref{kstiffC0chi0}) (identical to equation (3.1.26)
of \cite{hofmann}), see also
(\ref{smallpar}), (\ref{isorespk}), for the specific relation
between the coupling constant $k^{-1}$ and isolated susceptibility
$\chi(0)$ {\it with presence of the stiffness term} $C(0)$ in "the
zero frequency limit" within the FLDM. As shown in 
Sec.\ \ref{zerofreqlim} through (\ref{chicollfldm}) by using the
expansion in small parameter $kC$ (\ref{smallpar}) up to the
second order terms in $kC$, the isolated susceptibility
$\chi(0)$, see (\ref{respintr}) at $\om =0$, is related to the
coupling constant $k^{-1}$ by (\ref{kstiffC0chi0}) with the
stiffness term $C(0)$. The correction related to the stiffness
$C(0)$ appears in (\ref{chix}) in a 
higher order than
$A^{-1/3}$ because it is of the order of the small parameter $kC
\sim A^{-2/3}$, see (\ref{smallpar}) and discussion near this
equation. The zero frequency stiffness $C(0)$ is equal
approximately to the liquid drop one $C$ (\ref{stiffness0}) in the
FLDM for the considered enough large temperatures for which the
quantum shell effects can be neglected.

The derivatives of the quasistatic particle density in
(\ref{quasdens}), (\ref{Favfld}) and (\ref{couplfld}) can be
found from (\ref{denseq}), 
\bea\l{drdqdt} 
\left(\frac{\partial \rho }{\partial Q}\right)_T^{\tt X}
&=& -\left(\frac{\partial \rho 
}{\partial r}\right)^{\tt X} R_0 Y_{L0}({\hat r}),
\nonumber\\
\left(\frac{\partial \rho }{\partial T}\right)_Q^{\tt X} &=& 
\frac{1}{\rho_0^{\tt X}}\frac{\partial \rho_0^{\tt X} }{\partial T}
\rho^{\tt X}+\frac{R_0 }{3\rho_\infty} \frac{\partial \rho_\infty
}{\partial T} \left(\frac{\partial \rho }{\partial
r}\right)^{\tt X}, 
\eea
 for $Q=0$ with 
\bel{rho0x} 
\rho_0^{\tt X}=
\rho_\infty \left(1+\frac{6 b_{{}_{\! S}}^{\tt X} r_{{}_{\! 0}} }{K^{\tt X} R_0
}\right), 
\ee
 as in (\ref{rho0}). We emphasize
that the surface energy constant $b_{{}_{\! S}}^{\tt X}$ (or the
surface tension coefficient $\alpha^{\tt X}$) depends also on the
type of the process specified by index ${\tt X}$ as the
in-compressibility $K^{\tt X}$ because of the ${\tt X}$-dependence
of the particle density derivative in the integrand of
(\ref{tensionconst}) for the tension coefficient. The total
quasistatic energy is the sum of the volume  and surface parts
determined by the in-compressibility $K^{\tt X}$ and surface
$b_{{}_{\! S}}^{\tt X}$ constants, respectively. The in-compressibility
modulus $K^{\tt X}$ (responsible for the change of the volume
energy) is given by (\ref{incomprTdef}) for ${\tt X}=T$, and
(\ref{incompraddef}) for ${\tt X}$ is "ad", see also
(\ref{isotherk}), (\ref{Kadiabat}) or 
(\ref{isotherkexp}), (\ref{incompradexp}) of their more 
specific expressions for
nuclear matter.
The in-compressibility $K$ equals the adiabatic one
$K^{\varsigma}$ as shown through (\ref{dtdrhoexp}) and
(\ref{incompradexp}), $K=K_{\rm tot}(\om=0)=K^{\varsigma}$. In the
derivations of (\ref{drdqdt}), we took into account that 
$\rho_0^{\tt X}$ (\ref{rho0x}) does not depend on $Q$, and the
density $\rho_\infty$ (or $r_0$) is assumed to be approximately
independent of index "${\tt X}$" in (\ref{drdqdt}).

Substituting (\ref{drdqdt}) into (\ref{couplfld}) for the
coupling constant $k_{\tt X}^{-1}$, one writes 
\bea\l{couplfld1}
&&k_{\tt X}^{-1} = R_0^2 \int~{\rm d} \r \left\{{\partial V \over
{\partial r}} \frac{\partial \rho }{\partial r} Y_{L0}({\hat r})~\qquad
\right.\nonumber\\
&\times & \left.\left[Y_{L0}({\hat r}) -\frac{1}{3 \rho_\infty} 
\frac{\partial
\rho_\infty }{\partial T} \frac{\delta T }{\delta
Q}\right]\right\}_{Q=0}^{\tt X} \mathcal{O}_1.
\eea
The
first term proportional to the density in $\partial \rho/
\partial T$ of (\ref{drdqdt}) leads to small $A^{-1/3}$
corrections to the coupling constant $k_{\tt X}^{-1}$
(\ref{couplfld1}) with respect to the second component depending
on the coordinate derivative $\partial \rho/\partial r$. However,
all terms including these corrections related to the variation of
the temperature $\delta T$ in (\ref{couplfld}) [or
(\ref{couplfld1})] can be neglected as compared to the first
term in the square brackets. (It comes from the variation of the
collective variable $\delta Q$ up to the same relatively small
corrections of the order of $A^{-1/3}$.) Indeed, for the
isothermal case "${\tt X}=T$" one has it exactly by its
definition. For other "${\tt X}$" the quantity $\delta T/\delta
Q$ in (\ref{Favfld}) and (\ref{couplfld1}) with the density
(\ref{denseq}) can be transformed within the ES precision 
\bel{dtdqx} 
\left(\frac{\delta
T }{\delta Q}\right)^{\tt X} =\left(\frac{\delta T }{\delta
\rho} \frac{\partial \rho }{\partial Q}\right)^{\tt X}
=\left(\frac{\partial T }{\partial r}\right)^{\tt
X}R_0Y_{L0}({\hat r}).
\ee
 For
instance, for the constant entropy (adiabatic) condition $S=\int
{\rm d}r \rho \varsigma=S(\rho,T)={\rm const.}$, see
(\ref{entropydef}) with the quasistatic particle density
$\rho$ (\ref{denseq}), the derivative $\delta T/\delta Q$ can be
calculated through a variation of this density $\rho$ as shown in
the middle of (\ref{dtdqx}). In the quasistatic limit $\om
\to 0$ all quantities of the equilibrium state can be considered
also as a functional of the only density $\rho$ (\ref{denseq})
in the ESA 
and one has again
(\ref{dtdqx}). We have used already this property in the
derivation of the operator ${\hat F}(\r)$ for transformations of
the derivatives of a mean field $V$ in (\ref{foperl}). As noted
and used for the derivations in Appendix A.1,  
the
temperature $T(\r)$ is approximately independent of the spatial
coordinates $\r$ at equilibrium.  
Therefore, according to (\ref{dtdqx}),
the second terms in (\ref{Favfld}) and (\ref{couplfld1}),
which appear due to the temperature variation $\delta T$, turn
into zero with the FLDM precision. 

After the simple integration over the angles ${\hat r}$ in
(\ref{couplfld1}) for the coupling constant $k_{\tt X}^{-1}$,
one then arrives at
\bel{couplfld2}
k_{\tt X}^{-1}
= R_0^4 \int_0^\infty~{\rm d} r \left[ \frac{\delta V }{\delta
\rho} \left(\frac{\partial \rho }{\partial r}\right)^2
\right]_{Q=0}^{\tt X} \mathcal{O}_1.
\ee
According to (\ref{denseq}),
the integrands in (\ref{couplfld2}) contains the sharp bell
function ${\partial \rho/\partial r}$ of $r$. Therefore, the
integrals converges there to a small spatial region near the
effective nuclear surface defined as the positions of maxima of
this derivative at $r=R_0$ (Appendix D). We use these properties of the
integrand in the derivation of (\ref{couplfld2}) taking
smooth quantities as $r^2$ at the nuclear surface point $r=R_0$
off the integrals up to small corrections of the order of
$A^{-1/3}$ within the same ESA. [This is like for the
derivations of the boundary conditions
(\ref{bound1}), (\ref{bound2}), see
\cite{strutmagbr,magstrut,strutmagden}, and of (\ref{foperl})
for the operator ${\hat F}(\r)$.] In this way, we get the expansion
of the coupling constant $k_{\tt X}^{-1}$ (\ref{couplfld}), in
powers of the $A^{-1/3}$ with the leading term shown in the second
equation there, see (\ref{couplfld2}).

For the following derivations, we specify now the quasistatic
derivative $\left(\delta V / \delta \rho\right)^{\tt X}$ at $Q=0$
taken it from (\ref{gradrelmu}), 
\bel{derVderP1} 
{\bf \nabla}
V^{\tt X}~ =~ \frac{K^{\tt X} }{9 \rho_0^{\tt X}}~ {\bf \nabla}
\rho^{\tt X}, 
\ee
 where index ${\tt X}$ in ${\bf \nabla} f^{\tt
X}$ means the gradient of the quantity $f$ taken for the condition
marked by ${\tt X}$ as in the variation $\delta f^{\tt X}$. The
proportionality of the gradients in (\ref{derVderP1}) shows the
self-consistency within the ESA precision, see
\cite{magstrut} for more general relations of the self-consistency
in the FLDM.

Using (\ref{derVderP}) and (\ref{tensionconst}) in 
(\ref{couplfld2}) for the coupling
constants and (\ref{chix}) for the susceptibilities, one obtains
the identical results for these quantities with 
small corrections
of the order of $A^{-1/3}$, 
\bel{couplxchix} 
\chi^{\tt X}= -k_{\tt X}^{-1}\mathcal{O}_1=
\frac{K^{\tt X} b_{{}_{\! S}}^{\tt X} R_0^4 }{72 \pi
r_0^2 \rho_\infty^{\tt X} \mathcal{C}} \mathcal{O}_1.
\ee

We shall show now from (\ref{couplxchix}) that the adiabatic
susceptibility $\chi ^{\rm ad}$ and coupling constant $k_{\rm ad}^{-1}$
are equal to the isolated ($\chi(0)$) and quasistatic ($k^{-1}$)
ones, respectively, up to small $A^{-1/3}$ corrections within the
ESA.

As noted above, for the adiabatic ($K^{\varsigma}$) and
quasistatic ($K$) in-compressibility modula,  we got
$K^{\varsigma}=K$, see after (\ref{dtdrhoexp}).  
The surface energy constant $b_{{}_{\! S}}$ equals
the adiabatic one $b_{S}^{\rm ad}$, according to
(\ref{keybound}). Indeed, the volume energy per particle  is
also approximately the same for these cases, 
$b_{\mathcal{V}}^{\rm ad} = b_{{}_{\! \mathcal{V}}}$,
because of its relation  $b_{{}_{\! \mathcal{V}}} \approx K/18$ to the
in-compressibility modulus ($b_{\mathcal{V}}^{\rm ad}=K^{\varsigma}/18$)
within the ESA \cite{strutmagden}
and equivalence of the corresponding in-compressibility modula. The
functional derivative in (\ref{keybound}) is the quasistatic
chemical potential $\mu$ which does not depend obviously on the
type of the process ${\it X}$.
From (\ref{keybound}), one gets now $\alpha^{\rm ad}=\alpha$ for
the surface tension coefficient or
$b_{{}_{\! S}}=b_{S}^{\rm ad}$ for the surface energy constant. Namely,
this quantity should be identified with the experimental value
$b_{{}_{\! S}}=17-19~\mbox{MeV}$ in the FLDM computations. 
Thus, from (\ref{couplxchix}) one obtains the ergodicity condition
(\ref{ergodicity1}) for the susceptibilities within the ESA precision,
\bel{chi0adfld} 
\chi(0)=\chi^{\rm ad}=\frac{K b_{{}_{\! S}} r_0^5
}{54 \mathcal{C}} A^{4/3}, 
\ee
 up to small $A^{-1/3}$ corrections.
As seen from  (\ref{couplxchix}), one gets  also
$k^{-1}=k_{\rm ad}^{-1}$ for the coupling constants within the same
approximation , see (\ref{kfld}) for $k^{-1}$. The index "ad"
for the coupling constant will be omitted below in line of
\cite{hofmann}.

We are interested also in the discussion of the difference between
the susceptibilities $\chi^T$ and $\chi^{\rm ad}$. From
(\ref{chix}), one has 
\bel{chiTchiad}
\chi^T-\chi^{\rm ad}=\left(k_T^{-1}-k^{-1}\right) \mathcal{O}_1.  
 \ee 
It is useful to re-derive this
relation by applying Appendix A1 of \cite{hofmann} for the
specific Fermi-liquid drop thermodynamics, see
(\ref{ctcaddif}). As noted in Appendix A.4,  
it is
more convenient to use the Gibbs free energy $G$ instead of the
free energy $F$. As the derivations of the $\chi^T-\chi^{\rm ad}$ in
Appendix A1 of \cite{hofmann} do not contain any change in the
volume and pressure variables, we can use all formulas in A1 of
\cite{hofmann} here with the replace of the free energy $F$ by
the Gibbs free energy $G$, in particular (\ref{ctcaddif}).
There is a specific property of the FLDM with respect to the
microscopic shell model with the residue interaction of
\cite{hofmann}. The second derivatives of the Hamiltonian
$\langle\left(\partial ^2 {\hat H} / {\partial
Q^2}\right)_{Q=0}\rangle_0$ in equations (A.1.6) and (A.1.7) of
\cite{hofmann} 
depend in the FLDM on the type of the
process ${\tt X}$, isothermal and adiabatic one, relatively. The
first derivative of the Hamiltonian ${\partial H/\partial Q}$ in
equations (A.1.2) and (A.1.3), see \cite{hofmann}, is proportional to
the derivatives of the density ${\partial \rho/\partial Q}$ like
in the susceptibilities (A.1.8), 
\bel{dhdq} 
\left(\frac{\partial H
}{\partial Q}\right)^{\tt X} 
=\left(\frac{\delta V }{\delta
\rho} \frac{\partial \rho }{\partial Q}\right)^{\tt X}
=-\left(\frac{\partial V }{\partial r}\right)^{\tt
X}R_0Y_{L0}({\hat r}). 
\ee
We used also this self-consistent
dependence of mean potential $V$ through the particle density
$\rho$ in the derivations of the operator ${\hat F}(\r)$
(\ref{foperl}) in the FLDM: The derivatives of $V$ are
proportional to the ones of the density $\rho$ [see (\ref{derVderP1})], 
which both depend
on ${\tt X}$, i.e., whether we consider the latter for the fixed temperature
or entropy. Therefore, (A.1.6) and (A.1.7) with the definitions
(A.1.8) of \cite{hofmann} as applied to the FLDM, see
(\ref{ctcaddif}), should be a little modified to 
\bea\l{d2GEdq2}
\left(\frac{\partial^2 G }{\partial Q^2}\right)_T&=&
\left\langle\left(\frac{\partial ^2 {\hat H} }{\partial
Q^2}\right)_{Q=0}\right\rangle^T -\chi^T ,\nonumber\\
\left(\frac{\partial^2 E }{\partial Q^2}\right)_S&=&
\left\langle\left(\frac{\partial ^2 {\hat H} }{\partial
Q^2}\right)_{Q=0}\right\rangle^{\rm ad} -\chi^{\rm ad} .
\eea 
The derivatives of
the thermodynamic potential G are considered for the constant
pressure instead of the constant volume as used for the free
energy case. Similar calculations of the average value of the
second derivative of the Hamiltonian $\langle\partial ^2 {\hat H}/
\partial Q^2\rangle^{\tt X}$ as for the coupling constants lead
to
\bea\l{secderiv}
&&\left\langle\left(\frac{\partial ^2 {\hat H} }{\partial
Q^2}\right)_{Q=0}\right\rangle^{\tt X}= \int {\rm d} \r 
\left(\frac{\partial
^2 V }{\partial Q^2}\right)_{Q=0}^{\tt X} \rho \qquad\nonumber\\
&=& -R_0^4\int_0^{\infty}
{\rm d} r \left(\frac{\partial V }{\partial r}~ \frac{\partial \rho
}{\partial r}\right)_{Q=0}^{\tt X} \mathcal{O}_1. 
\eea
We integrated in the second equation of (\ref{secderiv}) by
parts. Taking then (\ref{denseq}) for the quasistatic density
$\rho_{\rm qs}$, one gets 
\bel{d2hdq2} 
\left\langle\left(\frac{\partial ^2
{\hat H} }{\partial Q^2}\right)_{Q=0}\right\rangle_0^{\rm qs}=
-k^{-1} \mathcal{O}_1
\ee 
with the
coupling constant $k$ (\ref{kfld}). Using the same transformations
of the thermodynamic derivatives as in Appendix A1 of
\cite{hofmann}, see (\ref{ctcaddif}),
from (\ref{d2GEdq2}) and (\ref{d2hdq2}) up
to relatively small $A^{-1/3}$ corrections of the ESA, one gets
\bea\l{chiTchiad1}
&&\chi^T-\chi^{\rm ad}= 
-\left[\left(\left(\frac{\partial^2 G}{\partial
T^2}\right)_Q\right)^{-1} \left(\frac{\partial^2 G}{\partial T\partial
Q}\right)^2\right]_{Q=0}\nonumber\\
&+& k_T^{-1}-k^{-1} \;.
\eea
Applying then the relation of the Gibbs free energy $G=A \mu$ to
the chemical potential $\mu$, we note that there is the factor $A^{-1}$ which
suppress much the contribution of the first term compared to
second one, $k_T^{-1}-k^{-1}$, see (\ref{couplxchix}),
\bel{ktk} 
k_T^{-1}-k^{-1}= \frac{r_0^5 A^{4/3} K b_{{}_{\! S}}  }{56
\mathcal{C}} \left(1-\frac{K^{T} }{K} \frac{b_{{}_{\! S}}^T 
}{b_{{}_{\! S}}}\right) \mathcal{O}_1.
\qquad\qquad 
\ee 
Moreover, the terms in
the square brackets of
(\ref{chiTchiad1}) are zero because the second derivative
$~\left(\partial \mu/\partial Q\right)_T~$ is zero within the
precision of the FLDM. To show this, let us take the equation as
for the temperature (\ref{dtdqx}) with the only replace of the
temperature $T$ by the chemical potential $\mu$. 
The above mentioned statement becomes
now obvious because the chemical
potential $\mu$ is a constant as function of the spatial
coordinates at the equilibrium as the temperature $T$
independently on the type of the process ${\tt X}$
within the FLDM.
As the result, we obtain the same
relation (\ref{chiTchiad}) with the difference of the coupling
constants shown in (\ref{ktk}).

We can evaluate the ratio of the surface energy coefficients
$b_{S}^T/b_{{}_{\! S}}$ of (\ref{couplxchix}) using in
(\ref{chiTchiad}) the fundamental relation
(\ref{chitchiskSkT}) for the ratio of the susceptibilities
$\chi^T/\chi^{\rm ad}$ in terms of the in-compressibility modula
$K/K^T$ ($K=K^{\varsigma}=K^{\rm ad}$), 
\bel{bstbs} 
\frac{b_{S}^T }{b_{{}_{\! S}}}= 
\frac{K }{K^T}~\left(2-\frac{K }{K^T}\right)
\approx 1+\frac{2 \pi^2 \tbar^2 }{3 \mathcal{G}_0}. 
\ee 
In the last
equation, we used the temperature expansions for the
in-compressibilities $K$ (\ref{incompradexp}) and $K^T$
(\ref{isotherkexp}). Thus, the surface energy constant $b_{S}^T$
for the constant temperature is larger than adiabatic (or
quasistatic) $b_{{}_{\! S}}$ and their difference is small as
$\tbar^2$.


\setcounter{equation}{0}

\section{Symmetry-energy density functional and boundary conditions}
\l{endenfun}

The nuclear energy, 
$E=\int \d \r\; \rho_{+}\;\epsi\left(\rho_{+},\rho_{-}\right)\;$, 
in the local density approach
\cite{brguehak,chaban,reinhard,bender,revstonerein,ehnazarrrein,pastore}
can be calculated through the energy density 
$\epsi\left(\rho_{+},\rho_{-}\right)$ per particle, 
\bea\l{enerden}
&&\epsi\left(\rho_{+},\rho_{-}\right) =
- b_{{}_{\! \mathcal{V}}} 
+ J \mathcal{I}^2 
+  
\frac{K}{18}\epsilon_{+}(\rho_{+}) -
J \mathcal{I}^2\epsilon_{-}(\rho_{+},\rho_{-}) +
\quad\nonumber\\
&& 
\left(\frac{\mathcal{C}_{+}}{\rho_{+}} +\mathcal{D}_{+} 
\right) 
\left(\nabla \rho_{+}\right)^2  
+ \left(\frac{\mathcal{C}_{-}}{\rho_{+}} + 
\mathcal{D}_{-} 
\right) 
\left(\nabla \rho_{-}\right)^2.\quad
\eea
Here, $\rho_{\pm}=\rho_n \pm \rho_p$ are the isoscalar, $\rho_{+}$, and 
isovector, $\rho_{-}$, particle densities, $\mathcal{I}=(N-Z)/A$ 
is the asymmetry parameter, 
$N$ and $Z$ are the neutron and proton numbers in the nucleus, $A=N+Z$. 
The particle separation energy  
$b_{{}_{\! \mathcal{V}}} \approx$ 16 MeV  
and  the symmetry energy constant of nuclear matter $~J \approx$ 
30 MeV are introduced also in (\ref{enerden}). 
The in-compressibility modulus of 
the symmetric nuclear matter 
 $K \approx 220-260$ MeV is shown in Table I of 
\cite{chaban,reinhard,BMRV}). 
Equation (\ref{enerden})
can be applied approximately to the
realistic Skyrme forces \cite{chaban,reinhard}, in particular
by neglecting small semiclassical $\hbar$ corrections 
and Coulomb terms \cite{brguehak,strtyap,strutmagbr,strutmagden,magsangzh}. 
$\mathcal{C}_{\pm}$ and $\mathcal{D}_{\pm}$ are constants defined by the 
basic Skyrme force parameters. The isoscalar 
surface energy-density part, independent explicitly 
of the density gradient terms, is determined by the dimensionless function  
$\epsilon_{+}(\rho_{+})$ satisfying
the standard saturation conditions \cite{strutmagden,magsangzh,BMRV}.
For the derivation of the explicitly analytical results, we 
use the quadratic approximation $\epsilon_{+}(\rho_{+}) = 
(1-\rho_{+}/\rho_\infty)^2=(1-w_{+})^2$, where $\rho_\infty\approx$ 0.16 fm$^{-3}$ 
is the density of infinite nuclear matter [see around 
(\ref{denseq})].
The isovector component can be simply evaluated as   
$\epsilon_{-}=1 - \rho_{-}^2/(\mathcal{I}\rho_{+})^2=1-w_{-}^2/w_{+}^2$.
In both these energies $\epsilon_{\pm}$, 
$w_{\pm}=\rho_{\pm}/(\mathcal{I}_{\pm} \rho_\infty)$ are the dimensionless
particle densities, $\mathcal{I}_{+}=1$ and 
$\mathcal{I}_{-}= \mathcal{I}$.
The isoscalar SO gradient terms in (\ref{enerden}) are defined with a
constant: 
$\mathcal{D}_{+} = -9m W_0^2/16 \hbar^2$, where
$W_0 \approx$100 - 130~ MeV$\cdot$fm$^{5}$ 
 and $\mathcal{D}_{-}$ is relatively small 
\cite{brguehak,chaban,reinhard}.

From the condition of the minimum energy $E$   
under the constraints
of the fixed particle number $A=\int \d \r\; \rho_{+}(\r)$ and
 neutron excess $N-Z= \int \d \r\; \rho_{-}(\r)$,
one arrives at the Lagrange equations for $\rho_{\pm}$ with the corresponding 
multipliers being the 
isoscalar and isovector  chemical potentials
with the surface corrections at the first order, 
$\Lambda_\pm \propto \mathcal{I}_\pm a/R \sim A^{-1/3}$ 
\cite{strutmagbr,strutmagden,magsangzh,BMRV}.

The isoscalar  and isovector  
particle densities $w_\pm$  
can be derived from (\ref{enerden})
first at the leading approximation in a small parameter 
$a/R$. For the isoscalar particle density $w_{+}=w_{+}(\xi)$ 
[$\xi$ is the distance
of the given point $\r$ from the ES in units of the diffuseness
parameter $a$ in the local ES coordinates, see (\ref{denseq}), 
$\xi = (r-R)/a$ for the spherical nuclei], 
one finds (Appendix B of \cite{BMRV} and
\cite{strutmagden,magsangzh}),
\bel{ysolplus}
\xi=-\int_{w_r}^{w}\d y\; \sqrt{\frac{1 +\beta y}{y\epsilon(y)}}\;,\qquad
\ee
below the turning point $~\xi(w=0)~$ and $~w=0~$ for $~\xi \geq \xi(w=0)~$,
$\beta=\mathcal{D}_{+}\rho_\infty/\mathcal{C}_{+}$ is the dimensionless SO
parameter (for simplicity of the notations, we omit the low index
``+'' in $w_{+}$).  
For $w_r=w(\xi=0)$,
one has the boundary condition,
$\d^2 w(\xi)/\d \xi^2=0$
at the ES ($\xi=0$):
\bel{boundeq}
\epsilon(w_r)+w_r(1 +\beta w_r) \left[\d \epsilon(w_r)/\d w\right]=0.
\ee
(see Appendix B of \cite{BMRV} where the specific solutions for $\xi(w)$
in the quadratic approximation for $\epsilon_{+}(w)$ in terms of 
elementary functions were derived).
For $\beta=0$ (i.e. without SO terms), it simplifies to
the solution $w(\xi)=\tanh^2\left[(\xi-\xi_0)/2\right]$ for
$\xi\leq \xi_0=2{\rm arctanh}(1/\sqrt{3})$
and zero for $\xi$ outside the nucleus ($\xi>\xi_0$).
For the same leading term of the isovector density, $w_{-}(w)$,
one approximately (for large enough constants $c_{sym}$ 
of all desired Skyrme forces 
\cite{chaban,reinhard})  
finds (Appendix A of \cite{BMRV}) 
\bel{ysolminus}
w_{-} =  w\;\left(1-\frac{\widetilde{w}^2(w) \left[1
+ \tilde{c}\widetilde{w}(w)\right]^2}{2\left(1+\beta\right)}
\right).
\ee
where
\bel{defpar}
\widetilde{w}=\frac{1-w}{c_{sym}}, \quad 
c_{sym}=a \sqrt{\frac{J}{\rho_\infty\vert\mathcal{C}_{-}\vert}},\quad
\widetilde{c}=\frac{\beta c_{sym}/2-1}{1+\beta},
\ee
and $a \approx 0.5 - 0.6$ fm is the 
diffuseness parameter [see (\ref{denseq})].

Simple expressions for the constants 
$b_S^{(\pm)}$ 
(\ref{bsplusminus}) can be easily
derived  in terms of
the elementary functions 
in the quadratic approximation to $\epsilon_{+}(w)$,
given explicitly in Appendix A \cite{BMRV}.
Note that in these derivations we neglected curvature terms 
and being of the same order shell corrections. 
The isovector energy terms were obtained within the ES 
approximation with high accuracy up to the product of two
small quantities, $\mathcal{I}^2$ and $(a/R)^2$.

Within the improved ES approximation accounting also 
for next order corrections in a small parameter 
$a/R$, we derived 
the macroscopic boundary conditions 
(Appendix B of \cite{BMRV})
\bea\l{macboundcond}
&&\delta \mathcal{P}_{\pm}\Big|_{ES} 
\equiv \left(\rho_\infty\;\mathcal{I}_{\pm}\;
\Lambda_{\pm}\right)_{ES} =  \delta P_{S}^{\pm},\nonumber\\ 
&&\mbox{where}\quad \delta P_{S}^{\pm} \equiv 2 \alpha_{\pm} 
\delta \mathcal{H} 
\eea
 are the
isovector and isoscalar surface-tension (capillary) pressures, 
$\delta \mathcal{H} \approx -\delta R_\pm/R_\pm^2$ are small 
variations of 
mean ES curvatures $\mathcal{H}$ (\ref{keybound}), 
$\delta R_\pm $
are radius variations (\ref{surface}),
and $\alpha_{\pm}$ are the tension coefficients, respectively,
\bea\l{sigma}
\alpha_{\pm}&=&b_S^{(\pm)}/4 \pi r_0^2,\quad b_{S}^{(\pm)} \approx
\frac{8 \pi}{a}\left(\rho_\infty \mathcal{I}_\pm\right)^2 
\mathcal{C}_{\pm}\nonumber\\
&\times&\int_{-\infty}^{\infty} \d \xi\;
\left(1 + \frac{\mathcal{D}_{\pm}\rho_\infty}{\mathcal{C}_{\pm}} w_{+}\right)
\left(\frac{\partial w_{\pm}}{\partial \xi}\right)^2.\quad
\eea
The conditions (\ref{macboundcond}) ensure 
the equilibrium through the equivalence
of the volume and surface pressure 
(isoscalar or isovector) variations, see detailed derivations 
in Appendix B of \cite{BMRV}.
As shown in Sec.\ III \cite{strutmagbr,strutmagden,kolmagsh}, 
the pressures 
$\delta \mathcal{P}_{\pm}$
can be obtained through moments of dynamical variations of the
corresponding phase-space distribution functions 
$\delta f_\pm(\r,\p,t)$ (\ref{planewave}) in the nuclear volume.

For the nuclear energy $E$  
in this improved ESA (Appendix C of \cite{BMRV}), 
one obtains
\bel{EvEs}
E \approx -b^{}_V\; A + J (N-Z)^2/A + E_S^{(+)} + E_S^{(-)},
\ee
with the following 
isoscalar (+) and isovector (-) surface energy components,
\bel{Espm}
 E_S^{(\pm)}= \alpha_{\pm}\mathcal{S}=b_{S}^{(\pm)} \mathcal{S}/(4\pi r_0^2),
\ee
and the ES area $\mathcal{S}$. 
These energies  
are determined by the
isoscalar and isovector 
surface energy constants $b_{S}^{(\pm)} \propto \alpha_{\pm}$ (\ref{sigma}) 
through the solutions 
for $w_{\pm}(\xi)$ taken 
 at the leading order in $a/R$.

For the energy surface coefficients
 $b_{S}^{(\pm)}$ (\ref{sigma}) with $\mathcal{D}_{-} \approx 0$] 
in the quadratic approximation  $\epsilon_{+}(w)=(1-w)^2$,
we finally arrived at the following explicit analytical expressions
in terms of
the Skyrme force parameters 
(Appendix C of \cite{BMRV})
\bel{bsplusminus} 
b_{S}^{(+)}=\frac{6 \mathcal{C}_{+}
\rho_\infty \mathcal{J}_{+}}{ r_0 a},\quad
b_{S}^{(-)}=k^{}_S \mathcal{I}^2,\quad
k_{{}_{\! S}}= 6 \rho_\infty \mathcal{C}_{-}\mathcal{J}_{-}/(r_0 a), 
\ee
where
\bea\l{Jp}
&&\mathcal{J}_{+}=\int_0^1 \d w\; \sqrt{w(1+\beta w)}\;(1-w)\nonumber\\
&=&\frac{1}{24}\;(-\beta)^{-5/2}\;
\left[\mathcal{J}_{+}^{(1)}\; \sqrt{-\beta(1+\beta)}
+\mathcal{J}_{+}^{(2)}\; \arcsin\sqrt{-\beta}\right],\nonumber\\
&&\mathcal{J}_{+}^{(1)}=3 + 4 \beta(1+\beta),\quad
\mathcal{J}_{+}^{(2)}=-3-6\beta,
\eea
and
\bea\l{Jm}
\mathcal{J}_{-}&=&-\frac{1}{1+\beta}\;
\int_0^1 \d w\; \sqrt{
w(1+\beta w)}\;(1-w)(1+\widetilde{c} \widetilde{w})^2
\nonumber\\
&=&\frac{\widetilde{c}^2}{1920 (1+\beta) (-\beta)^{9/2}}\;
\left[\mathcal{J}_{-}^{(1)}\left(c_{sym}/\widetilde{c}\right)\;
\sqrt{-\beta(1+\beta)}\right.\nonumber\\ 
&+& \left.
\mathcal{J}_{-}^{(2)}\left(c_{sym}/\widetilde{c}\right)\;
\arcsin\sqrt{-\beta}\right],
\nonumber\\
\mathcal{J}_{-}^{(1)}(z)&=& 105- 4 \beta \left\{95 +75 z +
\beta \left[119+10z (19+6z) \right.\right.\nonumber\\
&+& \left.\left. 8 \beta^2
\left(1+ 10z(1+z)\right)
+ 8 z \left(5 z (3 +2 z) -6\right)\right]\right\},
\nonumber\\
\mathcal{J}_{-}^{(2)}(z)&=&15 \left\{7+2\beta \left[
5 (3 + 2 z)\right.\right.\nonumber\\ 
&+& \left.\left. 8 \beta (1+z)
\left(3 +z +2 \beta (1+z)\right)\right]\right\},
\eea
see also (\ref{defpar})
for $\widetilde{c}$, $c_{sym}$ and $\widetilde{w}$.
For the limit $\beta \rightarrow 0$ from (\ref{Jp}) and (\ref{Jm}),
one has
$\mathcal{J}_{\pm} \rightarrow 4/15$. In the limit
$\mathcal{C}_{-} \rightarrow 0$,  one obtains $k^{}_S \rightarrow 0$.


\setcounter{equation}{0}
\section{POT calculations of the MI}
\l{semcalmi}

\subsection{Energy shell corrections}
\l{enshcor}

The energy shell corrections $\delta E$ can be expressed approximately  
through the oscillating level density component 
$\delta g_{{}_{\! \Gamma}}(\eps)$, 
averaged locally by using the convolution (folding) integral 
with a small averaging parameter $\Gamma$ of the 
Gaussian weight function
 \cite{strut,fuhi}.
As shown in  \cite{fuhi}, neglecting small corrections of the order of 
the squares of the Fermi energy shell fluctuations  
$(\delta \eps_{{}_{\! {\rm F}}})^2$ at 
$\Gamma \ll \hbar \Om \sim \eps_{{}_{\! {\rm F}}}/A^{1/3}$  (see (\ref{hom}), 
also \cite{strutmag}), one has
\bea\l{dedge}
&&\delta E = \int \d \eps\; n(\eps) 
\left(\eps-\eps_{{}_{\! {\rm F}}}\right)\; \delta g_{{}_{\! \Gamma}}(\eps),\nonumber\\
&& {\rm with} \quad  N=\int \d \eps\; n(\eps),\quad 
n(\eps)=\theta(\eps_{{}_{\! {\rm F}}}-\eps). 
\eea
Substituting 
\bel{avdden}
\delta g_{{}_{\! \Gamma}}(\eps)=
\Re \sum_{\rm PO} \delta g_{{}_{\! {\rm PO}}}(\eps)\;
\exp\left[-\left(\frac{t_{{}_{\! {\rm PO}}} \Gamma}{\hbar}\right)^2\right]
\ee
with (\ref{dlevdenscl}) for $\delta g_{{}_{\! {\rm PO}}}(\eps)$ into (\ref{dedge}),
one can expand a smooth action in exponent at the linear order,
\bel{actexpef}
S_{\rm PO}(\eps) \approx S_{\rm PO}(\eps_{{}_{\! {\rm F}}}) + t_{{}_{\! {\rm PO}}} 
\left(\eps-\eps_{{}_{\! {\rm F}}}\right),
\ee 
and pre-exponent 
amplitude at zero order  over $\eps$ near the Fermi energy
 $\eps_{{}_{\! {\rm F}}}$
[$t_{{}_{\! {\rm PO}}}= \partial S_{\rm PO}(\eps_{{}_{\! {\rm F}}})/\eps$]
(see a similar derivation of the averaged density $\delta g_{{}_{\! \Gamma}}(\eps)$
in \cite{strutmag,sclbook,migdalrev}). 
These expansions are valid for a small enough
width $\Gamma$ mentioned above to get a sharped bell-shaped Gaussian 
averaging function near $\eps_{{}_{\! {\rm F}}}$.  
Calculating then
simple Gaussian integrals over the energy $\eps$ by integration
by parts, one arrives at 
(\ref{descl}). In these derivations at the leading order in expansion in
$(\delta \eps_{{}_{\! {\rm F}}})^2$, we accounted for the zero value
originated by the lower limit $\eps=0$ in (\ref{dedge}) 
by using that
$t_{{}_{\! {\rm PO}}}(\eps)$ is relatively large at small but finite $\Gamma$.
Thus, one stays with the only contribution (independent of $\Gamma$)
 at the upper limit
$\eps=\eps_{{}_{\! {\rm F}}}$, in line of the basic concepts that
the energy shell correction is determined by the quantum s.p.\ states
near the Fermi surface \cite{fuhi,strut}. 
Similarly, the same result can be obtained
by using the Lorentzian weight function [the summand in 
(\ref{avdden}) is proportional to 
$\exp\left(-t_{{}_{\! {\rm PO}}} \Gamma/\hbar\right)$ 
in the Lorenzian
case, instead of the Gaussian exponent]. In this case, the 
local convolution averaging
of the oscillation level density component with the Lorentzian width
parameter $\Gamma$ 
is resulted in a formal shift of the energy 
$\eps \to \eps +i\Gamma$ ($\Gamma \ll \hbar \Om$). Thus, the 
strightforward
calculations by the residue method also gives (\ref{descl}).

\subsection{Derivation of the rigid-body MI }
\l{derrigmi}

\subsubsection{TF COMPONENT }
\l{tfmi}

We substitute approximately the Green's function
$\langle G_{0}\rangle_{\Gamma_p}$, locally averaged over the momentum $p$
by using the Gaussian weight function with a finite small width 
$\Gamma_p$, into $\Theta^{00}$ [see (\ref{thetaxnnp}) at $\nu=\nu'=0$] 
instead of $G_0$,
\bea\l{G0av}
\left\langle G_{0}\right\rangle_{\Gamma_p} &=& 
\frac{1}{\Gamma_p\sqrt{\pi}}\int_{-\infty}^\infty \d p'\; G_0(s,p')\;
\exp\left[-\left(\frac{p'-p}{\Gamma_p}\right)^2\right]
\nonumber\\
&\approx&
G_0(s,p)\; \exp\left[
-\frac{s^2 \Gamma_p^2}{4\hbar^2}\right],
\eea
as in \cite{strutmag} for the level density.
Transforming then
the integration variables $\r_1$ and $\r_2$ to the canonical 
average $\r$ and
difference $\s=\s_{{}_{\! 12}}$ ones (\ref{newcoord}), 
for the corresponding locally averaged MI component
$\left\langle \Theta_x^{00} \right\rangle_{\Gamma_p}$, one approximately gets  
\bea\l{thetaxnnp00}
&&\left\langle \Theta_x^{00} \right\rangle_{\Gamma_p}
\approx\frac{d_sm^2}{4\pi^3}\int \d \eps \;n(\eps)
\int \d \r \int \frac{\d \s}{s^2}\;
\ell_x\left(\r+\frac{\s}{2}\right)\quad
\nonumber\\
&\times&
\ell_x\left(\r-\frac{\s}{2}\right)\;\sin\left(\frac{2 s p}{\hbar}\right)\;
\exp\left[- \frac{s^2 \Gamma_p^2}{2\hbar^2}\right].\quad
\eea
For simplicity, we omit here and below the 
index in $s_{{}_{\! 12}}$ within this 
Appendix E.2  because it will not interfer with
different notations. As shown below (at the end of this Appendix  E.2a), 
the final result for $\langle \Theta_x^{00} \rangle_{\Gamma_p}$
(\ref{thetaxnnp00}) does not depend approximately on $\Gamma_p $,
that looks as a plateau of the SCM (without correction polynomials).
Within the NLLLA (\ref{nllla}), 
used already in  (\ref{G0})
\cite{gzhmagsit,gzhmagsit,mskbPRC2010} after the averaging over 
the phase-space variables, the main contribution is given
by small distance $s_{{}_{\! 12}}$ with respect to
the wave length $\hbar/p_{{}_{\! {\rm F}}}$ of the particle near the Fermi surface.
In this approximation at the leading zero order, due to the 
exponential cut-off factor decreasing with $s$ and $\Gamma_p$,
one may expand smooth classical quantities in $sp/\hbar$ in the argument
of exponent and pre-exponent amplitude factors in (\ref{thetaxnnp00})
 at the leading 
order, in particular, applying 
\bel{l2}
\ell_x\left(\r+s/2\right)\;\ell_x\left(\r-\s/2\right) \approx
\ell_x^2(\r) \approx p^2 r_{\perp x}^2.
\ee
In (\ref{thetaxnnp00}), we integrate over $\s$ in the spherical 
coordinates, $\d \s=s^2\d s\; \sin\theta_s\d \theta_s\;\d \varphi_s$,
with the polar axis $z_s$
directed along $\p(\r)$ (see Fig.\ \ref{fig14}). Then, 
for the ${\rm CT}_0$ momentum $\p(\r)$, i.e., 
along the $\s_{{}_{\! 12}}$, 
one takes into account that the integrand and 
limits of the integration  
over angles $\theta_s$,
and $\varphi_s$ are constants independent of 
other variables. 
Therefore, this integration over all angles
gives simply $4 \pi$, and we arrive at 
\bea\l{thetaxnnpint00}
\left\langle \Theta_x^{00} \right\rangle_{\Gamma_p}
&\approx&\frac{d_sm^2}{\pi^2}\int \d \r \;r_\perp^2\int_0^{\eps_{{}_{\! {\rm F}}}} 
\d \eps \; \left[{\tt I}_{00}(s_{\rm max},\eps,\Gamma_p)\right.
\nonumber\\
&-& \left.{\tt I}_{00}(0,\eps,\Gamma_p)\right].
\eea
Here, we exchanged the order of integrations over $\eps$
and $\r$. The remaining indefinite integral ${\tt I}_{00}(s,\eps,\Gamma_p)$ 
over $s$ as function of $s$, $\eps$ and $\Gamma_p$
can be approximately (within the NLLLA) taken
analytically, 
\bea\l{sinttf}
&&{\tt I}_{00}(s,\eps,\Gamma)=\int \d s\; \sin\left(\frac{2 s p}{\hbar}\right)\;
\exp\left[-\frac{s^2 \Gamma^2}{2\hbar^2}\right] 
\nonumber\\
&=& 
\sqrt{\frac{\pi}{2}}\;\frac{i \hbar}{2 \Gamma}\;
\exp\left[-2 \left(\frac{p}{\Gamma}\right)^2\right]\;
\left[{\rm erf}\left(\frac{ip \sqrt{2}}{\Gamma}+
\frac{\Gamma s}{\hbar\sqrt{2}}\right)\right.
\nonumber\\ 
&+&\left.
{\rm erf}\left(\frac{ip \sqrt{2}}{ \Gamma}-
\frac{\Gamma s}{\hbar \sqrt{2}}\right)\right],
\eea
where 
${\rm erf}(z)$
is the standard error function, ${\rm erf}(z)=
(2/\sqrt{\pi})\int_0^z \d t\; \exp(-t^2)$.
This integral, taken at the upper limit
$s=s_{max}$, is rather a complicated function
of $\r$, especially near the ES of the potential well. 
However, the Gaussian factor in the integrand 
with any small but 
a finite Gaussian parameter $\Gamma_p$,
\bel{avcondgp}
\hbar/R \ll \Gamma_p \ll p_{{}_{\! {\rm F}}},
\ee
removes the oscillating contribution
arising from the upper limit $s_{\rm max}$ ($R$ is the mean 
nuclear radius). The reason is due to the exponential 
asymptotics at a large argument
$s$, such as
\bea\l{int00as}
&&\exp\left[-\frac{\Gamma_p^2 s_{\rm max}^2}{2 \hbar^2}\right], \qquad
{\rm or} 
\nonumber\\
&&\exp\left[-\frac{2 p^2}{\Gamma_p^2}\right],\quad 
{\rm at}\quad p\sim p_{{}_{\! {\rm F}}}, 
\eea
or even strongly as the product of these exponents.
Then, according to another asymptotics for small $s \rightarrow 0$,
\bea\l{int00as0}
{\tt I}_{00}(s,\eps,\Gamma) &=& -\frac{\hbar}{2p} + 
\frac{i\; \sqrt{2 \pi}\;\hbar}{\Gamma}\;
\exp\left(-\frac{p^2}{2\Gamma^2}\right)\nonumber\\
&+&\frac{p s^2}{\hbar}\left\{1 + 
\mathcal{O}\left[\left(\frac{ps}{\hbar}\right)^2\right]\right\},
\eea
we are left with the only constant contribution from the lower
limit $s=0$, independent of $s$ and of
the Gaussian averaging parameter $\Gamma_p$ satisfying the conditions 
(\ref{avcondgp}),
 \bel{sinttffin}
{\tt I}_{00}(s,\eps,\Gamma) \approx -\hbar/(2 p).
\ee
Finally, from (\ref{thetaxnnpint00}) and (\ref{sinttffin})
one obtains 
\bea\l{TFrig}
\left\langle\Theta_x^{00}\right\rangle&=&
\frac{d_sm^2}{2\pi^2 \hbar^3}\left\langle\int \d \eps \;n(\eps)
\int \d \r \; r_{\perp x}^2 \; p(\r)\right\rangle\nonumber\\
&=& d_s m \int \d \r\; r_{\perp x}^2\;\rho_{{}_{\! {\rm TF}}}(\r)=
\Theta_{x,{\rm TF}}^{\rm (RB)}.
\eea
We used also the expression for the TF particle density through
$G_0$ (\ref{G0}),
\bel{tfpartden}
\rho_{{}_{\! {\rm TF}}}(\r) = 
-\frac{1}{\pi} \Im \int_0^{\eps_{{}_{\! {\rm F}}}} \d \eps\; G_0\Big|_{s\rightarrow 0}=
\frac{m}{2 \pi^2\hbar^3}\int_0^{\eps_{{}_{\! {\rm F}}}} \d \eps\; p(\r).
\ee
Similarly, using the Lorentzian weight function for the averaging in 
(\ref{G0av}) instead of the Gaussian one,
\bel{G0avlor}
\left\langle G_{0}\right\rangle_{\Gamma} =
\frac{1}{\pi}\int_{-\infty}^\infty \d p'\; 
\frac{\Gamma G_0(s,p')}{\left(p'-p\right)^2 + \Gamma^2}
=G_0(s,p+i\Gamma),
\ee
one obtains the same result (\ref{TFrig}) independently of the choice of
the averaging function ($\Gamma=\Gamma_p$ in this Appendix E.2a). 
In these derivations, we used
the residue technics for the analytical evaluations of the integrals,
that means formally the replace of such a local averaging by the shift of the
momentum, $p \rightarrow p +i \Gamma_p$ [see (\ref{thetaxnnp}) 
at $\nu=\nu'=0$ and (\ref{l2})],
\bea\l{theta00avlor}
&&\left\langle \Theta_x^{00} \right\rangle_{\Gamma}\approx
\frac{d_sm^2}{\pi^2} \Im \int \d \r\; \left(y^2+z^2\right)
\nonumber\\ 
&\times&\int_0^{\eps_{{}_{\! {\rm F}}}} 
\d \eps \int_{0}^{s_{\rm max}} \d s\;
\exp\left[ \frac{2 i s \left(p+i \Gamma\right)}{\hbar}\right]
\approx \frac{d_s m \hbar}{2 \pi^2}
\nonumber\\
&\times& \int \d \r \left(y^2+z^2\right)\;\int_0^{p_{{}_{\! {\rm F}}}}
\d p\; p^2 \left\{1 - \exp\left[-\frac{2 \Gamma s_{\rm max}}{\hbar}\right]\right.
\nonumber\\
&\times& \left. 
\left[\frac{\Gamma}{p}\; 
\sin\left(\frac{2 p s_{\rm max}}{\hbar}\right) +
\cos\left(\frac{2 p s_{\rm max}}{\hbar}\right)\right]\right\}. 
\eea
Again, according to (\ref{avcondgp}), the second strongly 
oscillating term of the integrand coming
from the upper limit $s=s_{\rm max}$ in the last
line can be neglected as exponentially small, instead of 
the Gaussian behavior above. Then, 
we are left with the main first TF term [see (\ref{TFrig})] 
independent of $\Gamma_p$, as in the case of the Gaussian averaging.  

\bigskip
\subsubsection{MI SHELL CORRECTIONS}
\l{mishcor}

To average the oscillating component $\delta \Theta_x^{01}$ of the sum 
(\ref{thetaxsum})  (see 
(\ref{thetaxnnp}) at $n=0$ and $n'=1$) over the phase space variables,
one may use the Green's function 
$\left\langle G_0\right\rangle_\Gamma$ (\ref{G0av}), locally averaged
with a Gaussian weight 
instead of $G_0$, and similarly, instead of $G_1$ \cite{strutmag},
\bea\l{G1av}
&&\left\langle G_1\right\rangle_\Gamma=
\frac{1}{\Gamma \sqrt{\pi}} \int \d \eps'\; G_1\left(\r_1,\r_2,\eps'\right)\;
\exp\left[-\frac{(\eps'-\eps)^2}{\Gamma^2}\right]=\qquad
\nonumber\\
&=&
\sum_{CT_1}\mathcal{A}_{CT_1}\;
\exp\left[\frac{i}{\hbar} S_{CT_1}\! -\! \frac{i\pi}{2}\sigma_{{}_{\! CT_1}}
\!- \! i \phi_d 
\! -\! \frac{t_{{}_{\! CT_1}}^2 \Gamma^2}{4\hbar^2}\right].\,\,\,
\eea
Transforming also the integration variables ${\bf r}_1$ and ${\bf r}_2$
to the canonical ones (\ref{newcoord}) at zero temperature, one finds
\bea\l{dthetax01new}
&&\left\langle\delta \Theta_x^{01}\right\rangle=-\frac{d_s m}{\pi^2 \hbar^2}
\sum_{{\rm CT}_1} \Im\int_0^{\varepsilon_{{}_{\! {\rm F}}}} \d \eps\;
\int \d \r \int \frac{\d \s}{s} \;
\ell_x\left(\r-\frac{\s}{2}\right)
\nonumber\\
&\times& \ell_x\left(\r+\frac{\s}{2}\right)
\;\cos\left[\frac{s}{\hbar} p(\r)\right]\;
\exp\left[-\frac{s^2 \Gamma_p^2}{4 \hbar^2}\right]
\nonumber\\
&\times& \mathcal{A}_{{\rm CT}_1}\left(
\r-\frac{\s}{2},\r+\frac{\s}{2};\eps\right)\;
\exp\left[\frac{i}{\hbar} S_{{\rm CT}_1}\left(
\r-\frac{\s}{2},\r+\frac{\s}{2};\eps\right)\right. 
\nonumber\\
&-&\left.
i \frac{\pi}{2} \sigma_{{}_{\! CT_1}}- i \phi_{\rm d}
- \frac{t_{{\rm CT}_1}^2\Gamma^2}{4 \hbar^2}\right].
\eea
We shall put $\Gamma$ and $\Gamma_p$ to be zero in the final 
expressions
for this average $\left\langle\delta \Theta_x^{01}\right\rangle$, as far as
$\Gamma$ is much smaller than the distance between gross shells 
$\hbar \Om$ (\ref{hom}) and $\Gamma_p$ satisfies inequalities
 (\ref{nllla}).
Expanding then the action phase of the second exponent and its 
pre-exponent factors 
in small ${\bf s} p/\hbar$ up to first nonzero terms 
(i.e., 
up to the first and zeroth order ones, respectively), due to 
the first sharp-peaked exponential Gaussian factor in the second line of 
(\ref{dthetax01new}), 
one applies (\ref{l2}) and 
\bea\l{ampactoc}
\mathcal{A}_{{\rm CT}_1}\left(\r - \frac{\s}{2},\r + \frac{\s}{2}; \eps\right)
&\approx& \mathcal{A}_{{\rm CCT}_1}\left(\r,\r; \eps\right),\qquad
\nonumber\\
S_{{\rm CT}_1}\left(\r - \frac{\s}{2},\r + \frac{\s}{2}; \eps\right)
&\approx& S_{{\rm CCT}_1}\left(\r,\r; \eps\right) +\p \s.\qquad
\eea
With these expansions in (\ref{dthetax01new}), 
for the integration over $d\s=s^2 \d s\; \d x_s\;\d \varphi_s$ 
in (\ref{dthetax01new}), we use the 
same spherical coordinate system ($s, \theta_s, \varphi_s$)
with the polar axis $z_s$ directed again along the momentum vector 
$\p(\r)=(\p_1 +\p_2)/2$, 
$x_s=\cos \theta_s$ (Fig.\ \ref{fig14}). The integral over 
the azimuthal angle $\varphi_s$ gives simply $2 \pi$ due to the azimuthal
symmetry. The integration limits  over $x_s$ 
can be considered as from -1 to 1 within the NLLLA (\ref{nllla})
(neglecting thus the dependence of limits 
for the integration over angles $\theta_s$ on $s_{\rm max}$ 
and $\r$),
one approximately finds from (\ref{dthetax01new}) 
\bea\l{dthetax01new1}
\left\langle \delta \Theta_{x}^{01} \right\rangle
&\approx&  \frac{2d_s m}{\pi \hbar}\;
\Im \sum_{{\rm CCT}_1} \int \d \r \;r_{\perp x}^2
\int_0^{\varepsilon_{{}_{\! {\rm F}}}} \d \eps\;
\mathcal{A}_{{\rm CCT}_1}
\quad\nonumber\\
&\times&
\exp\left[\frac{i}{\hbar} S_{{\rm CCT}_1} - \frac{i \pi}{2}\sigma_{{}_{\! {{\rm CCT}_1}}} - i \phi_{d}
\right]\;I_{01},  
\eea
where 
\bea\l{int01}
I_{01}&=&\int_0^{s_{\rm max}} \d s\; s\; \int_{-1}^{1} \d x_s\; 
\cos\left[s p(\r)/\hbar\right]\;
\exp\left[-\frac{s^2 \Gamma_p^2}{4 \hbar^2}\right]\nonumber\\
&\times&\exp\left[\frac{i}{\hbar} s p(\r)\; x_s\right]\;
\exp\left[-\frac{t_{{}_{\! CT_1}}^2\Gamma^2}{4 \hbar^2}\right]
\nonumber\\
& \approx&  \left[{\tt I}_{00}\left(s_{\rm max},\eps,
\frac{\Gamma_p}{\sqrt{2}}\right)
-{\tt I}_{00}\left(0,
\eps,\frac{\Gamma_p}{\sqrt{2}}\right)\right]
\nonumber\\
&\times& \exp\left[-\frac{t_{{}_{\! CCT_1}}^2\Gamma^2}{4 \hbar^2}\right]
\approx \frac{\hbar}{2p}\;
\exp\left[-\frac{t_{{}_{\! CCT_1}}^2\Gamma^2}{4 \hbar^2}\right].
\eea
The sum runs all of ${\rm CCT}_1$s (closed ${\rm CT}_1$s).
Taking then the integral over the angle variable $x_s$ 
in the NLLLA (\ref{nllla}), one then integrate
over the
modulus $s$ within integration limits
from $0$ to $s_{\rm max}$.  Note that with the approximation 
$t_{{}_{\! CT_1}} \approx t_{{}_{\! CCT_1}}$, due to $\Gamma \ll \hbar \Omega$
(but significantly larger
than a distance between neighboring energy levels), this integral is 
reduced to $s=0$ and $s_{\rm max}$ boundaries of 
${\tt I}_{00}(s,\eps,\Gamma_p/\sqrt{2})$ (\ref{sinttf}), 
see the third line in (\ref{int01}).
Calculating approximately the integral over $s$ as 
in the subsection E.2a of this Appendix, and using the 
same asymptotics (\ref{int00as})
at large upper integration limit $s=s_{\rm max}$ 
and
(\ref{int00as0}) at small lower one $s=0$, one obtains
the nonzero contribution only from the lower integration limit $s=0$
as in the previous subsection E.2a.
Other contributions of the upper limit $s_{\rm max}$ can be neglected 
because the integrand over $s$ contains rapidly
oscillating functions, and after a local averaging in 
the phase space variables (even with a small but finite Gaussian
averaging parameter), they  exponentially disappear under the 
condition (\ref{avcondgp})
for $\Gamma_p$ as in the calculations of 
the Thomas-Fermi MI component
(Appendix E.2a). Finally, by making use of (\ref{int01}) 
in (\ref{dthetax01new1}), 
one obtains
\bea\l{dthetax01new2}
&&\left\langle \delta \Theta_{x}^{01}\right\rangle \approx 
-\frac{d_s m}{\pi}
\Im \sum_{{\rm CCT}_1} \int \d \r\; r_{\perp x}^2 \int_0^{\eps_{{}_{\! {\rm F}}}} \d \eps\;
\mathcal{A}_{{\rm CCT}_1}
\nonumber\\
&\times&
\exp\left[\frac{i}{\hbar} S_{{\rm CCT}_1} - 
\frac{i \pi}{2}\sigma_{{}_{\! {{\rm CCT}_1}}} - i \phi_{d}
-\frac{t_{{\rm CCT}_1}^2\Gamma^2}{4\hbar^2}\right]
\nonumber\\
&=&d_s m\int \d \r \;\left(z^2+y^2\right)\; \delta \rho_{\rm scl}(\r)=
\delta \Theta_{x,{\rm scl}}^{\rm (RB)}.
\eea
In these derivations, we used
(\ref{rperpcoord}) for 
the perpendicular coordinate $r_{\perp x}$,
and (\ref{dTxrigSCL}) for the oscillating shell component 
$\delta \Theta_{x\; {\rm scl}}^{\rm RB}$ of the semiclassical MI 
(\ref{rigmomsplit}). This component is 
related to the oscillating shell part $\delta \rho_{\rm scl}(\r)$
[see (\ref{ddenpart}) and (\ref{Gct}) with a closed ${\rm CT}_1$] 
in the semiclassical particle
density (\ref{denpartscl}). 
Like in the previous subsection of Appendix E.2, we obtain the 
same result (\ref{dthetax01new2}) 
by using the Lorentzian weight function for the local average
($\eps \rightarrow \eps+i \Gamma$). Indeed, using its definition
(\ref{G0avlor}) for both Green function components $G_0$ and $G_1$,
and performing the same integrations in the NLLLA (\ref{nllla}),
one gets
\bea\l{dthetax01new2lor}
&&\langle \delta \Theta_x^{01}\rangle 
=-\frac{m d_s}{\pi}\int \d \r \;\left(y^2 +z^2\right)\Im
\sum_{{\rm CCT}_1}\int \d \eps 
\mathcal{A}_{{\rm CCT}_1} 
\nonumber\\
&\times&\exp\left[\frac{i}{\hbar}\; S_{{\rm CCT}_1}
-\frac{i \pi}{2}\sigma_{{}_{\! {\rm CCT}_1}} -i \phi_{\rm d} -
\frac{\Gamma t_{{}_{\! {\rm CCT}_1}}}{\hbar}\right]
\nonumber\\
&\times& \left\{1 + \exp\left(-\frac{\Gamma_ps_{\rm max}}{\hbar}\right)
\left[\frac{\Gamma_p}{2p}\; \sin\left(\frac{2 p s_{\rm max}}{\hbar}\right)
\right.\right.
\nonumber\\
&-&\left.\left.\cos\left(\frac{2 p s_{\rm max}}{\hbar}\right)\right]\right\}.
\eea
As transparently seen from this explicit expression,  
one has exponential disappearance of the oscillating contributions
on the upper integration limit $s_{\rm max}$ 
under the conditions (\ref{avcondgp})
for $\Gamma_p$, see the second term in figure brackets of the last two lines
of (\ref{dthetax01new2lor}). Therefore, 
the first constant term 
in these brackets (coming from the lower integration limit $s=0$) 
yields immediately the finite $\Gamma \rightarrow 0$ rigid-body limit 
(\ref{dthetax01new2}) for $\langle \delta \Theta_x^{01}\rangle$.

Using analogous analytical calculations of the other terms 
$\langle\delta \Theta_{x}^{10}\rangle$ and 
$\langle\delta \Theta_{x}^{11}\rangle$ [see (\ref{thetaxnnp})], 
one finds the essentially different integrals over $s$, such as
\bea\l{int10} 
I_{10}&=&\int_0^{s_{\rm max}} \d s\;
\sin^2 (ps/\hbar)\;
\exp\left[-\frac{s^2 \Gamma_p^2}{4  \hbar^2}-
\frac{t_{{\rm CT}_1}^2\Gamma^2}{4 \hbar^2}\right]
\nonumber\\
&\approx& \exp\left[-
\frac{t_{{\rm CT}_1}^2\Gamma^2}{4 \hbar^2}\right]\;
\int_0^{s_{\rm max}} \d s\;
\left[1-\cos\left(\frac{2ps}{\hbar}\right)\right]
\nonumber\\
&\times&
\exp\left[-\frac{s^2 \Gamma_p^2}{4 \hbar^2}\right],
\eea
 and 
\bea\l{int11} 
I_{11}&=&\int_0^{s_{\rm max}} \d s\; s\; \sin(2 p s/\hbar)\;
\exp\left[-\frac{s^2 \Gamma_p^2}{4 \hbar^2}-
\frac{t_{{\rm CT}_1}^2\Gamma^2}{4 \hbar^2}\right]
\nonumber\\
&\approx& \exp\left[-
\frac{t_{{\rm CCT}_1}^2\Gamma^2}{4 \hbar^2}\right]\;
\int_0^{s_{\rm max}} \d s\;
s\;\sin\left(\frac{2ps}{\hbar}\right)
\nonumber\\
&\times&
\exp\left[-\frac{s^2 \Gamma_p^2}{4 \hbar^2}\right],
\eea
respectively.   
Integrating analytically in (\ref{int10}) and (\ref{int11}),
one can see that any contributions coming from the upper limit 
$s_{\rm max}$ exponentially disappear as shown in (\ref{int00as}) 
(with the formal replace $\Gamma_p$ by $\Gamma_p/\sqrt{2}$) as above.
However, in contrast to the calculations 
of $\left\langle \Theta_x^{00}\right\rangle$ 
and $\left\langle \Theta_x^{01}\right\rangle$, the
 contributions from the lower 
integration limit
$s=0$ turn into zero too, according to the asymptotics at the 4th order in 
distance $s$ in units of the wave-length $\hbar/p$:
\bea\l{intas1011}
I_{10} &=& \frac{2 p^2 s^3}{3 \hbar^2} \exp\left[-
\frac{t_{{}_{\! CCT_1}}^2\Gamma^2}{4 \hbar^2}\right]\left[1 +
\mathcal{O}\left(\frac{ps}{\hbar}\right)\right],\qquad
\nonumber\\
I_{11} &=& \frac{2 p s^3}{3 \hbar} \exp\left[-
\frac{t_{{}_{\! CCT_1}}^2 \Gamma^2}{4 \hbar^2}\right]\left[1 +
\mathcal{O}\left(\frac{ps}{\hbar}\right)\right].\qquad 
\eea
Therefore, 
in addition to (\ref{dthetax01new2}), independently of the
weight function for averaging, the two components associated with 
integrals (\ref{int10}) and (\ref{int11})
do not contribute at both integration limits within 
the NLLLA, as explained above. Thus, for  
all nonzero terms of the oscillating part of the MI,
$\left\langle \delta \Theta_x\right\rangle$,
one finally approximately arrives
at the same rigid-body MI shell component 
in the NLLLA (\ref{nllla}). 
This result does not depend on the choice of the weight (Gaussian
and Lorentzian) functions for the local averaging over the phase space.

\end{appendix}


\vspace{0.2cm}

\newpage
\centerline{{\bf CAPTIONS TO FIGURES AND TABLE:}}
\bigskip

{\bf Fig.\ 1.} Inertia correction $(M(0)-M_{\rm LD})/M_{\rm LD}$
(\ref{masscorr}) are shown by solid; dashed line shows the effective
damping parameter $\eta^2$, see (\ref{eta}), $\eps_{{}_{\! {\rm
F}}}=40~\mbox{MeV}$, $r_{{}_{\! 0}}=1.2~\mbox{fm}$,
$\mathcal{F}_1=-0.6$, $\tau_{{}_{\! 0}}=2.2\cdot
10^{-21}~\mbox{Mev}^2\cdot \mbox{sec}$ as in \cite{magkohofsh};
$b_{{}_{\! S}}=17~\mbox{MeV}$, $L=2$, ${\it
\Gamma}_0=33.3~\mbox{MeV}$, $c=20~\mbox{MeV}$.

\vspace{0.2cm}

{\bf Fig.\ 2}. Response function $\mbox{Im}\chi_{QQ}^{\rm coll}$
[the sum of the response-function ($s^{(n)}$) branches over $n$
($n=0,1$) in (\ref{chicollfldm}) without $k^2$ factor] versus
frequency $\om$ in units $\Omega=v_{{}_{\! {\rm F}}}/R$ for
different temperatures $T$ shown by numbers; $\mathcal{F}_0=-0.2$;
other parameters are the same as in Fig.\ \ref{fig1}.

\vspace{0.2cm}

{\bf Fig.\ 3.} Response function for the hydrodynamic collision
regime and heat pole of Fig.\ \ref{fig2} for small frequencies and
temperature $T=6~\mbox{MeV}$. Numbers $n=0~ \mbox{and/or}~ 1$ show
the sum of the response-function ($s^{(n)}$) branches or one of them
(the latter curves coincide).

\vspace{0.2cm}

{\bf Fig.\ 4.} Stiffness coefficient $C$ in units of the LDM value
$C_{\rm LD}$ versus temperature $T$; full squares are obtained by
fitting for the second (hydrodynamic-sound) response function peak
in (\ref{chicollfldm}) as explained in sections \ref{transprop} and 
\ref{discuss}; joined
open squares are the same for the third (Fermi-surface-distortion)
peak; open circles are the heat pole stiffness; thick solid shows
$C(0)=C_{\rm LD}$ (\ref{C0}); parameters are the same as in Figs.\
\ref{fig1}, \ref{fig2}.

\vspace{0.2cm}

{\bf Fig.\ 5.} Mass parameter $M$ in units of the LDM value $M_{\rm
LD}$ versus temperature; thin and thick solids show $M(0)/M_{\rm
LD}$ and unit LDM value; other notations and parameters are the same
as in Fig.\ \ref{fig4}.

\vspace{0.2cm}

{\bf Fig.\ 6.} Temperature dependence of the friction $\gamma$ in
$\hbar$ units; thick solid is the LDM value
$\gamma(0)=\gamma_{{}_{\! {\rm LD}}}$ (\ref{friction0}),
(\ref{gamma0}); dashed line is $\gamma(0)$ for $c=\infty$; other
notations and parameters are the same as in Fig.\ \ref{fig4}.

\vspace{0.2cm}

{\bf Fig.\ 7.} Friction $\gamma$ (logarithmic scale, $\hbar$ units)
as function of temperature; short dashed is asymptotical heat pole
friction $\gamma^{\rm hp}/\hbar$ (\ref{frictionhpm}); other
notations and parameters are the same as in Figs.\ \ref{fig4} and
\ref{fig6}.

\vspace{0.2cm}

{\bf Fig.\ 8.} Contribution of the ``heat pole'' to friction for the
non-ergodic system: for the fully drawn curve $\Gamma_T$ is
evaluated for $c=20~$MeV and for dashed curve $1/c=0~$; as reference
values, the result of the wall formula (line with stars) and the
contribution from the non-diagonal matrix elements (line with
squares) are shown (after \cite{hofivyam,hofmann}).

\vspace{0.2cm}

{\bf Fig.\ 9.} Isovector $w_{-}$ [(\ref{ysolminus})] particle
density versus $\xi$ with and without ($\beta=0$) SO terms for the
Skyrme force SLy7 as a typical example; the isoscalar $w$ [see
(\ref{ysolplus}) at $\epsilon=(1-w)^2$] is also shown by solid lines
(after \cite{BMRV}).

\vspace{0.2cm}

{\bf Fig.\ 10.} Isovector particle densities $w_{-}(\xi)$
(\ref{ysolminus})  as functions of $\xi$ within the quadratic
approximation to $\eps(w)$ for several critical Skyrme forces
\cite{chaban,reinhard} in the logarithmic scale (after \cite{BMRV}).

\vspace{0.2cm}

{\bf Fig.\ 11.} Semiclassical moments of inertia $\Theta_x$ (divided
by
  $\hbar^2$ and expressed in Mev$^{-1}$) as functions of the mass
  number $A$. Extended Thomas--Fermi results correspond to black full dots
  whereas those obtained upon neglecting the  spin degrees of freedom are
  represented by crosses. Thomas--Fermi MI are plotted as open circles.
  Plus signs refer, finally, to the Inglis cranking dynamical MI
  (after \cite{bartelnp}).

\vspace{0.2cm}

{\bf Fig.\ 12.} Equilibrium deformations of $^{90}$Zr in the
($\beta,\gamma$)
  plane for different angular momentum values $I$ (after \cite{bartelpl}).

\vspace{0.2cm}

{\bf Fig.\ 13.} Variational ETF moment of inertia $\Theta_{ETF}$
  (in $\hbar^2$ MeV$^{-1}$) of $^{90}$Zr as function
  of the rotational energy $\hbar \om$ (in MeV)
  (after \cite{bartelpl}).

\vspace{0.2cm}

{\bf Fig.\ 14.} Trajectories connecting points $\r_1$ and $\r_2$
without (CT$_0$; solid line) and with (CT$_1$; dashed line)
reflections; the initial $\p_1^{(0)}$ and $\p_1^{(1)}$, and the
final $\p_2^{(0)}$ and $\p_2^{(1)}$ momenta of a particle at these
points; $\s_{12}=\r_2-\r_1$; polar axises $z$ and $z_s$ and the
corresponding angles $\theta_1$ and $\theta_2$ are shown,
respectively.

\vspace{0.2cm}

{\bf Fig.\ 15.} Shell-structure free energy $\delta F$ (in HO units
$\hbar \om^{}_0$) as function of the particle number variable
$A^{1/3}$ for the critical deformations $\eta=1$, $1.2$ and $2$ at a
temperature of $T = 0.1 \hbar \om_{{}_{\! 0}}$; The SCM smoothing
parameters are $\gamma/\hbar \om_{{}_{\! 0}} =1.5-2.5$, and $M=4-8$.
Thin dots show the contribution of the 3D orbits, and the thin
dashed curves the EQ orbit contributions (after \cite{mskbPRC2010}).

\vspace{0.2cm}

{\bf Fig.\ 16.} Moment of inertia shell correction $\delta \Theta_x$
(in the same units as in Fig.\ \ref{fig15} for the perpendicular
rotation as function of the particle number variable, $A^{1/3}$,
temperatures $T=0.1$ and $0.2 \hbar \omega_{{}_{\! 0}}$. The thin
dotted line shows the contribution of 3D orbits, the thin dashed
line the contribution of EQ orbits for a temperature $T=0.1 \hbar
\omega_{{}_{\! 0}}$, and broad dashed line the one of EQ orbits for
$T=0.2 \hbar \omega_{{}_{\! 0}}$ (after \cite{mskbPRC2010}).

\vspace{0.2cm}

{\bf TABLE 1.} Isovector energy $k_S$
and coefficients $\mathcal{C}_{-}$
for several Skyrme forces
\cite{chaban,reinhard};
$D(A)$ is the mean IVGDR energy constants
for particle numbers $A=50-200$ within the FLDM and last
within the hydrodynamic (Steinwendel-Jensen) model; experimental data is about
80 MeV (after \cite{BMRV}).

\clearpage

\begin{table*}[pt]
\begin{tabular}{|c|c|c|c|c|c|c|c|c|c|c|}
\hline
& SkM$^*$ & SkM & SIII & SGII & RATP &SkP &
T6 & Ski3 & SLy5 & SLy7\\
\hline
$\mathcal{C}_{-}$ & -4.79 &-4.69 & -5.59 & -0.94 & 13.9  &-20.2
 &  0 & 12.6 & -22.8 & -13.4 \\
$k_S$  & -0.77 &-1.90 & -0.52  & -0.21 & 1.42 & -1.93
    & 0  & 4.88 & -6.96 & -6.32 \\
 $D_{HD}$  & 73-82 & 71-76 & 79-104 & 74-77 & 87 & 70-69
   & 86-88 & 105-100 & 76-84 &  77-89  \\
 $D_{{}_{\! FLDM}}$   & 85-86 & 85-86 & 82 & 82 & 90-89 & 87
   & 88 & 101-106 & 79-83 &  81-84  \\
\hline
\end{tabular}

\vspace{0.5cm}
TABLE\ I.
\end{table*}
%

\begin{figure}
\begin{center}
\includegraphics[width=0.8\columnwidth,clip=true,angle=-90]{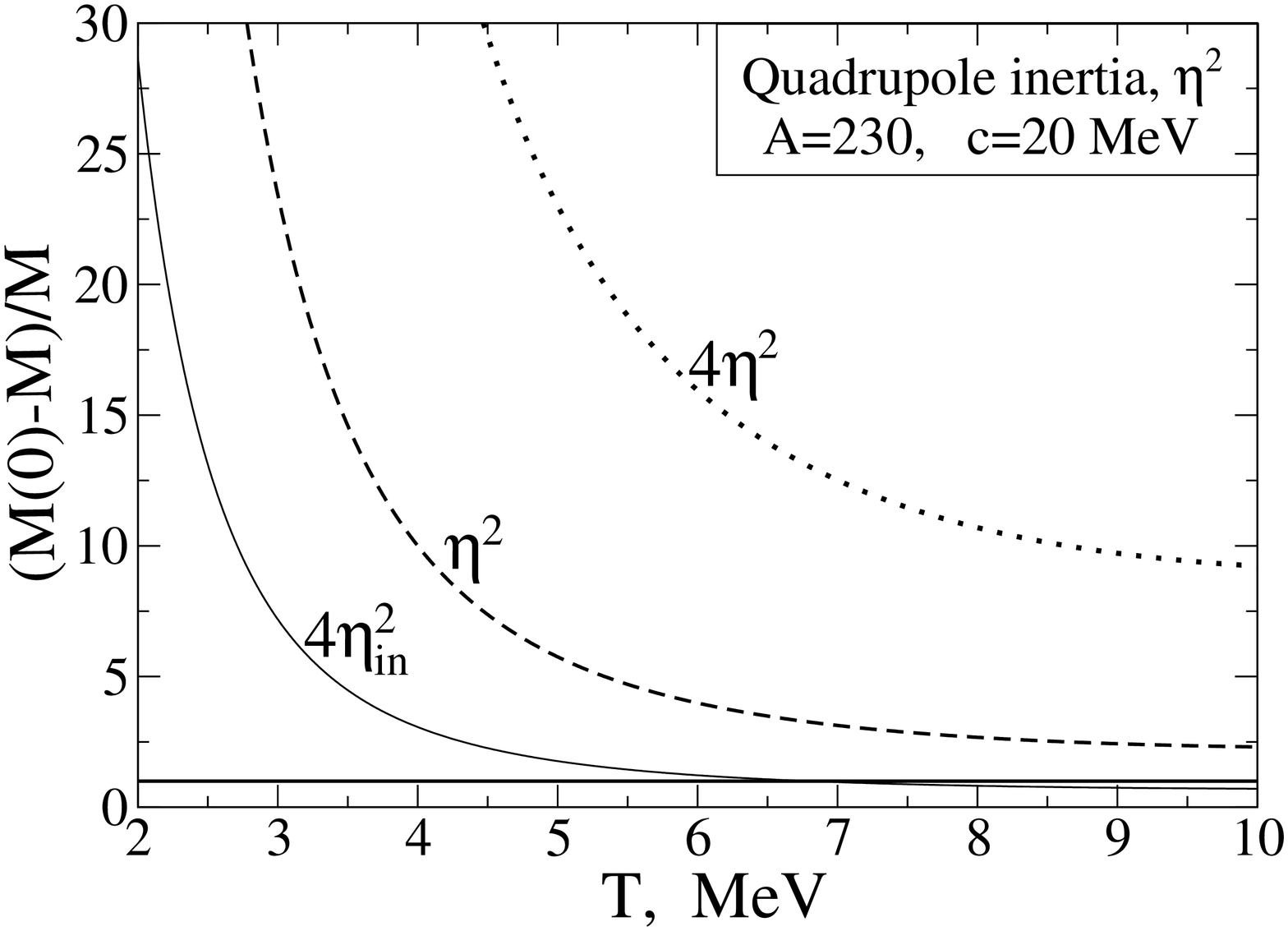}
\end{center}
\vspace{-1.0cm}
\caption{
}
\label{fig1}
\end{figure}
\begin{figure}
\begin{center}
\includegraphics[width=0.8\columnwidth,clip=true,angle=-90]{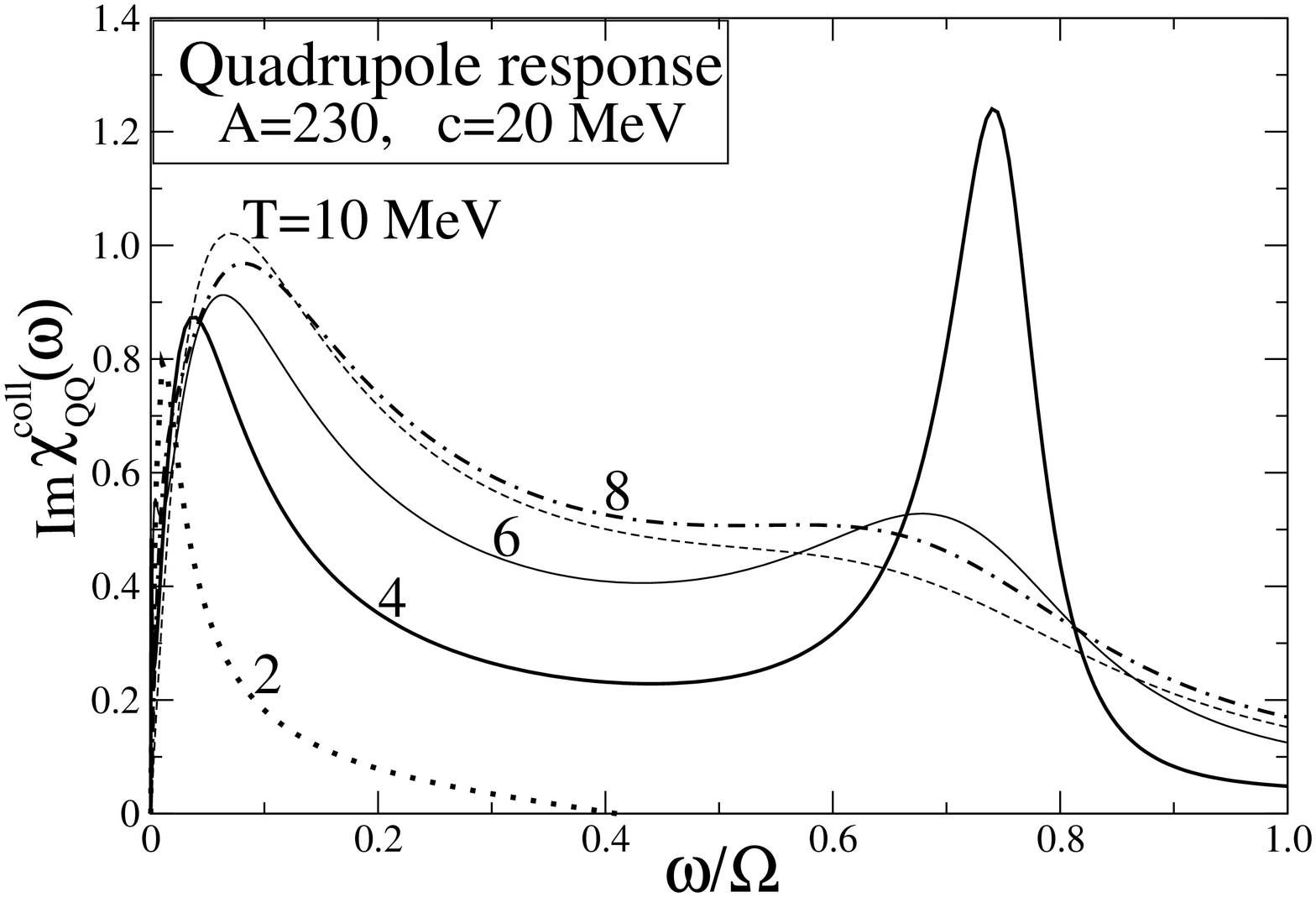}
\end{center}
\vspace{-1.0cm}
\caption{
}
\label{fig2}
\end{figure}
\begin{figure}
\begin{center}
\includegraphics[width=0.8\columnwidth,clip=true,angle=-90]{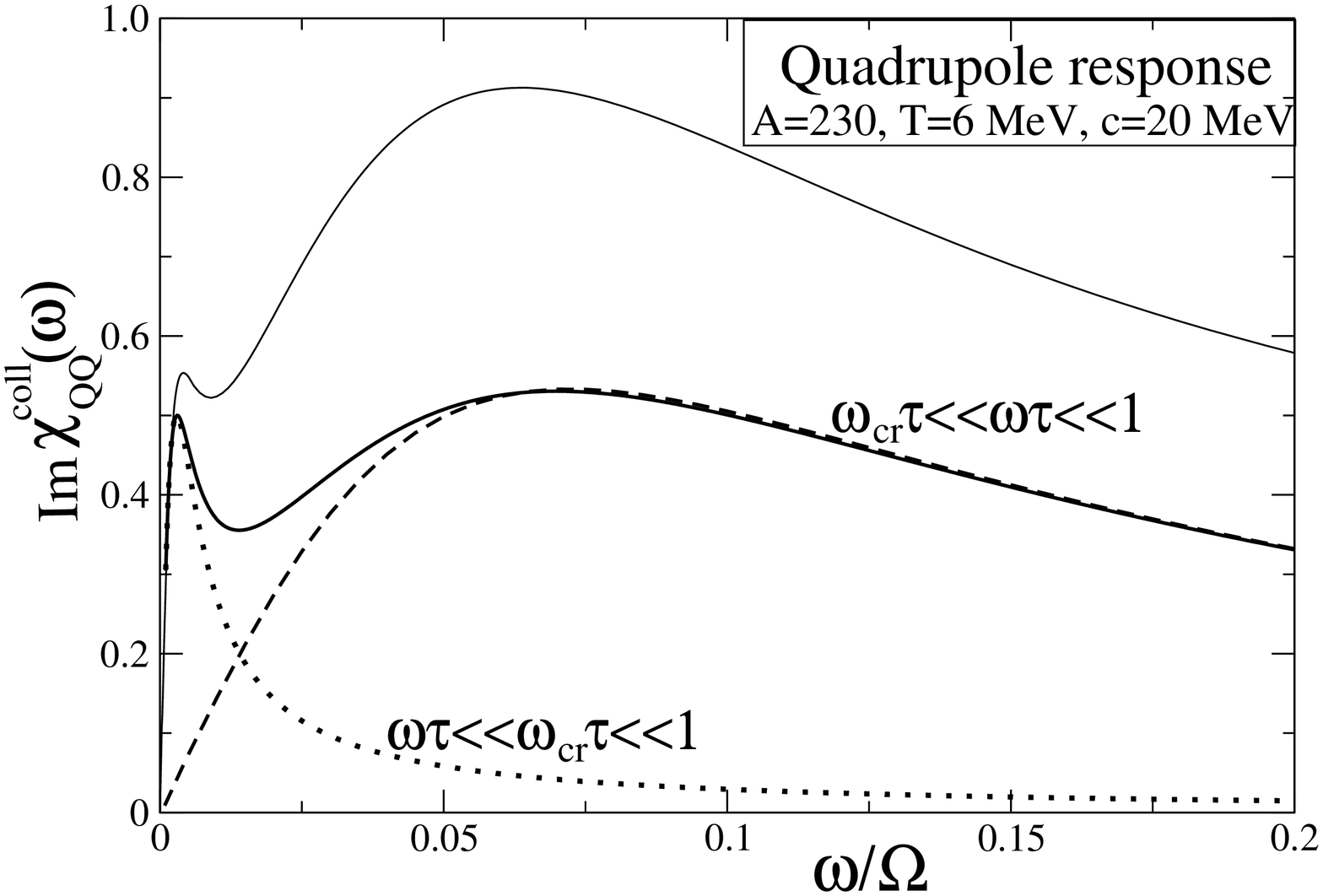}
\end{center}
\vspace{-1.0cm}
\caption{
}
\label{fig3}
\end{figure}
\begin{figure}
\begin{center}
\includegraphics[width=0.8\columnwidth,clip=true,angle=-90]{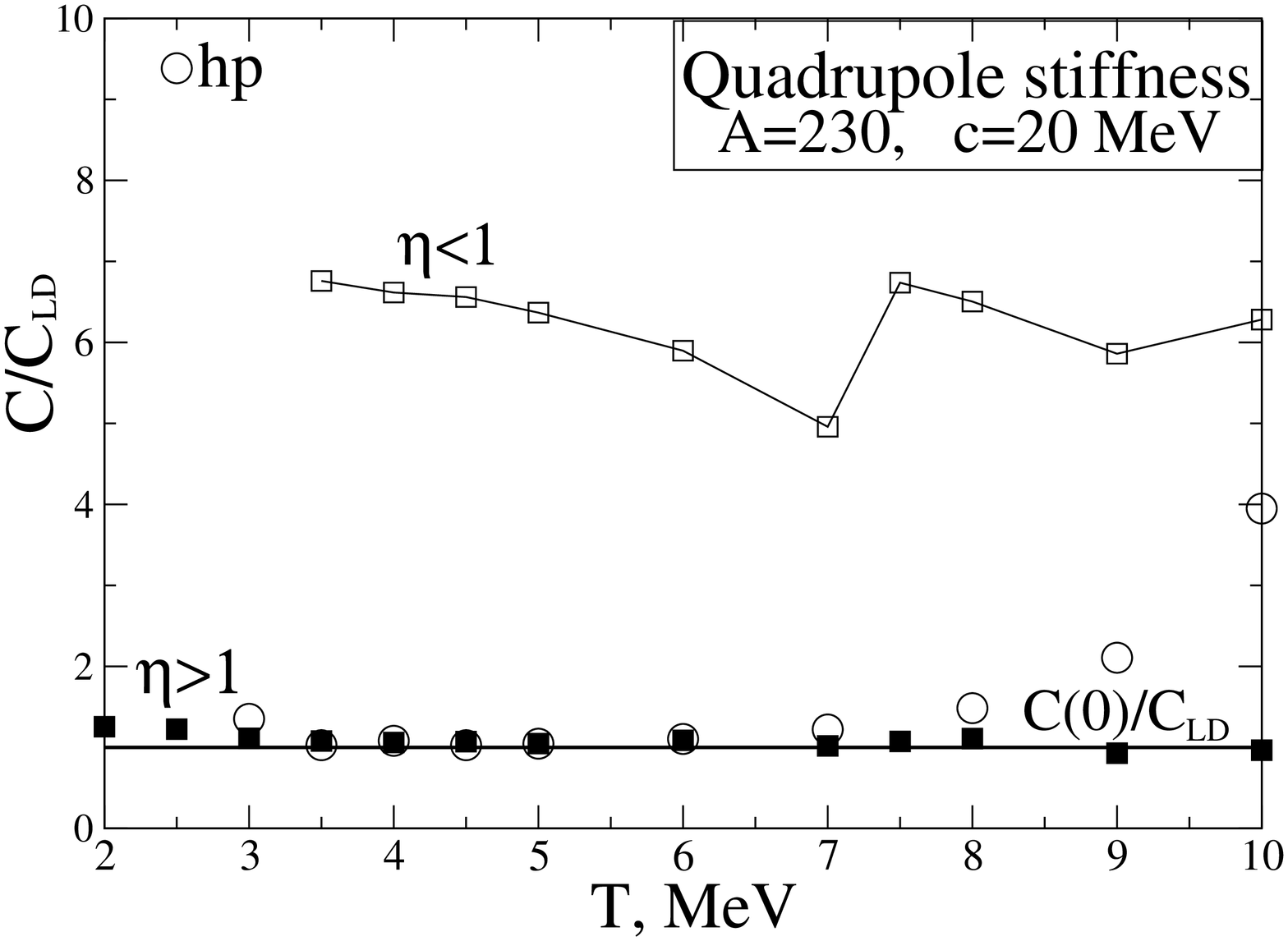}
\end{center}
\vspace{-1.0cm}
\caption{
}
\label{fig4}
\end{figure}
\begin{figure}
\begin{center}
\includegraphics[width=0.8\columnwidth,clip=true,angle=-90]{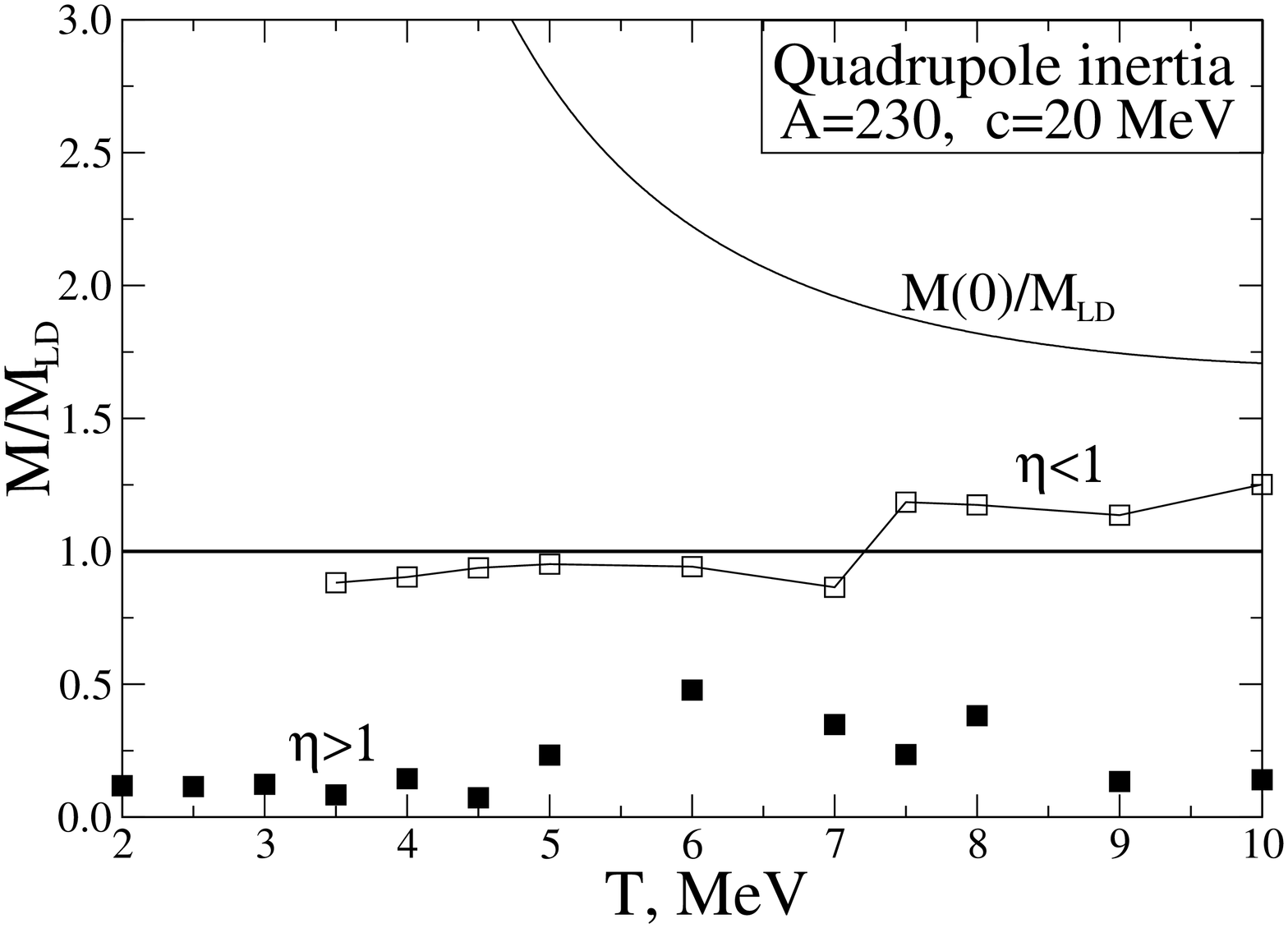}
\end{center}
\vspace{-1.0cm}
\caption{
}
\label{fig5}
\end{figure}
\begin{figure}
\begin{center}
\includegraphics[width=0.8\columnwidth,clip=true,angle=-90]{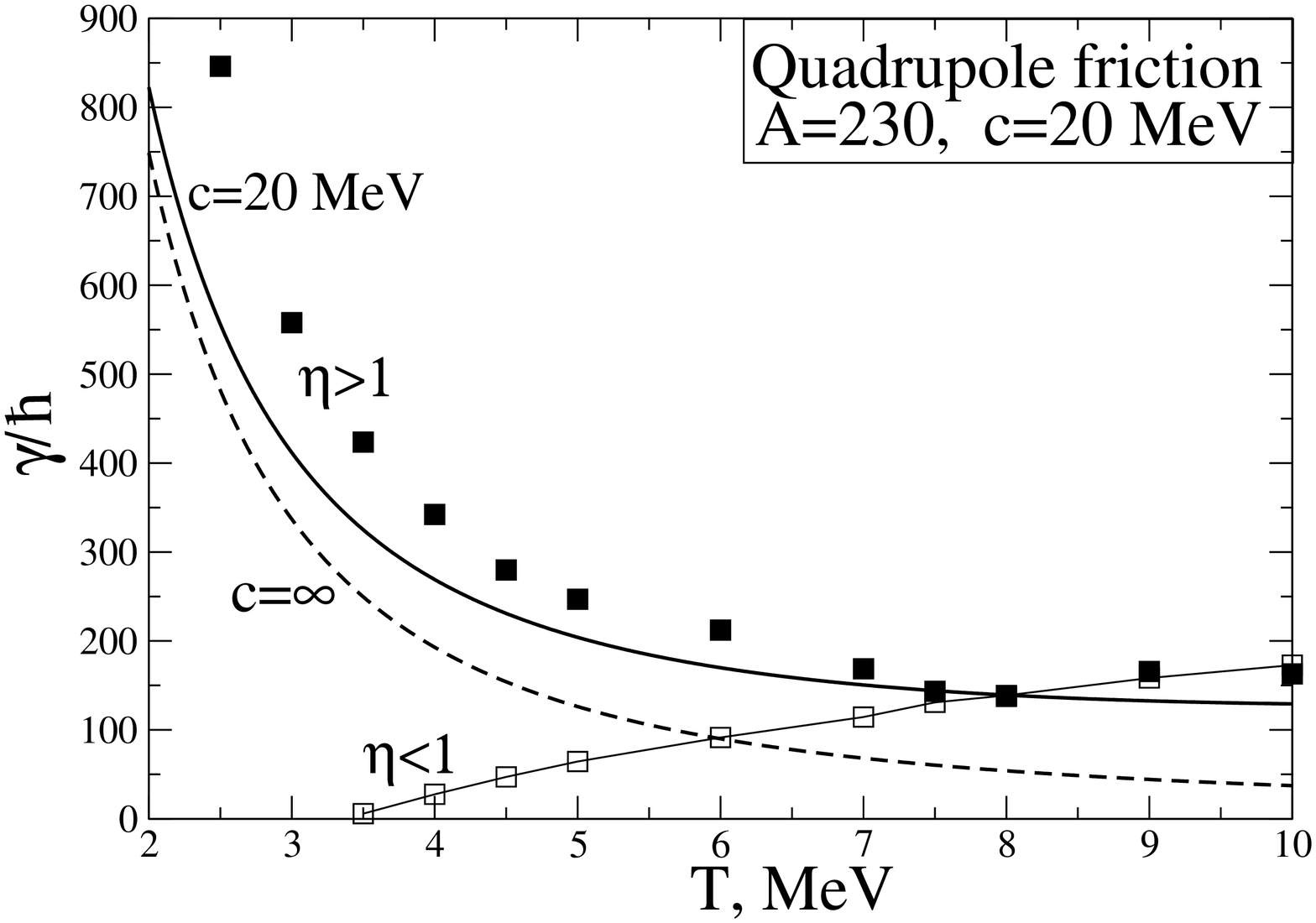}
\end{center}
\vspace{-1.0cm}
\caption{
}
\label{fig6}
\end{figure}
\begin{figure}
\begin{center}
\includegraphics[width=0.8\columnwidth,clip=true,angle=-90]{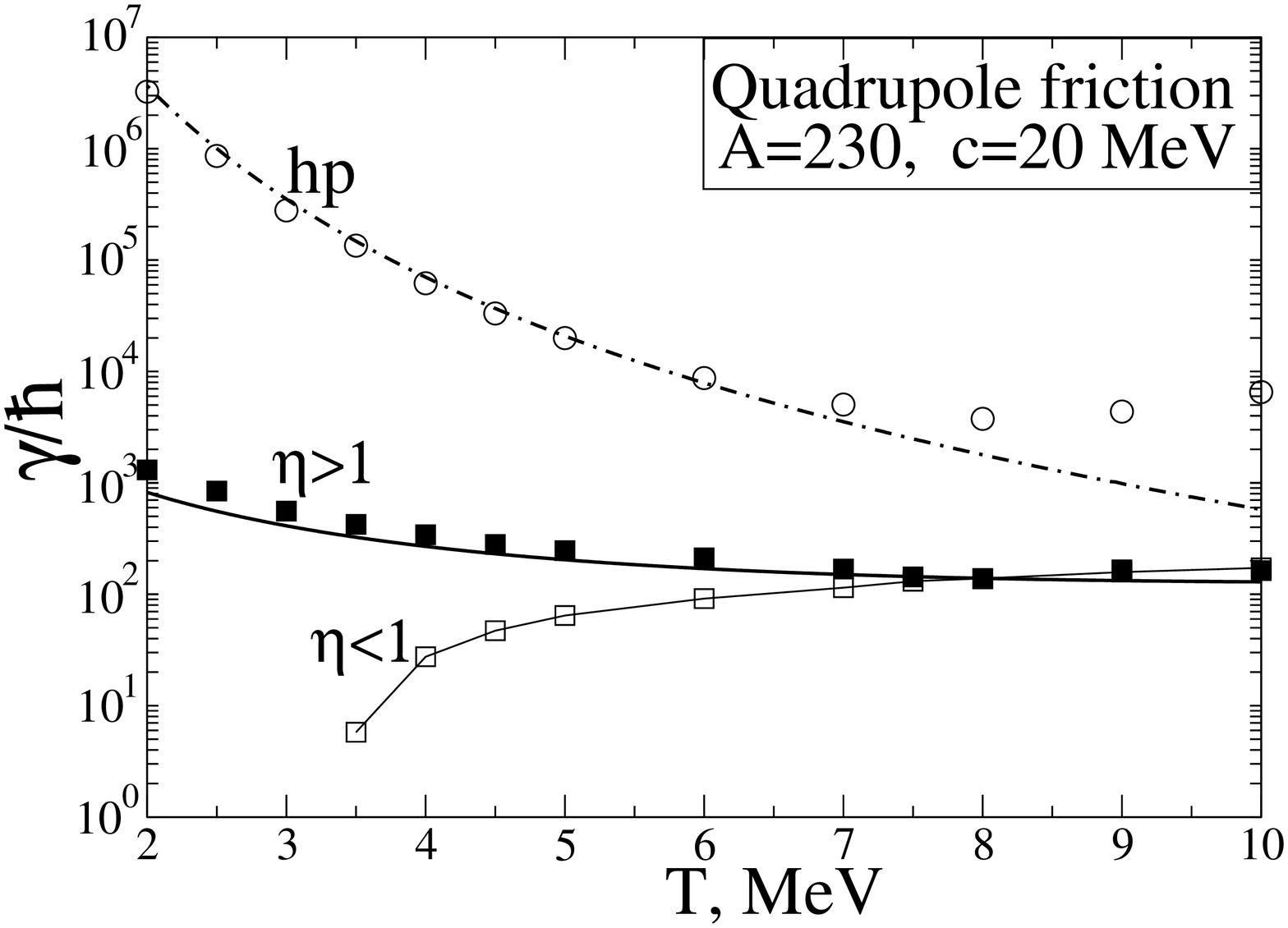}
\end{center}
\vspace{-1.0cm}
\caption{
}
\label{fig7}
\end{figure}
\begin{figure}
\begin{center}
\includegraphics[width=0.8\columnwidth,clip=true]{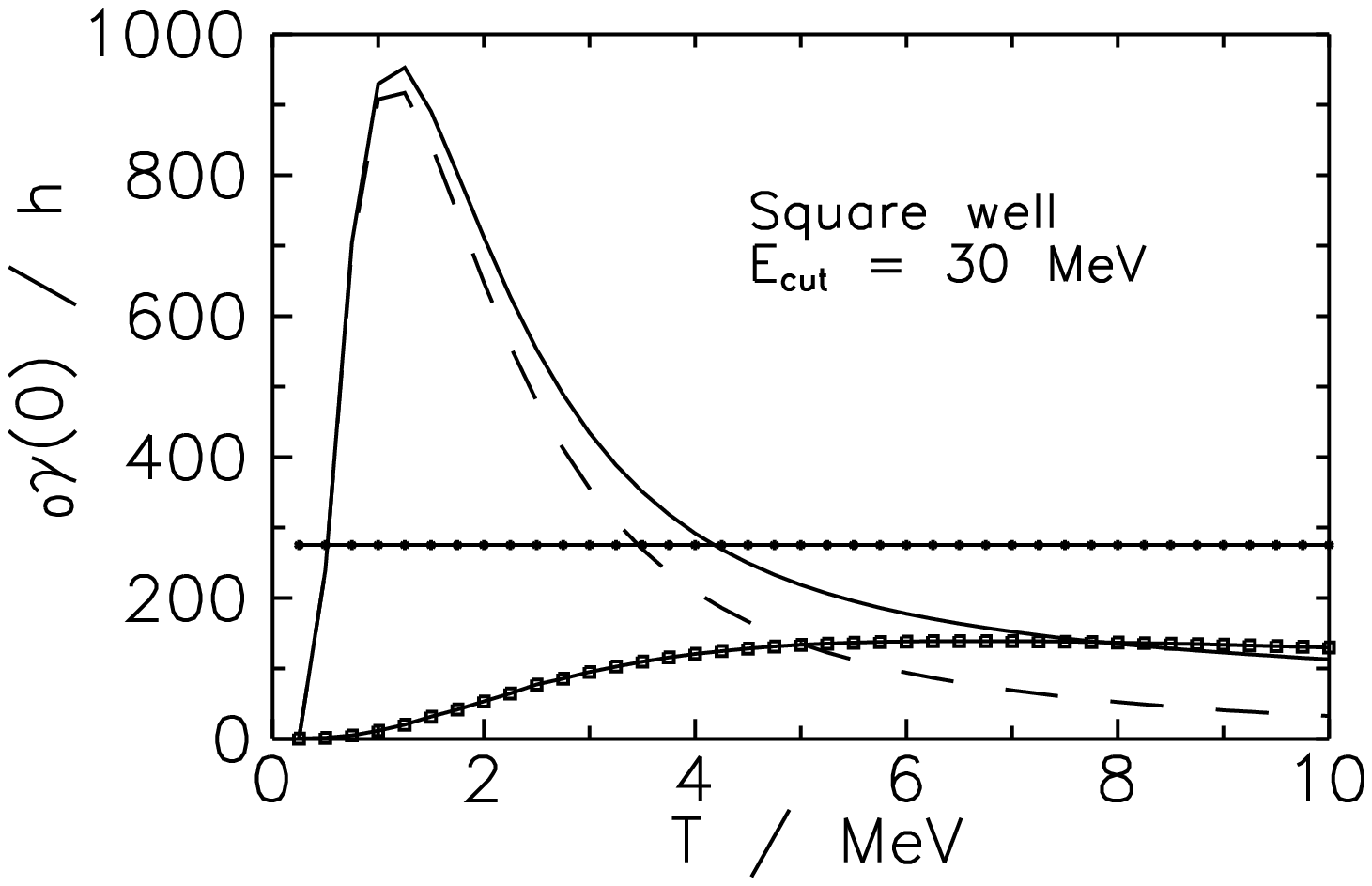}
\end{center}
\vspace{-0.7cm}
\caption{
}
\label{fig8}
\end{figure}
\begin{figure}
\begin{center}
\includegraphics[width=0.8\columnwidth,clip=true]{fig09.eps}
\end{center}
\vspace{-1.0cm}
\caption{
}
\label{fig9etf}
\end{figure}
\begin{figure}
\begin{center}
\includegraphics[width=0.8\columnwidth,clip=true]{fig10.eps}
\end{center}
\vspace{-1.0cm}
\caption{
}
\label{fig10etf}
\end{figure}
\begin{figure}
\begin{center}
\includegraphics[width=0.8\columnwidth,clip=true]{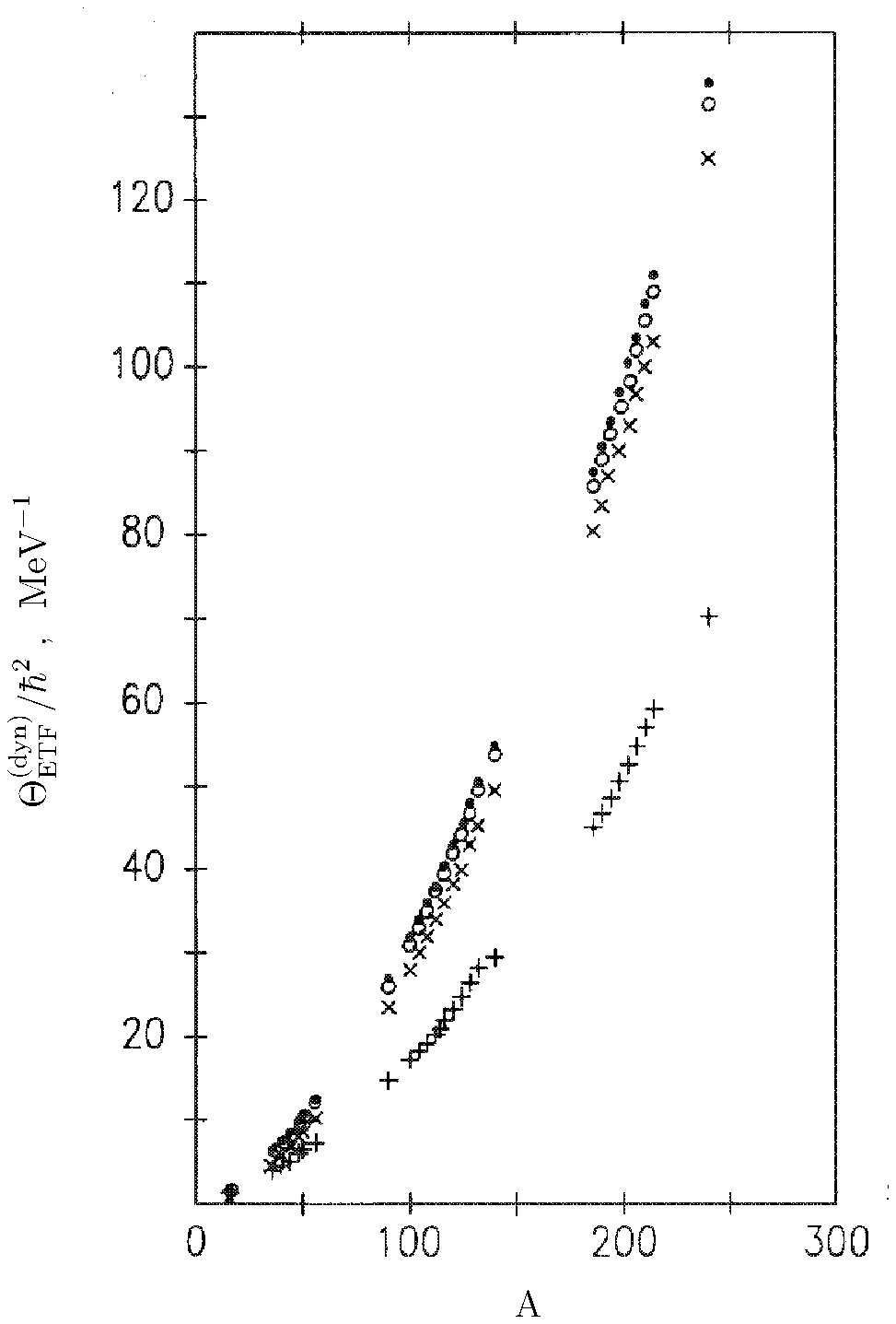}
\end{center}
\vspace{-0.5cm}
\caption{
}
\label{fig11}
\end{figure}
\begin{figure}
\begin{center}
\includegraphics[width=0.8\columnwidth,clip=true]{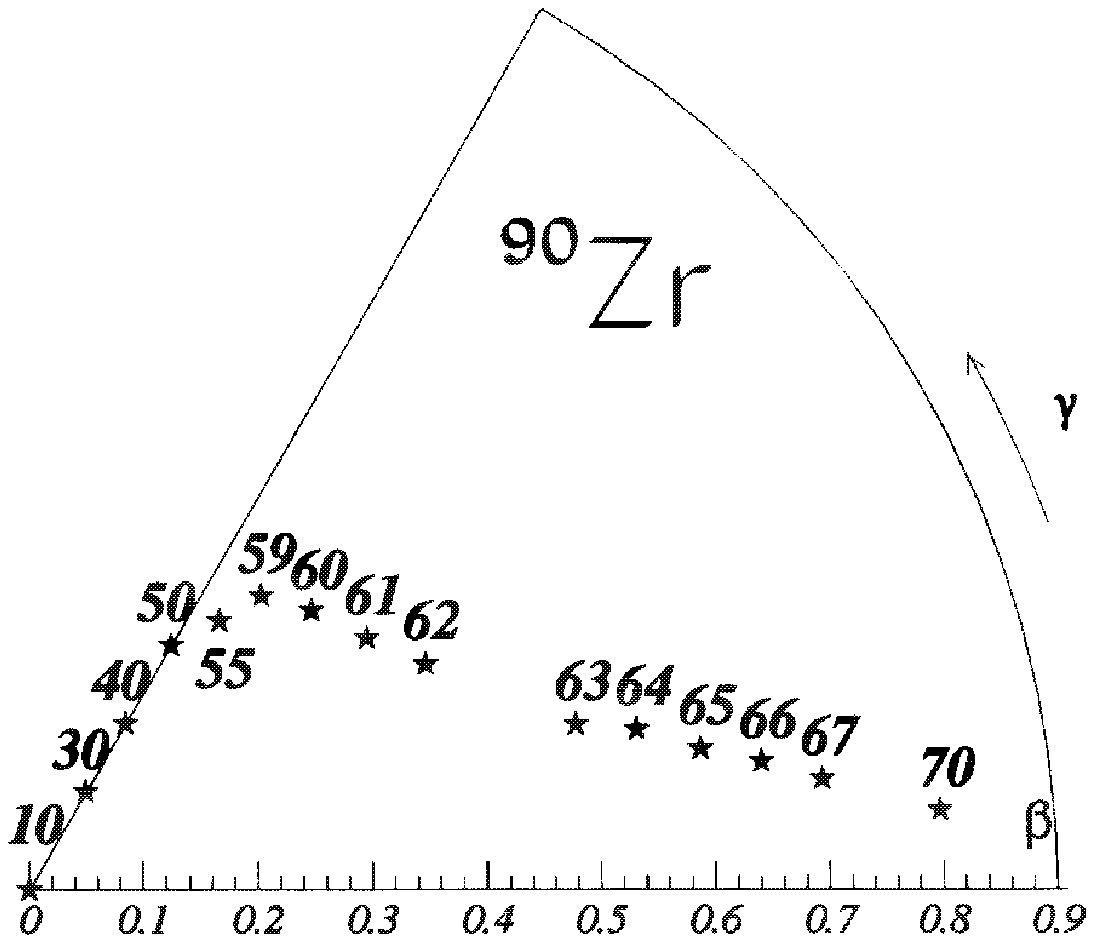}
\end{center}
\vspace{-0.5cm}
\caption{
}
\label{fig12}
\end{figure}
\begin{figure}
\begin{center}
\includegraphics[width=0.8\columnwidth,clip=true]{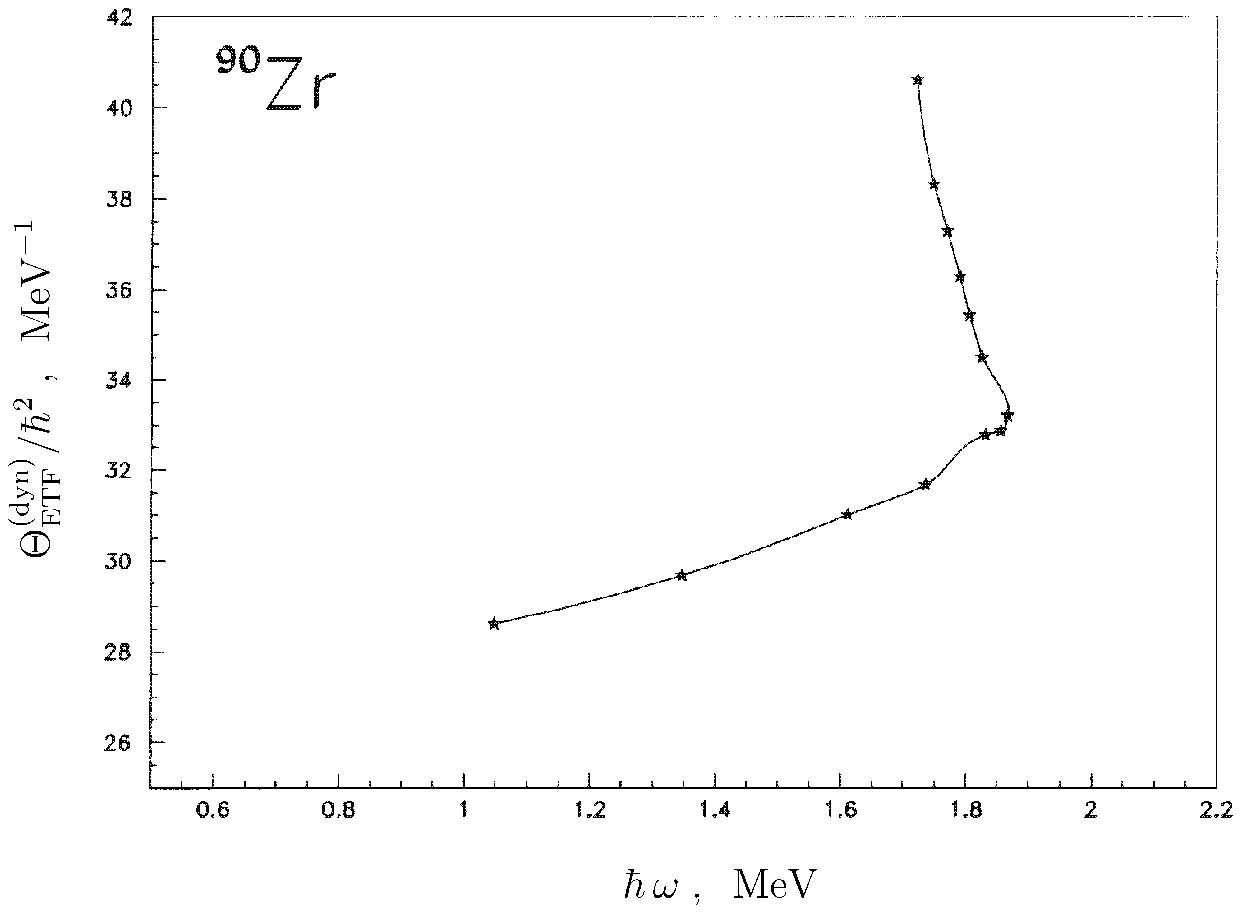}
\end{center}
\vspace{-0.5cm}
\caption{
}
\label{fig13}
\end{figure}
\begin{figure}
\begin{center}
\includegraphics[width=0.45\textwidth,clip=true]{fig14.eps}
\caption{
}
\label{fig14}
\end{center}
\end{figure}
\begin{figure}
\begin{center}
\includegraphics[width=0.8\textwidth,clip=true]{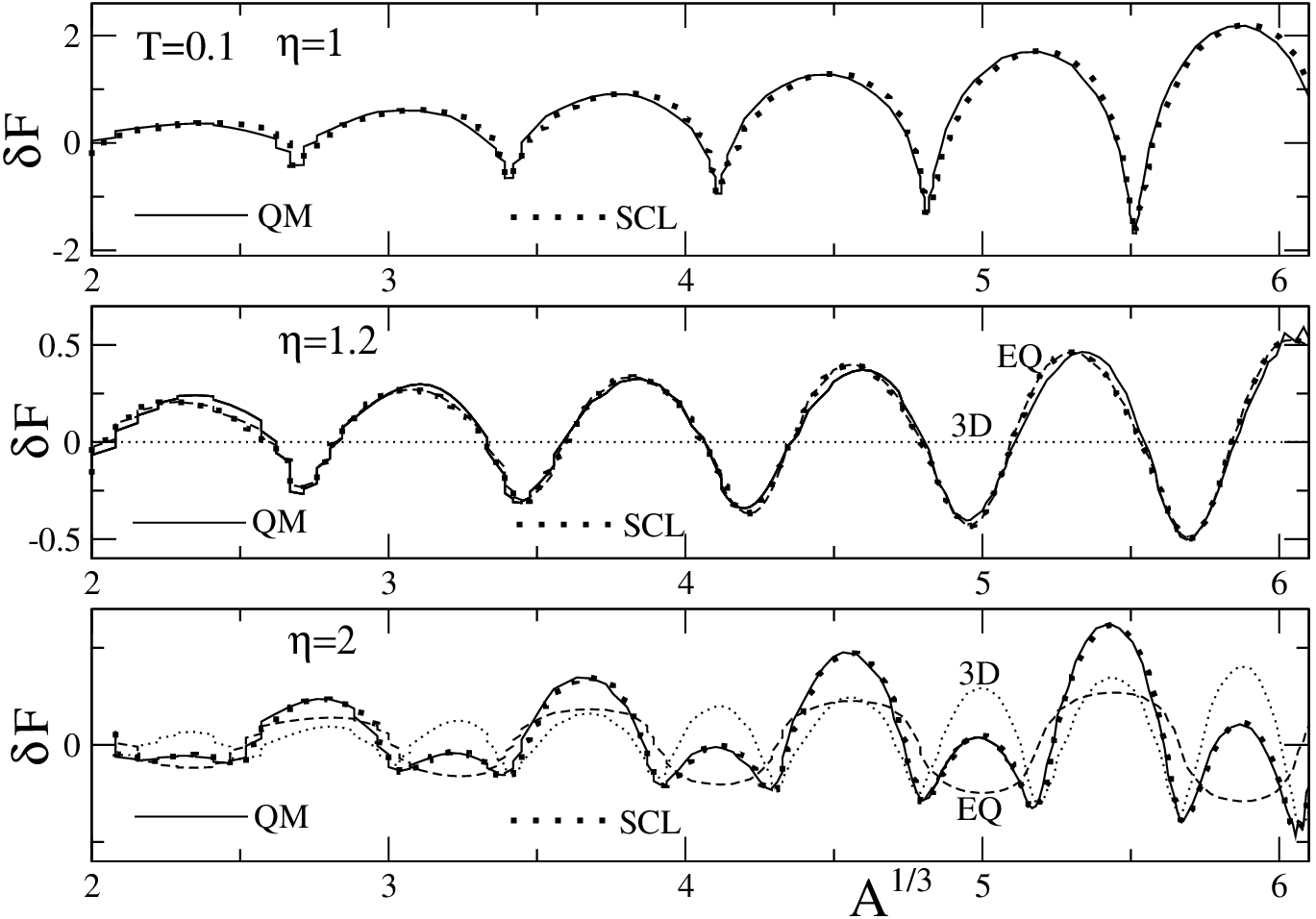}
\caption{
}
\label{fig15}
\end{center}
\end{figure}
\begin{figure}
\begin{center}
\includegraphics[width=0.8\textwidth,clip=true]{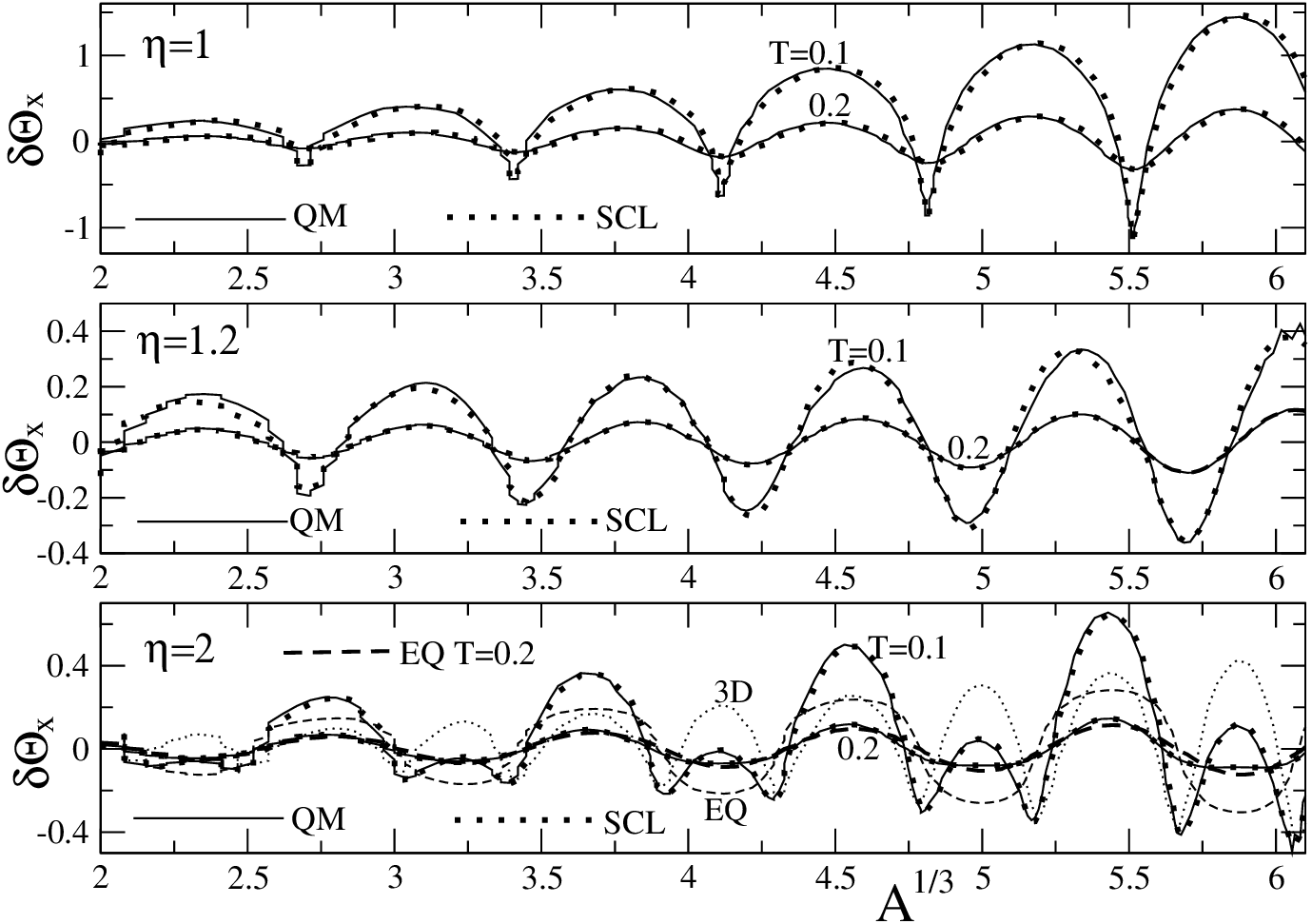}
\caption{
}
\label{fig16}
\end{center}
\end{figure}

\end{document}